\documentclass[%
reprint,
superscriptaddress,
frontmatterverbose, 
nofootinbib,
amsmath,amssymb,
aps,
prstab,
prstper,
floatfix,
]{revtex4-2}

\usepackage{graphicx}
\usepackage{dcolumn}
\usepackage{bm}
\usepackage{hyperref}
\usepackage[mathlines]{lineno}
\usepackage[caption=false]{subfig}
\usepackage{ulem} 
\usepackage{comment}
\usepackage{orcidlink}

\usepackage[
scale=0.7, marginratio={1:1, 2:3}, ignoreall,
text={7in,10in},centering,
margin=0.24in, 
total={6.5in,8.75in}, top=1.2in, left=0.6in,right = 0.6in, includefoot, 
]{geometry}

\usepackage{setspace}
\hypersetup{
	colorlinks=true, linkcolor=blue, 
    urlcolor=blue, 
	citecolor=green, 
    linktocpage =true
}
\usepackage{url}
\usepackage{multirow}
\usepackage{standalone}
\usepackage{cancel}
\usepackage{ulem}
\usepackage{colortbl}
\usepackage{enumitem}

\usepackage{tablefootnote}

\newcommand{\gv}[1]{{\textcolor{red}{\sf{[George: #1]}} }}

\begin{document}

\preprint{APS/123-QED}




\title{Universal description of the Neutron Star's surface and its key global properties: \\ A Machine Learning Approach for nonrotating and rapidly rotating stellar models}

\author{\large Grigorios Papigkiotis \orcidlink{0009-0008-2205-7426}}
\email{gpapigki@auth.gr}
\affiliation{\large Department of Physics, Aristotle University of Thessaloniki,\\ Thessaloniki 54124, Greece}
\author{\large Georgios Vardakas \orcidlink{0000-0003-1352-2062}}
\email{g.vardakas@uoi.gr}
\affiliation{\large Department of Computer Science \& Engineering, University of Ioannina, Ioannina 45110, Greece}
\author{\large Aristidis Likas \orcidlink{0000-0003-3170-5428}}
\email{arly@cs.uoi.gr}
\affiliation{\large Department of Computer Science \& Engineering, University of Ioannina, Ioannina 45110, Greece}
\author{\large Nikolaos Stergioulas \orcidlink{0000-0002-5490-5302}}
\email{niksterg@auth.gr}
\affiliation{\large Department of Physics, Aristotle University of Thessaloniki,\\ Thessaloniki 54124, Greece}

\date{\today}

\begin{abstract}
Neutron stars provide an ideal theoretical framework for exploring fundamental physics when nuclear matter surpasses densities encountered within atomic nuclei. Despite their paramount importance, uncertainties in the equation of state (EoS) have shrouded their internal structure. For rotating neutron stars, the shape of their surface is contingent upon the EoS and the rotational dynamics. This work proposes new universal relations regarding the star's surface, employing machine-learning techniques for regression. More specifically, we developed highly accurate universal relations for a neutron star's eccentricity, the star's ratio of the polar to the equatorial radius, and the effective gravitational acceleration at both the pole and the equator. Furthermore, we propose an accurate theoretical formula for $(d\log R(\mu)/d\theta)_{\max}$. This research addresses key astronomical aspects by utilizing these global parameters as features for the training phase of a neural network. Along the way, we introduce new effective parameterizations for each star's global surface characteristics. Our regression methodology enables accurate estimations of the star's surface $R(\mu)$, its corresponding logarithmic derivative $d\log R(\mu)/d\theta$, and its effective acceleration due to gravity $g(\mu)$ with accuracy better than $1 \%$. The analysis is performed for an extended sample of rotating configurations constructed using a large ensemble of 70 tabulated hadronic, hyperonic, and hybrid EoS models that obey the current multimessenger constraints and cover a wide range of stiffnesses. Above all, the suggested relations could provide an accurate framework for the star's surface estimation using data acquired from the NICER X-ray telescope or future missions, and constrain the EoS of nuclear matter when measurements of the relevant observables become available.


\end{abstract}
\maketitle
\section{Introduction}
Neutron stars (NSs) are unparalleled cosmic laboratories for investigating fundamental physics and gaining profound insight into the nature of the densest compact objects in the Universe. Distinguished by high nuclear density within the stellar core, rapid rotation, and remarkable compactness, NSs embody an astrophysical system described by the General Theory of Relativity (GR) and serve as a joint research area for relativistic astrophysics and nuclear physics.

One of the foremost challenges in nuclear astrophysics is to unravel the properties of ultradense and cold nuclear matter beyond the nuclear saturation density, denoted as $\rho_0 \approx 2.8\times 10^{14} \ \mathrm{g/cm^3} $. These properties remain inadequately understood. Within this context, establishing the relation between the structure of an NS, its global attributes (such as mass and radius), and the underlying microphysics—specifically, the equation of state (EoS)—is imperative for comprehending and validating diverse astrophysical scenarios. Presuming that nuclear matter within the star is appropriately described by a perfect fluid, the intrinsic microphysical properties are encapsulated by a barotropic EoS, representing the relation between the pressure and energy density of matter \cite{rezzolla_physics_2018,steiner2010equation,lattimer2011neutron, ozel_masses_2016,burgio2021modern,suvorov2024premerger,compose2022compose,oertel2017equations,lattimer2021neutron,chatziioannou2024neutron,miller2019psr, miller2021radius}. 

The accurate determination of NS properties, including masses and equatorial radii, is anticipated to unveil insights into the EoS \cite{baym2018hadrons,miller2019psr, miller2021radius,suvorov2024premerger,bauswein2019equation,drischler2021limiting,lattimer2021neutron,kashyap2022numerical,pang2021nuclear,imam2024implications}. While accurate measurements of the former have been obtained through observations of orbital parameters in double pulsar systems using radio astronomy, determining the latter remains challenging \cite{ozel2012surface, coleman2016observational, rezzolla_physics_2018,pang2021nuclear,fonseca2021refined,chatziioannou2024neutron}. Considerable efforts have been made toward this scope, mainly through the spectroscopic observations of quiescent NSs \cite{heinke2006hydrogen,webb2007constraining, guillot2011neutron,bogdanov2016neutron,van2024simultaneous} and X-ray bursters \cite{ozel2009mass, ozel2012surface,ozel2016dense,chatziioannou2024neutron,parikh2021uv}. However, current inferences regarding the equatorial radii carry substantial systematic errors \cite{ozel2012surface, coleman2016observational,miller2019psr, miller2021radius,riley2021nicer, pang2021nuclear,fonseca2021refined,chatziioannou2024neutron,choudhury2024nicer, lattimer2021neutron,guver2016systematic}.

In recent years, a wealth of observational data on NSs has emerged using diverse methods, including gravitational wave ground-based detectors such as Advanced LIGO \cite{aasi2015advanced} and Advanced Virgo \cite{acernese2014advanced}, as well as measurements in the electromagnetic band, such with the NICER mission, that focuses on observing X-ray pulse emissions emanating from {\it hot spots} on the surfaces of NSs \cite{gendreau2012neutron, arzoumanian2014neutron, gendreau2017searching,watts2016colloquium, miller2019psr, miller2021radius, riley2021nicer, choudhury2024nicer, kurpas2024detection, chatziioannou2024neutron}. These astrophysical observations have led to numerous efforts to narrow down and constrain the EoS, resulting in a better understanding of the fundamental particle interactions occurring at the NSs' interiors. These endeavors include utilizing NICER measurements of mass and radius \cite{riley2019nicer, miller2019psr, miller2021radius,riley2021nicer,pang2021nuclear,fonseca2021refined,choudhury2024nicer, kurpas2024detection, chatziioannou2024neutron}, measuring tidal deformability through gravitational waves (GWs) \cite{van2017upper, hinderer2008tidal, binnington2009relativistic, damour2009relativistic, chatziioannou2020neutron, dietrich2021interpreting,bauswein2019equation,gamba2023resonant,ripley2024constraint,williams2024phenomenological,carson2019future, huxford2024accuracy,francesco2023nuclear}, and applying joint constraints, as seen in \cite{raaijmakers2020constraining, biswas2022constraining, biswas2021impact, biswas2022bayesian, traversi2020bayesian, xie2019bayesian, biswas2021towards, dietrich2020multimessenger, landry2020nonparametric, raaijmakers2021constraints,bauswein2019equation,ho2023new,ripley2024constraint,williams2024phenomenological,carson2019future, huxford2024accuracy,francesco2023nuclear}. In particular, the detection of the binary NS merger event GW170817 \cite{abbott2017gw170817, abbott2019properties} has prompted additional studies in this field \cite{jiang2019equation, fattoyev2018neutron, most2018new, abbott2018gw170817, landry2018constraints, annala2018gravitational, lim2018neutron, kumar2019inferring,ho2023new,ripley2024constraint,williams2024phenomenological,huxford2024accuracy}.

At the same time, the nuclear physics community has developed a wide variety of EoS models. These models differ in terms of the assumed composition of the NS interior, the nucleon interaction properties, and the methods used to tackle the associated many-body problem. In each case, the derivation of the NS properties directly depends on the specific properties of the chosen EoS \cite{steiner2010equation,lattimer2011neutron, zdunik2013maximum, ozel_masses_2016,haensel_neutron_2007,compose2022compose}. Given the observational data on the macroscopic properties of NSs, common approaches based on Bayesian statistics have been implemented to infer the EoS, see e.g. \cite{miller2021radius,raaijmakers2020constraining, biswas2022constraining, biswas2021impact, biswas2022bayesian, traversi2020bayesian, xie2019bayesian, biswas2021towards, dietrich2020multimessenger, al2021combining, kruger2023rapidly}. In addition, recent studies have employed Artificial Neural Networks (ANNs) to reconstruct the EoS based on global properties of NSs \cite{fujimoto2020mapping, fujimoto2018methodology, fujimoto2018methodology, ferreira2021unveiling,farrell2023deducing,krastev2023deep}. Notably, Morawski \& Bejger \cite{Morawski2020} investigated the application of ANNs supported by the autoencoder architecture, while Sona et al. \cite{soma2022neural, soma2023reconstructing} employed ANNs to represent the EoS in a model-independent manner, leveraging the unsupervised automatic differentiation framework. Several Machine-Learning (ML) techniques have also been employed to explore the NS EoS. For instance, Lobato et al. \cite{lobato2022cluster} used a clustering method to identify patterns in mass-radius curves, while \cite{lobato2022unsupervised} investigated correlations between different dense matter EoSs using unsupervised ML techniques. Furthermore, efforts have been made to derive nuclear matter properties from NS EoS and observations using deep neural networks, as demonstrated in \cite{ferreira2022extracting, krastev2023deep}.

While NS global parameters are related to the EoS, the pursuit of investigating EoS-insensitive (universal) relations between stellar parameters has seen significant progress. 
Ravenhall and Pethick \cite{ravenhall1994neutron} initially introduced a relation between the normalized moment of inertia $I/MR_{\mathrm{eq}}^2$ and stellar compactness $C=M/R_{\mathrm{eq}}$, emphasizing its apparent EoS insensitivity. Subsequent modifications to this relation were made by Lattimer and Prakash \cite{lattimer2001neutron} and Bejger and Haensel \cite{bejger2002moments} and then used by Lattimer and Schutz \cite{lattimer2005constraining} to estimate the NS's radius based on combined mass and moment of inertia measurements of a pulsar in a binary system. Similar relations were also proposed by Breu and Rezzola \cite{breu_maximum_2016} for both slowly and rapidly rotating equilibrium configurations. Laarakkers and Poisson \cite{Laarakkers:1997hb} demonstrated a quadratic dependence of the quadrupole moment $Q$ on the star's angular momentum $J$, while Pappas and Apostolatos \cite{Pappas:2012ns, Pappas:2012qg} proposed a cubic dependence of the spin octupole $S_3$ on $J$, suggesting a Kerr-like behavior for the moments. Urbanec et al. \cite{urbanec2013quadrupole} identified a universal relation in slow-rotating NSs, expressing the reduced quadrupole moment $\bar{Q} = -MM_2/J^2$ as a quantity inversely proportional to compactness. Yagi and Yunes extended these findings, highlighting a new universal relation between the normalized quadrupole moment $\bar{Q}$ and stellar compactness in both slow-rotating NSs and quark stars \cite{yagi2017approximate}.

Yagi and Yunes also discovered relations involving the reduced moment of inertia $I$, quadrupole moment $Q$, and tidal love number $\lambda$ with accuracy better than 1\% in the slow-rotation limit assuming small tidal deformations \cite{yagi2013love}. Bauböck et al. \cite{baubock2013relations} conducted related work using the Hartle-Thorne approximation, offering insights that serve as a valuable tool for resolving degeneracies in modeling GW signals from inspiraling binaries \cite{yagi_i-love-q_2013a, maselli2013equation}. Although these relations show promise, challenges arise in the case of rapid rotation. For example, Doneva et al. \cite{doneva2013breakdown} revealed a weakening in EoS independence between the moment of inertia $I$ and the quadrupole moment $Q$ under rapid rotation. Pappas and Apostolatos \cite{pappas2014effectively} restored universality by introducing the dimensionless angular momentum $\chi = J/M^2$ as a spin parameter instead of the rotation frequency. This insight, extended by Chakrabarti et al. \cite{chakrabarti2014q}, underscores the significance of employing dimensionless quantities to establish universal relations for rapidly rotating NSs. However, there is still some confusion on this matter. In particular, Konstantinou and Morsink \cite{Konstantinou:2022vkr} discuss the potential loss of universality concerning the equatorial radius of NSs. Nevertheless, this may be due to an unfavorable selection of parameters, which is similar to the case outlined in \cite{doneva2013breakdown}.

Taking a slightly different direction, Pappas and Apostolatos \cite{pappas2014effectively}, along with Stein et al. \cite{stein2014three}, explored EoS-insensitive three-parameter relations ($M,\chi,\bar{Q}$) applicable to NSs. Subsequently, these relations were extended to include quark stars by Yagi et al. \cite{yagi2014effective} and Chatziioannou et al. \cite{chatziioannou2014toward}, incorporating higher-order multipole moments. In addition,  recent progress in formulating universal relations has incorporated innovative data science approaches. For instance, Papigkiotis and Pappas \cite{papigkiotis2023universal} utilized supervised ML techniques to derive EoS-insensitive relations for NS global parameters, while Manoharan and Kokkotas \cite{manoharan2023finding} employed statistical data analysis methods. Efforts also extended to provide theoretical justifications, linking universality to the homologous isodensity profiles within stars and the stiffness of ultradense EoSs \cite{yagi_i-love-q_2013a,stein2014three,martinon2014rotating}. These endeavors enrich our understanding of NS properties, affirming the impact of nuclear matter properties at low-mass densities and emphasizing the role of homologous structures in both Newtonian and general relativistic scenarios \cite{yagi_i-love-q_2013a,stein2014three, yagi2014love}. Additionally, proposals by Martinon et al. \cite{martinon2014rotating} and Sham et al. \cite{sham2015unveiling} suggest that ultradense stiff EoSs, when treated as expansions around the incompressible limit, contribute to the EoS independence observed at higher compactnesses, where nuclear matter approaches the limit of an incompressible fluid. The two pictures are complementary to each other, since a homologous and almost constant ellipticity profile describes isodensity surfaces in the incompressible limit.

Above all, measuring the NS bulk properties is one of the most rigorous tests of our comprehension of matter under extreme conditions. In that direction, several methods developed involve analyzing the X-ray flux emission originating from {\it hot spots} on the stellar surface \cite{gendreau2012neutron, arzoumanian2014neutron, gendreau2017searching,bogdanov2008thermal,miller2015determining,mereghetti2010x,chatziioannou2024neutron,silva2019neutron,yunes2022gravitational,parikh2021uv,bogdanov2016neutron,choudhury2024nicer}. Detected as pulsations from the NICER telescope, the shape of this radiation (i.e., its profile) carries valuable information about the star's surface properties and the surrounding spacetime, see e.g. \cite{watts2016colloquium, watts2019constraining}. This information enables the simultaneous inference of the mass and the equatorial radius $R_{\mathrm{eq}}$ at a precision level of $5\%–10\%$. Recent announcements have revealed measurements of the mass and the equatorial radius of the isolated millisecond pulsars $\mathrm{J}0030+0451$ \cite{miller2019psr, riley2019nicer,raaijmakers2019nicer,jiang2020psr}, and $\mathrm{J}0740 + 6620$ \cite{miller2021radius,riley2021nicer,fonseca2021refined,biswas2021impact,raaijmakers2021constraints,salmi2024radius,dittmann2024more,cromartie_relativistic_2020}. Furthermore, it should be highlighted that for PSR $\mathrm{J}0740+6620$, the most recent equatorial radius estimate \cite{salmi2024radius,dittmann2024more} based on the NICER data exhibits strong consistency with the prior ones \cite{cromartie_relativistic_2020,miller2021radius,riley2021nicer}, aligning well within the reported confidence intervals. PSR $\mathrm{J}0437-4715$, the nearest and most luminous known millisecond pulsar, provides supplementary constraints on mass and radius, building upon the findings from previous measurements \cite{choudhury2024nicer,reardon2024neutron}.
However, the determination of mass and radius for PSR $\mathrm{J}1231-1411$ is complicated by its complex surface emission geometry, including non-antipodal {\it hot spots}, and the weak interpulse feature in its X-ray pulse profile \cite{salmi2024nicer}. Reliable inferences from the NICER and XMM-Newton data require constrained radius priors, highlighting the sensitivity of the results to these assumptions and the challenges of accurately modeling the observational data. Furthermore, additional properties, such as the star's moment of inertia, can be deduced using quasi-equation-of-state insensitive relations, see e.g. \cite{silva2021astrophysical}. To the best of our knowledge, expected inferences from the observation of one another PSR: $\mathrm{J}2124-3358$ is anticipated to be released in the near future \cite{bogdanov2019constraining}.

Many of these primary targets as well as those investigated through spectroscopic observations have moderate spins of a few hundred $\mathrm{Hz}$. For instance, rotation-powered X-ray pulsars observed by NICER are relatively slowly rotating, with spin frequencies ranging from $174$ $\mathrm{Hz}$ (e.g., PSR $\mathrm{J}0437-4715$) to $346$ $\mathrm{Hz}$ (e.g., PSR $\mathrm{J}0740+6620$). In contrast, accretion-powered pulsars exhibit much higher rotation rates, such as IGR $\mathrm{J}00291+5934$ which rotates at $599$ $\mathrm{Hz}$ (the fastest AMXP known) \cite{galloway2005discovery,Patruno2021}. Moreover, the spin frequencies of quiescent low-mass X-ray binaries (LMXBs) and X-ray bursters are largely uncertain. However, examples such as $4\mathrm{U} \ 1608-522$, with a spin frequency of $620$ $\mathrm{Hz}$, demonstrate the potential for substantial rotation rates \cite{muno2002frequency,tarana2008integral}. Table (\ref{tab:indicative_psrs_table}) provides an indicative summary of the typical parameters for the NS configurations previously discussed, including the range of the dimensionless spin parameter $\sigma = \Omega^2 R_{\mathrm{eq}}^3/GM$, derived from plausible values for masses and radii. For PSRs $\mathrm{J}0437-4715$ and $\mathrm{J}0740+6620$ the parameters displayed are based on \cite{bogdanov2019constraining,fonseca2021refined,dittmann2024more,salmi2024radius}. In contrast, for $\mathrm{J}00291+5934$ and $4\mathrm{U} \ 1608-522$, typical and reasonable NS mass and radius values are employed to extract an estimate for the reduced spin.

\begin{table}[!ht]
    \footnotesize
    \caption{\label{tab:indicative_psrs_table} Indicative parameters for different populations of X-ray NSs.}
    \begin{ruledtabular}
        \begin{tabular}{c|c|c|c|c}
            {NS} & $M \ [M_{\odot}]$  & $R_{\mathrm{eq}} \ [\mathrm{km}]$&  $f$ [$\mathrm{Hz}$] & spin parameter $\sigma$\\
    
            \hline
             PSR $\mathrm{J}0437-4715$ & $1.44$ & $15.3^{+2.0}_{-1.6}$ &  $174$ & $\sim [0.016, 0.032]$    \\
             \hline
             PSR $\mathrm{J}0740+6620$ & $2.08^{+0.07}_{-0.07}$& $12.49^{+1.28}_{-0.88}$& $346$ & $\sim [0.028, 0.043]$ \\
             \hline
             IGR $\mathrm{J}00291+5934$ & $\sim 1.4$& $\sim 10-14$ & $599$ &  $\sim [0.076,0.209]$\\
             \hline
             $4\mathrm{U} \ 1608-522$ & $\sim 1.4$ &  $\sim 10-14$ & $620$ & $\sim [0.082,0.224]$ \\
        \end{tabular}
    \end{ruledtabular}
\end{table}



At the frequencies of a few hundred $\mathrm{Hz}$ or higher, gravitational effects are influenced not only by the mass and radius but also by additional parameters, including the NS's quadrupole moment and the oblateness of its surface, see e.g. \cite{morsink2007oblate, baubock2013narrow}. Leveraging these observations to measure the masses and radii of NSs with the precision necessary for constraining their EoS necessitates taking these non-negligible effects into account. These electromagnetic observations combined with GW inferences on the tidal deformability from NS binaries will considerably improve our understanding of the NS EoS, see e.g. \cite{raithel2019constraints, raaijmakers2019nicer, raaijmakers2020constraining, jiang2020psr, zimmerman2020measuring, dietrich2020multimessenger, chatziioannou2020neutron, essick2020direct,chatziioannou2024neutron,suvorov2024premerger,bauswein2019equation,ripley2024constraint,pang2021nuclear,bogdanov2016neutron,parikh2021uv, miller2015determining,silva2019neutron,yunes2022gravitational,bogdanov2016neutron,choudhury2024nicer,raaijmakers2021constraints,carson2019future, huxford2024accuracy,francesco2023nuclear}.

Mainly, the shape of the NS depends on the EoS and the rotation frequency parametrization. In the direction of describing the surface properties of a rotating star, notable contributions have been made by Morsink et al. \cite{morsink2007oblate} and AlGendy and Morsink \cite{algendy2014universality}, who introduced EoS-insensitive formulas for surface estimation associated with the equatorial radius $R_{\mathrm{eq}}$. These proposals hinge on the assumption that the fitting coefficients are contingent upon both the stellar compactness $C$ and the dimensionless spin parameter $\sigma = \Omega^2R^3_{\mathrm{eq}}/GM$. A comprehensive review of these coefficients across a wider spectrum of NS models was conducted by Silva et al. \cite{silva2021surface}. In their work, a new novel fitting formula based on an elliptical isodensity approximation \cite{1993ApJS...88..205L} was also proposed. Furthermore, regarding the effective gravitational acceleration on the star's surface, AlGendy and Morsink \cite{algendy2014universality} proposed universal relations applicable to both slowly and rapidly rotating NSs. The coefficients for the relative fitting functions also rely on $C$ and $\sigma$ parameters. However, it is important to note that their analysis was based on a limited sample of EoS models.

Driven by the aforementioned motivation, this work focuses on the determination, with high accuracy, of the NS's surface, and its effective gravitational acceleration for a wide range of rotating stellar models. In order to make that possible, we deployed an ANN architecture for regression. Crucially, this approach is designed to be independent of the specific EoS chosen. The choice of an ANN model lies in its ability to discover complex patterns and relations within data \cite{hornik1989multilayer}. Additionally, the capacity of neural networks for learning from large datasets allows them to generalize well to unseen data, enhancing their predictive accuracy.

Along the way, we have formulated several new EoS-insensitive relations, serving as valuable global quantities that characterize the NS configuration. Moreover, these astrophysical parameters play a crucial role in modeling various aspects of the star's surface. The derivation of these auxiliary universal relations was performed by utilizing supervised ML techniques, including Cross-Validation and least-squares regression. Such techniques facilitated the identification of the most suitable functional forms that validate the correlated data. The primary investigation encompasses a diverse range of rotating models, ranging from static configurations to frequencies reaching close to the mass-shedding (Kepler) limit \footnote{Astrophysical mechanisms, such as the r-mode instability \cite{andersson1999relevance,andersson2000r}, may limit the spin to a fraction of the Kepler limit.}. Specifically, we propose new accurate universal relations about the star's polar radius, and the star's eccentricity (oblateness), encompassing various observable parameters, such as the stellar compactness $C$ and the reduced spin $\sigma$. In addition, we explore non-conventional relations, including the theoretical universal description of the star's surface maximum value of the logarithmic derivative $d\log R(\mu)/d\theta$ as well as those concerning the effective acceleration due to gravity at the star's pole and equator. With these new insights, we are able to acquire the essential information for modeling the star's surface for arbitrary rotation, leading the trained ANN model to make precise predictions at $0.25\%$ accuracy for each particular NS model within our test ensemble. In addition, the designed ANN architecture provides a universal estimation of the surface's logarithmic derivative $d\log R(\mu)/d\theta$, and the star's effective acceleration at the surface with an accuracy of order less than $1\%$.

The plan of the paper is as follows. Initially, in Sec. \ref{sec:num_data}, we briefly introduce the numerical setup and the associated sample of EoSs used to estimate our equilibrium NS-models ensemble. Then, in Sec. \ref{sec:surface_localization}, we present the enthalpy-based method utilized for the accurate localization of the NS's surface. This section also covers the numerical framework for the estimation of the logarithmic derivative and the surface effective acceleration due to gravity. Following this, in Sec. \ref{sec:univ_rel_surf_prop}, we present the corresponding new fitting functions proposed regarding the investigated NS's surface properties, whereas in Sec. \ref{sec:univ_rel_with_ANNs}, we present the formulation for the precise inference of the quantities relating to the star's surface itself. In Sec. \ref{sec:conclusion}, we summarize our findings and present our concluding remarks. Finally, in Appendix \ref{sec:ML_part}, we introduce the ML framework employed to extract our results. We introduce the methodology used for linear regression and cross-validation, accompanied by the designed ANN architecture for training and testing purposes. Then, in Appendix \ref{app:eos_tables}, we incorporate the tables with the EoS-ensemble that we have used. Lastly, in Appendix \ref{sec:errors_violin_plots}, we provide a further overview regarding the inference in the test set for the new ANN fitting functions investigated in Sec. \ref{sec:univ_rel_with_ANNs}. Unless stated otherwise, we set $G = c = 1$.

\section{\label{sec:num_data} NUMERICAL SCHEME AND EoS DATA}
The primary uncertainties in NS bulk properties arise from unknown particle interactions in high-density regions. Beyond the nuclear saturation density $\rho_0$, typical of standard symmetric nuclear matter, NS's structure and composition become increasingly uncertain \cite{haensel_neutron_2007,lattimer2021neutron}. Different EoSs exhibit markedly distinct bulk properties, directly influencing the construction of static and rotating NS sequences. The EoS is crucial for describing the macroscopic properties of NS physics, serving as a key input for solving Einstein's field equations. In this work, we generated a comprehensive ensemble of equilibrium NS configurations, incorporating various EoSs. These models include nonrotating and uniformly rotating configurations, covering frequencies from the static case up to a high rotational frequency of $\sim 1.87 \ \mathrm{kHz}$ (larger than encountered in known pulsars). The nonrotating solutions were obtained through the integration of hydrostatic equilibrium equations in spherical symmetry \cite{oppenheimer_massive_1939}, while for the rotating ones,  we have used the RNS code \cite{rns, stergioulas1994comparing}, which integrates the nonlinear elliptic-type field equations alongside the hydrostationary equilibrium equation \cite{friedman_rotating_2013, cook1992spin}.

Specifically, we consider the stellar matter as a perfect fluid exhibiting local isotropy and described by the energy-momentum tensor \cite{friedman_rotating_2013,rezzolla2013relativistic},
\begin{equation}
\label{Tmn}
T^{a\beta}=\left(\epsilon+P\right)u^au^{\beta}+Pg^{a\beta},
\end{equation}
where $u^a$ represents the fluid four-velocity, $g^{\alpha\beta}$ is the metric tensor, and $\epsilon$ and $P$ are scalar quantities denoting the fluid's total energy density and pressure, respectively. 

For nonrotating NSs, we adopt a spherically symmetric metric 
\begin{equation}
\label{ds2_stat}
ds^2=- e^{2\nu(r)}dt^2+e^{2\lambda(r)}dr^2+r^2\left(d\theta^2+\sin^2\theta d\phi^2\right).
\end{equation}
Here, $\nu(r)$ and $\lambda(r)$ are time-independent metric functions of the radial coordinate $r$, following Birkhoff's theorem. The time independence of the metric tensor implies that the matter within the NS is in hydrostatic equilibrium. Equilibrium configurations are determined as solutions of the Tolman-Oppenheimer-Volkoff (TOV) equations \cite{oppenheimer_massive_1939,haensel_neutron_2007}, given by
\begin{subequations}
	\label{eq:tov_system}
	\begin{align}
	\frac{dm}{dr} & = 4\pi r^2 \epsilon(r),\\
	\frac{d\nu}{dr}& =\frac{m(r)+4\pi r^3 P(r)}{r(r-2m(r))},\\
	\frac{d P}{d r}& =-\left(\epsilon(r) +P(r)\right)\frac{d\nu}{dr},
	\end{align}
\end{subequations}
where the mass-function $m(r)$ is identified as $m(r) = \frac{r}{2}\left(1-e^{-2\lambda (r)}\right)$. The TOV equations are supplemented by a cold, ultra-dense, and barotropic ($\epsilon=\epsilon(P)$) nuclear matter EoS, establishing a relation between energy density and pressure \cite{friedman_rotating_2013, rezzolla2013relativistic, haensel_neutron_2007, rezzolla_physics_2018}. It is important to emphasize that Eqs. (\ref{ds2_stat}) and (\ref{eq:tov_system}) provide the general definitions incorporating the Schwarzschild radius $r$. However, the RNS code utilizes the quasi-isotropic coordinate $\tilde{r}$ \cite{friedman_rotating_2013}, which becomes the isotropic coordinate in the nonrotating limit.

To address the uncertainty of the EoS at ultrahigh densities in $\beta$ equilibrium, numerous models have been proposed in the literature, based on diverse many-body nonrelativistic and relativistic theories \cite{dutra2014relativistic,dutra2012skyrme,compose2022compose,lattimer2021neutron,oertel2017equations,dutra2014relativistic,haensel_neutron_2007,chatziioannou2024neutron}. Nonrelativistic approaches include nuclear effective interaction forces (EI), cluster energy functionals (CEF), density functionals (NR DF), and unified Scyrme-Hartree-Fock nuclear forces (SHF) \cite{haensel_neutron_2007}. Relativistic methods encompass the relativistic mean-field theory (RMF), relativistic density functionals (RDF), chiral perturbation theory (chPT), perturbative Brueckner-Bethe-Goldstone quantum theory (BBG), Brueckner-Hartree-Fock approximation with continuous choice for the auxiliary single-particle potential (BHF), chiral mean-field theory models (CMF), nonperturbative functional renormalization group approach (NP-FRG), Nambu-Jona-Lasinio (NJL) model \cite{yu_self-consistent_2020} within the mean-field approximation (NJL-MF), and a two-flavor quark-meson truncation in the local potential approximation (LPA) including vector interactions within the nonperturbative functional renormalization group approach (NP-FRG) \cite{haensel_neutron_2007, malik_gw170817_2018}.

Incorporating these theoretical considerations, we employed realistic EoS models sourced from the CompOSE \cite{compose,compose2022compose} database. The utilized EoS ensemble includes hadronic, hyperonic, and hybrid models, presented in tabulated form, providing a comprehensive description of the NS's interior, encompassing both the crust and the stellar core. The extensive list, consisting of 70 cold EoSs employed for each NS category can be found in Tables (\ref{tab:hadronic},\ref{tab:hyperonic},\ref{tab:hybrid}), displayed in Appendix \ref{app:eos_tables}. To the best of our knowledge, this work utilizes the most comprehensive EoS ensemble that has been employed to derive NS universal relations.

All EoSs are assumed to follow $\beta$ equilibrium and zero temperature conditions. In addition, various individual families are further subdivided based on the specific physical theory employed to describe the EoS data. Furthermore, each EoS is accompanied by detailed information on the matter composition within the star's core, as well as essential NS properties, including the nonrotating maximum mass, the corresponding equatorial radius, and the equatorial radius of a $1.4 \ M_{\odot}$ configuration.

The selected EoS models listed in the Tables (\ref{tab:hadronic},\ref{tab:hyperonic},\ref{tab:hybrid}) satisfy the stipulated constraints defined by the lower bounds on the maximum nonrotating mass for PSR $\mathrm{J}0348+0432$ ($M = 2.01^{+0.04}_{-0.04} \ M_{\odot}$) \cite{demorest2010two, antoniadis_massive_2013} and PSR $\mathrm{J}0740+6620$ ($M = 2.14^{+0.20}_{-0.18} \ M_{\odot}$) within a $2\sigma$ credible interval \cite{cromartie_relativistic_2020} (first measurement). It is worth noting that Fonseca et al. \cite{fonseca2021refined} provided an improved estimate of the initially reported PSR $\mathrm{J}0740+6620$ mass, determining it as $M = 2.08^{+0.07}_{-0.07} \ M_{\odot}$ with $1\sigma$ credibility. This refinement was achieved using additional radio data and leveraging the relativistic Shapiro time delay. Furthermore, the selected ensemble of EoSs complies with this constraint within the reported credible intervals. Despite that, it is important to point out that PSR J0740 rotates at $346 \ \mathrm{Hz}$, allowing it to support a slightly higher maximum mass than the nonrotating case. This increase, however, is minimal—likely no more than $1\%$, depending on the star's equatorial radius. Currently, the $1\sigma$ uncertainty in the mass measurement estimated by \cite{fonseca2021refined} is approximately $3\%$, rendering the rotational mass increase negligible at this stage. Nevertheless, it is expected that future observations will reduce these error margins, thus the effect of rotation on the maximum mass could become more significant.

From the binary neutron star merger perspective, these EoS models also yield nonrotating maximum mass NS with a radius $R_{{M_{max}}} \geq 9.60^{+0.14}_{-0.03} \ \mathrm{km}$, as indicated by the GW170817 NS-NS merger analysis \cite{bauswein_neutron-star_2017, friedman_astrophysical_2020-2}. Additionally, none of the selected EoS models surpass a maximum mass of $2.33 \ M_{\odot}$ within the $2\sigma$ bound, assuming that the final remnant of GW170817 was a black hole \cite{dietrich2020multimessenger, rezzolla_using_2018}. It is crucial to highlight that all these EoS models verified the physical acceptability conditions (see e.g., \cite{papigkiotis2023universal, haensel_neutron_2007}, for a review), ensuring $\beta$-equilibrium.

\begin{figure}[!ht]
	\includegraphics[width=0.47\textwidth]{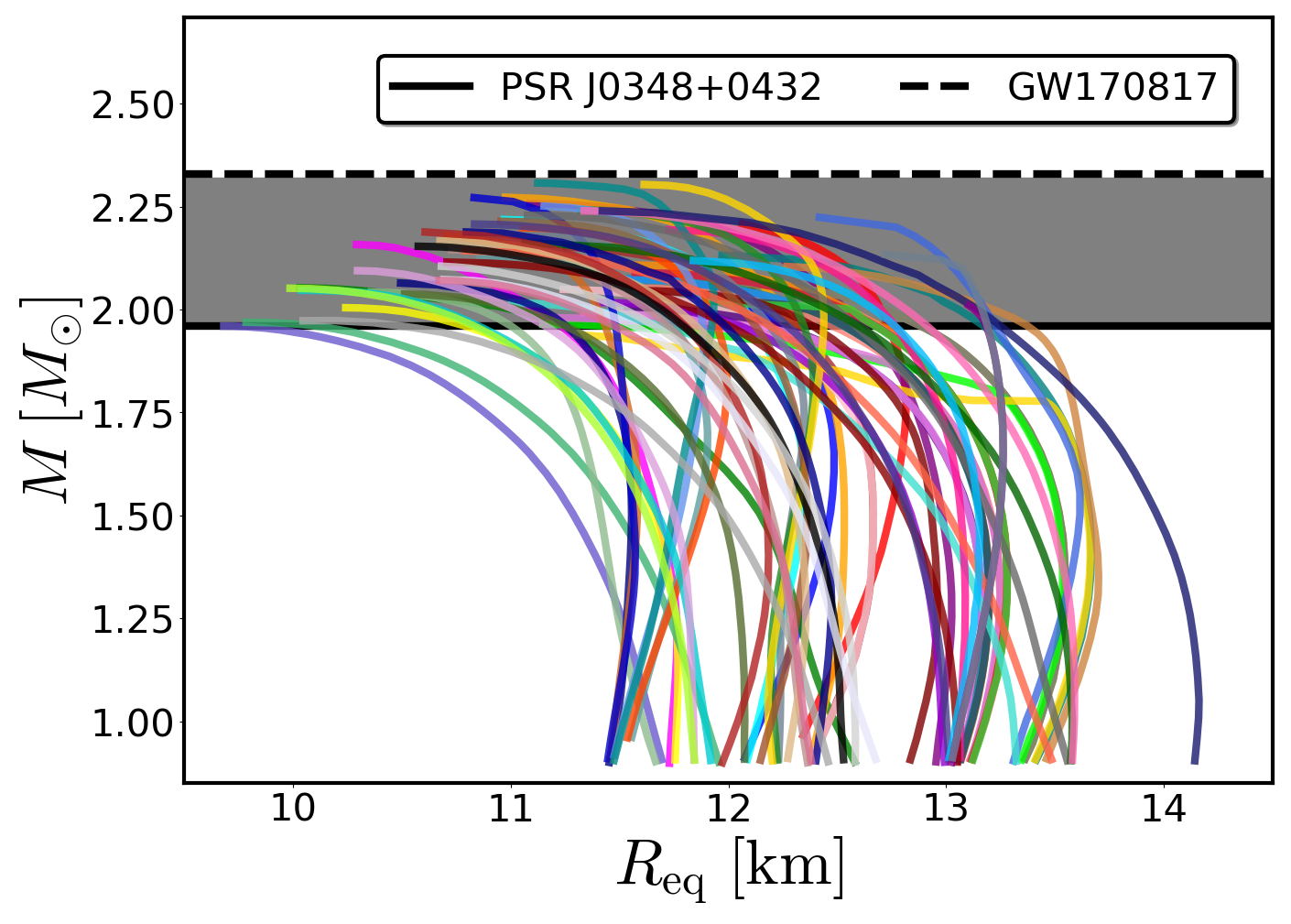}
	\caption{\label{fig:M_R} {$M=M(R_{\mathrm{eq}})$} diagram for sequences of nonrotating NS configurations. The various colors are representative of different EoSs as indicated by the legend provided in Fig. \ref{fig:color_band}, located in Appendix \ref{app:eos_tables}. This color-to-EoS mapping remains consistent across all subsequent fit figures presented in the Sec. \ref{sec:univ_rel_surf_prop}.}
\end{figure}

In Fig. \ref{fig:M_R}, we illustrate the Mass-Radius curves for the cold EoS ensemble employed in this work. Each curve presented corresponds to a nonrotating NS sequence.

Most EoSs assume $npe\mu$-particle composition in the stellar core, while others incorporate other exotic matter components such as hyperons (Table \ref{tab:hyperonic}) or quarks (Table \ref{tab:hybrid}). The $M-R_{\mathrm{eq}}$ relation for an EoS is presented only up to the star's maximum mass. It is noteworthy that horizontal lines represent the $2\sigma$ lower and upper range for the mass of one of the two most massive known radio pulsars, PSR $\mathrm{J}0348+0432$ ($M = 2.01^{+0.04}_{-0.04} \ M_{\odot}$) \cite{antoniadis_massive_2013} (solid line, lower limit), and the maximum mass of GW17087 final remnant \cite{dietrich2020multimessenger, rezzolla_using_2018} (dashed line, upper limit). Recently, a revised mass measurement for PSR $\mathrm{J}0348+0432$ has been reported, lowering its estimated mass to $1.8 \ M_{\odot}$ \cite{saffer2024lower}. Regardless, this updated lower mass does not affect the mass threshold that an EoS must support utilized in this work, as the mass limit of PSR $\mathrm{J}0740+6620$ continues to provide essentially the same constraint.

Nevertheless, most stellar objects commonly experience rotation, and sometimes, this rotational motion can be quite rapid. A rotating compact object is characterized by its mass $M$ and its angular momentum $\mathrm{J}$ \cite{friedman_rotating_2013}. In this case, the spacetime is considered to be stationary, axisymmetric, and asymptotically flat. These assumptions are formulated mathematically by introducing two Killing vectors, $t^{\alpha}$ and $\phi^{\alpha}$. In quasi-isotropic coordinates ($\tilde{r},\theta$), the stationary metric satisfying these criteria is described by the line element \cite{friedman_rotating_2013, butterworth_structure_1976}: 
\begin{align}
\label{eq:ds_rot_2}
\nonumber ds^2=&-e^{(\gamma+\rho)}dt^2+e^{(\gamma-\rho)}\tilde{r}^2\sin^2\theta(d\phi-\omega dt)^2\\
&+e^{2a}(d\tilde{r}^2+\tilde{r}^2d\theta^2),
\end{align}
where the metric functions $\gamma, \rho, \omega, \ \mathrm{and} \ \alpha$ are functions of the $(\tilde{r}, \theta)$.
For a uniformly rotating stellar configuration, the star's angular velocity $\Omega$ as defined by an observer at infinity remains constant. The equation governing hydrostatic equilibrium for a stationary, axisymmetric, and uniformly rotating NS is given by \cite{friedman_rotating_2013},
\begin{equation}
\label{hydro_uniform}
\frac{\nabla_a P}{\epsilon + P}=\nabla_{a} \ln u^t,
\end{equation}
where
\begin{equation}
    \label{u_a}
    u^\alpha=u^t(t^\alpha+\Omega\phi^\alpha),
\end{equation}
is the four-velocity of a perfect-fluid element, expressed in terms of the timelike and spacelike Killing vectors $t^\alpha$ and $\phi^\alpha$ while,  
\begin{equation}
\label{u_t}
u^t=\frac{e^{-(\rho+\gamma)/2}}{\sqrt{1-(\Omega-\omega)^2\tilde{r}^2\sin^2(\theta) \ e^{-2\rho}}},
\end{equation}
follows from the normalization condition $u^\alpha u_\alpha = -1$ and the metric (\ref{eq:ds_rot_2}).

To solve the nonlinear Einstein field equations along with the hydrostationary equilibrium equation \cite{friedman_rotating_2013, paschalidis2017rotating}, various numerical methods have been developed \cite{wilson1972models, bonazzola1974exact-2, friedman1989implications, komatsu_rapidly_1989-2, komatsu_rapidly_1989_II-2, stergioulas1994comparing}. Solutions are obtained through numerical integration on a discrete grid, employing a combination of integral and finite differences techniques \cite{komatsu_rapidly_1989-2}. Notably, Komatsu, Eriguchi, and Hachisu (KEH) \cite{komatsu_rapidly_1989-2, komatsu_rapidly_1989_II-2} and Cook, Shapiro, and Teukolsky (CST) \cite{friedman_rotating_2013} employed an iterative numerical method, utilizing integral representation with Green's functions. In this work, the numerical integration for the equations of structure and field equations is performed using the RNS code \cite{rns, stergioulas1994comparing}, which is based on the aforementioned methods.

Assuming a perfect fluid, the {RNS} code solves for the NS's interior (matter and spacetime) and exterior spacetime on a discrete grid with the radial coordinate $\tilde{r}$ compactified and equally spaced in the interval $s\in [0,1]$, using 
$s\equiv \tilde{r}/(\tilde{r}+\tilde{r}_{\mathrm{eq}})$. The angular coordinate $\mu = \cos(\theta)$ is equally spaced in the interval $\mu \in[0,1]$. Here, $\tilde{r}_{\mathrm{eq}}$ corresponds to the coordinate radius of the star's surface at the equator. The computational grid is structured so the star's center is at $s = 0$, the surface at $s=1/2$, and infinity at $s=1$. The equatorial plane is at $\mu = 0$, and the pole is at $\mu=1$. It is worth noting that, in the equatorial plane, half of the grid is assigned to the star's interior, while the other half is allocated to the vacuum exterior. In this work, a grid size of $\mathrm{MDIV}\times \mathrm{SDIV}=261\times 521$ is employed, where MDIV is the number of points in the angular $\mu$ direction, while SDIV is the number of points in the compactified radial $s$ direction.

For a given EoS under uniform star rotation, the {RNS} code yields unique equilibrium solutions by specifying the central energy density $\epsilon_c$ and the axial ratio $\tilde{r}_{\mathrm{pole}}/\tilde{r}_{\mathrm{eq}}$ between the polar and equatorial coordinate radii. Stellar models are then computed along sequences varying the central energy density and axial ratio \cite{friedman_rotating_2013, stergioulas1994comparing}. The numerical computation includes the determination of the star's metric functions within both interior and exterior regions, encompassing the fluid configuration and various equilibrium quantities. Across our EoS ensemble, we computed a diverse sample of static, relatively slowly and rapidly rotating NSs, covering central densities $\epsilon_c \sim [3.928\times10^{14}-3.029\times10^{15}] \ \mathrm{g/cm^3}$ and masses from $\sim 0.9 \ M_{\odot}$ up to the star's maximum mass $M_{\mathrm{max}}$. Our comprehensive sample comprises $40015$ rotating NS models with frequencies spanning from a few hundred $\mathrm{Hz}$ ($f\sim 190.27 \ \mathrm{Hz}$) up to the $\mathrm{kHz}$ ($f\sim 1.87\ \mathrm{kHz}$) limit, and $2679$ nonrotating equilibrium ones. This ensemble of NSs stands as an adequate dataset for a thorough investigation.

\section{\label{sec:surface_localization} Numerical SURFACE LOCALIZATION AND EFFECTIVE GRAVITY IN ROTATING STARS}
When a numerical NS solution is provided, we can estimate the star's coordinate surface $\tilde{r}_{s}(\theta)$ by identifying the loci where the pressure vanishes. Then, the star's circumferential radius $R(\mu)$ is determined in the {RNS} computational grid as a function of the cosine of the colatitude $\theta$ and the compactified radial coordinate $s$ as:
\begin{equation}
\label{R_circ}
R(\mu)=\tilde{r}_s(\mu)\ e^{(\gamma_s-\rho_s)/2},
\end{equation}
where $\mu$ has already been defined,
$\gamma_s\equiv\gamma(\tilde{r}_s),\  \textrm{and} \ \rho_s\equiv\rho(\tilde{r}_s)$ are the metric functions computed in the radial coordinate $\tilde{r}_s = \tilde{r}_s({\mu})$ \cite{butterworth_structure_1976, cook_spin-up_1992, komatsu_rapidly_1989-2}. Utilizing this formulation, we can express the ratio between the star's polar and equatorial radius as 
\begin{equation}
   \label{R_p/R_e}
   \mathcal{R} \equiv \frac{R_{\mathrm{pole}}}{R_{\mathrm{eq}}} = \frac{R(\mu=1)}{R(\mu=0)}.
\end{equation}
Additionally, the ellipticity $\epsilon_s$ and eccentricity $e$ of the meridional cross-section of the star are defined as \cite{silva2021surface},
\begin{align}
\label{ellipticity}
\mathrm{\epsilon_s} & \equiv 1- \mathcal{R} \\
\label{eccentricity}
e & \equiv \sqrt{1-\mathcal{R}^2}.
\end{align}

\begin{figure}[!th]
	\includegraphics[width=0.46\textwidth]{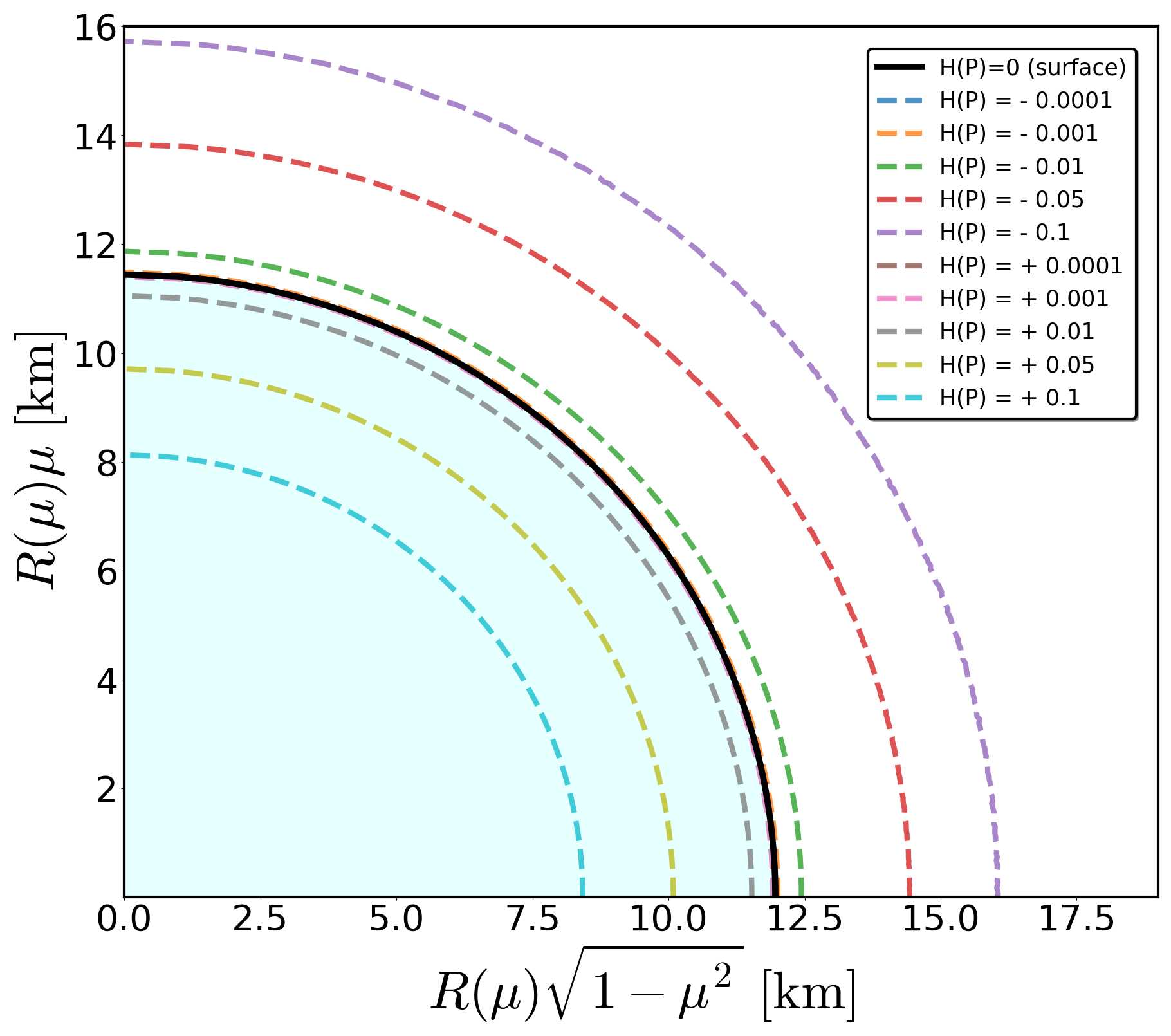}
	\caption{\label{fig:H_p_curves} Indicative contours of constant enthalpy per unit mass, $H$, for a NS configuration with central energy density $\epsilon_c=9.875\times10^{14} \ \mathrm{g/cm^3}$, mass $M=1.439 \ M_{\odot}$, and rotational frequency $f = 454.0 \ \mathrm{Hz}$ constructed using EoS SLY4. The solid black line represents the star's surface $R(\mu)$ at $H(P)=0$, while the dashed contour curves within the stellar surface have $H(P)>0$, and those outside have $H(P)<0$. This particular stellar object corresponds to model 2 with physical parameters highlighted in Table (\ref{tab:indicative_propert}).}
\end{figure}

To accurately estimate the NS surface solution, we employ an enthalpy-based method, specifically leveraging the first integral of the hydrostatic equilibrium equation (\ref{hydro_uniform}). For the barotropic EoS case discussed, defining the enthalpy per unit mass as \cite{friedman_rotating_2013}:
\begin{equation}
    \label{enthalpy}
    H(P)\equiv\int_{0}^{P}\frac{dP'}{\epsilon(P')+P'},
\end{equation}
the first integral of hydrostationary equilibrium (\ref{hydro_uniform}) takes the form 
\begin{equation}
\label{surf_1}
H(P)-\ln(u^t)=\textrm{const}=\left(\frac{\rho+\gamma}{2}\right)_{\mathrm{pole}},
\end{equation}
with the right-hand side term evaluated at the star's coordinate pole ($\tilde{r}_{\mathrm{pole}}=\tilde{r}(\mu=1)$). Notably, the enthalpy goes to zero along the star's surface, whereas it remains positive in the interior region. In the exterior, the enthalpy function (\ref{surf_1}) can still be calculated and has negative values, although it no longer describes fluid properties, but is just a combination of spacetime properties. In addition, the {RNS} code conveniently provides the polar redshift value as,
\begin{equation}
    \label{zp}
    z_{\mathrm{pole}}=\exp \left(-\frac{\rho_{\mathrm{pole}}+\gamma_{\mathrm{pole}}}{2}  \right)-1
\end{equation}
Therefore, the coordinate surface of the star $\tilde{r}_s(\mu)$ is determined by locating the points where $H(P)=0$, where the constant term in Eq.(\ref{surf_1}) is equivalently substituted by $-\ln(1+z_{\mathrm{pole}})$. In this way, using Eq.(\ref{u_t}), we solve numerically \footnote{For the numerical solution, we used a modification of the Powell hybrid method as implemented in the Python SciPy library \cite{2020SciPy-NMeth}: \url{https://docs.scipy.org/doc/scipy/reference/generated/scipy.optimize.root.html\#scipy.optimize.root}.} the equation
\begin{equation}
\label{H_r_mu}
u^t(\tilde{r},\mu) - (1+z_{\mathrm{pole}} )= 0,
\end{equation}
seeking in a sequence of values for $\mu$ within the range $[0,1]$, for the values of $\tilde{r}$ that the relation (\ref{H_r_mu}) is satisfied. This yields $\tilde{r}_s(\mu)$, and subsequently, the circumferential radius can be determined using Eq.(\ref{R_circ}). For the selected grid $\mathrm{MDIV}\times \mathrm{SDIV}$, we acquire 261 data points for $R(\mu)$ through this numerical method, where each data point corresponds to the respective $\mu_i \in [0,1]$ value. As an example, in Fig. \ref{fig:H_p_curves}, we present the contours of constant enthalpy per unit mass $H$ for an NS model characterized by $\epsilon_c=9.875\times10^{14} \ \mathrm{g/cm^3}$, a mass of $1.439 \ M_{\odot}$, and a rotational frequency of $f = 454.00 \ \mathrm{Hz}$. We have to note that the construction of this NS configuration is based on the SLY4 EoS. The solid black line delineates the surface where $H(P)=0$, while the indicative colored dashed lines represent different enthalpy contour curves relevant to the star's interior and exterior, respectively. In addition, it is important to emphasize that, under the isodensity approximation \cite{1993ApJS...88..205L}, the contour curves of constant enthalpy exhibit a consistent polar-to-equatorial radius ratio across the entire star. Numerical deviations from this property, determined by comparing the values associated with the illustrated enthalpy curves, are found to be less than  $1\%$.
\begin{figure}[!ht]
	\includegraphics[width=0.46\textwidth]{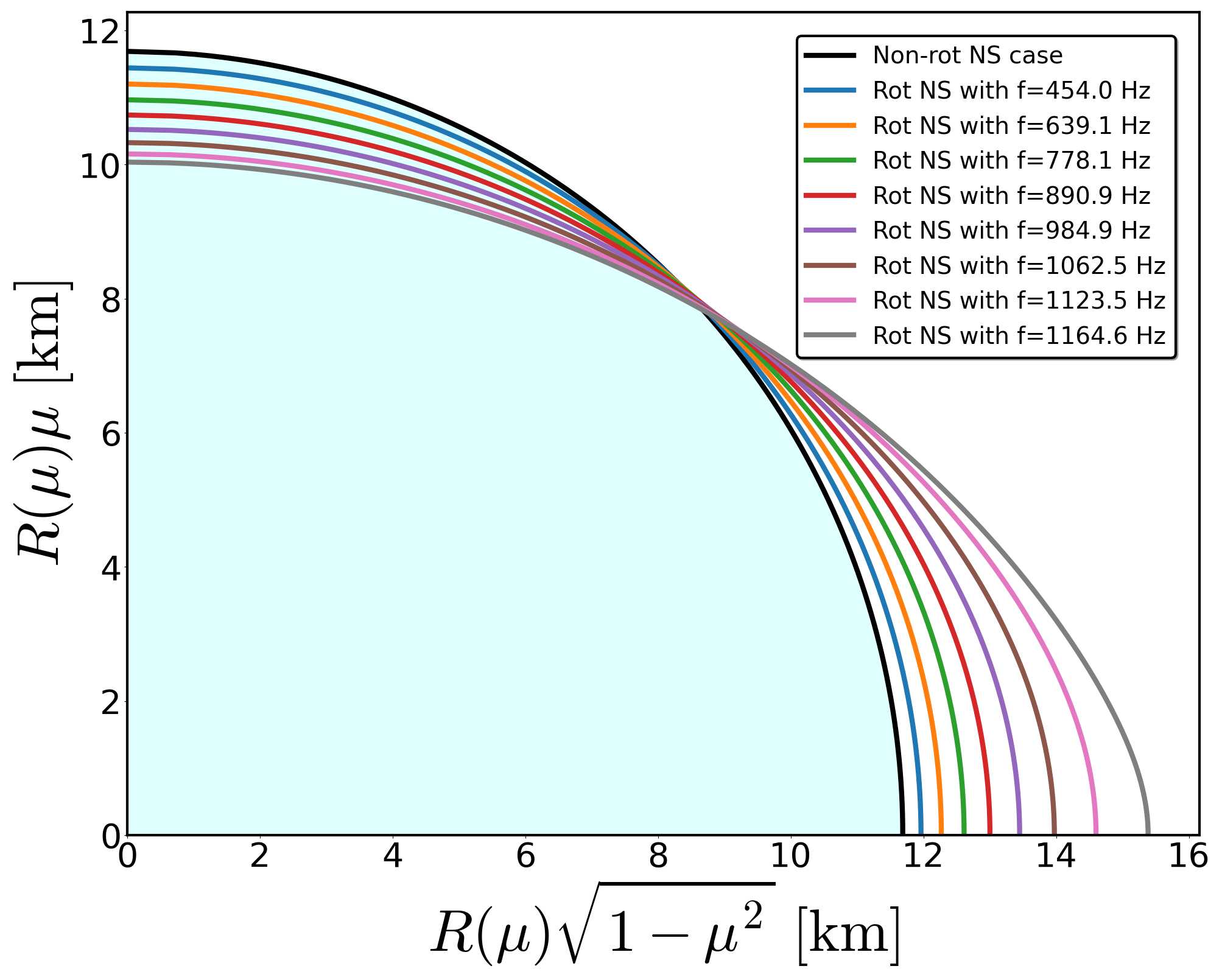}
	\caption{\label{fig:R(mu)_rep1} 
     Deformation of the NS's shape induced by rotation. The solid black curve corresponds to the surface of a nonrotating NS configuration, while the various colored ones correspond to surfaces associated with rapidly rotating NS models with the same central energy density $\epsilon_c=9.875\times10^{14} \ \mathrm{g/cm^3}$. The shaded cyan region represents the configuration of a nonrotating NS, characterized by an equatorial radius of $R_{\mathrm{eq}} = 11.688 \ \mathrm{km}$ and a mass of $M = 1.404 \ M_{\odot}$. The properties of the NS models presented are summarized in Table (\ref{tab:indicative_propert}).
    }
\end{figure}
\begin{figure}[!ht]
    \includegraphics[width=0.46\textwidth]{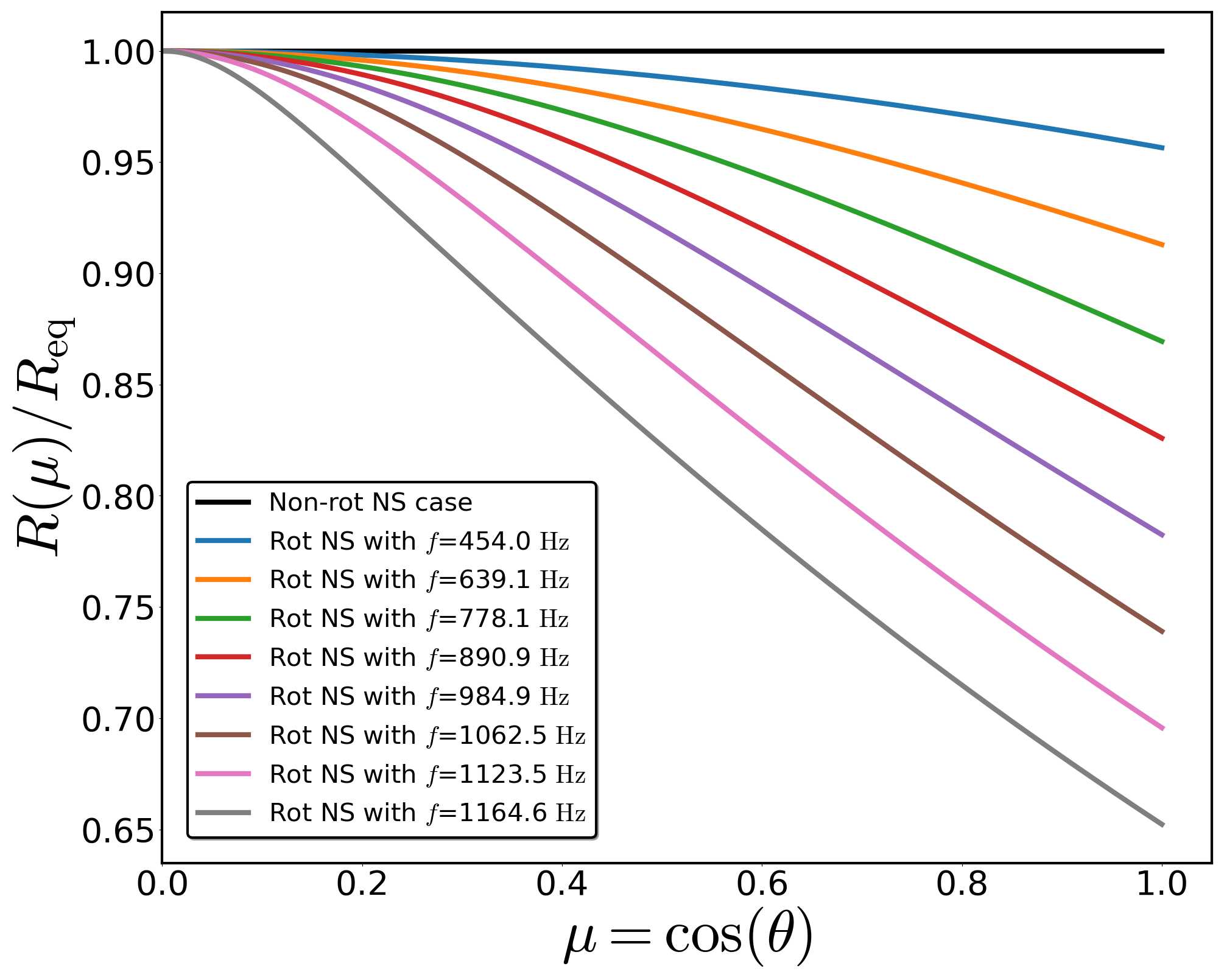}
	\caption{\label{fig:R(mu)_rep2} 
    Illustration of the normalized circumferential radius $R(\mu)/R_{\mathrm{eq}}$ as a function of the cosine of the colatitude $\theta$. The solid black curve corresponds to the static case, while the various colored ones correspond to rapidly rotating NS configurations with central energy density $\epsilon_c=9.875\times10^{14} \ g/\mathrm{cm^3}$. The models have been constructed by evenly spacing varying the polar-to-equatorial coordinate axial ratio ($\tilde{r}_{\mathrm{pole}}/\tilde{r}_{\mathrm{eq}}$) by a fixed step. The physical parameters of each NS model depicted are presented in Table (\ref{tab:indicative_propert}). 
    }
\end{figure}
\begin{table*}[th] 
    \caption{\label{tab:indicative_propert}Indicative NS models and their properties. They are obtained using the EoS SLY4 and correspond to a sequence of fixed central energy density $\epsilon_c=9.875\times10^{14} \ g/\mathrm{cm^3}$ stars with increasing rotational frequency. The columns represent the gravitational mass, the equatorial and polar radii, the stellar compactness $C=M/R_{\mathrm{eq}}$, the axes ratio $\tilde{r}_{\mathrm{pole}}/\tilde{r}_{\mathrm{eq}}$, the star's rotational frequency, the dimensionless spin $\sigma = \Omega^2R^3_{\mathrm{eq}}/GM$, the eccentricity $e$, the reduced effective acceleration at the equator, and, finally, the reduced effective acceleration at the pole.}
    \begin{ruledtabular}
    \begin{tabular}{|c|c|c|c|c|c|r|c|c|c|c|}
        Model & $M \ [M_{\odot}]$ & $R_{\mathrm{eq}} \ [\mathrm{km}]$ & $R_{\mathrm{pole}} \ [\mathrm{km}]$ & $C \ [-]$ & $\tilde{r}_{\mathrm{pole}}/\tilde{r}_{\mathrm{eq}}$ & $f$ ($\mathrm{Hz}$) & $\sigma \ [-]$ & $e \ [-]$ & $g_{\mathrm{eq}}/g_0$ & $g_{\mathrm{pole}}/g_0$ \\
        \hline
        1 & 1.404 &  11.688 & 11.688 &  0.1773 & 1.000 & 0.0 & 0.000& 0.000& 1.000 & 1.000 \\
        \hline
        2 & 1.439 &   11.963 & 11.442 & 0.1774 & 0.950 & 454.0 & 0.073 & 0.292 & 0.952 & 1.065 \\
        \hline
        3 & 1.476 &	12.269 &	11.201& 0.1775 &	0.900 &	639.1 & 0.152 & 0.408 & 0.896 & 1.136  \\
        \hline
        4 & 1.518 &	12.613 &	10.965 & 0.1775& 0.850 &	778.1 & 0.238	& 0.494&	0.830 & 1.215 \\
        \hline
        5 & 1.563 &	13.002 &	10.738 & 0.1773 	& 0.800 &	890.9 & 0.332 &	0.564 &	0.753 & 1.303 \\
        \hline
        6 & 1.612 &	13.449 &	10.524 &0.1768  & 0.750 &	984.9 & 0.436 &	0.623 &	0.661 & 1.405 \\
        \hline
        7 & 1.663 &	13.973 &	10.327 & 0.1756& 0.700 &	1062.5 & 	0.551 &	0.674 & 0.548 &	1.525\\
        \hline
        8 & 1.713 &	14.601 & 10.158 & 0.1731& 0.650 & 1123.5 & 0.683 &	0.718 &	0.406 & 1.672\\
        \hline
        9&  1.753 &	15.386 & 10.036 & 0.1681& 0.600 & 1164.6 & 0.839 & 0.758  & 0.218 &1.868\\
    \end{tabular}
    \end{ruledtabular}  
\end{table*}

Expanding on the use of the EoS SLY4, we present diverse surface representations of NS models, all sharing the same fixed central energy density ($\epsilon_c=9.875\times10^{14} \ g/\mathrm{cm^3}$) but different rotation frequencies, as illustrated in figures (\ref{fig:R(mu)_rep1}) and (\ref{fig:R(mu)_rep2}). The corresponding properties of these indicative stellar models are presented in Table (\ref{tab:indicative_propert}). Throughout this sequence of models, coordinate axial ratios (polar-to-equatorial) were evenly distributed, covering a range of rotational frequencies from the static case to $1164.55 \ \mathrm{Hz}$. The impact of the increasing rotation frequency on the star's shape is evident in Fig. \ref{fig:R(mu)_rep1}, where the star flattens at the poles while bulging out in the equator. Hence, rotation heightens oblateness, causing the star's shape to deviate from the spherical static case. This methodology for estimating the surfaces $R(\mu)$ of numerically simulated NS models and their associated features extends across the entire sample of compact objects, encompassing the whole set of cold EoSs for dense matter included in our ensemble.

\begin{figure}[!ht]
    \includegraphics[width=0.46\textwidth]{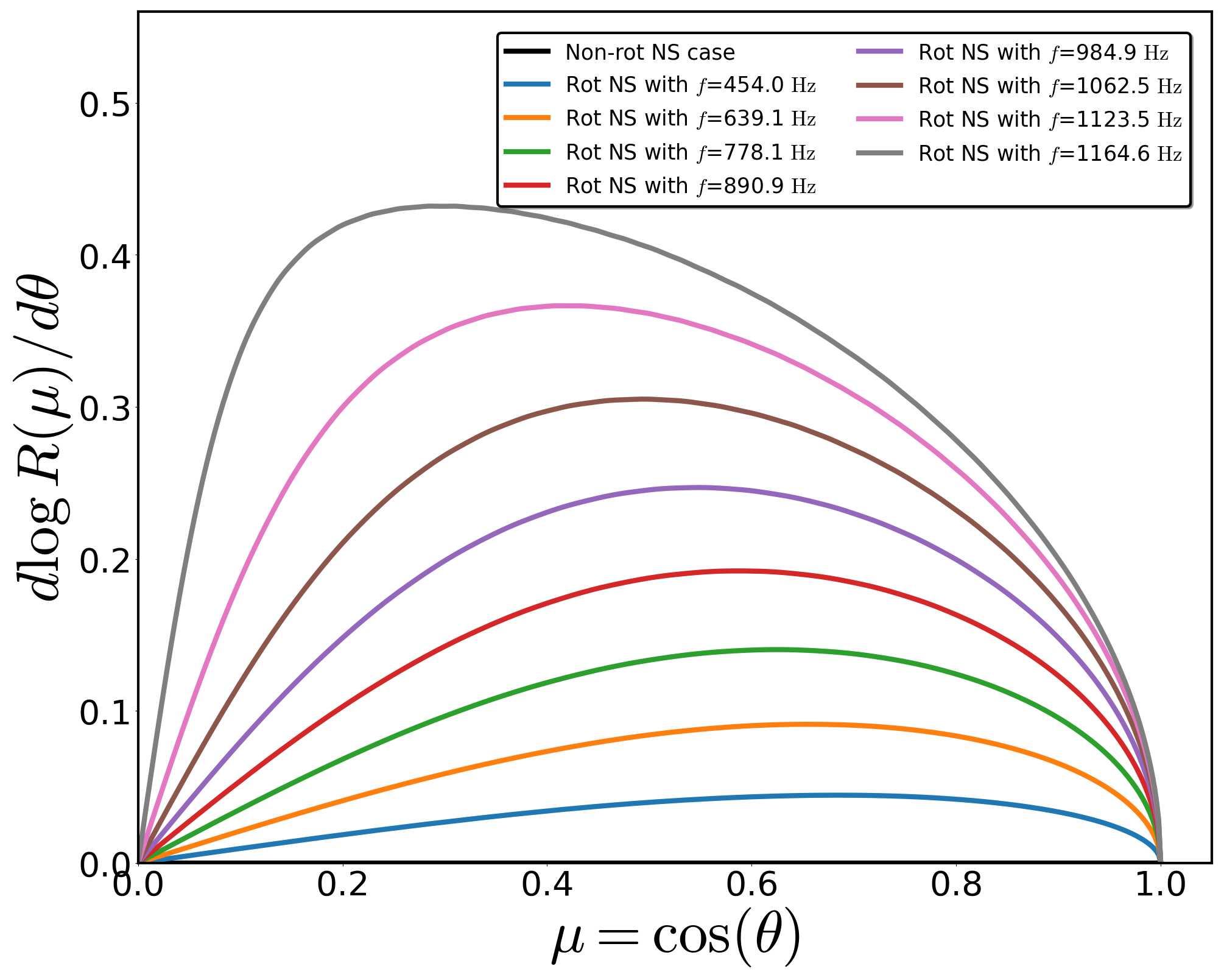}
	\caption{\label{fig:log_der_bench_models} Logarithmic derivative representation given as a function of the cosine of the colatitude $\theta$ for the NS benchmark models summarized in Table (\ref{tab:indicative_propert}). The diverse colored curves illustrate an enhanced deviation from sphericity as the rotational frequency increases.}
\end{figure}

In our endeavor to provide an accurate representation of the oblate shape of the rotating NS, we introduce the angle $\varphi$ that defines the inclination between the vector normal to the surface, $\bf{n}$, and the radial direction $\bf{r}$ as, 
\begin{equation}
    \label{varph}
    \cos(\varphi) = \left[1+\left(\frac{d \log R(\mu)}{d \theta}\right)^2 \right]^{-1/2}.
\end{equation}
The normal vector to the surface can be expressed in terms of the radial $\mathbf{r}$ and tangential $\boldsymbol{\theta}$ unit vectors in spherical coordinates as $\mathbf{n} = \cos(\varphi) \mathbf{r} + \sin(\varphi) \boldsymbol{\theta}$. Therefore, in order to describe the oblate shape of the rotating stellar object as precisely as possible, we need to estimate both $R(\mu)$ and the logarithmic derivative expressed as,
\begin{equation}
\label{log_der}
  \frac{d \log R(\mu)}{d \theta} = -\left(1-\mu^2 \right)^{1/2} \frac{1}{R(\mu)} \frac{d R(\mu)}{d \mu},  
\end{equation}
with the latter (Eq.(\ref{log_der})) being a measure of the deviation from the sphericity of the star’s surface and subject to the constraints
\begin{equation}
    \label{log_der_constr}
    \left[\frac{d \log R(\mu)}{d \theta}\right]_{\mu = 0} = \left[\frac{d \log R(\mu)}{d \theta}\right]_{\mu = 1} = 0.
\end{equation}
Although our NS models ensemble does not include Keplerian configurations, it is worth noting that at the equator this constraint does not hold when the star is exactly at the mass-shedding limit and a cusp forms.
In any case, the logarithmic derivative plays a crucial role in computing the beaming angle $a_e$ for a photon emitted at the surface of the NS, as highlighted in \cite{baubock2012ray,cadeau2007light}. For the NS benchmark models catalog provided in the Table (\ref{tab:indicative_propert}), we illustrate in Fig. \ref{fig:log_der_bench_models} the logarithmic derivative (\ref{log_der}) as a function of $\mu = \cos(\theta)$ for various rotation frequencies.
This estimation, supplemented with the aforementioned constraints (\ref{log_der_constr}), is performed by using the central sixth-order numerical finite differencing formula. As depicted in Fig. \ref{fig:log_der_bench_models}, an increase in rotational frequency aligns with a heightened deviation from sphericity for the NS's surface.

Lastly, NSs exhibit an extraordinary intensity of gravitational acceleration on their surfaces, establishing an environment where gravity is markedly pronounced. More specifically, considering hydrostatic equilibrium, the fluid's acceleration $a_{\alpha}$ corresponding to the case of the four-velocity (\ref{u_t}) can be expressed as
\begin{equation}
    \label{acceleration}
  a_{\alpha} = -\nabla_{\alpha}\ln u^t,  
\end{equation}
which is normal to the star's surface \cite{friedman_rotating_2013}. The magnitude of the  acceleration is identified as the effective acceleration due to gravity and is given by
\begin{equation}
    \label{effective_g}
    g = a = \left(g^{\alpha \beta} a_{\alpha}a_{\beta} \right)^{1/2}.
\end{equation}

In the absence of rotation, the metric exterior of the line element (\ref{ds2_stat}) to the star’s surface (at radius $R\equiv R_{\mathrm{eq}}$) is given by the Schwarzschild solution \cite{friedman_rotating_2013}. Plugging this metric into Eqs. (\ref{u_a}), (\ref{acceleration}, and (\ref{effective_g}), the acceleration due to gravity is given by
\begin{equation}
    \label{g_0}
    g_0 = \frac{M}{R^2} \left(1 - \frac{2M}{R}\right)^{-1/2}.
\end{equation}

In the case of a rotating star, the three-velocity of a fluid element as measured by a zero angular momentum observer (ZAMO) at infinity is given by \cite{friedman_rotating_2013,osti_4483768,butterworth_structure_1976}
\begin{equation}
    \label{3_velocity}
    V = (\Omega-\omega)\tilde{r}\sin(\theta)e^{-\rho}.
\end{equation}
Given the definition (\ref{3_velocity}) for the three-velocity, the acceleration vector (\ref{acceleration}) in the coordinate system described by the metric (\ref{eq:ds_rot_2}) is 
\begin{equation}
    a_{\alpha} = \frac{1}{2}\frac{\partial (\rho+\gamma)}{\partial x^{\alpha}} - \left(\frac{V}{1-V^2}\right)\frac{\partial V}{\partial x^{\alpha}}.
\end{equation}
Therefore, the coordinate-independent effective acceleration due to gravity (\ref{effective_g}) is then
\begin{equation}
    \label{eff_grav_at_surf}
    g = e^{-a}\left[\alpha_{\tilde{r}}^{2} + \left(\frac{\alpha_{\theta}}{\tilde{r}}\right)^2\right]^{1/2},
\end{equation}
where all relevant quantities are evaluated on the star's surface \cite{cumming2002hydrostatic}. In this context, the dimensionless effective gravity $g/g_0$ can be determined for a rapidly rotating star, with $g_0$ denoting the effective gravity on the surface of a nonrotating stellar object possessing the same mass and equatorial radius as the spinning one. Due to the chosen normalization, the effective gravity of a nonrotating star is denoted by a horizontal line at $g/g_0 = 1.00$.

\begin{figure}[!ht]
    \includegraphics[width=0.46\textwidth]{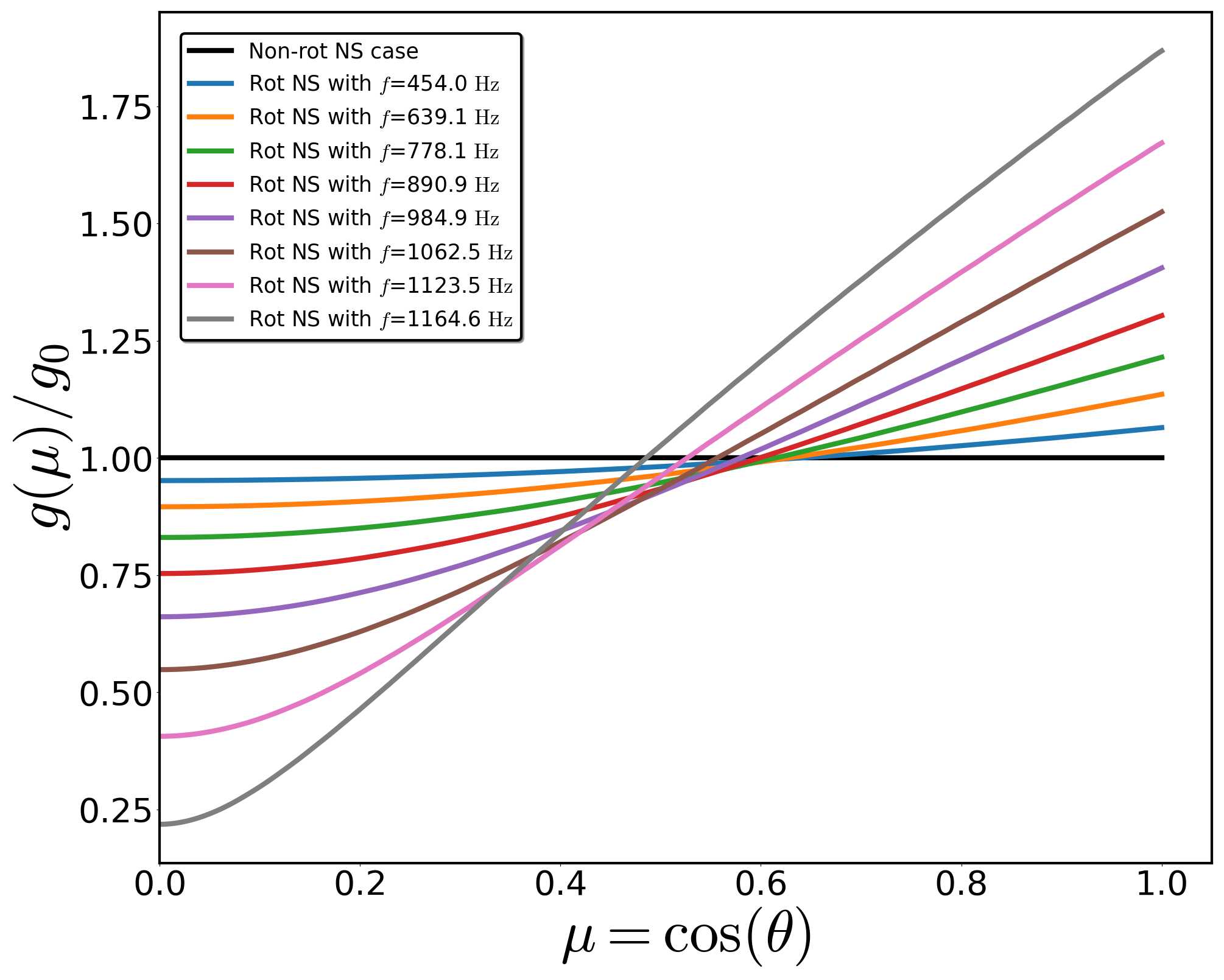}
	\caption{\label{fig:g_mu_indicative} Normalized effective gravity $g(\mu)/g_0$ as a function of the cosine of the colatitude $\theta$. The lines correspond to the indicative benchmark models computed with EoS SLY4, identified by their corresponding rotational frequency values. The nonrotating NS model in this sequence is labeled by the horizontal black line, denoted $g(\mu)/g_0) = 1.00$. For the rotating models, the effective gravity is smaller at the star's equator ($\mu=0$) and larger at the star's pole ($\mu=1$). As the rotational frequency increases, there is an increase in the deviation from the static case. The physical parameters, including the values for $g_{\mathrm{eq}}/g_0$, and $g_{\mathrm{pole}}/g_0$ are presented in Table (\ref{tab:indicative_propert}).}
\end{figure}

In Fig. \ref{fig:g_mu_indicative}, we present the normalized effective gravity at the NS's surface as a function of $\mu = \cos(\theta)$ for the catalog of benchmark models provided in Table (\ref{tab:indicative_propert}). The nonrotating configuration is highlighted by a horizontal black line with $g/g_0 = 1.00$. In general, the effective acceleration varies across the NS's surface, being minimal at the equator ($\mu=0$) and maximal at the pole ($\mu=1$), reflecting the oblate shape induced by rotation. This leads to heightened acceleration at the pole and diminished magnitude at the equator. For the rotating case, the higher the star's angular velocity, the larger the deviation from the static case. Finally,  in the mass-shedding (Keplerian) limit, it is worth noting that the effective gravity at the star's equator tends to zero, as highlighted in \cite{algendy2014universality}. 

\section{\label{sec:univ_rel_surf_prop} \bf PART I: UNIVERSAL RELATIONS for the GLOBAL PROPERTIES OF THE STAR'S SURFACE}
In this section, we will present our results which are either improvements on the established universal relations or entirely new relations. Then, in the next section (\ref{sec:univ_rel_with_ANNs}), we will explain why these quantities are essential for precise inference of the star's surface.

The organization of our findings is as follows: Sec. \ref{sec:poly_methods_1} is dedicated to proposing new EoS-insensitive relations for the star's global polar-to-equatorial ratio, $\mathcal{R} = R_{\mathrm{pole}}/R_{\mathrm{eq}}$, the star's eccentricity $e$, and the maximum value of the logarithmic derivative $(d \log R(\mu)/ d \theta)_{\mathrm{max}}$. Subsequently, in Sec. \ref{sec:poly_methods_2}, we also suggest new universal relations for the effective acceleration due to gravity at both the star's pole and the star's equator. In each case, we utilized the least squares regression method for fitting, complemented by the leave-one-out cross-validation (LOOCV) evaluation process. Also, we employed evaluation measures throughout all examined regression models to estimate and assess each model's performance on the corresponding validation set. The interested reader can refer to Appendix \ref{sec:ML_part} for specific details about the fitting formulation and the methodology employed for the evaluation.

We deem a relation between selected parameters to be ``universal'' when the relative errors in the validation set are $\lesssim \mathcal{O} (10 \%)$. The cross-validation evaluation method that we used ensures that the suggested EoS-insensitive relations possess generalization ability beyond the training data. This critical aspect sets our models apart from other proposed fits that can be found in the literature that lack this form of validation evaluation, thus carrying an elevated risk of overfitting at the training set. In summary, our analysis ensures that the new fitting functions stand a better chance of performing well within the corresponding relative errors on new, previously unseen data.

\subsection{\label{sec:poly_methods_1} RELATIONS FOR $\mathcal{R}$, $e$ AND $(d \log R(\mu)/d\theta)_{\mathrm{max}}$}
The relation between the NS's polar and equatorial radius $\mathcal{R}$ (\ref{R_p/R_e}) is not solely influenced by the star's rotation but is also contingent on the star's internal structure, determined by the unknown EoS. Therefore, investigating relations that are insensitive to the EoS and involve the ratio of polar to equatorial radius is quite significant. 

\begin{figure}[!htb]
    \includegraphics[width=0.46\textwidth]{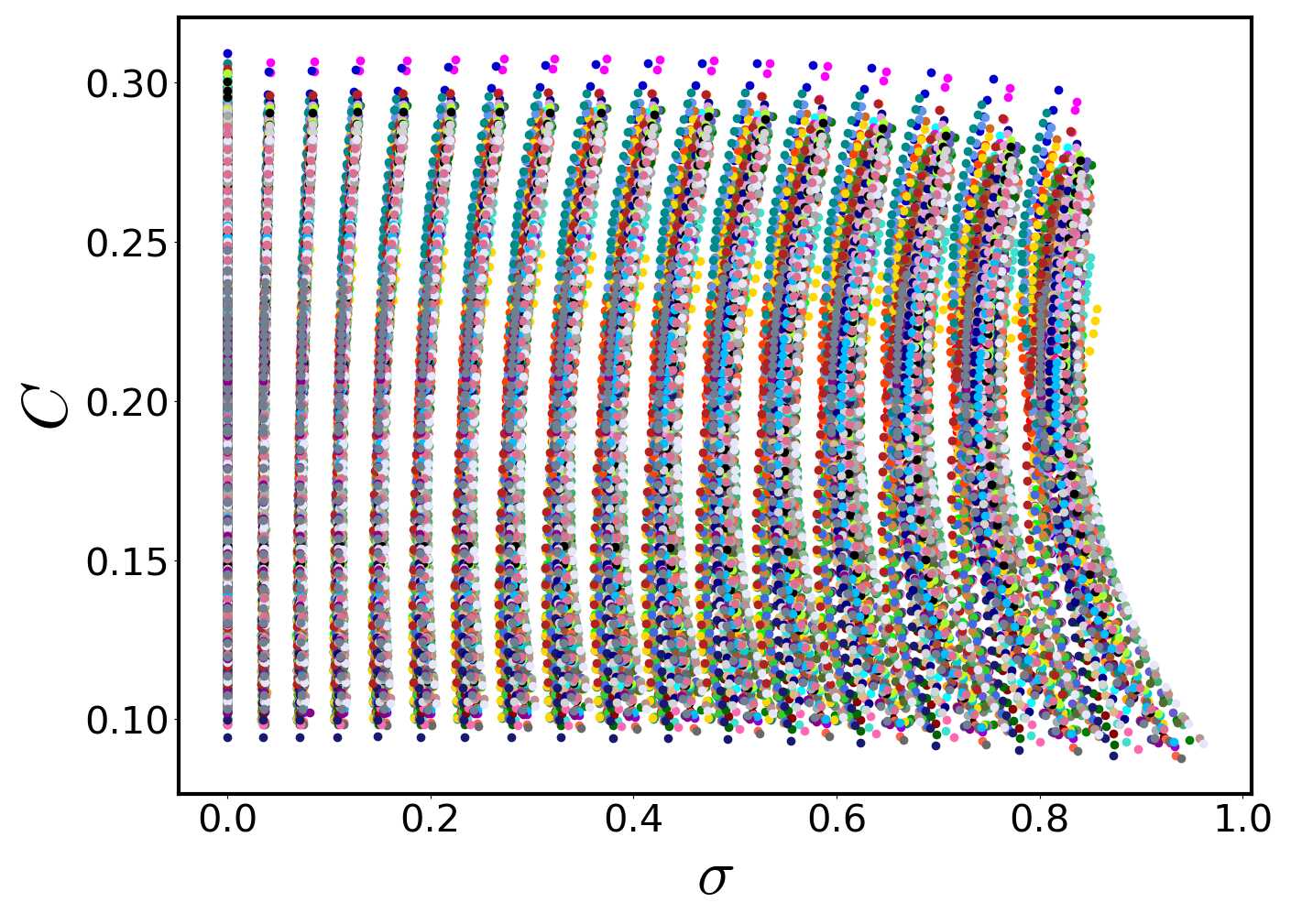}
    \caption{\label{fig:C-sigma} Distribution of the $C-\sigma$ parameter space, encompassing a wide range of rotation rates and degree of stiffness. Each color represents the EoS mapping as highlighted in Fig. \ref{fig:color_band} of Appendix \ref{app:eos_tables}.}
\end{figure}

In that direction, to investigate effective universal relations for the $\mathcal{R}$ ratio, we have considered the dimensionless stellar compactness $C = M/R_{\mathrm{eq}}$ and the dimensionless spin $\sigma = \Omega^2R^3_{\mathrm{eq}}/GM$ referring to the star's rotation, both as parameters. For our sample of NS models, these parameters are in the respective ranges, $0.0876 \leq C \leq 0.3095$, and $0.000 \leq \sigma \leq 0.961$. In addition, it is well-established that the choice of parameters influences the EoS-independent behavior among observable quantities \cite{rezzolla_physics_2018}. In Fig. \ref{fig:C-sigma}, we illustrate, for completeness, the $C-\sigma$ representation for each EoS selected from our ensemble.
Therefore, having our choice of feature parameters established, we investigate the relation that connects the $\mathcal{R}$ ratio with the parameters $C$ and $\sigma$.

The regression model $\mathcal{R}(C,\sigma)$ that best describes the data has the functional form, 
\begin{equation}
        \label{eq:R_c_sigma}
        \mathcal{R}(C,\sigma)=\sum_{n=0}^{4}\sum_{m=0}^{4-n}\hat{\mathcal{A}}_{nm} \ C^n \ \sigma^m.
\end{equation}
Compared to other regression functions examined, this is the less complicated model with the optimal evaluation measures at LOOCV. The corresponding results for an investigation of different polynomial models are highlighted in Table (\ref{tab:R_c_sigma_tab}).
\begin{table}[!h]
    \small
    \caption{\label{tab:R_c_sigma_tab} Indicative list of LOOCV evaluation measures (the definition of each quantity is presented in the Appendix \ref{sec:train_test}) for the $\mathcal{R}(C,\sigma)=\sum_{n=0}^{\kappa}\sum_{m=0}^{\kappa-n}\hat{\mathcal{A}}_{nm} \ C^n \ \sigma^m$ parametrization, where $\kappa$ is the highest order of the polynomial function.}
    \begin{ruledtabular}
        \begin{tabular}{ccccccc}
            MAE & Max Error & MSE & $d_{\text{max}}$ & MAPE & Exp Var & $\kappa$ \\
            $\times 10^{-3}$ & $\times 10^{-2}$ & $\times 10^{-4}$ & ($\%$) & $\times 10^{-1}$ ($\%$)&  &  \\
            \hline
            11.254  & 6.830 & 1.963  & 10.83 & 14.02  & 1.0 & 1 \\
            3.819  & 2.785 & 0.252 & 4.42 & 4.90  & 1.0 & 2 \\
            3.213 & 1.955& 0.193 & 3.12 & 4.21  & 1.0 & 3 \\
           {\bf 3.112}  & {\bf1.876} & {\bf 0.187} & {\bf 2.80}  & {\bf 4.09}  & {\bf 1.0} & {\bf 4} \\
            3.089 & 1.998 & 0.185 & 2.95  & 4.06  & 1.0 & 5 \\
            3.067 & 2.007  & 0.183 & 2.96  & 4.02  & 1.0 & 6 \\
            3.043 & 2.063 & 0.181 & 3.15  & 3.99  & 1.0 & 7 \\
            3.014 & 2.013 & 0.179 & 3.08  & 3.95  & 1.0 & 8 \\
        \end{tabular}
    \end{ruledtabular}
\end{table}

\begin{figure}[!ht]
  \centering
  \includegraphics[width=0.46\textwidth]{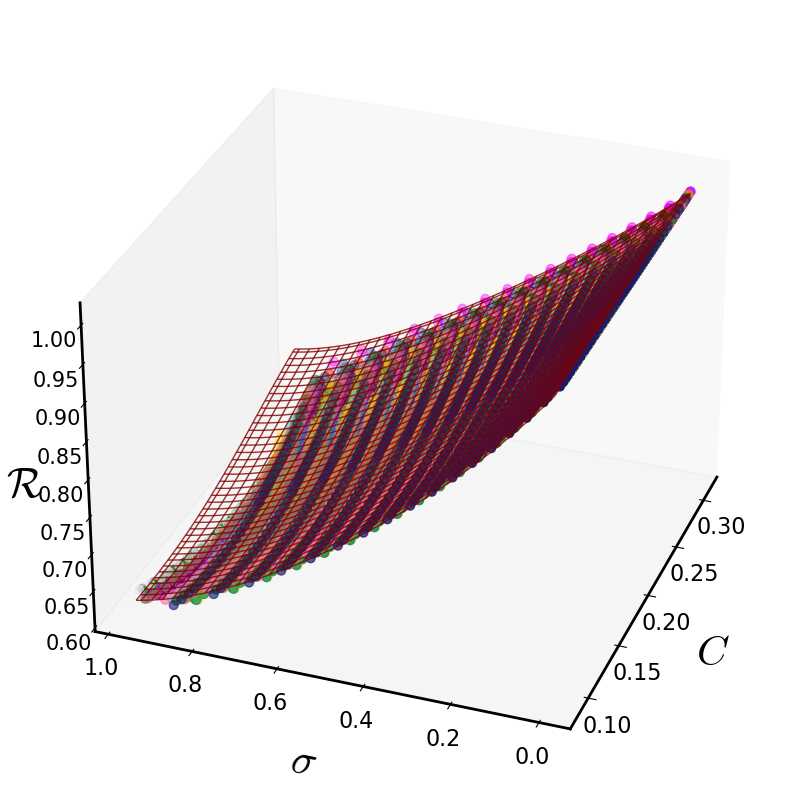}\hfill
  \includegraphics[width=0.46\textwidth]{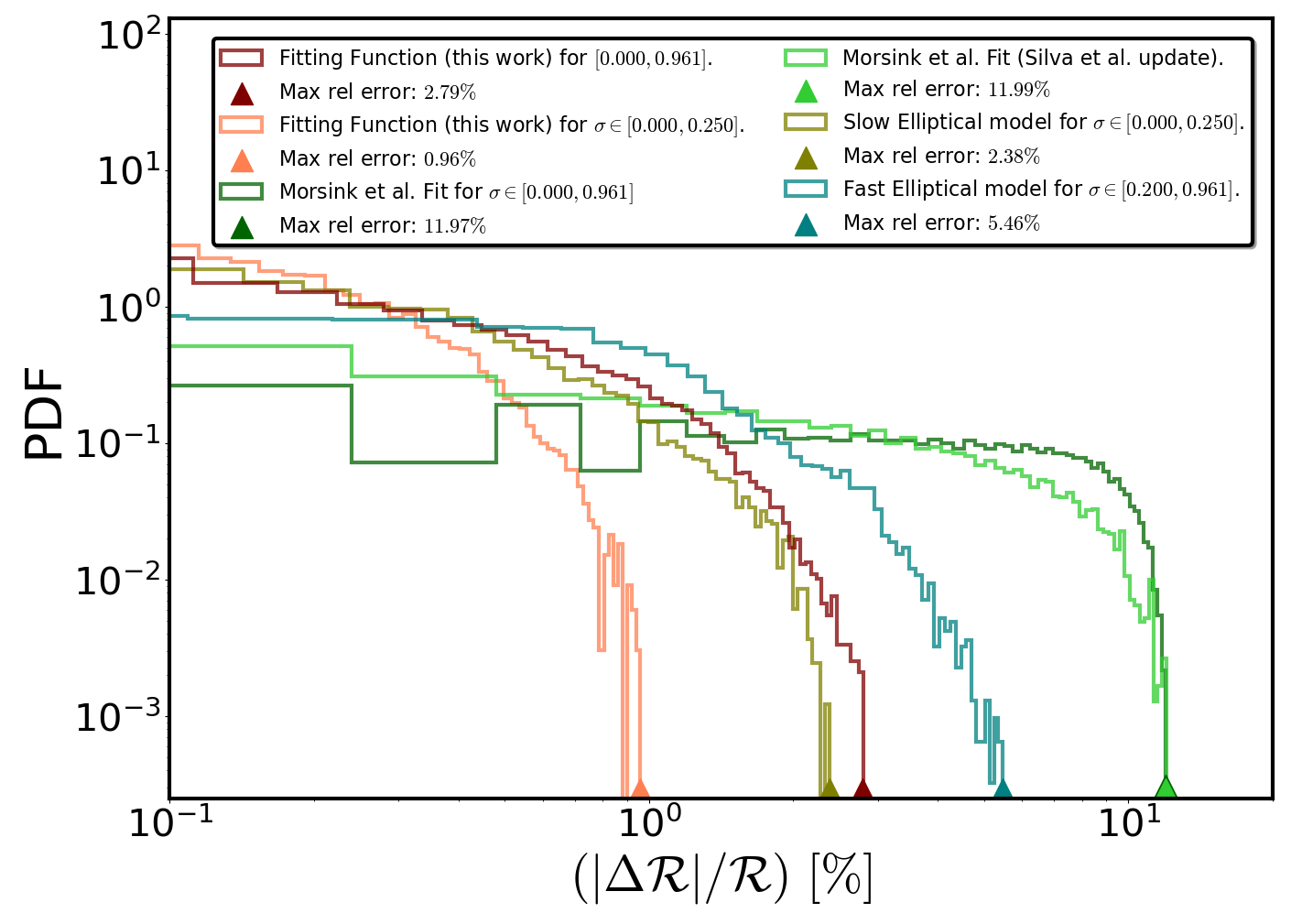}
  \caption{\label{R_C_sigma_fit} $\mathcal{R}$ ratio as a function of the dimensionless parameters $C$, and
$\sigma$ (Top panel) and probability density function (PDF) distributions for the different fits and the assumed ranges of rotation (Bottom panel). In the top panel, the maroon grid corresponds to the regression polynomial formula (\ref{eq:R_c_sigma}). In the bottom panel, the absolute relative errors to the fit are given as ($100\% (|\Delta \mathcal{R}|)/\mathcal{R} = 100\% (|\mathcal{R}_{\mathrm{fit}} - \mathcal{R}|)/\mathcal{R}$) in logarithmic scale. Furthermore, additional relative error distributions for fitting functions proposed in the literature are illustrated with distinct colors, providing a basis for comparative analysis.}
\end{figure}

From the model-fit evaluation, the polynomial function’s (\ref{eq:R_c_sigma}) (best fit) parameters $\hat{\mathcal{A}}_{nm}$ are presented in Table (\ref{tab:R_c_sigma_optimizers}).
\begin{table}[!h]
    \caption{\label{tab:R_c_sigma_optimizers} Optimal $\hat{\mathcal{A}}_{nm}$ regression coefficients for the $\mathcal{R}(C,\sigma)$ parametrization (\ref{eq:R_c_sigma}). For this model, the coefficient of determination is exceptionally high with a value of $R^2 = 0.9983$.}
    \begin{ruledtabular}
        \begin{tabular}{cccc}
            $\hat{\mathcal{A}}_{00}$ & $\hat{\mathcal{A}}_{01}$ & $\hat{\mathcal{A}}_{02}$ & $\hat{\mathcal{A}}_{03}$ \\
            0.942328& -0.617711& 0.544639&-0.440968 \\
            \hline\hline    
            $\hat{\mathcal{A}}_{04}$ & $\hat{\mathcal{A}}_{10}$ & $\hat{\mathcal{A}}_{11}$ & $\hat{\mathcal{A}}_{12}$ \\
            0.196118&1.296632 &-1.458921 &-0.226904 \\
            \hline\hline    
            $\hat{\mathcal{A}}_{13}$ & $\hat{\mathcal{A}}_{20}$ & $\hat{\mathcal{A}}_{21}$ & $\hat{\mathcal{A}}_{22}$ \\
            0.527775& -10.45611& 8.668382& -2.506686\\
            \hline\hline
            $\hat{\mathcal{A}}_{30}$ & $\hat{\mathcal{A}}_{31}$ & $\hat{\mathcal{A}}_{40}$ & \\
            36.131881&-7.524662 &-45.301523 & 
        \end{tabular}
    \end{ruledtabular}
\end{table}
The surface fit and the corresponding relative error histogram for the whole ensemble of the NS models examined are presented in Fig. \ref{R_C_sigma_fit}.

Based on the histogram presented in Fig. \ref{R_C_sigma_fit} (bottom panel), it is evident that the relative errors between the regression model (\ref{eq:R_c_sigma}) and the actual $\mathcal{R}$ values are $\leq 2.79 \%$ for all EoSs and all rotating models in the $[0, 0.961]$ range for $\sigma$. Relative deviations $> 1 \%$ correspond to $2222$ stellar models mainly due to the Hyperonic and Hybrid EoSs utilized with equatorial radius $R_{\mathrm{eq}}\in [12.63, 19.41] \ \mathrm{km}$, reduced spin $\sigma \in [0.284, 0.873]$, independently of the star’s compactness. Furthermore, when we restrict to the range $\sigma \leq 0.25$, the maximum percentage error is only $0.96\%$. Therefore, the $\mathcal{R}(C,\sigma)$ regression formula (\ref{eq:R_c_sigma}) corresponds to a well-behaved EoS insensitive relation for all the NS models considered.

By setting $\mu = 1$ in the proposed formulas found in the literature \cite{morsink2007oblate, silva2021surface, algendy2014universality} designed for estimating the star's surface $R(\mu)$, one can deduce the corresponding value for the polar-to-equatorial radius $\mathcal{R}$. To facilitate a thorough comparison, we applied both the $R(\mu=1)$ fit proposed by Morsink et al. \cite{morsink2007oblate, silva2021surface} and the ``slow elliptical'' and ``fast elliptical'' fits proposed by Silva et al.\cite{silva2021surface}. The ``slow elliptical'' fit pertains to NS models with $\sigma \leq 0.25$, while the ``fast elliptical'' fit describes NS configurations with $\sigma \geq 0.20$, as elaborated in \cite{silva2021surface}. From Fig. \ref{R_C_sigma_fit} (bottom panel), it is evident that our new fit Eq.(\ref{eq:R_c_sigma}) achieves a higher accuracy than previous fits in the literature, for all rotational regimes examined.

An alternative method for estimating the surface $R(\mu)$ of slowly rotating NSs with $\sigma \leq 0.1$ has been proposed by AlGendy and Morsink \cite{algendy2014universality}, and is currently employed in pulse profile modeling by the NICER collaboration \cite{bogdanov2019constraining}. This fitting function was further examined and its coefficients were updated by Silva et al. \cite{silva2021surface}. Within our ensemble, consisting of 7639 slowly rotating NS models with $\sigma \leq 0.1$, we can compare the outcome of the updated Algendy and Morsink fits to our regression model (\ref{eq:R_c_sigma}).
\begin{figure}[!htb]
    \includegraphics[width=0.46\textwidth]{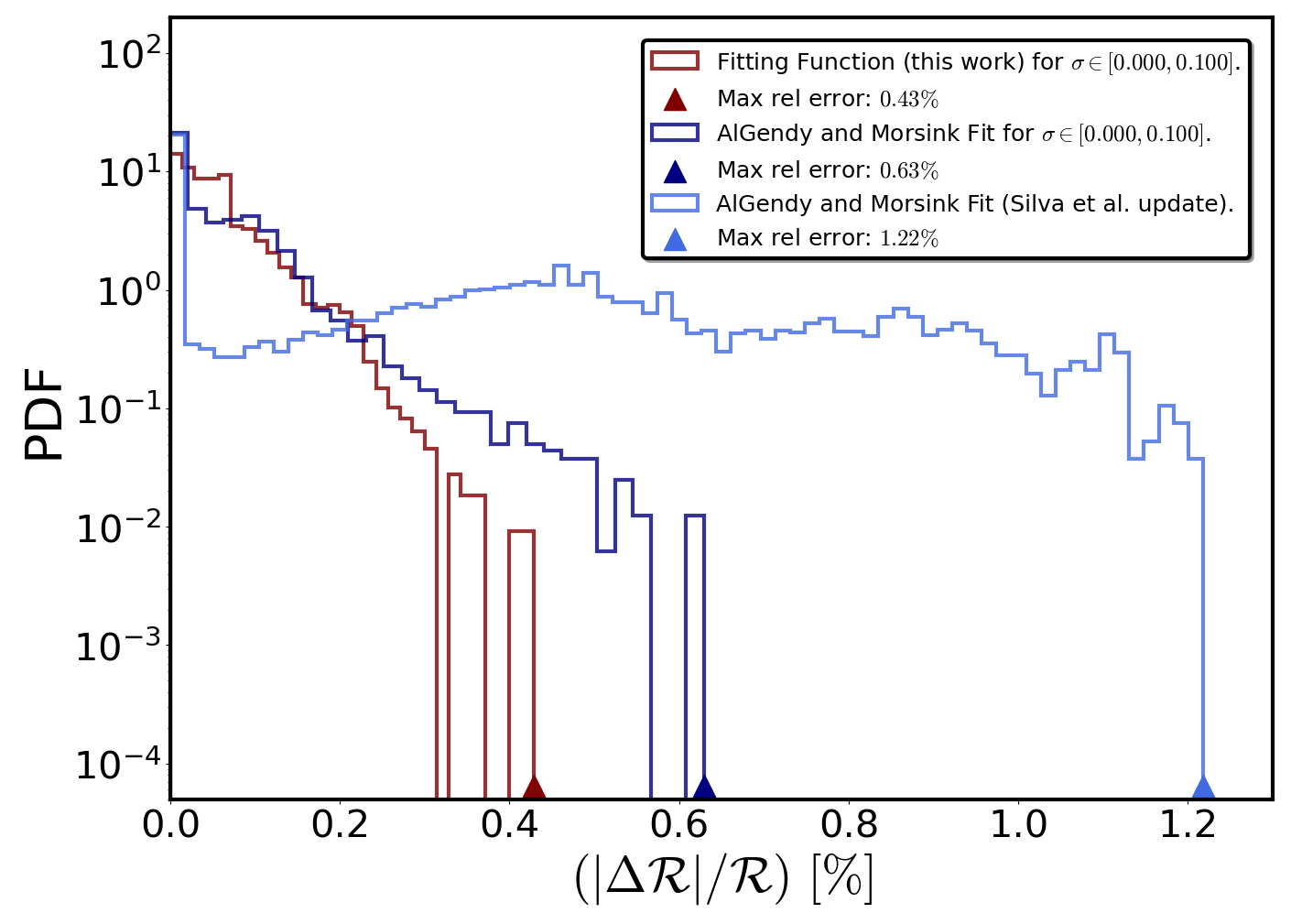}
    \caption{\label{fig:R_reg_vs_alg} Same as Fig. \ref{R_C_sigma_fit} (bottom panel), but for rotating models limited in the range $\sigma\in [0, 0.1]$. }
\end{figure}

Examining the distribution of errors for slowly rotating models presented in Fig. \ref{fig:R_reg_vs_alg}, it is {\it apparent} that our new fit Eq.(\ref{eq:R_c_sigma}) adeptly reproduces the majority of data values, showcasing an absolute relative error of less than $0.43\%$. This is comparable (and somewhat better) than the results obtained using the original AlGendy and Morsink fitting function \cite{algendy2014universality} and the corresponding one with coefficients provided by Silva et al. \cite{silva2021surface}. This discrepancy primarily arises from the broader ensemble of predominantly slowly rotating NS models used in the original AlGendy and Morsink study, while Silva et al. concentrated on a more restricted set of slowly rotating stellar configurations in this $\sigma$ range. Nevertheless, for all fits the relative differences remain relatively small.

\begin{figure}[!htb]
    \includegraphics[width=0.46\textwidth]{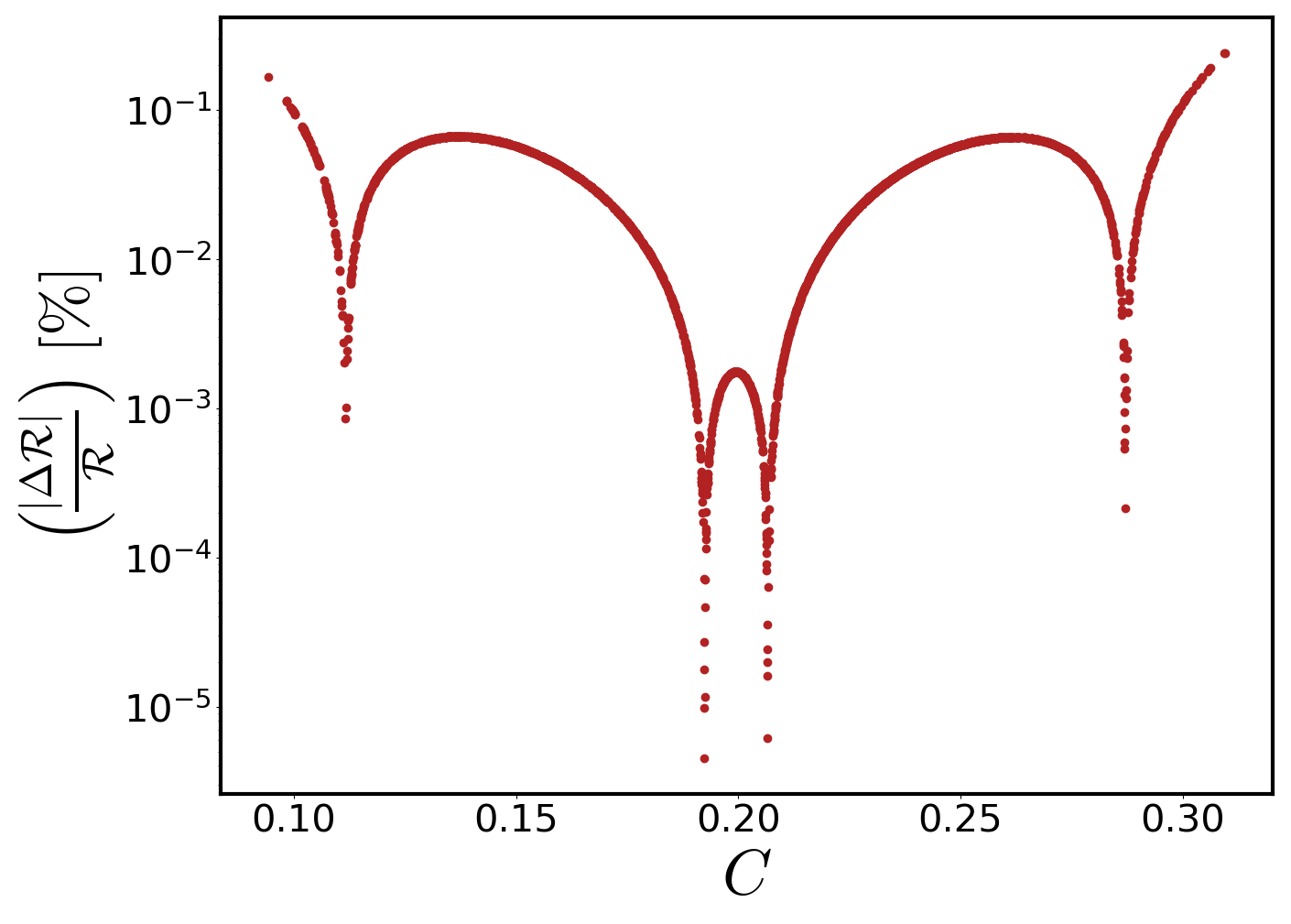}
    \caption{\label{fig:R_C_static} Absolute relative deviation $\left(|\Delta \mathcal{R}|/\mathcal{R}\right) \ [\%]$ as a function of the stellar compactness $C$ for nonrotating NS configurations.}
\end{figure}

As an additional demonstration of the success of Eq.(\ref{eq:R_c_sigma}), we present in Fig. \ref{fig:R_C_static} the distribution of the nonrotating NS models derived from the relation $\mathcal{R}(C,\sigma = 0)$ (\ref{eq:R_c_sigma}) as a function of the stellar compactness $C$, incorporating the corresponding data from all of our EoSs.
As we can see in Fig. \ref{fig:R_C_static}, the theoretical prediction (\ref{eq:R_c_sigma}), setting $\sigma = 0$, reproduces the $R_{\mathrm{pole}} = R_{\mathrm{eq}}$ constraint with accuracy $\leq 0.24 \%$ for all the nonrotating NS models considered.

In addition, another universal relation that would be useful to investigate is one that links directly the star's eccentricity $e$ (\ref{eccentricity}) with the parameters $C$ and $\sigma$. For nonrotating NS models, the star's eccentricity should be zero. Recognizing that $e$ is a parameter of interest for rotating NSs, we decided to focus our analysis on stellar models that rotate with frequencies in the range of $0.1902 \lesssim f \ [\mathrm{kHz}] \lesssim 1.871$ and have stellar parameters that range from $0.0876 \lesssim C \lesssim 0.3075$, and $0.0328 \lesssim \sigma \lesssim 0.9612$. This ensemble includes 40015 stellar models out of the total 42694 that were used previously.

The surface-formula $e(C,\sigma)$ that optimally describes the data has the functional form,
\begin{equation}
    \label{eq:e_c_sigma}
    e(C,\sigma)=\sum_{n=0}^{5}\sum_{m=0}^{5-n}\hat{\mathcal{B}}_{nm} \ C^n \ \sigma^m.
\end{equation}
This is the least complicated regression function that provided a satisfactory fit; i.e., among the different polynomial functions we examined, there were higher order ($\kappa > 5$) functions that gave better evaluation scores at LOOCV compared to the selected $\kappa = 5$ fit, as can be seen in Table (\ref{tab:e_c_sigma_tab}). However, choosing polynomial models that were too complicated was not worth the slight improvement of the fit quality.
\begin{table}[!h]
	\small
	\caption{\label{tab:e_c_sigma_tab} Indicative list of LOOCV evaluation measures for the $e(C,\sigma)=\sum_{n=0}^{\kappa}\sum_{m=0}^{\kappa-n}\hat{\mathcal{B}}_{nm} \ C^n \ \sigma^m$ parametrization.}
	\begin{ruledtabular}
		\begin{tabular}{ccccccc}
			MAE & Max Error & MSE & $d_{\text{max}}$ & MAPE & Exp Var & $\kappa$ \\
            $\times 10^{-3}$ & $\times 10^{-3}$ & $\times 10^{-4}$ & ($\%$) & ($\%$)&  &  \\
			\hline
			40.421 & 156.100& 23.978 & 61.98 & 9.79 & 1.0 & 1 \\
			13.682 & 49.544 & 2.877 & 23.57 & 3.32 & 1.0 & 2 \\
			7.252 & 37.523 & 0.802 & 11.37 & 1.69 & 1.0 & 3 \\
			5.109 & 21.952 & 0.384 & 6.14 & 1.08 & 1.0 & 4 \\
			{\bf 4.652} &{\bf 19.808} & {\bf 0.326} & {\bf 4.58} & {\bf 0.92} & {\bf 1.0} & {\bf 5} \\
			4.365 & 19.091 & 0.295 & 3.99 & 0.84 & 1.0 & 6 \\
			4.357 & 18.747 & 0.294 & 3.90 & 0.83 & 1.0 & 7 \\
			4.302 & 18.830 & 0.289 & 3.76 & 0.82 & 1.0 & 8 \\
		\end{tabular}
	\end{ruledtabular}
\end{table}

\begin{figure}[!htb]
    \includegraphics[width=0.46\textwidth]{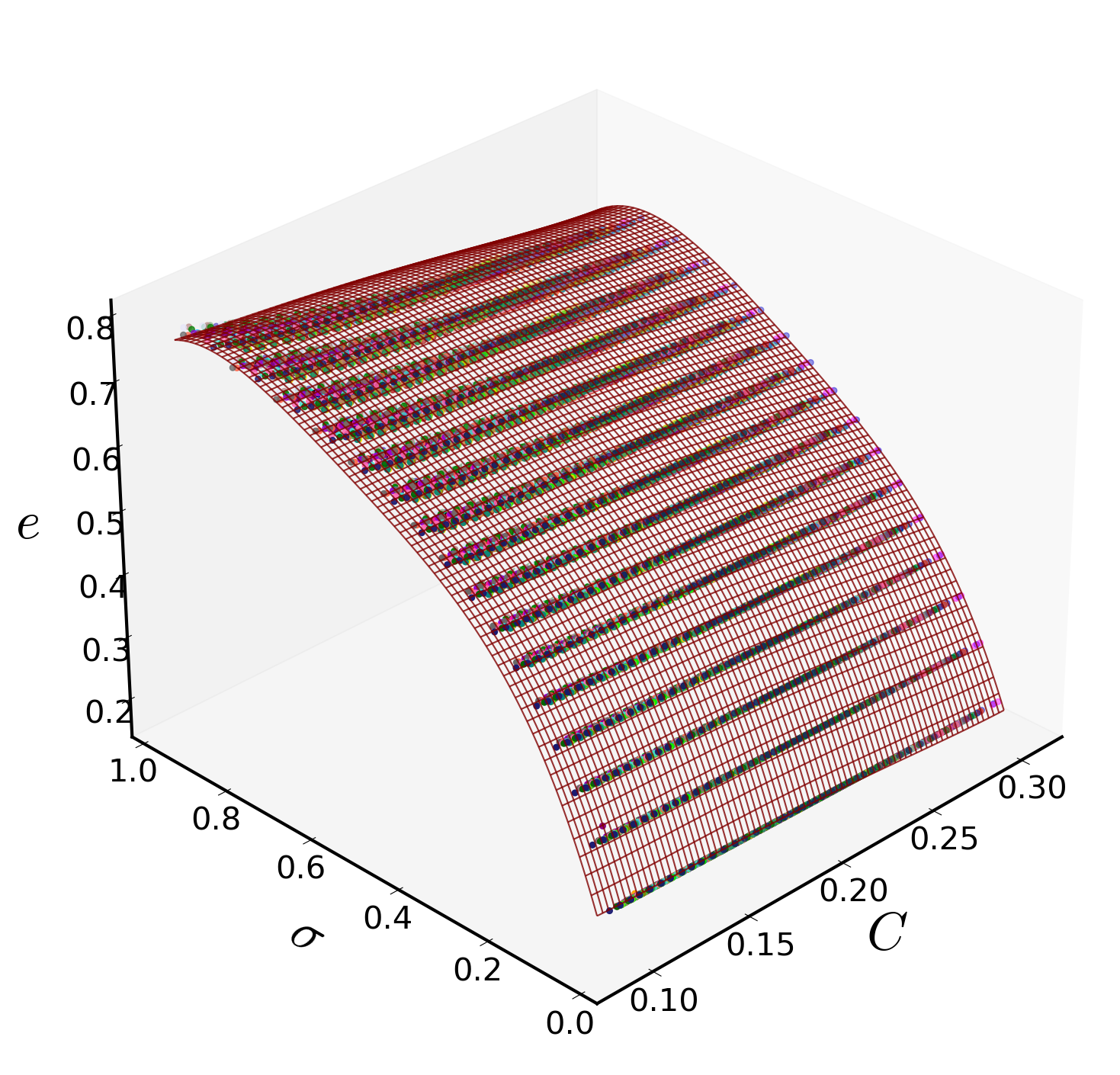}\hfill
    \includegraphics[width=0.46\textwidth]{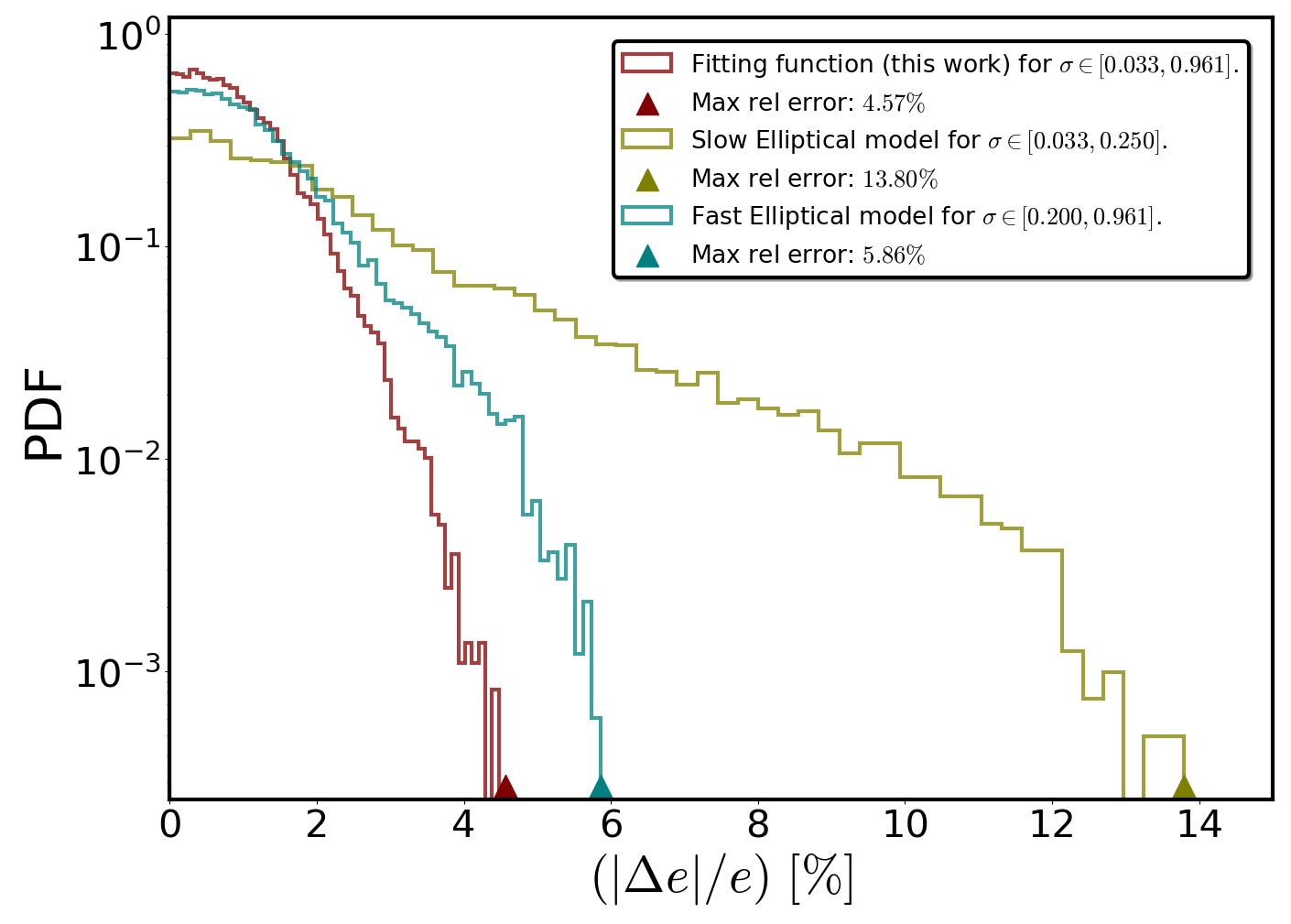}
    \caption{\label{fig:e_C_sigma_fit} Top panel: Rotating NS's eccentricity $e$ as a function of the dimensionless parameters $C$, and $\sigma$ and relative deviations histogram in logarithmic scale for an extended sample of models. The analytic surface shown as a maroon grid corresponds to the regression polynomial formula (\ref{eq:e_c_sigma}). Bottom panel: The relative errors to the fit are given as ($100\% (|\Delta e|)/e = 100\% (|e_{\mathrm{fit}} - e|)/e$) and are shown for our fitting function (\ref{eq:e_c_sigma}), while the others correspond to associated models found in the literature. }
\end{figure}

From the surface-fit evaluation, the regression model's (\ref{eq:e_c_sigma}) parameters $\hat{\mathcal{B}}_{nm}$ are presented in Table (\ref{tab:e_c_sigma_optimizers}).
\begin{table}[!h]
    \caption{\label{tab:e_c_sigma_optimizers} $\hat{\mathcal{B}}_{nm}$ regression coefficients for the $e(C,\sigma)$ parametrization (\ref{eq:e_c_sigma}). For this functional form, the coefficient of determination is exceptionally high with a value of $R^2 = 0.9987$.}
    \begin{ruledtabular}
        \begin{tabular}{cccc}
            $\hat{\mathcal{B}}_{00}$ & $\hat{\mathcal{B}}_{01}$ & $\hat{\mathcal{B}}_{02}$ & $\hat{\mathcal{B}}_{03}$   \\
            0.182561 & 3.042299& -8.712805& 15.471220\\
            \hline\hline	
            $\hat{\mathcal{B}}_{04}$ & $\hat{\mathcal{B}}_{05}$ & $\hat{\mathcal{B}}_{10}$ & $\hat{\mathcal{B}}_{11}$  \\
            -13.854751& 4.714505& -1.525336& -1.332937\\
            \hline\hline
            $\hat{\mathcal{B}}_{12}$ & $\hat{\mathcal{B}}_{13}$ & $\hat{\mathcal{B}}_{14}$ & $\hat{\mathcal{B}}_{20}$ \\
            2.754990 & -6.446261 & 4.848852& 14.900646\\
            \hline\hline
            $\hat{\mathcal{B}}_{21}$ & $\hat{\mathcal{B}}_{22}$ & $\hat{\mathcal{B}}_{23}$ & $\hat{\mathcal{B}}_{30}$ \\
            9.197133& 8.083461& -9.033159& -67.794879\\
            \hline\hline
            $\hat{\mathcal{B}}_{31}$ & $\hat{\mathcal{B}}_{32}$ & $\hat{\mathcal{B}}_{40}$ & $\hat{\mathcal{B}}_{41}$ \\
            -57.474308& 12.934989& 137.191421& 68.05573\\
            \hline\hline
            $\hat{\mathcal{B}}_{50}$ & & & \\
            -99.173163& & & \\
        \end{tabular}
    \end{ruledtabular}
\end{table}
The surface evaluation fit (\ref{eq:e_c_sigma}) and the corresponding absolute relative error histogram are presented in Fig.  \ref{fig:e_C_sigma_fit}.

We observe that, the $e(C,\sigma)$ parametrization gives relative deviations between the fit (\ref{eq:e_c_sigma}) and the observed data values that are $\leq 4.57 \%$. To elaborate further on the fit quality, it is worth mentioning that only 3070 rotating models out of the total 40015 exhibit relative errors $\gtrsim 2\%$. For the stellar configurations with the larger deviations, it should also be noted that there is no specific pattern regarding the variance of errors concerning the EoSs or the $(R_{\mathrm{eq}}, C, \sigma)$ parameters characterizing the investigated parameter space. Therefore, the regression formula (\ref{eq:e_c_sigma}) corresponds to a well-behaved EoS-insensitive relation, which provides accurate results for the majority of the rotating stellar models considered. 

Finally, as an additional demonstration of the accuracy of equation (\ref{eq:e_c_sigma}), we incorporated in Fig. \ref{fig:e_C_sigma_fit} (bottom panel) the relative error distribution for both the ``slow elliptical'' and the ``fast elliptical'' $e(C,\sigma)$-fits using the coefficients presented in \cite{silva2021surface}. In all cases, the suggested formula (\ref{eq:e_c_sigma}) has higher accuracy than the already established universal relations.

One could also look for EoS-insensitive relations that can describe properties of NSs that are not directly observable. One such interesting quantity would be the maximum value of the logarithmic derivative $d \log R(\mu)/d\theta$ (\ref{log_der}). For example,  as illustrated in Fig. \ref{fig:log_der_bench_models}, it is apparent that this quantity varies among different NS models, and its variation is influenced by the star's rotational frequency. 

For the ensemble of rotating NS models that we used previously, we explored a universal relation that relates $(d \log R(\mu)/d\theta)_{\mathrm{max}}$ with the parameters $C$, $\sigma$, and $\mathcal{R}$. We observed that including the polar-to-equatorial ratio, $\mathcal{R}\in[0.626,0.981]$, as a feature is crucial for reducing errors and improving the accuracy of data capture. The fitting hyper-surface that best reproduces the data has the functional form,
\begin{equation}
        \small
        \label{eq:dlogR_c_sigma_Rp_Re}
        \left( \frac{d \log R(\mu)}{d \theta} \right)_{\mathrm{max}}=\sum_{n=0}^{3}\sum_{m=0}^{3-n}\sum_{q=0}^{\small 3-(n+m)}\hat{\mathcal{C}}_{nmq} \ C^n \ \sigma^m \ \mathcal{R}^q.
\end{equation}
Once more, this is the less complicated function among those investigated, avoiding the complexity of higher-order functions (i.e., polynomials with orders higher than $\kappa = 3$). Despite its simplicity, it provides a satisfactory fit, and any potential enhancement from adopting a higher-order polynomial function is only marginal and not worth the effort. The corresponding results for an indicative list of regression models examined are shown in Table (\ref{tab:dlogR_C_sigma_R}). Therefore, from the best model fit evaluation, the associated coefficients are presented in Table (\ref{tab:logR_C_sigma_R_optimizers}).

\begin{table}[!h]
    \caption{\label{tab:dlogR_C_sigma_R} Indicative list of LOOCV evaluation measures for the 
     $\sum_{n=0}^{\kappa}\sum_{m=0}^{\kappa-n}\sum_{q=0}^{\kappa-(n+m)}  \hat{\mathcal{C}}_{nmq} \ C^n \sigma^m  \mathcal{R}^q$ parametrization.}
    \begin{ruledtabular}
       \begin{tabular}{ccccccc}
            MAE & Max Error & MSE & $d_{\text{max}}$ & MAPE & Exp Var & $\kappa$ \\
             $\times 10^{-3}$& $\times 10^{-3}$& $\times 10^{-6}$ & ($\%$) & ($\%$)&  &  \\
            \hline
             3.774 & 26.318 & $23.757$ & 79.08 & 4.05 & 1.0 & 1 \\
             0.669 & 21.541 & $1.259$ & 5.23 & 0.40 & 1.0 & 2 \\
             {\bf 0.497} & {\bf 7.817 }&  ${ \bf 0.783}$ & {\bf 3.23} & {\bf 0.23} & {\bf 1.0} & {\bf 3} \\
             0.438 & 7.156 & $0.690$ & 5.80 & 0.18 & 1.0 & 4 \\
             0.416 & 6.877 & $0.653$ & 2.61 & 0.17 & 1.0 & 5 \\
             0.403 & 6.840 & $0.632$ & 3.91 & 0.16 & 1.0 & 6 \\
             0.395 & 6.831 & $0.614$ & 2.63 & 0.16 & 1.0 & 7 \\
             0.382 &  6.846 & $0.599$ & 3.54 & 0.14 & 1.0 & 8 \\
        \end{tabular}
    \end{ruledtabular}
\end{table}

\begin{figure}[!htb]
    \includegraphics[width=0.46\textwidth]{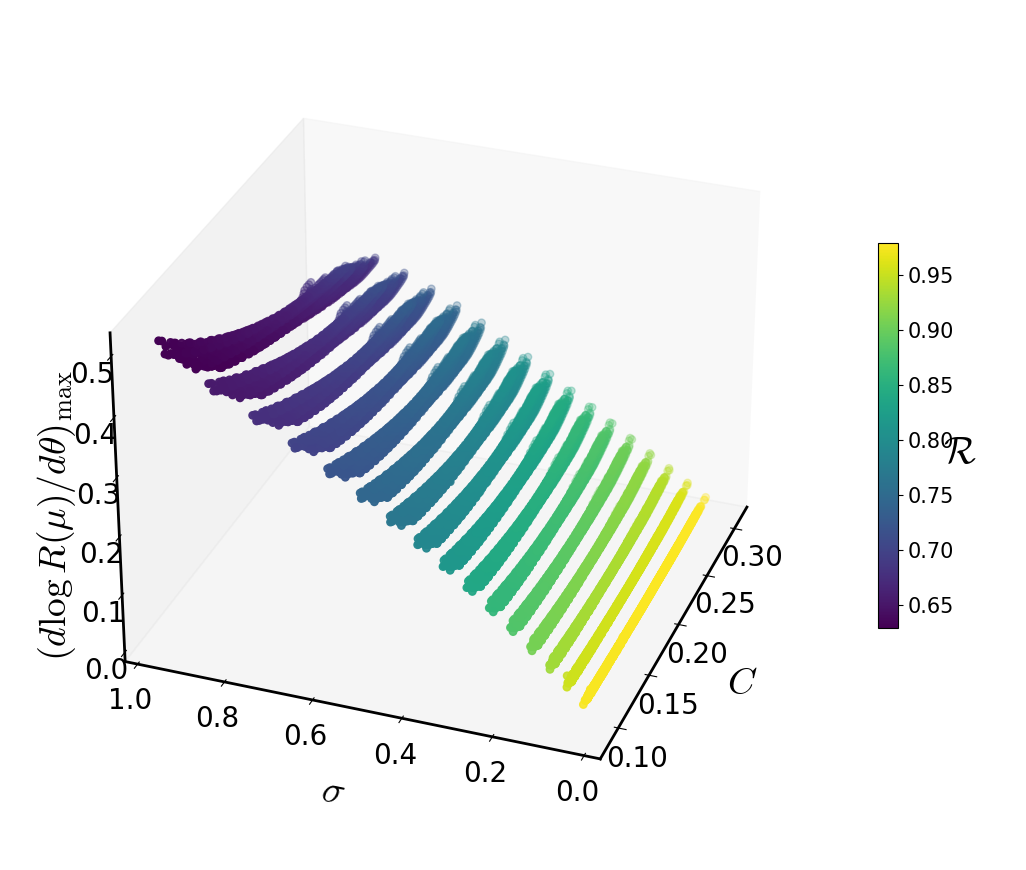}\hfill
    \includegraphics[width=0.46\textwidth]{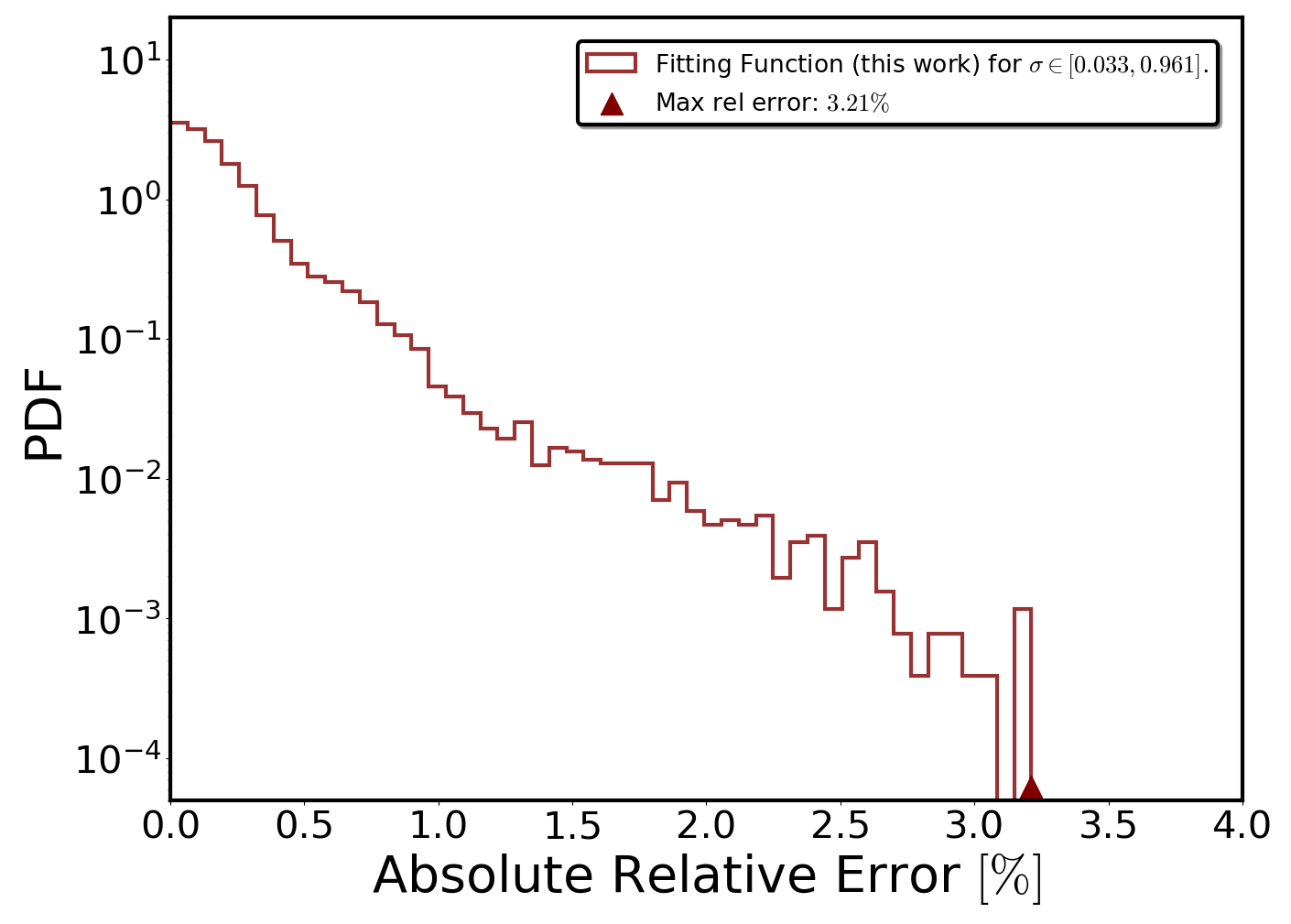}
    \caption{\label{fig:log_der_max_fit} Top panel: Numerical values of the $(d \log R(\mu)/d\theta)_{\mathrm{max}}$ as a function of the dimensionless parameters $C$, and $\sigma$. In this illustration, the colored variation of the data points corresponds to the $\mathcal{R}$ ratio dependence, as highlighted in the accompanying vertical color bar. Bottom panel: Distribution of the absolute values of the relative error between the numerical values and the regression formula (\ref{eq:dlogR_c_sigma_Rp_Re}) for the sample of rotating models described in the text. 
    }
\end{figure}

\begin{table}[!h]
    \caption{\label{tab:logR_C_sigma_R_optimizers} $\hat{\mathcal{C}}_{nmq}$ regression parameters for the logaritmic derivative at maximum parametrization (\ref{eq:dlogR_c_sigma_Rp_Re}). For this formula, the coefficient of determination is exceptionally high with a value of $R^2 = 0.99995$.}
    \begin{ruledtabular}
        \begin{tabular}{cccc}
            $\hat{\mathcal{C}}_{000}$ & $\hat{\mathcal{C}}_{001}$ & $\hat{\mathcal{C}}_{002}$ & $\hat{\mathcal{C}}_{003}$ \\
             -38.441438 & 131.341937& -146.858523 & 53.966304\\
            \hline\hline    
            $\hat{\mathcal{C}}_{010}$ & $\hat{\mathcal{C}}_{011}$ & $\hat{\mathcal{C}}_{012}$ & $\hat{\mathcal{C}}_{020}$ \\
            58.559062 & -130.455829 & 72.135318 & -27.241808\\
            \hline\hline    
            $\hat{\mathcal{C}}_{021}$ & $\hat{\mathcal{C}}_{030}$ & $\hat{\mathcal{C}}_{100}$ & $\hat{\mathcal{C}}_{101}$ \\
            30.488008 & 4.079915& -42.134981 & 87.095661\\
            \hline\hline
            $\hat{\mathcal{C}}_{102}$ & $\hat{\mathcal{C}}_{110}$ & $\hat{\mathcal{C}}_{111}$ & $\hat{\mathcal{C}}_{120}$\\
            -45.056782 & 38.466379 & -39.679675 & -8.676257\\
            \hline\hline
            $\hat{\mathcal{C}}_{200}$ & $\hat{\mathcal{C}}_{201}$ & $\hat{\mathcal{C}}_{210}$ & $\hat{\mathcal{C}}_{300}$\\
            -2.231085 & 2.574342 & 0.661519 & -0.403277\\
        \end{tabular}
    \end{ruledtabular}
\end{table}
The data distribution and the corresponding relative errors histogram are highlighted in Fig. \ref{fig:log_der_max_fit}. In this representation, the colored variation of the data points presented corresponds to different values of the $\mathcal{R}$ ratio, as indicated in the accompanying vertical colored bar.

From Fig. \ref{fig:log_der_max_fit} (bottom panel), it is evident that the regression formula (\ref{eq:dlogR_c_sigma_Rp_Re}) reproduces the data with accuracy better than $ 3.21 \%$. Notably, when applying this formula, only 266 out of the entire set of rotating NS models show relative deviations $> 1\%$. There is no discernible pattern linking this small subset of stellar configurations mainly to a specific EoS or category of EoSs. However, for completeness, it is worth noting that these models correspond to parameters within the range $C\in[0.094,0.238], R_{\mathrm{eq}}\in[11.594,16.766] \ \mathrm{km},\sigma\in[0.033,0.844]$, and $\mathcal{R}\in[0.639,0.978]$. Most importantly, the fitting function (\ref{eq:dlogR_c_sigma_Rp_Re}) stands as an accurate universal relation for all considered NS configurations. The significance of this relation will become evident in Sec. \ref{sec:univ_rel_dlogR}, especially concerning the universal determination of the logarithmic derivative as a function of the cosine of the colatitude $\theta$.

\subsection{\label{sec:poly_methods_2} RELATIONS FOR $g_{\mathrm{pole}}$ AND $g_{\mathrm{eq}}$}

The hotspot models used in the analysis of NICER data assume that the NS's atmosphere is described by a Hydrogen \cite{heinke2006hydrogen}. A key aspect of these models is the dependence of atmospheric properties on the star's effective gravity, which varies significantly between the poles and the equator due to the star's rapid rotation (see e.g., Fig \ref{fig:g_mu_indicative} for a review). At the poles, the effective gravity is stronger, while at the equator, it is weaker because of the centrifugal force. This variation highlights the importance of a universal description for surface gravity, which is crucial for accurately modeling the NS's atmosphere and interpreting the observed emission patterns.

\begin{figure}[!htb]
    \includegraphics[width=0.46\textwidth]{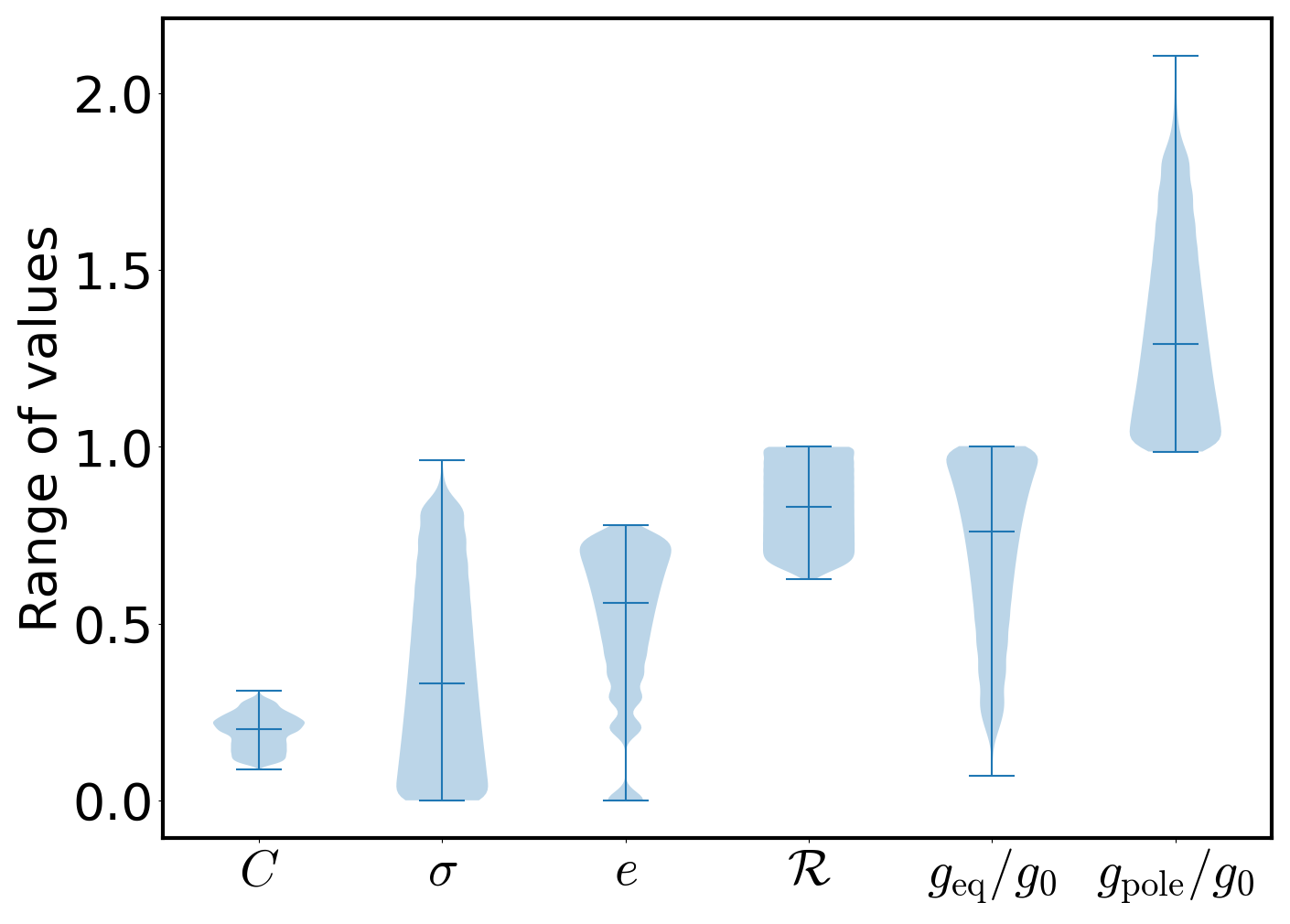}
    \caption{\label{fig:violin_plot} Range of values and data density representation for each parameter employed. For each feature illustrated on the horizontal axis, the associated range of data values corresponds to the whole sample of NS configurations used.}
\end{figure}

Drawing from the above, we now turn our attention to the universal estimation of the effective acceleration due to gravity, both at the star's pole and at the star's equator. In this pursuit, we reexamine the EoS-insensitive relations proposed in \cite{algendy2014universality} concerning the parameters $C\in[0.0876,0.3095]$, $\sigma \in [0.0000,0.9612]$ and explore potential enhancements. Regarding parameters, we also explore alternative possibilities by incorporating the star's eccentricity $ e\in[0.000,0.780]$ as a feature. Fig. \ref{fig:violin_plot} presents the corresponding violin plot illustrating the range of values for each associated quantity employed. These values correspond to the whole sample of 42694 stellar models contained in our ensemble.

We first look for a better parametrization in the relation that links the effective gravity at the star's pole by utilizing the $C$, and $\sigma$ as feature parameters. The polynomial function that best describes the data has the functional form,
\begin{equation}
        \label{eq:gpole_c_sigma}
        {g_{\mathrm{pole}}}(C,\sigma)=g_0\sum_{n=0}^{4}\sum_{m=0}^{4-n}\hat{\mathcal{D}}_{nm} \ C^n \ \sigma^m.
\end{equation}
Again, this is the simplest regression model compared to the others we tested that gave a satisfactory fit. Higher order polynomial functions with $\kappa > 4$ gave better evaluation measures at LOOCV from those demonstrated in Table (\ref{tab:g_pole_C_sigma}) and correspond to the $\kappa = 4$ fitting function. However, the improvement from selecting a more complex model is only marginal and not worth the effort.
\begin{table}[!h]
    \small
    \caption{\label{tab:g_pole_C_sigma} Indicative list of LOOCV evaluation measures for the $g_{\mathrm{pole}}(C,\sigma) = g_0\sum_{n=0}^{\kappa}\sum_{m=0}^{\kappa-n}\hat{\mathcal{D}}_{nm} \ C^n \ \sigma^m$ parametrization.}
    \begin{ruledtabular}
       \begin{tabular}{ccccccc}
            MAE & Max Error & MSE & $d_{\text{max}}$ & MAPE & Exp Var & $\kappa$ \\
            $\times 10^{-3}$ & $\times 10^{-2}$ & $\times 10^{-4}$ & ($\%$) &  $\times 10^{-1}$ ($\%$)&  &  \\

            \hline
              15.713 & 14.612 & 4.301 & 6.94 & 11.77 & 1.0 & 1 \\
              5.539 & 7.819 & 0.719 & 4.10 & 3.85 & 1.0 & 2 \\
              4.535 & 6.201 & 0.556 & 3.25 & 3.03 & 1.0 & 3 \\
              {\bf 4.320} & {\bf 5.377} & {\bf 0.524} & {\bf 3.07} & {\bf 2.86} & {\bf 1.0} & {\bf 4} \\
              4.268 &  5.141 & 0.511 & 3.06 & 2.83 & 1.0 & 5 \\
              4.222 & 5.316 & 0.500 & 2.97 &  2.80 & 1.0 & 6 \\
              4.171 & 5.559 & 0.487 & 3.10 & 2.78 & 1.0 & 7 \\
              4.090 & 5.789 & 0.475 & 3.22 & 2.72 & 1.0 & 8 \\
        \end{tabular}
    \end{ruledtabular}
\end{table}
Therefore, from the fit evaluation, the model's optimizers $\hat{\mathcal{D}}_{nm}$ are presented in Table (\ref{tab:g_pole_c_sigma_optimizers}).
\begin{table}[!h]
    \caption{\label{tab:g_pole_c_sigma_optimizers} $\hat{\mathcal{D}}_{nm}$ regression coefficients for the $g_{\mathrm{pole}}(C,\sigma)$ parametrization (\ref{eq:gpole_c_sigma}). For this model, the coefficient of determination is exceptionally high with a value of $R^2 = 0.9991$.}
    \begin{ruledtabular}
        \begin{tabular}{cccc}
            $\hat{\mathcal{D}}_{00}$ & $\hat{\mathcal{D}}_{01}$ & $\hat{\mathcal{D}}_{02}$ & $\hat{\mathcal{D}}_{03}$ \\
            0.908111 & 2.018696& 0.553202& -0.800025\\
            \hline\hline    
            $\hat{\mathcal{D}}_{04}$ & $\hat{\mathcal{D}}_{10}$ & $\hat{\mathcal{D}}_{11}$ & $\hat{\mathcal{D}}_{12}$ \\
            0.488087& 2.018696&-2.790572 &-1.469351 \\
            \hline\hline    
            $\hat{\mathcal{D}}_{13}$ & $\hat{\mathcal{D}}_{20}$ & $\hat{\mathcal{D}}_{21}$ & $\hat{\mathcal{D}}_{22}$ \\
            1.466061& -15.689925& 11.971482& 1.116029\\
            \hline\hline
            $\hat{\mathcal{D}}_{30}$ & $\hat{\mathcal{D}}_{31}$ & $\hat{\mathcal{D}}_{40}$ & \\
            52.068673&-23.257769 &-62.804547 & \\
        \end{tabular}
    \end{ruledtabular}
\end{table}

\begin{figure}[!thb]
    \includegraphics[width=0.46\textwidth]{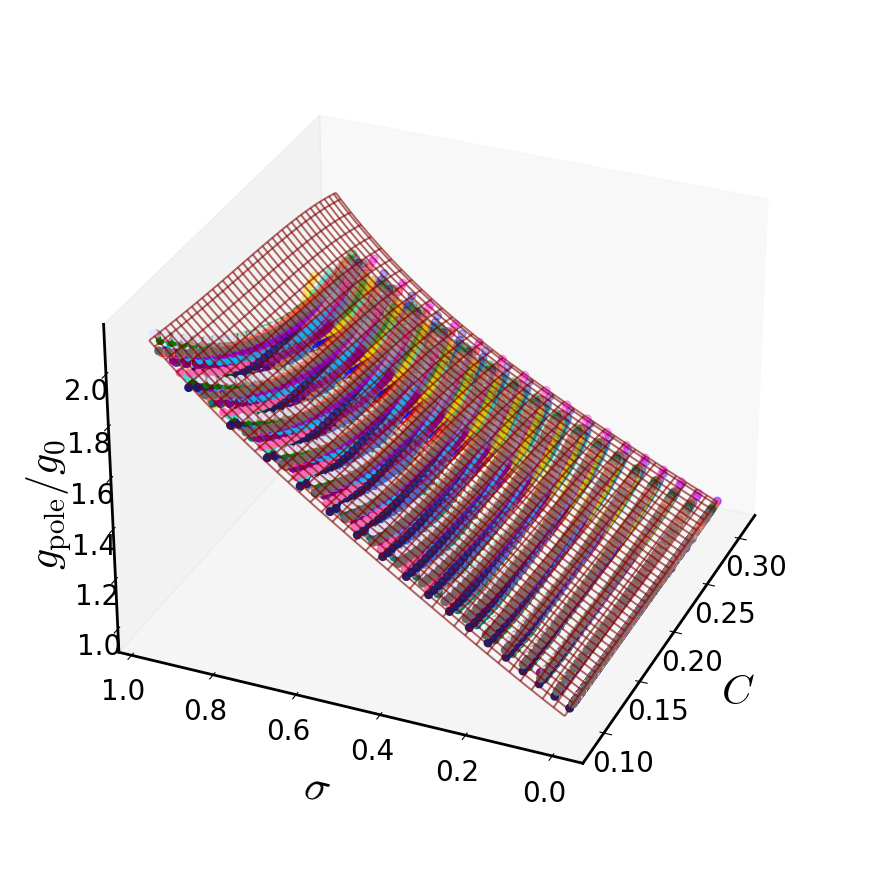}\hfill
    \includegraphics[width=0.46\textwidth]{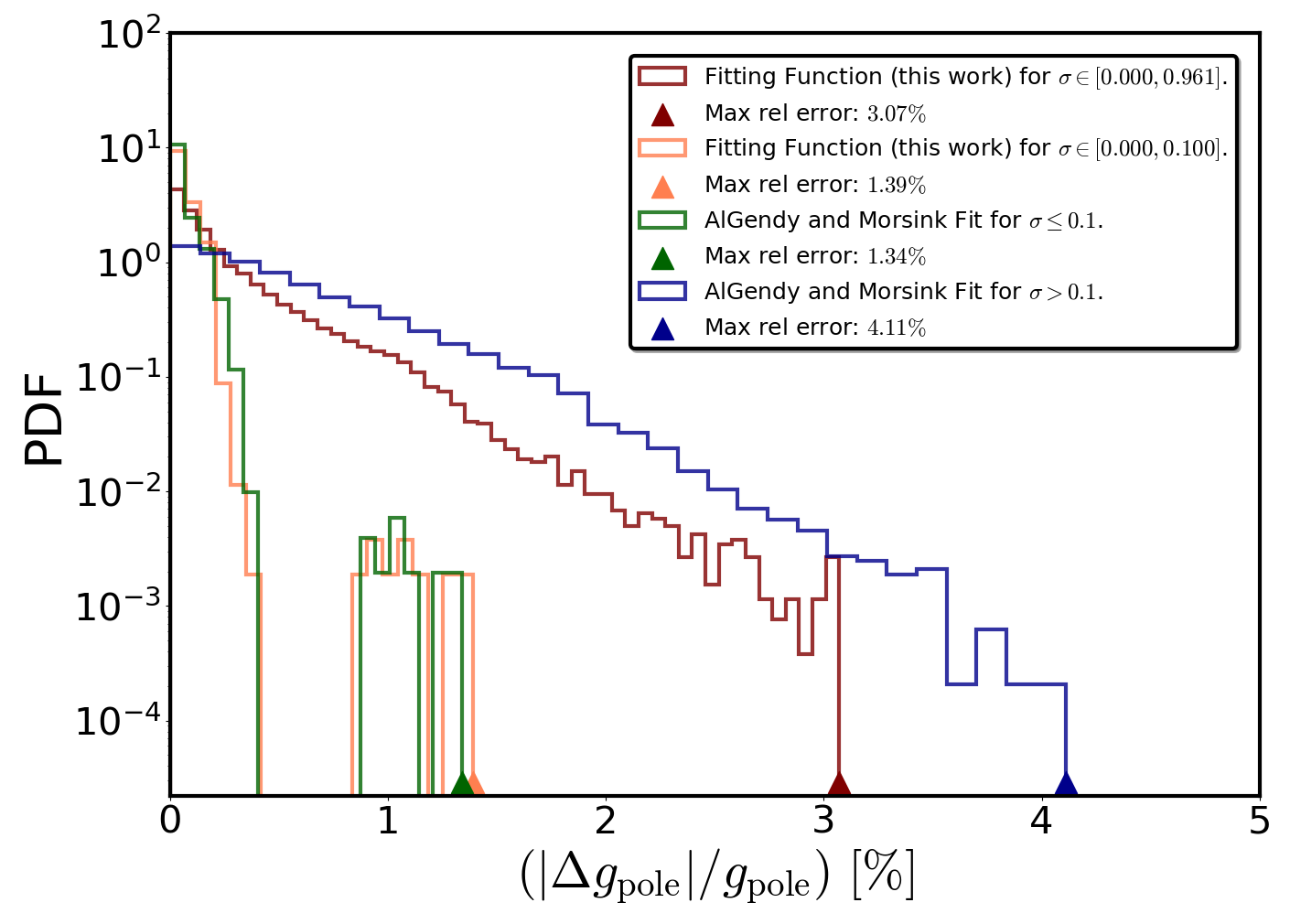}
    \caption{\label{fig:gp_fit} Top panel: $g_{\mathrm{pole}}/g_0$ as a function of the dimensionless parameters $C$ and $\sigma$.
    The plotted analytic surface which is shown with maroon grid lines
    corresponds to the fitting function (\ref{eq:gpole_c_sigma}). Bottom panel: Distribution of the absolute value of the relative deviations of the fit ($100\% (|\Delta g_{\mathrm{pole}}|)/g_{\mathrm{pole}} = 100\% (|g_{\mathrm{pole}, \mathrm{fit}} - g_{\mathrm{pole}}|)/ g_{\mathrm{pole}}$) given in logarithmic scale. The maroon and coral-colored histograms are the relative error predictions {associated with our} function, while the other two correspond to the fits proposed in the literature. }    
\end{figure}

The fitting function (\ref{eq:gpole_c_sigma}) that optimally reproduces the data and the corresponding relative deviations histogram are presented in Fig. \ref{fig:gp_fit}. Employing this parametrization for $g_{\mathrm{pole}}$, the relative errors between the regression model (\ref{eq:gpole_c_sigma}) and the data are $\leq 3.07\%$ (universality). It is important to highlight that, regardless of rotation and compactness, only 1190 NS configurations out of the total, having an equatorial radius in the range $R_{\mathrm{eq}}\in[11.53,18.38] \ \mathrm{km}$ have relative deviations $> 1\%$. The large majority of these stellar models correspond to Hybrid EoS models.

In addition, Fig. \ref{fig:gp_fit} (bottom panel) shows the comparison of our new regression formula (\ref{eq:gpole_c_sigma}) with methods already proposed, both in the case of slowly and rapidly rotating NSs \cite{algendy2014universality}. In the case of slowly rotating NS models with $\sigma \leq 0.1$, our fit yields results that closely align ($\lesssim 1.39 \%$) to the corresponding ones presented in \cite{algendy2014universality} when both are evaluated for the corresponding slowly-rotating data sample. However, when it comes to rapidly rotating NS models, our trained regression model has a higher accuracy when compared to the AlGendy and Morsink $g(\mu)$-formula for $\mu = 1$ (star's pole). Therefore, the surface fit (\ref{eq:gpole_c_sigma}) provides an accurate universal description for the effective gravity at the star's pole for each value of stellar compactness $C$ and rotation rate $\sigma$ within the parameter space. 

Now, we focus on investigating a universal description that connects the effective gravity at the star's equator with the parameters $C$, $\sigma$, and $e$. It is essential to highlight that adding eccentricity as a feature is crucial for minimizing deviations and enhancing the accuracy of data capture. The regression formula that best reproduces the data has the functional form,
\begin{align}
        \label{eq:g_eq_c_sigma_e}
        {g_{\mathrm{eq}}}(C,\sigma, e)=g_0\sum_{n=0}^{3}\sum_{m=0}^{3-n}\sum_{q=0}^{\small 3-(n+m)}\hat{\mathcal{E}}_{nmq} \ C^n \ \sigma^m \ e^q.
\end{align}

This is the least complicated fitting function that gave a satisfactory fit. It should be noted that higher order polynomial models gave better evaluation scores at LOOCV from those presented in the Table (\ref{tab:g_eq_C_sigma_e}) for the $\kappa = 3$ fitting function. However, selecting too complex models was not worth the slight improvement in the fit quality. From the best hyper-surface fit evaluation, the associated model coefficients are presented in Table (\ref{tab:g_eq_c_sigma_e_optimizers}).

\begin{table}[!h]
    \caption{\label{tab:g_eq_C_sigma_e} Indicative list of LOOCV evaluation measures for the $
    \sum_{n=0}^{\kappa}\sum_{m=0}^{\kappa-n}\sum_{q=0}^{\kappa-(n+m)}  \hat{\mathcal{E}}_{nmq} \ C^n \sigma^m  e^q$ parametrization.}
    \begin{ruledtabular}
        \begin{tabular}{ccccccc}
            MAE & Max Error & MSE & $d_{\text{max}}$ & MAPE & Exp Var & $\kappa$ \\
             $\times 10^{-3}$& $\times 10^{-2}$&  $\times 10^{-5}$&  ($\%$) & $\times 10^{-1}$($\%$) &  &  \\
            \hline
             12.457 & 9.099 & 23.452 & 60.24 & 23.56 & 1.0 & 1\\
             3.238 & 2.810 & 1.993 & 40.54 & 6.43 & 1.0 & 2 \\
             {\bf 0.760} & {\bf 1.291} & {\bf 0.123} & {\bf 4.34 }& {\bf 1.70} & {\bf 1.0} & {\bf 3} \\
             0.646 & 1.295 & 0.101 & 3.84 & 1.47 & 1.0 & 4 \\
             0.597 & 1.301 & 0.093 & 3.45 & 1.37 & 1.0 & 5 \\
             0.582 & 1.312 & 0.090 & 2.86 & 1.34 & 1.0 & 6 \\
             0.573 & 1.305 & 0.088 & 2.77 & 1.32 & 1.0 & 7 \\
             0.560 & 1.309 & 0.086 & 2.78 & 1.29 & 1.0 & 8 \\
        \end{tabular}
    \end{ruledtabular}
\end{table}

\begin{table}[!h]
    \caption{\label{tab:g_eq_c_sigma_e_optimizers} $\hat{\mathcal{E}}_{nmq}$ regression parameters for the $g_{\mathrm{eq}}(C,\sigma, e)$ parametrization (\ref{eq:g_eq_c_sigma_e}).  For this hyper-surface fit, the coefficient of determination is exceptionally high with a value of $R^2 = 0.99998$.}
    \begin{ruledtabular}
        \begin{tabular}{cccc}
            $\hat{\mathcal{E}}_{000}$ & $\hat{\mathcal{E}}_{001}$ & $\hat{\mathcal{E}}_{002}$ & $\hat{\mathcal{E}}_{003}$ \\
             0.995124 & -0.029767& 0.832182& 0.289041\\
            \hline\hline    
            $\hat{\mathcal{E}}_{010}$ & $\hat{\mathcal{E}}_{011}$ & $\hat{\mathcal{E}}_{012}$ & $\hat{\mathcal{E}}_{020}$ \\
            -1.691758 & -0.758367& 0.230731& 0.532801\\
            \hline\hline    
            $\hat{\mathcal{E}}_{021}$ & $\hat{\mathcal{E}}_{030}$ & $\hat{\mathcal{E}}_{100}$ & $\hat{\mathcal{E}}_{101}$ \\
            0.369276& -0.221009& 0.068663& 0.141318\\
            \hline\hline
            $\hat{\mathcal{E}}_{102}$ & $\hat{\mathcal{E}}_{110}$ & $\hat{\mathcal{E}}_{111}$ & $\hat{\mathcal{E}}_{120}$\\
            -2.032738& 2.331226& 2.630904&-4.035776 \\
            \hline\hline
            $\hat{\mathcal{E}}_{200}$ & $\hat{\mathcal{E}}_{201}$ & $\hat{\mathcal{E}}_{210}$ & $\hat{\mathcal{E}}_{300}$\\
            -0.28468& 0.128888& 1.205922&0.33807 \\
        \end{tabular}
    \end{ruledtabular}
\end{table}

\begin{figure}[!htb]
    \includegraphics[width=0.46\textwidth]{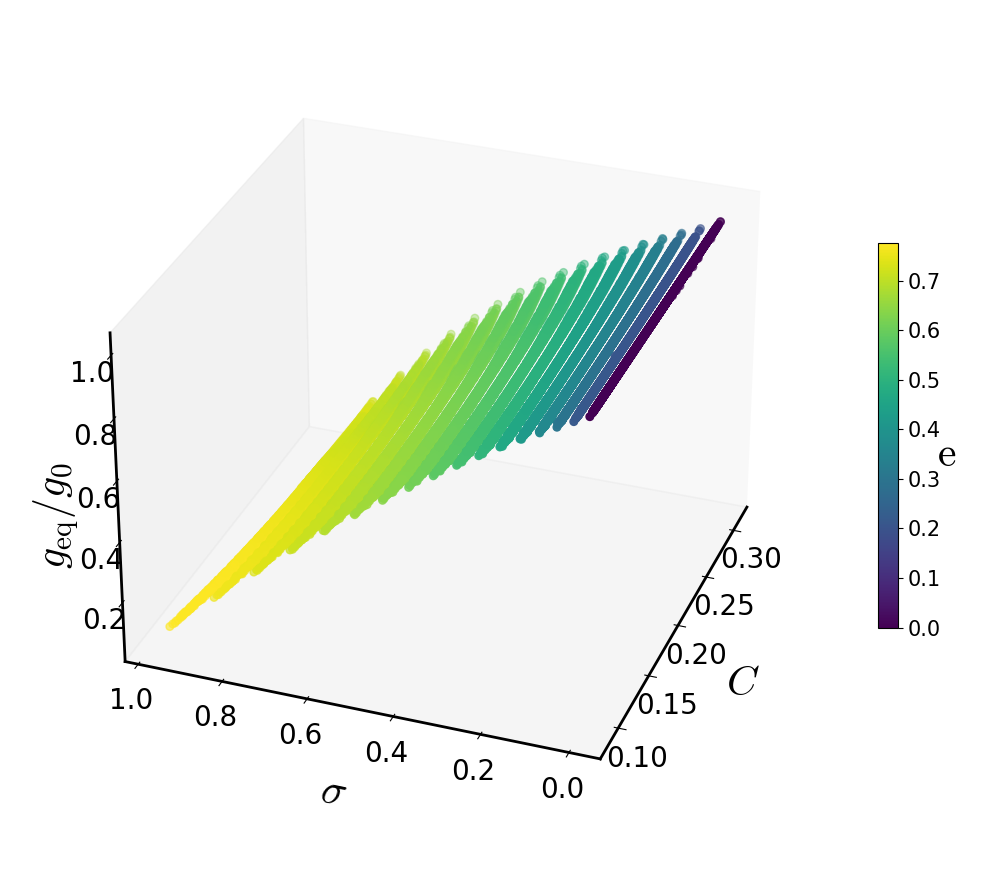}\hfill
    \includegraphics[width=0.46\textwidth]{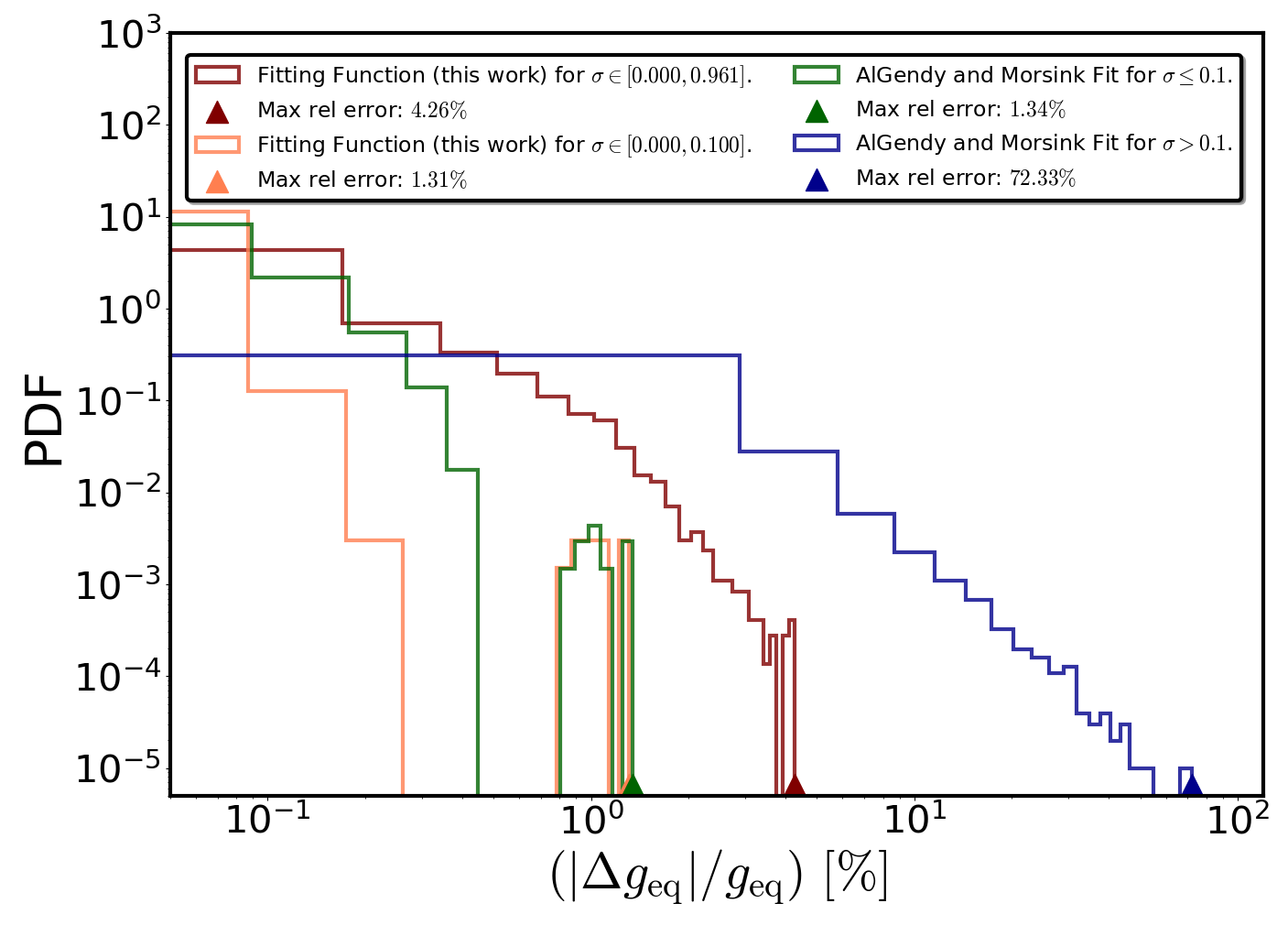}
    \caption{\label{fig:g_e_fit} Top Panel: $g_{\mathrm{eq}}/g_0$ as a function of the dimensionless parameters $C$, and $\sigma$. In this illustration, the colored variation of the presented data points corresponds to the star's eccentricity $e$ dependence, as highlighted in the accompanying vertical colored bar. Bottom panel: Absolute relative error distribution histogram derived using the regression formula (\ref{eq:dlogR_c_sigma_Rp_Re}) for the sample of NS models used. The horizontal axis is presented in a logarithmic scale. Also, the maroon and coral-colored histograms correspond to the absolute relative deviation predictions derived utilizing our fitting formula, while the others correspond to the associated models established in the literature.}

\end{figure}

The surface and distribution of the corresponding absolute values of the relative deviations histogram are presented in the two panels of Fig. \ref{fig:g_e_fit}. The varying colors of the data points correspond to different values of the star's eccentricity, as indicated in the accompanying vertical colored bar.

The regression formula (\ref{eq:g_eq_c_sigma_e}) verifies the data with accuracy better than $4.26 \%$. Most importantly, when applying this model, only 1107 out of the total ensemble of NS models have relative deviations $ >1\%$. Particularly for these stellar configurations, it is important to note that no clear pattern emerges in the variance of relative errors concerning either the EoSs or the EoS categories as well as the parameters $(C, \sigma, e)$ defining the examined parameter space. Therefore, the fitting formula (\ref{eq:g_eq_c_sigma_e}) is an accurate EoS-insensitive relation for all considered NS models. Fig. \ref{fig:g_e_fit} (bottom panel) also shows the comparison between our regression model (\ref{eq:gpole_c_sigma}) and those already suggested in the literature, both in the case of slowly and rapidly rotating NSs \cite{algendy2014universality}. In the case of slowly rotating configurations, our fitting function yields comparable results ($\lesssim 1.31 \%$) to the associated formula given in \cite{algendy2014universality} when both are evaluated using the particular slowly rotating data sample. In addition, when it comes to rapid rotation, it is evident that our trained regression model is more accurate than the corresponding AlGendy and Morsink $g(\mu)$-formula obtained by setting $\mu = 0$ (star's equator). Therefore, in all cases, our new fitting function (\ref{eq:gpole_c_sigma}) provides a robust EoS insensitive representation of the effective gravity at the star's equator.

\section{\label{sec:univ_rel_with_ANNs}Part II: Global inference of the star's surface USING ARTIFICIAL NEURAL NETWORKS}
Having obtained the data for the surface $R(\mu)$, the logarithmic derivative $d \log R(\mu)/ d\theta$, and the effective acceleration due to gravity at the star's surface $g(\mu)$ for each NS configuration included in our ensemble, we can now proceed to derive the associated regression model for each quantity of interest.

For the fitting procedure, we employed the feed-forward network architecture illustrated in Fig. \ref{fig:ANN_fig} of Appendix \ref{sec:ML_part}, following the steps, the formulation, and the optimization process outlined in Sec. \ref{sec:train_test}. At this point, it is worth mentioning that we consistently construct additional data points for each of the essential quantities (target-labels $\hat{z}$) using Hermite interpolation \cite{nozawa1998construction} as a pre-processing step.

More specifically, for each star, as highlighted in Sec. \ref{sec:num_data}, we have 261 (MDIV) data points for $R(\mu)$. By utilizing Hermite interpolation as a supplementary step to augment more data points, we aim to feed the network with additional information, thus enhancing the model's learning ability.
Through this method, we generate an extra data point $R(\tilde{\mu})$ for each centered $\tilde{\mu} = (\mu_i + \mu_{i+1})/2$ value in the range $\mu_i \leq \tilde{\mu} \leq \mu_{i+1}$. Covering the entire range of MDIV grid points $\mu \in [0,1]$, we produce $260$ synthetic data points for $R(\mu)$. This process was applied to each stellar configuration in the ensemble. As a result, each star is characterized by a total of 521 data points associated with its surface. Having acquired data on the star's surface, we apply the same methodology to generate additional synthetic data for the logarithmic derivative and the effective acceleration due to gravity at the star's surface. In addition, we have to note that the complete dataset contains $521 \ \mathrm{(data \ points/star)}\times 42694 \ \mathrm{stars} = 22243574 \ \mathrm{data \ points}$. For the NS models associated with each EoS, we partition $80 \%$ of the data for training and set aside the remaining $20 \%$ for testing, adhering to the way outlined in Sec. \ref{sec:train_test}. It should be emphasized that this vast amount of training and test data employed in this work play a crucial role in the neural network's training process, inference, and generalization ability.

\begin{figure}[!htb]
    \includegraphics[width=0.46\textwidth]{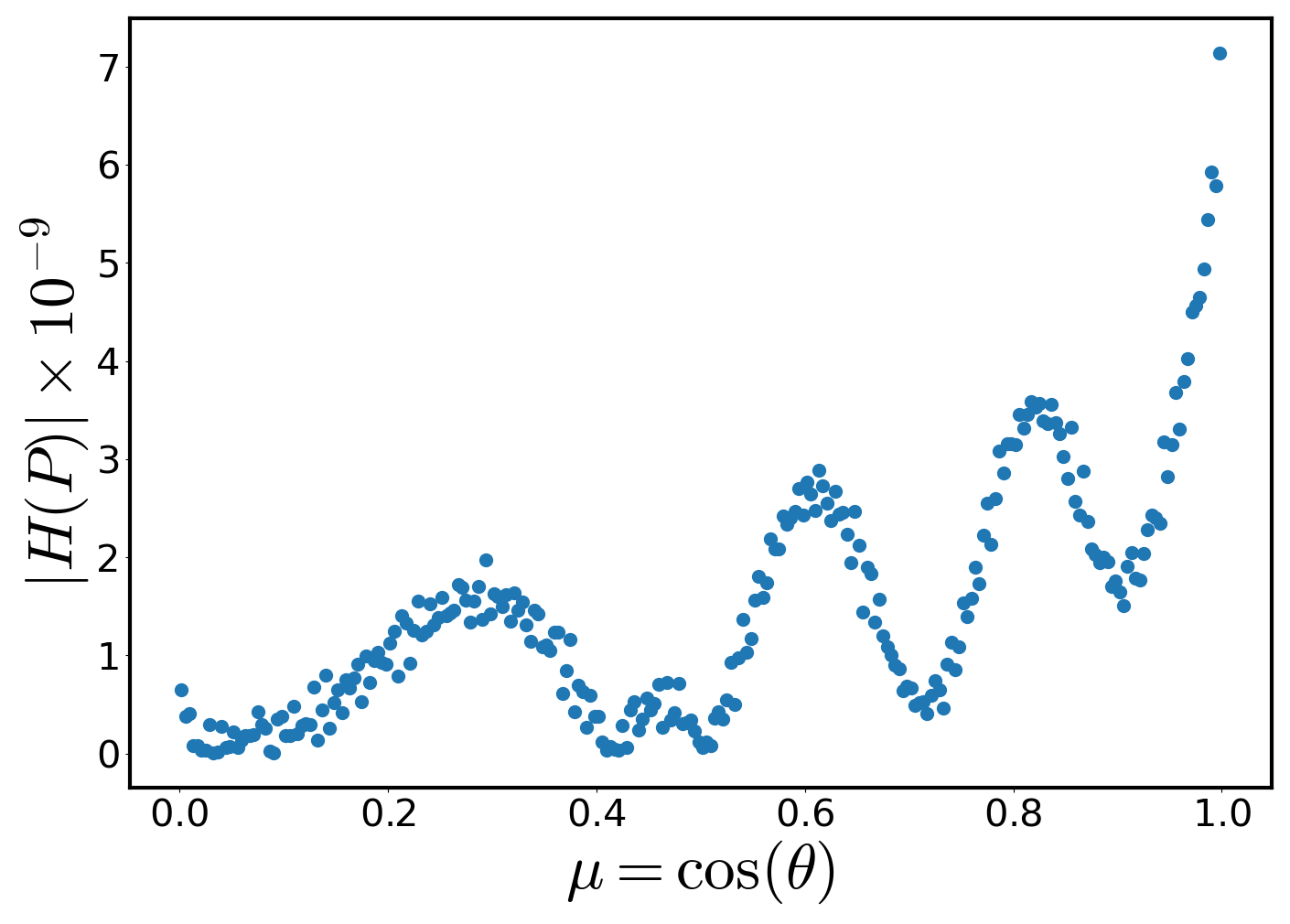}
    \includegraphics[width=0.46\textwidth]{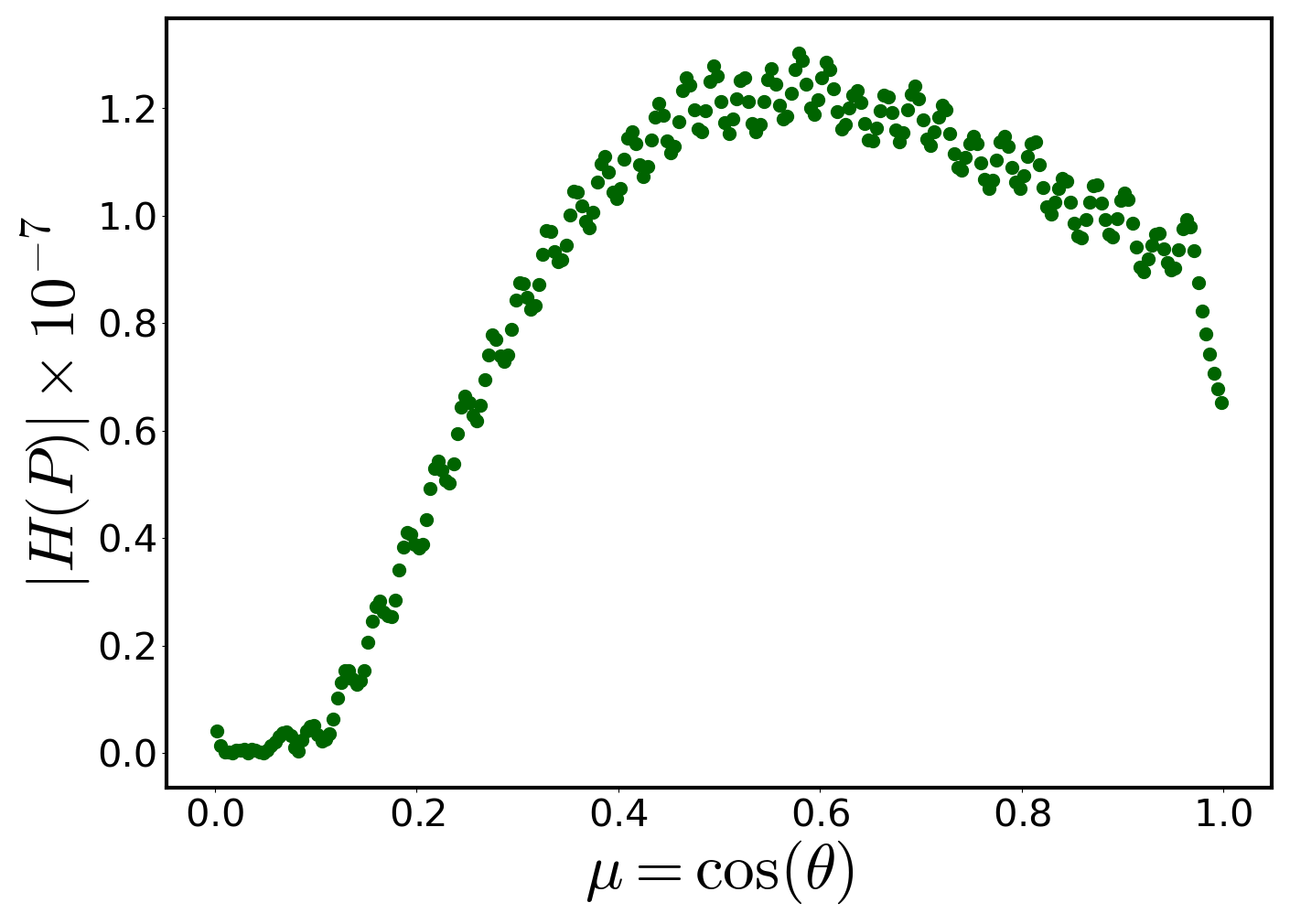}
    \caption{\label{fig:H(p)_root_ver} Top panel: Distribution of the numerical verification of the $H(P) = 0$ (Eq.(\ref{H_r_mu})) condition for the synthetic $R(\mu)$-data generated using Hermite interpolation. In the vertical axis, we present the range of $|H(P)|$ values corresponding to the NS benchmark model 2 shown in Table (\ref{tab:indicative_propert}). Bottom panel: Same as the top panel for the NS benchmark model 9 associated with the higher rotation frequency as shown in Table (\ref{tab:indicative_propert}).}
\end{figure}

For the benchmark NS model 2 presented in Table (\ref{tab:indicative_propert}), we provide an additional proof of concept in Fig. \ref{fig:H(p)_root_ver} (top panel), illustrating the distribution of the numerical verification of Eq.(\ref{H_r_mu}) for the synthetic data derived for $R(\mu)$. It is noteworthy to highlight that the condition $H(P) = 0$ is satisfied with an order of magnitude of $\lesssim 10^{-9}$. 
Furthermore, for the NS configuration with the higher rotation frequency, as presented in Table (\ref{tab:indicative_propert}) (model 9), Eq.(\ref{H_r_mu}) is numerically verified with an accuracy of approximately $\mathcal{O} (10^{-7})$ as shown in Fig. \ref{fig:H(p)_root_ver} (bottom panel). In this framework, it should be highlighted that a consistent pattern emerges across all NS configurations in our ensemble: the order of magnitude for the $H(P) = 0$ numerical verification for synthetic data varies systematically from $\mathcal{O} (10^{-9})$ to $\mathcal{O} (10^{-7})$ as the rotation frequency increases. In any case, the synthetic data effectively correspond to numerical solutions associated with the star's surface.

In practical applications like pulse profile modeling \cite{bogdanov2019constraining,gendreau2017searching,watts2019constraining,chatziioannou2024neutron} or the cooling tail method \cite{suleimanov2017direct,suleimanov2020observational,chatziioannou2024neutron}, we suggest an EoS-insensitive regression model that accurately describes the NS surfaces $R(\mu)$ across a relatively wide range of compactness values ($0.0876 \leq C \leq 0.3095$) and rotations ($0 \leq \sigma \leq 0.9612$) for each EoS within our ensemble \cite{algendy2014universality, morsink2007oblate, yagi2017approximate,silva2021surface}. In that direction, we also generate precise regression models for the logarithmic derivative $d \log R(\mu)/d \theta$ and the star's effective gravity $g(\mu)$ estimation, irrespective of the EoS. Most importantly, our regression models reproduce the data on the test set with high accuracy, thus demonstrating a state-of-the-art generalization ability beyond the training data.

The demonstration of the methodology and results will be ordered in the following way: Sec. \ref{sec:univ_rel_R} is dedicated to proposing an EoS-independent new relation for the star's circumferential radius $R(\mu)$, Sec. \ref{sec:univ_rel_dlogR} introduces a new relation for the logarithmic derivative, while Sec. \ref{sec:univ_rel_g} presents a new relation for the effective acceleration due to gravity $g(\mu)$ at the star's surface. It should be noted that except for the parameters $C$ and $\sigma$ commonly utilized in the literature, the proposed neural network architecture indirectly utilizes the star's polar radius as an additional parameter. The selection of this additional feature is sufficient for the model to excel in inference capabilities and generalization, achieving results in a way that is not substantially influenced by the EoS. Both indicative code {examples} for our fits and the trained model's optimal parameters $\theta^{\star}$ (weights) will be available in the following GitHub repository:\href{https://github.com/gregoryPapi/Universal-description-of-the-NS-surface-using-ML.git}{https://github.com/gregoryPapi/Universal-description-of-the-NS-surface-using-ML.git}. 

\subsection{\label{sec:univ_rel_R}UNIVERSAL ESTIMATION OF THE STAR'S SURFACE USING ANN}
To employ the designed ANN architecture for estimating the star's surface, we first consider the relation between the star's circumferential radius $R(\mu)$ and some input features $\tilde{x}$, encompassing the parameters $\mu, C, \sigma,$ and $e$.

Along a given sequence of data points associated with the star's surface, we then proceed to estimate the normalized circumferential radius given as,
\begin{equation}
  \label{R_mu_norm}  
  \large
  \hat{z}_{1} =
  \begin{cases}
    \  \ \frac{R(\mu) - R_{\mathrm{pole}}}{R_{\mathrm{eq}}-R_{\mathrm{pole}}}, \ \ \sigma \neq 0 \\
    \\
    \ \ \frac{R(\mu)}{R_{\mathrm{eq}}},  \ \  \sigma = 0.
  \end{cases}
\end{equation}
Depending on the star's rotation, this transformation is performed individually for each star included in our ensemble. In the case of nonrotating NSs we have $R(\mu)=R_{\mathrm{eq}}$, while for the rotating ones, with this linear transformation, we map the interval $[R_{\mathrm{pole}}, R_{\mathrm{eq}}]$ into the unit interval $[0,1]$. It is important to emphasize that possessing precise knowledge of the interval limits, not only enhances the effectiveness of the ANN model during the training process but also contributes to a significant reduction of the expected model's relative errors. Consequently, the choice of the sigmoid function (\ref{sigmoid}) placed on the output layer is a natural choice and guarantees that, the model's inference will consistently fall within the $[0,1]$ range, while at the same time enhancing the model's training during the learning processes (faster convergence and superior loss minimization).

\begin{figure}[!htb]
    \includegraphics[width=0.46\textwidth]{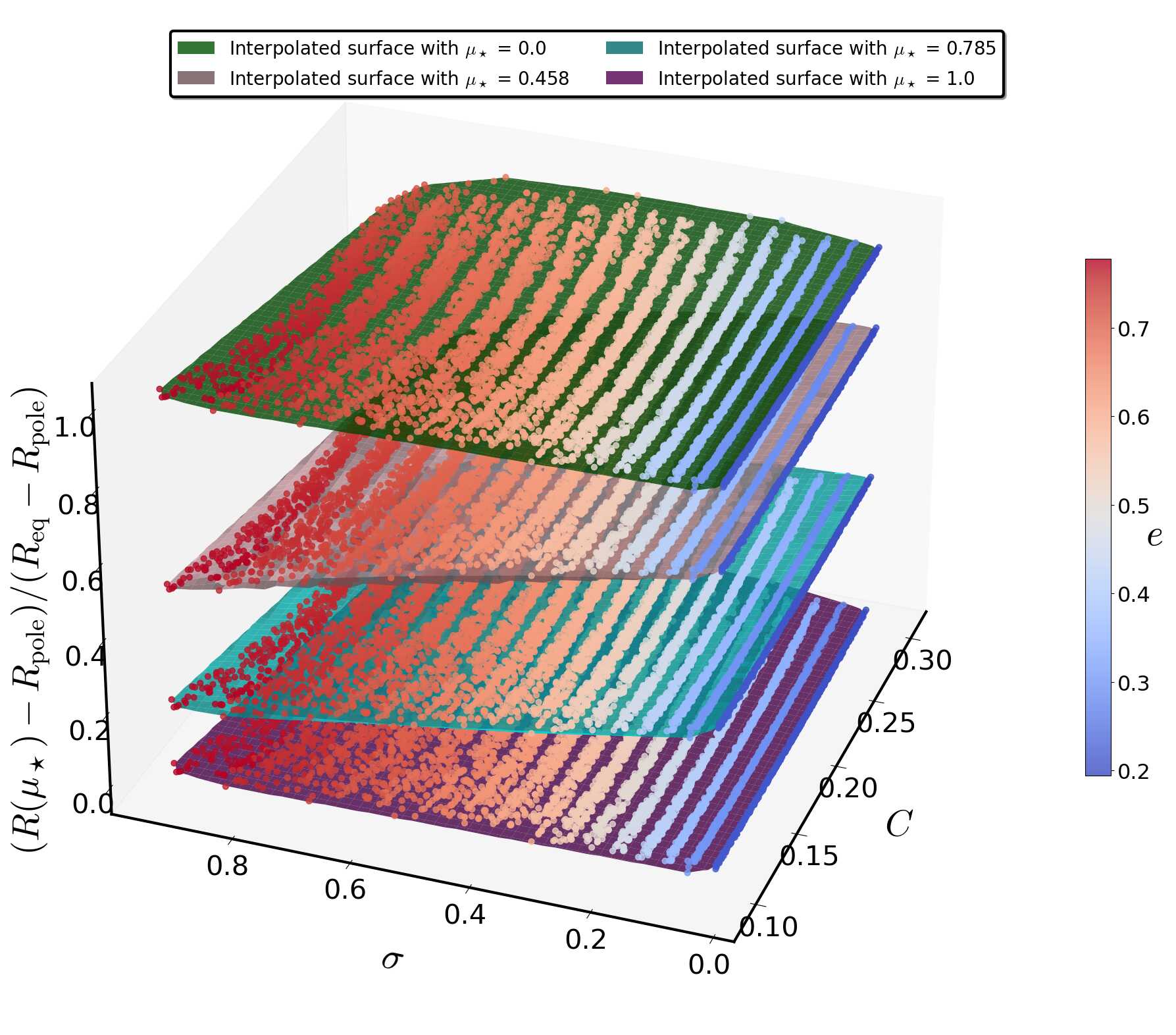}
    \caption{\label{fig:R(mu)_univ_illustration} Universal representation: Normalized radius $(R(\mu_\star)-R_{\mathrm{pole}})/(R_{\mathrm{eq}}-R_{\mathrm{pole}})$ as a function of the star's global parameters $C$ and $\sigma$ for a discrete array of $\mu$ values $\mu\in [0,1]$ moving from the rotating star's equator ($\mu = 0$) towards the star's pole ($\mu = 1$). Each colored numerically interpolated surface corresponds to an assigned $\mu_{\star}$ value, while the vertical colored bar represents the star's eccentricity $e$. }
\end{figure}

\begin{figure}[!htb]
    \includegraphics[width=0.46\textwidth]{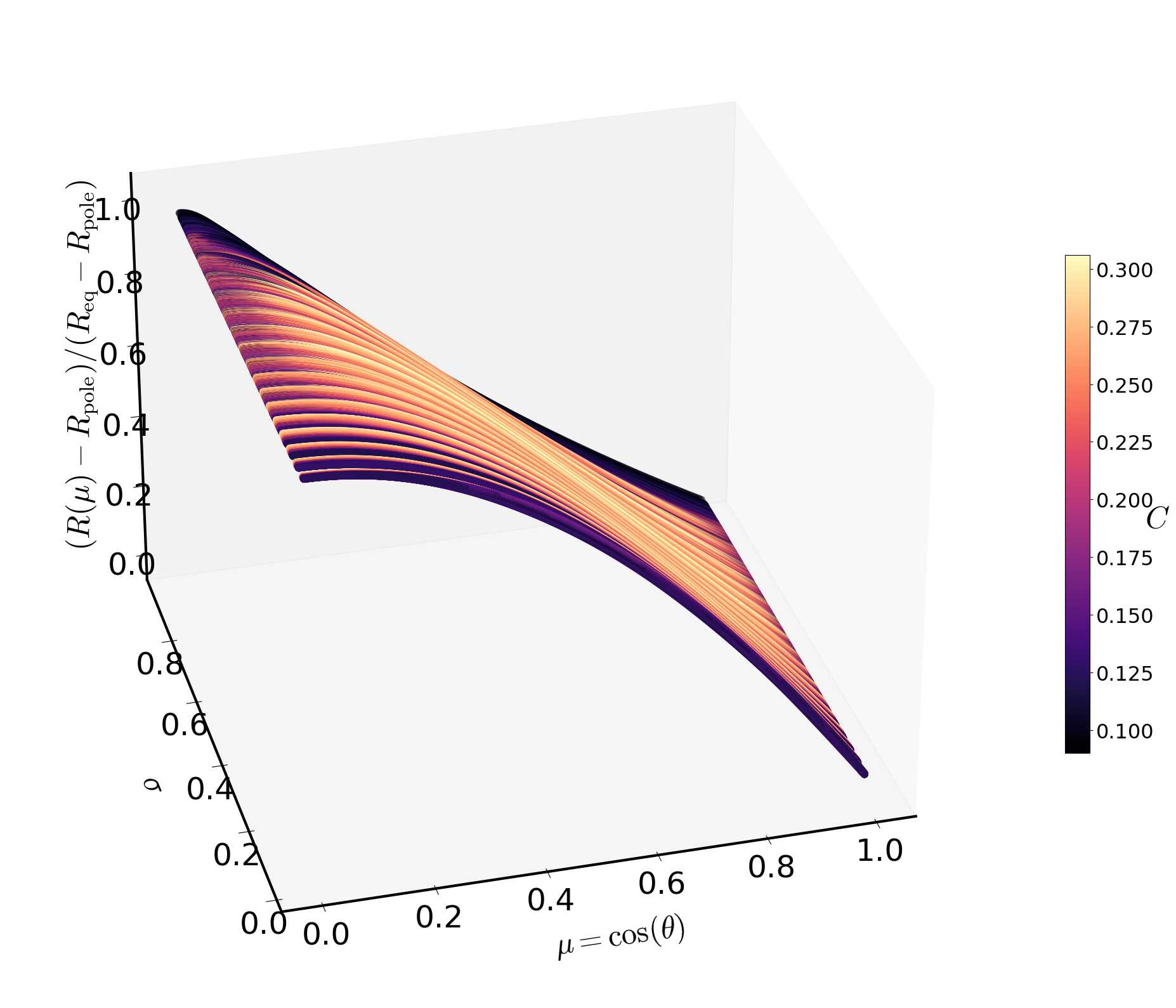}
    \caption{\label{fig:R(mu)_univ_illustration_2} EoS-insensitive relation: Normalized radius $(R(\mu)-R_{\mathrm{pole}})/(R_{\mathrm{eq}}-R_{\mathrm{pole}})$ as a function of 
    the angular position parameter on the star $\mu = \cos(\theta)$, and the reduced spin $\sigma$.
    The vertical colored bar represents the star's stellar compactness $C$.}
\end{figure}

For the rotating case, moving from the star's equator towards the star's pole, the proposed transformation (\ref{R_mu_norm}) is EoS-independent, aligning all data approximately onto a single universal plane for each specific value of the star's colatitude $\theta$. This representation is exemplified in Fig. \ref{fig:R(mu)_univ_illustration}, which displays the data distribution for an indicative discrete array of $\mu$ values $\mu \in [0,1]$ across the entire parameter space. In this illustration, the colored planes represent regions in space with fixed $\mu = \mu_\star$, while the vertical colored bar indicates the star's eccentricity. Overall, the data for each EoS form a universal hyperstructure within the parameter space. Furthermore, Fig. \ref{fig:R(mu)_univ_illustration_2} is an alternative representation of the universal parametrization (\ref{R_mu_norm}) for each $\mu$ within the parameter space, showing the data for each EoS in the test set over the full range of $\mu$ values across different rotation rates. 

\begin{figure}[!ht]
    	\centering
    	\includegraphics[width=0.46\textwidth]{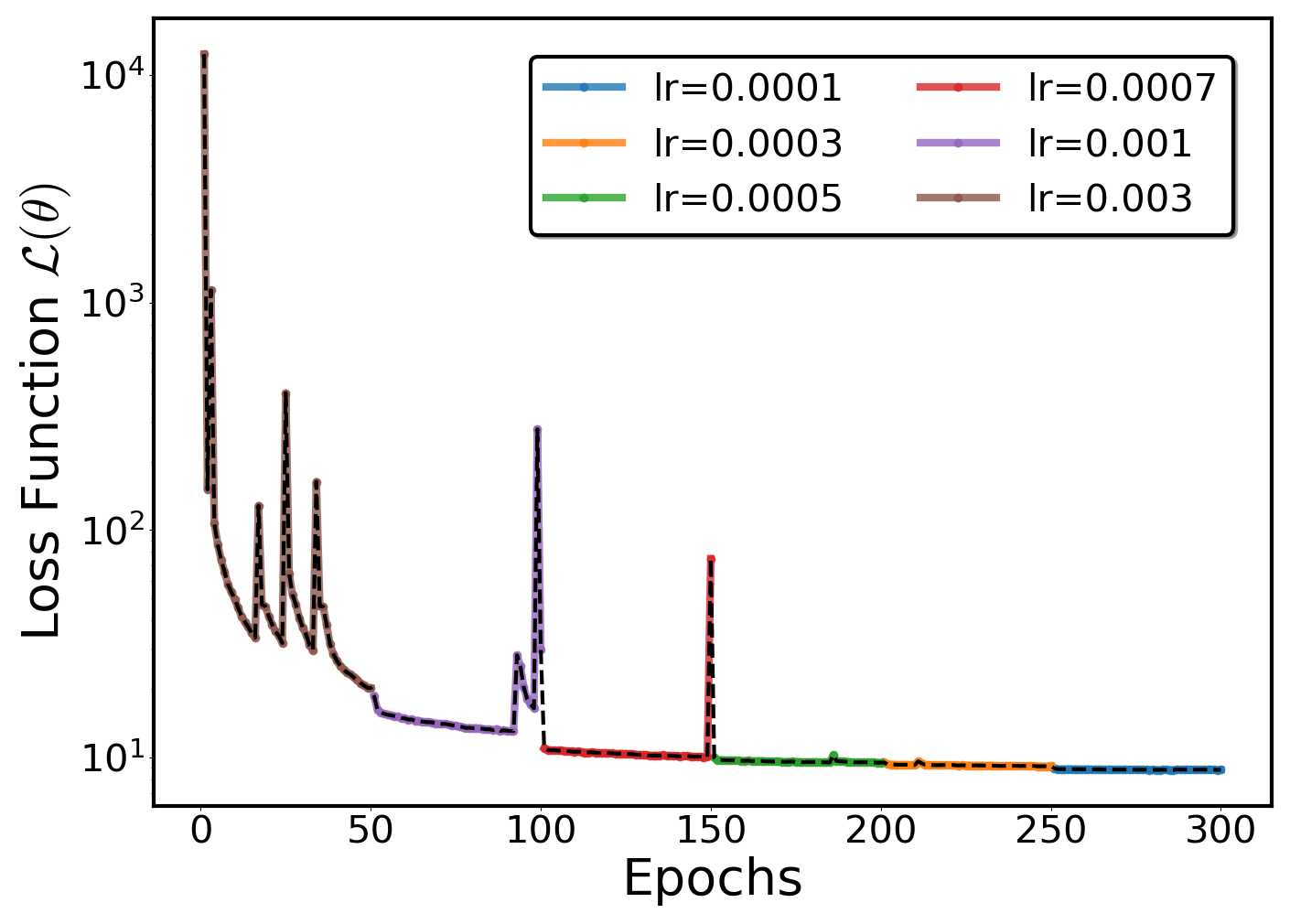}
        \caption{\label{fig:surf_loss} Normalized circumferential radius estimation using ANN: Illustration of the Loss function minimization results derived during the ANN model training. Each colored curve corresponds to the specific learning rate employed. The whole training process is carried out for 300 epochs.}
\end{figure}

Having obtained the scaled target data $\hat{z}_{1}$ and defining the train and test sets, we proceed to train the ANN model $\hat{F}_{\theta}(|\mu|, C, \sigma, e)$ (\ref{fig:ANN_fig}) to derive the optimal parameters $\theta^{\star}$ through the optimization process described in Sec. \ref{sec:train_test}. In Fig. \ref{fig:surf_loss}, we demonstrate the reduction of the loss function $\mathcal{L}(\theta)$ over 300 epochs of model training.

The regression model $R(\mu)$ with the optimal $\theta^{\star}$ parameters has the functional form,
\begin{equation}
    \label{eq:R_mu_fit}
    R(\mu) = R_{\mathrm{pole}} + (R_{\mathrm{eq}}-R_{\mathrm{pole}}) \hat{F}_{\theta^{\star}}(|\mu|, C, \sigma, e).
\end{equation}
The choice of $|\mu|$ as an input variable is intended to enforce $\mathbb{Z}_2$ mirror symmetry across the star's surface, ensuring that $R(\mu) = R(-\mu)$ along the rotation axis. The proposed fitting formula accurately reproduces the star's surface for an arbitrary order spin-induced deformation $\sigma$. Thus, it offers a universal fitting representation for the radius of the oblate star's shape at each specific $\mu$ value of interest. In addition, it is noteworthy that in the non-rotating limit ($\sigma = 0$), our formula perfectly adheres to the consistency condition $R(\mu) = R_{\mathrm{pole}} = R_{\mathrm{eq}}$, which is associated with spherical stars, for all values of $\mu$. 

\begin{figure}[!ht]
    	\centering
    	\includegraphics[width=0.46\textwidth]{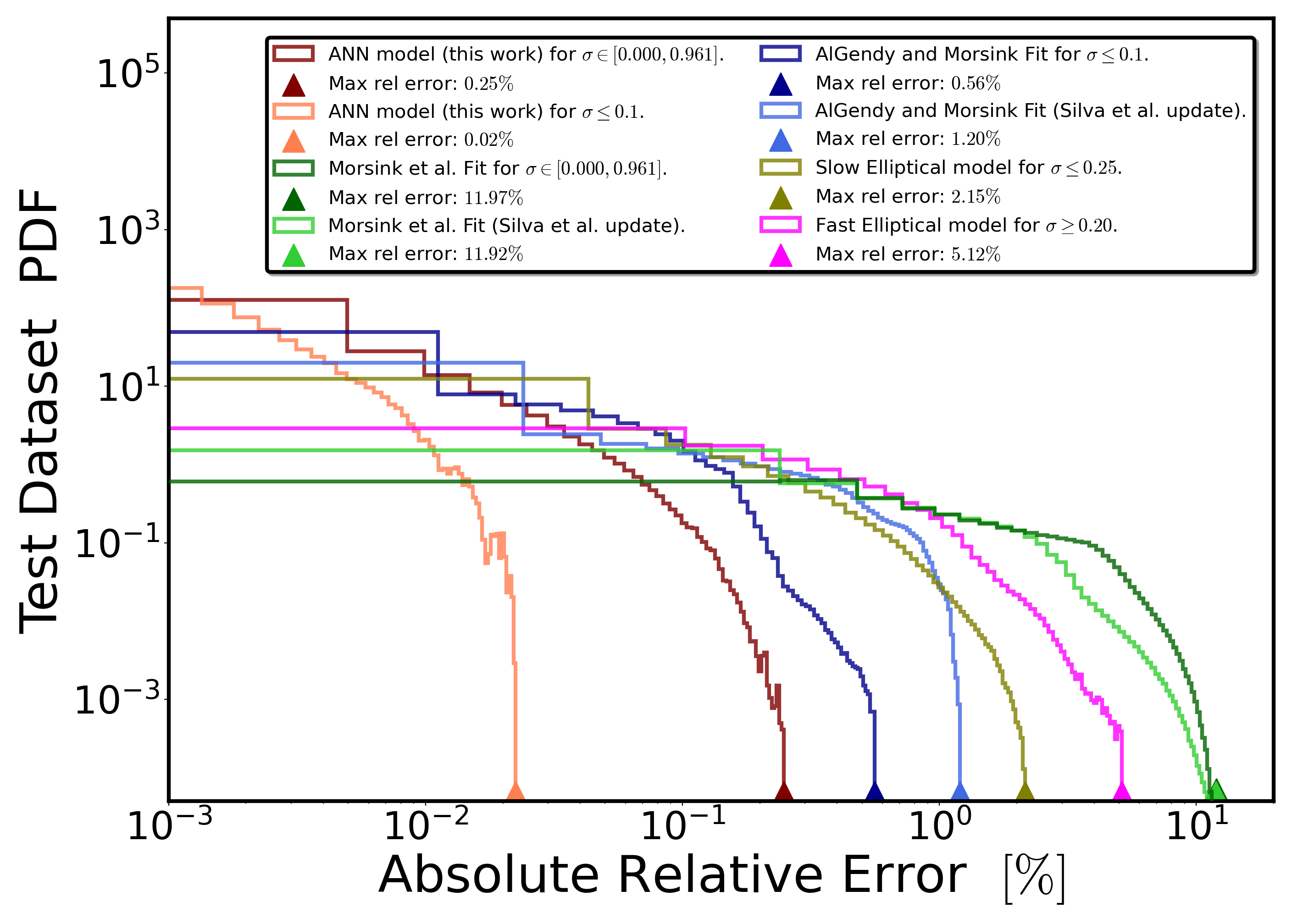}
        \caption{\label{fig:Rmu_comp} 
        Colored histograms depicting the distribution of absolute relative errors $100 \% \times(R(\mu)_{\mathrm{fit}}-R(\mu))/R(\mu)$ providing a basis for comparison between our ANN model's results with fitting functions established in the literature. Both axes are given in a logarithmic scale.
        Each histogram is referred to the test set, while the colored triangles denote the absolute maximum relative deviation produced by each functional form used to verify the test data.} 
\end{figure}

The evaluation measures for the model defined by equation (\ref{eq:R_mu_fit}) on the test set are presented in Table (\ref{tab:R_mu_eval_meas}).
\begin{table}[!h]
    \footnotesize
    \caption{\label{tab:R_mu_eval_meas} Evaluation measures for the parametrization given by equation (\ref{eq:R_mu_fit}) on the test set.}
    \begin{ruledtabular}
        \begin{tabular}{ccccccc}
            MAE & Max Error & MSE & $d_{\text{max}}$ & MAPE & Exp Var & $R^2$ \\
            $\times 10^{-3}$ & $\times 10^{-2}$ & $\times 10^{-6}$ & ($\%$) & $\times 10^{-5}$ ($\%$) & &   \\
            \hline
            1.149  & 3.333 & 5.865 & 0.25 & 8.98& 0.99999 & 0.99999  \\
        \end{tabular}
    \end{ruledtabular}
\end{table}
Additionally, Fig. \ref{fig:Rmu_comp} presents histogram distributions showcasing relative errors on the test set, offering a comparative analysis between our model and those proposed in \cite{silva2021surface,morsink2007oblate,algendy2014universality}.

Employing this $R(\mu)$ parametrization, the fitting formula (\ref{eq:R_mu_fit}) accurately reproduces the data on the test set, exhibiting a remarkable precision with a relative error of less than $0.25\%$. 
We can observe that the majority of the model predictions exhibit a relative error of less than $0.1 \%$. Therefore, for each EoS included in our ensemble, the associated ANN formulation highlights an excellent generalization ability. An important consideration at this point is the evaluation of the maximum relative deviation exhibited by the proposed regression model (\ref{eq:R_mu_fit}) in the test set, both across EoS categories and for individual EoSs. In Appendix \ref{sec:regression_violin_R_mu}, we present the violin plots illustrating the distribution of absolute fractional differences for each case, providing a comprehensive view of the model’s performance and its variability. The hybrid EoS class exhibits the largest surface deviations, around $ 0.25 \%$ ($d_{\mathrm{max}}$), while the hadronic and hyperonic categories show maximum relative errors of approximately $ 0.20 \%$ and $0.16 \%$, respectively. The EoS models associated with these values are the Holographic V-QCD model APR intermediate \cite{akmal_equation_1998, jokela2019holographic, ishii2019cool, ecker2020gravitational, jokela2021unified}, the EI-CEF-Scyrme model SKb \cite{danielewicz_symmetry_2009,gulminelli_unified_2015,kohler_skyrme_1976}, and the SU$(3)$-CMF model DS(CMF)$-1$ \cite{bennour_charge_1989,gulminelli_unified_2015,dexheimer_gw190814_2021,dexheimer_novel_2010,dexheimer_proto-neutron_2008,dexheimer_reconciling_2015,dexheimer_tabulated_2017} (see e.g., Appendix \ref{app:eos_tables} and Fig. \ref{fig:R_mu_violins} for a review).


In addition, Fig. \ref{fig:Rmu_comp}
provides the relative error comparison between our regression model (\ref{eq:R_mu_fit}) and the surface fitting functions that are available in the literature for both slowly and rapidly rotating NS configurations \cite{silva2021surface,morsink2007oblate,algendy2014universality}. For slowly rotating neutron star models with $\sigma \leq 0.1$, the AlGendy and Morsink fitting function \cite{algendy2014universality}, along with the updated coefficients provided by Silva et al. \cite{silva2021surface}, yields relative errors of $\leq 0.56 \%$ and $\leq 1.20 \%$, respectively. This difference mainly stems from the ensemble of predominantly slowly rotating NS models utilized in the initial AlGendy and Morsink work, while Silva et al. focused on a more limited set of slowly rotating stellar configurations. In contrast, our model achieves a significantly lower error margin of $\leq 0.02 \%$. Therefore, regardless of the star's rotation, the suggested formula (\ref{eq:R_mu_fit}) for estimating the star's circumferential radius has higher accuracy than the already established fitting relations. Hence, the regression Formula (\ref{eq:R_mu_fit}) offers a substantial improvement in accurately estimating the star's surface, irrespective of the EoS. 

\begin{figure*}[!ht]
        \centering
    	\includegraphics[width=0.9\textwidth]{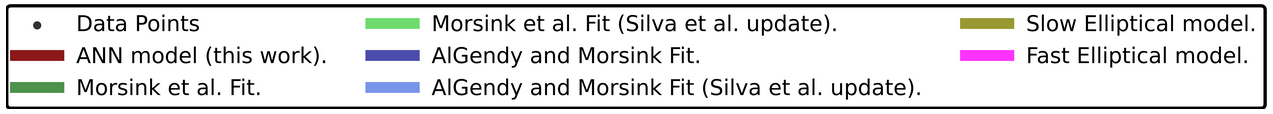}\\
        \centering
    	\includegraphics[width=0.32\textwidth]{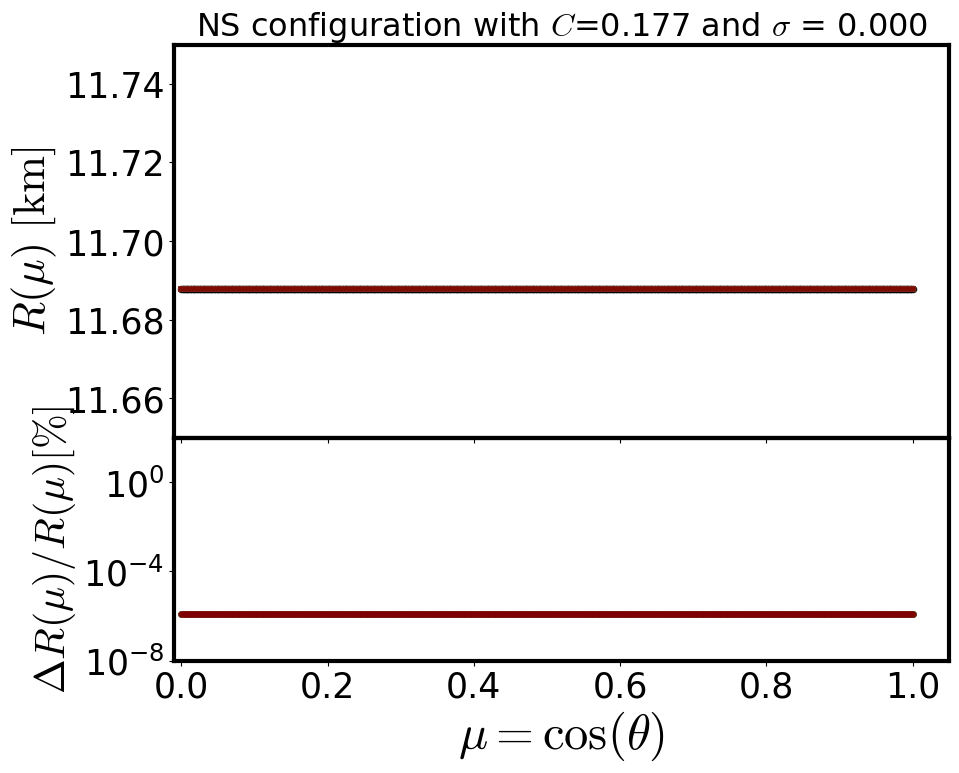}
        \includegraphics[width=0.32\textwidth]{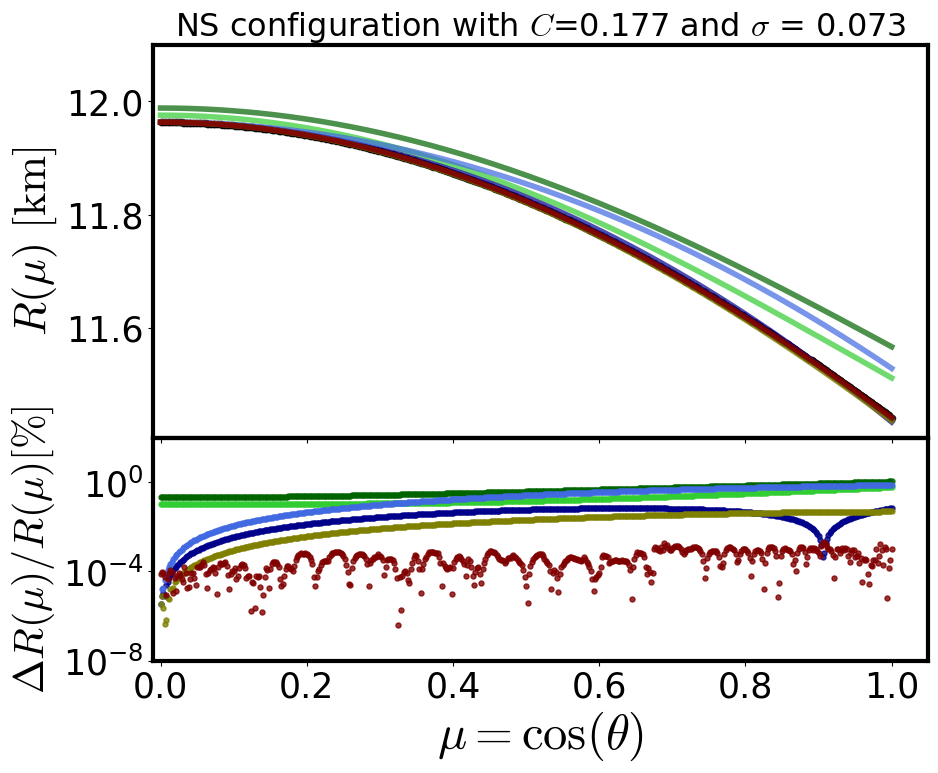}
        \includegraphics[width=0.32\textwidth]{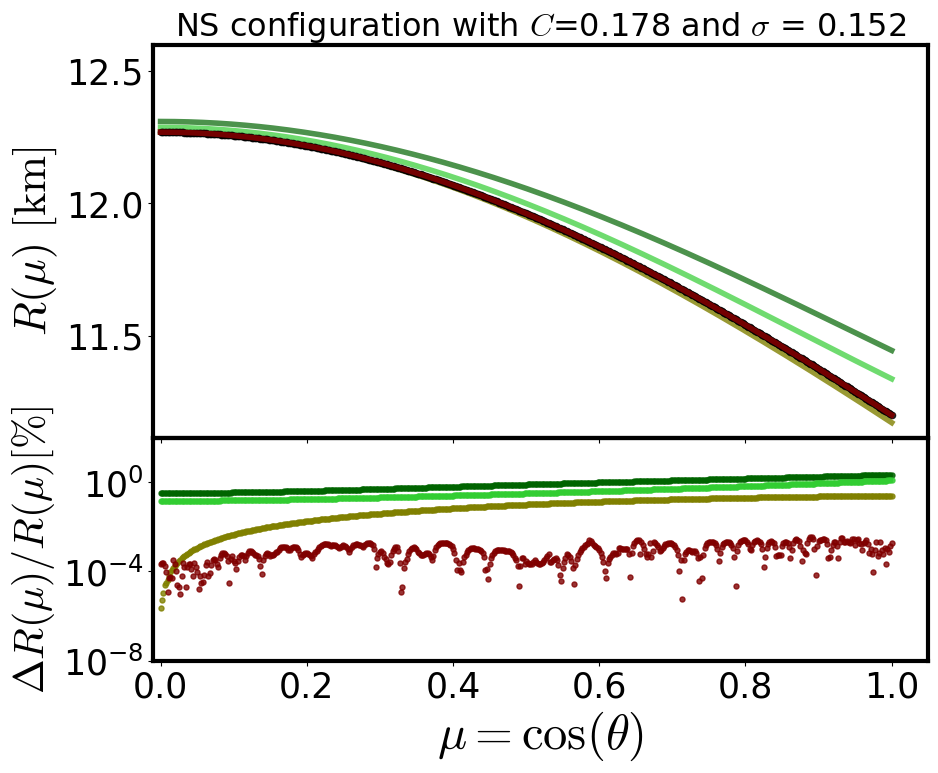}
        \includegraphics[width=0.32\textwidth]{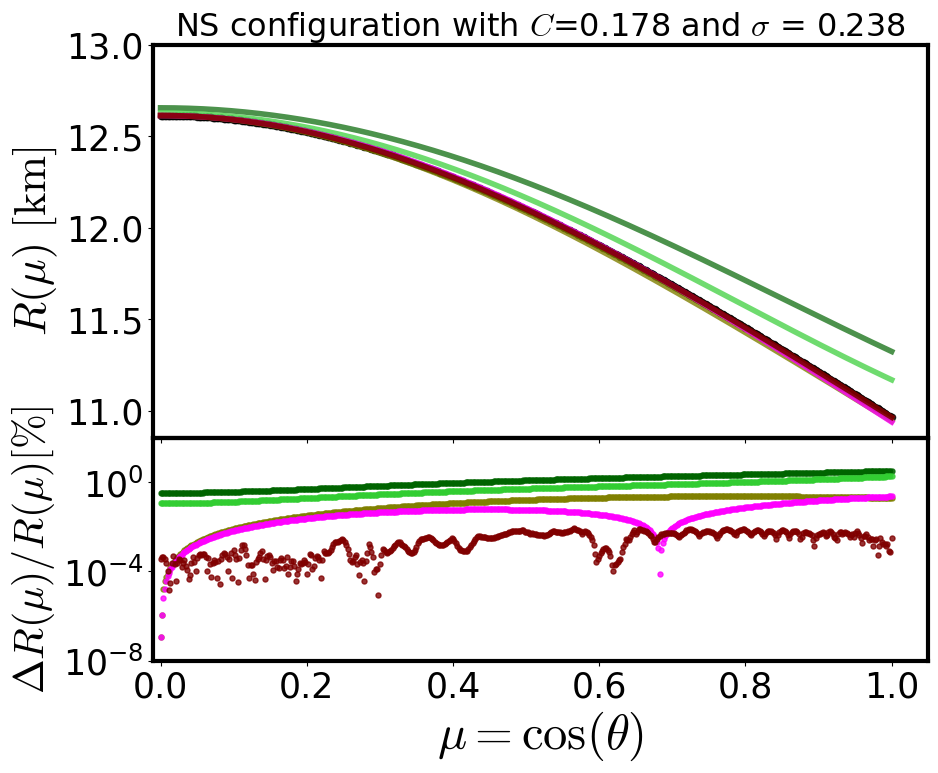}
        \includegraphics[width=0.32\textwidth]{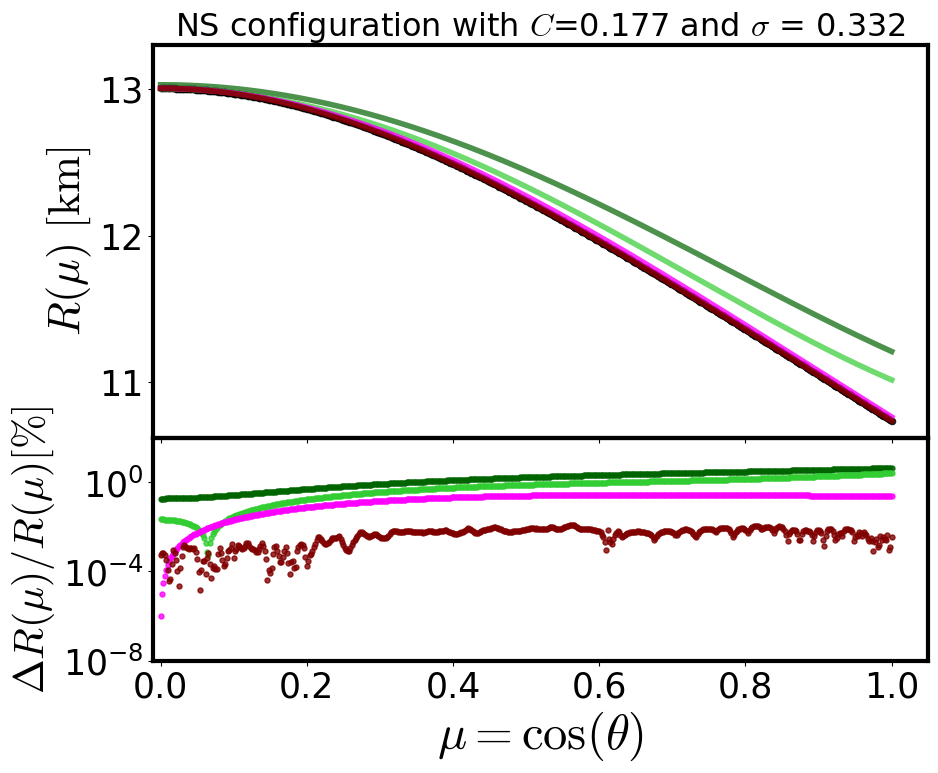}
        \includegraphics[width=0.32\textwidth]{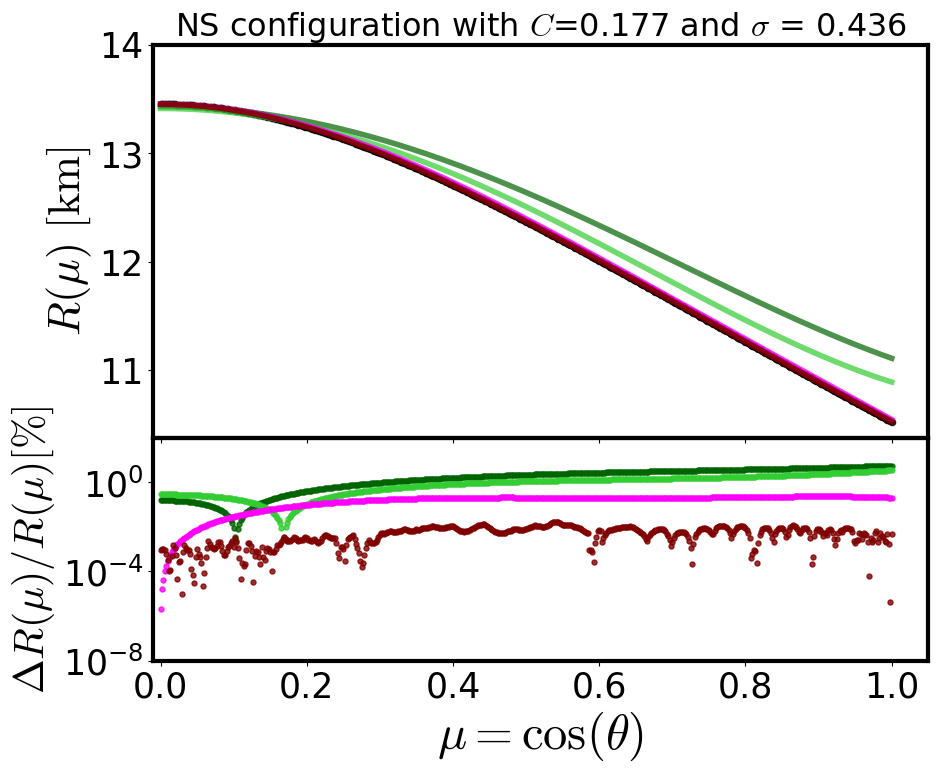}
        \includegraphics[width=0.32\textwidth]{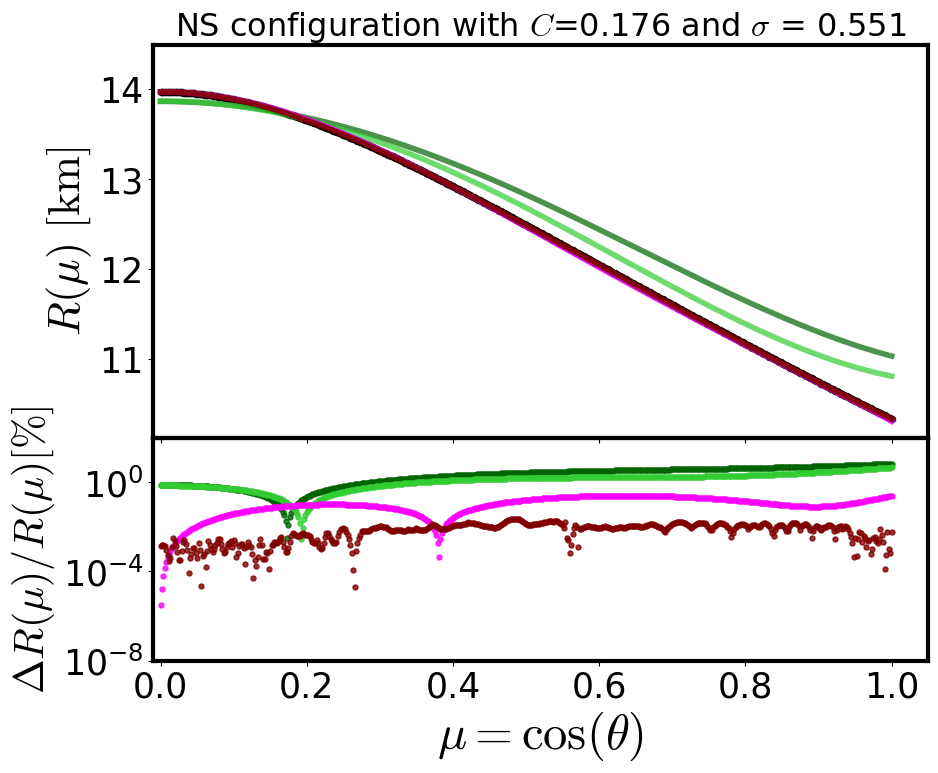}
        \includegraphics[width=0.32\textwidth]{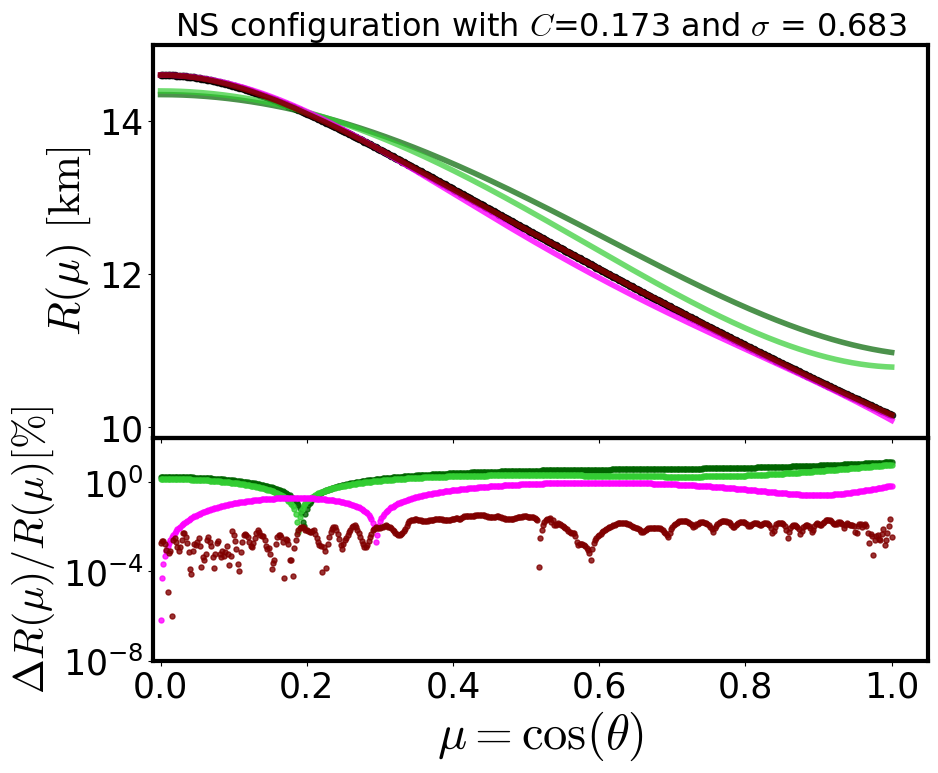}
        \includegraphics[width=0.32\textwidth]{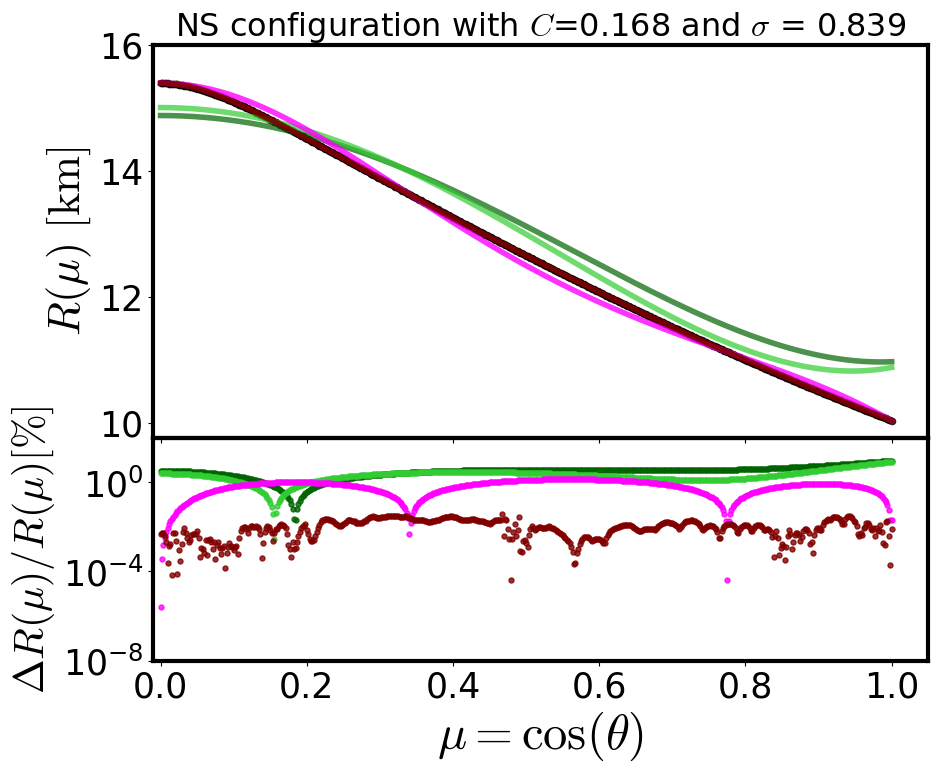}
        \caption{\label{fig:indicative_surfaces}Star's surface and relative errors given in logarithmic scale vs. angular position for the catalog of NS models presented in Table (\ref{tab:indicative_propert}). Each plot is for a different NS configuration and compares our results to previous results in the literature. The bottom panel in each plot shows the relative error for each fitting formula. Our ANN fit achieves the lowest relative errors, which are also independent of the rotation rate, in contrast to other fits.}         
\end{figure*}

As a comprehensive illustration of the effectiveness of the fitting function (\ref{eq:R_mu_fit}), Fig. \ref{fig:indicative_surfaces} presents various surfaces along with their corresponding relative errors for the whole selection of NS models outlined in Table (\ref{tab:indicative_propert}). In all cases, we incorporate existing fitting functions proposed in the literature, providing a basis for relative comparison. Each method accurately captures the static case, as expected. In the case of slowly rotating stars, the AlGendy and Morsink fit (with the contribution of Silva et al.), along with the slow elliptical fit accurately describe the star's equator at $\mu=0$. This accuracy also extends to the slow and fast elliptical fits as the rotation rate increases. Nevertheless, our fitting function excels in accurately reproducing the data. Unlike other fitting functions, our method mitigates the error accumulation as the model's predictions move towards the star's pole, thanks to effective normalization (\ref{R_mu_norm}).   

From an observational standpoint, having knowledge of the star's mass and its associated rotation frequency, one requires data for $R_{\mathrm{eq}}$ and $R_{\mathrm{pole}}$ to estimate the star's surface using the regression model (\ref{eq:R_mu_fit}). In this context, the previously derived EoS-insensitive relations for $R_{\mathrm{pole}}$ (\ref{eq:R_c_sigma}) and the star's eccentricity $e$ (\ref{eq:e_c_sigma}) play a crucial role in determining the star's surface with the suggested fitting function (\ref{eq:R_mu_fit}). These polynomial functions provide information with high accuracy when the equatorial radius $R_{\mathrm{eq}}$ is determined through observations (e.g., for NICER's telescope NS targets) or provided via a universal relation (see e.g., the relation suggested in \cite{papigkiotis2023universal}).

\subsection{\label{sec:univ_rel_dlogR}UNIVERSAL ESTIMATION OF THE SURFACE'S LOGARITHMIC DERIVATIVE USING ANN}
Apart from $R(\mu)$, the logarithmic derivative is fundamental in accurately modeling the oblate shape of a rotating NS configuration. At this point, we now proceed to employ the designed feed-forward network to predict the logarithmic derivative at the surface of the star. In this context, we utilize the parameters $\mu, C, \sigma$, and $\mathcal{R}$ as the model's input features $\tilde{x}$. In this case, we have to note the polar-to-equatorial radius $\mathcal{R}$ is a better feature than the star's eccentricity $e$ that was used in the Sec. \ref{sec:univ_rel_R}.

For a given NS configuration with spin parameter $\sigma$ within the parameter space, after obtaining the maximum value of the logarithmic derivative $\left(d\log R(\mu)/d\theta \right)$, we proceed to estimate the normalized logarithmic derivative denoted as 
\begin{equation} 
\label{dlogR_norm}
\hat{z}_{2} = \left(\frac{d\log R(\mu)}{d\theta}\right)/\left(\frac{d\log R(\mu)}{d\theta}\right)_{\mathrm{max}}.
\end{equation}
It is essential to note that the maximum value varies from one NS model to another. Recognizing that the interval boundaries, according to the constraints (\ref{log_der_constr}), should be zero in each case, the relation (\ref{dlogR_norm}) (min-max scaling) is devised to map the scaled logarithmic derivative interval associated with each NS model into the unit interval $[0,1]$. As previously, this transformation ensures a consistent representation of the data, facilitating the training process and optimizing the performance of the neural network.

\begin{figure}[!htb]
    \includegraphics[width=0.46\textwidth]{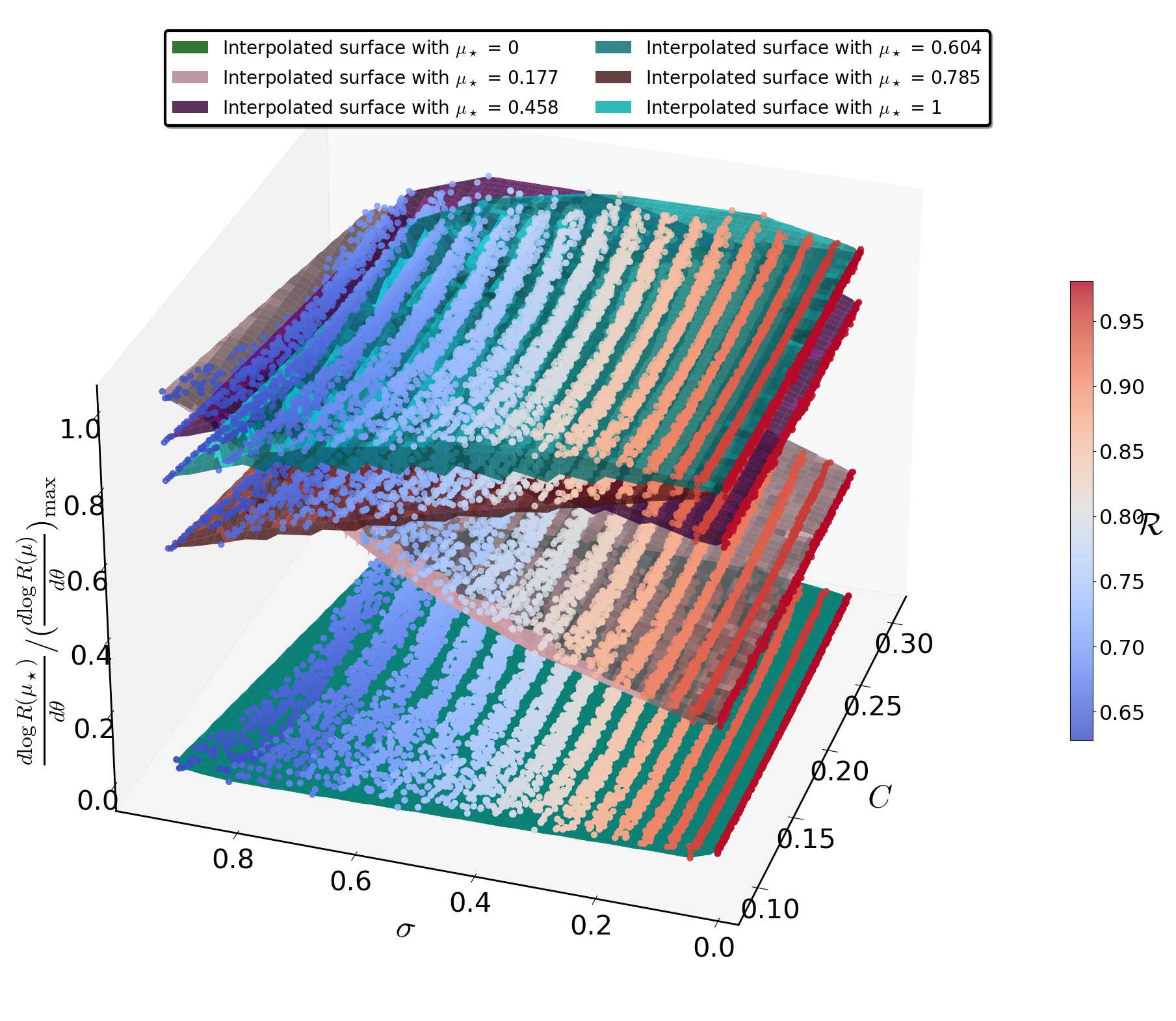}
    \caption{\label{fig:dR(mu)_univ_illustration} Universal representation: Normalized logarithmic derivative $ \frac{d\log R(\mu_\star)}{d\theta}/(\frac{d\log R(\mu)}{d\theta})_{\mathrm{max}}$ as a function of the star's global parameters $C$, and $\sigma$ for a discrete array of $\mu\in [0,1]$ values excluding the star's equator and pole which are clearly defined due to constraints (\ref{log_der_constr}). Each colored, numerically interpolated plane corresponds to an assigned $\mu_\star$ value, while the vertical colored bar represents the star's polar-to-equatorial ratio $\mathcal{R}$.}
\end{figure}

\begin{figure}[!htb]
    \includegraphics[width=0.46\textwidth]{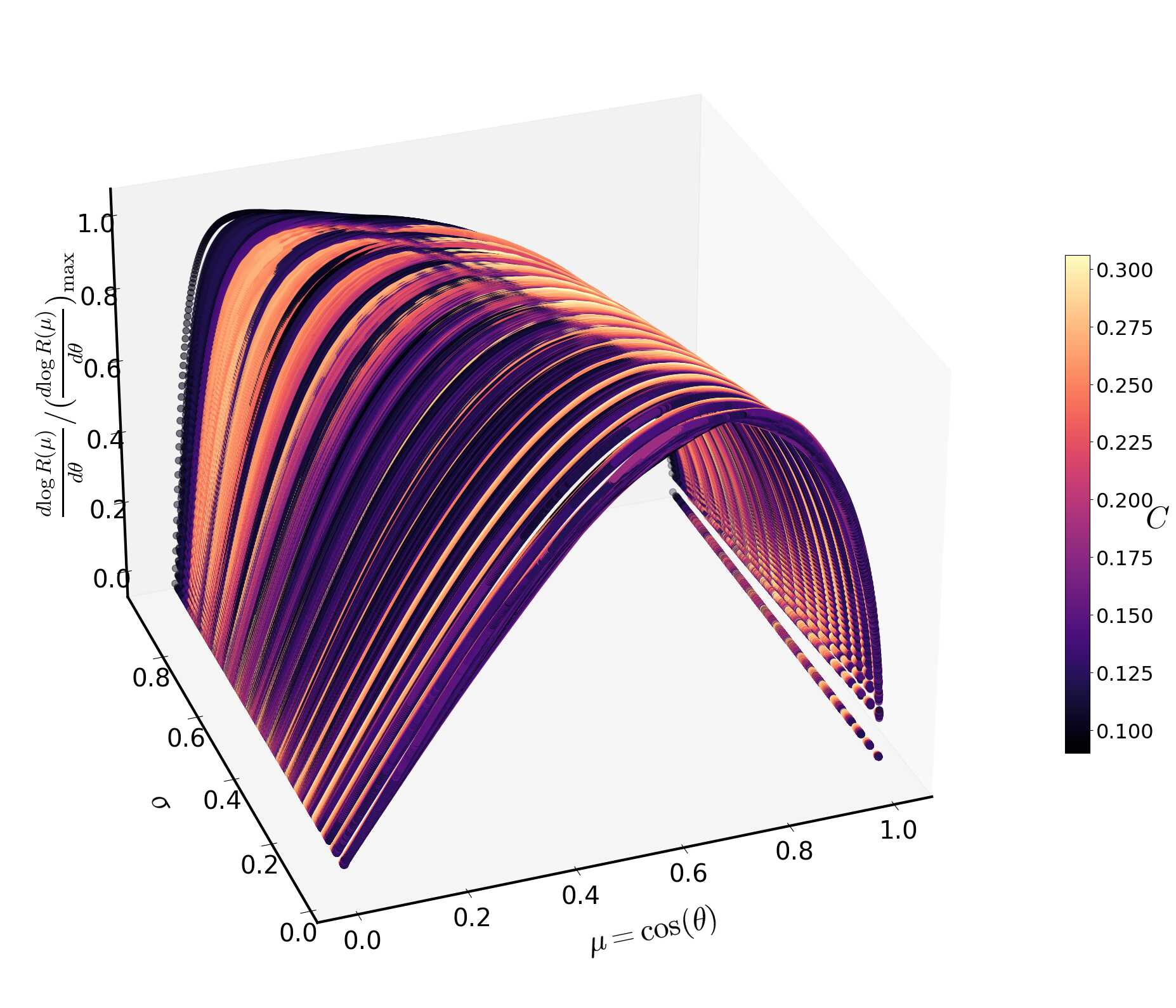}
    \caption{\label{fig:dR(mu)_univ_illustration_2} Universal parametrization: Effective normalization  (\ref{dlogR_norm}) shown as a function of the angular position  $\mu = \cos(\theta)$, and the reduced spin $\sigma$. The vertical colored bar represents the star's stellar compactness $C$. }
\end{figure}

Especially for the rotating case,  this transformation aligns all data points onto a single universal plane for each specific colatitude value, $\theta$. This behavior is depicted in Fig. \ref{fig:dR(mu)_univ_illustration}, showcasing the data distribution for a representative discrete set of $\mu$ values within the range $\mu \in [0,1]$ across the entire parameter space. In this illustration, the colored planes correspond to regions in space with fixed $\mu = \mu_\star$, while the vertical colored bar denotes the star's polar-to-equatorial ratio $\mathcal{R}$. In addition, Fig. \ref{fig:dR(mu)_univ_illustration_2} demonstrates a different morphology of the EoS-insensitive parametrization (\ref{dlogR_norm}) for each $\mu$ within the parameter space, showing the data in the test dataset over the full range of $\mu$ values across different rotation rates.

\begin{figure}[!ht]
    \centering
    \includegraphics[width=0.46\textwidth]{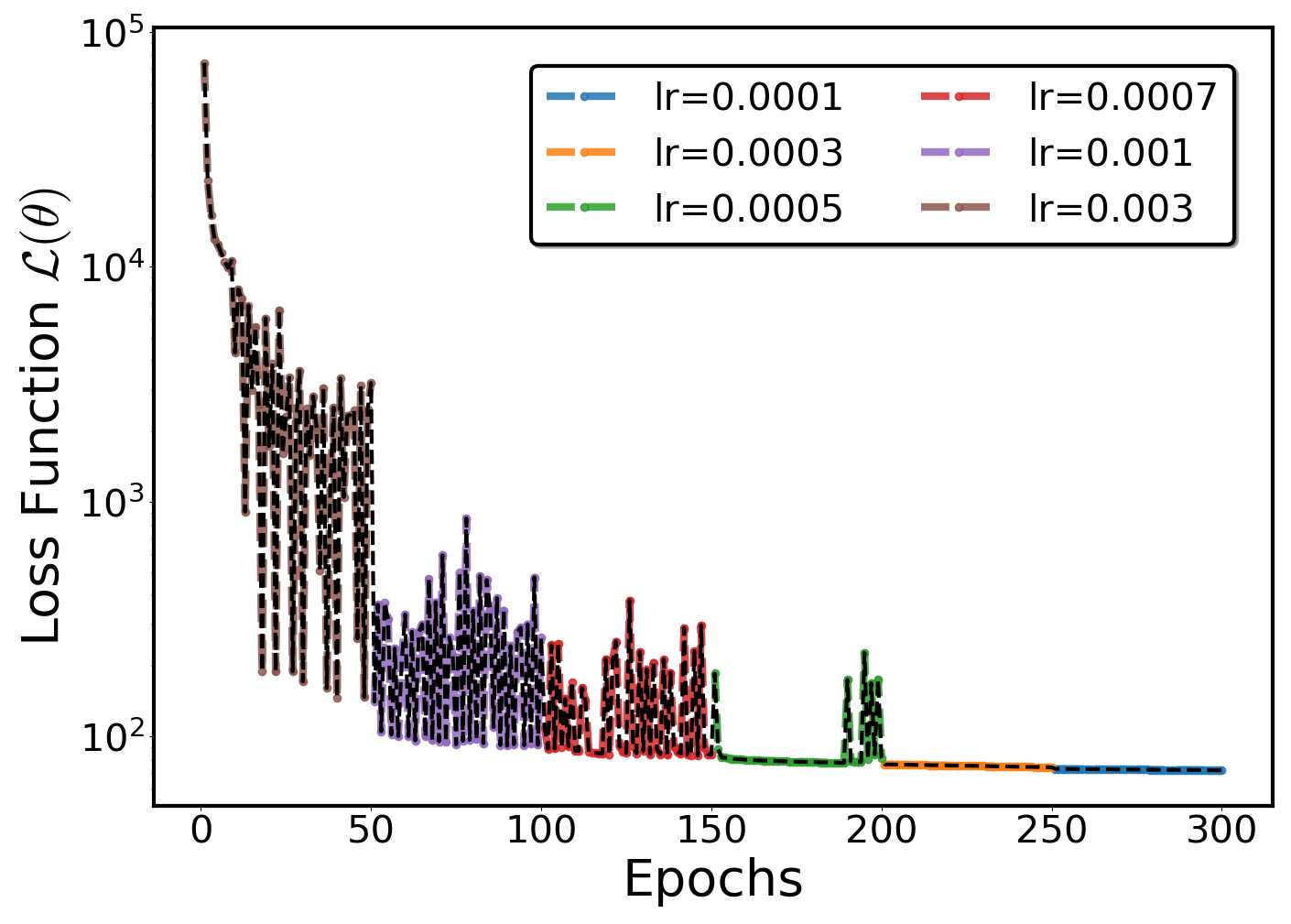}    
    \caption{\label{fig:dlogR_opt_process} Minimization of the Loss function (\ref{logR_loss}) during the model training. Each colored curve corresponds to the corresponding learning rate employed. The whole training process is carried out for 300 epochs. }
\end{figure}

After scaling the data for each particular NS configuration included within our ensemble, we proceed to train the ANN model $\hat{\mathcal{F}}_{\theta}(\mu, C, \sigma, \mathcal{R})$ to derive the best parameters $\theta^{\star}$ through optimization. At this point, it is crucial to emphasize that, to effectively capture the intricacies of data specifically at the interval boundaries and enhance the learning process, our training procedure incorporates a specialized form of the loss function given as, 
\begin{equation}
    \label{logR_loss}
    \mathcal{L}(\theta) = \sum_{i=1}^{n} w_i||z_i - \hat{z}_i||^2,
\end{equation}
where $w_i$ is a weight factor taking values
\begin{equation*}
w_i = \begin{cases} 
        10, &\text{for the star's pole and equator respectively}  \\
        1, & \text{otherwise.}
     \end{cases}
\end{equation*}
In Fig. \ref{fig:dlogR_opt_process}, we present the minimization of the loss function $\mathcal{L}(\theta)$ (\ref{logR_loss}) over 300 epochs of model training.

The regression model $d\log R(\mu)/d\theta$ with the optimal $\theta^{\star}$ parameters has the functional form
\begin{equation}
    \label{dlogR_reg_fit}
    \left(\frac{d\log R(\mu)}{d\theta}\right) = \left(\frac{d\log R(\mu)}{d\theta}\right)_{\mathrm{max}} \ \hat{\mathcal{F}}_{\theta^{\star}}(\mu, C, \sigma, \mathcal{R}).
\end{equation}
The proposed regression formula accurately reproduces the star’s deviation from sphericity as measured by the logarithmic derivative for each value of the spin-induced deformation $\sigma$ within the parameter space.    

The evaluation measures for the fitting formula (\ref{dlogR_reg_fit}) on the test set are highlighted in Table (\ref{tab:dlogR_mu_eval_meas}). 
\begin{table}[!ht]
\caption{\label{tab:dlogR_mu_eval_meas} Evaluation measures for the parametrization given by equation (\ref{dlogR_reg_fit}) on the test set.}
    \begin{ruledtabular}
        \begin{tabular}{ccccc}
            MAE & Max Error & MSE  & Exp Var & $R^2$ \\
            $\times 10^{-4}$ & $\times 10^{-3}$ & $\times 10^{-7}$  &   \\
            \hline
            2.658  & 8.360 & 3.187 &  0.99999 & 0.99999  \\
        \end{tabular}
    \end{ruledtabular}
\end{table}
We have to note that we intentionally omit the $d_{\mathrm{max}}$ and $\mathrm{MAPE}$ evaluation measures from this table (\ref{tab:dlogR_mu_eval_meas}). This decision is based on the observation that these evaluation functions tend to produce extremely high values for static cases where $R(\mu)$ is constant or for rotating stars where the interval boundaries are constrained by Eq.(\ref{log_der_constr}). In both cases, the associated derivative in the denominator coincides with zero.

\begin{figure}[!ht]
    	\centering
        \includegraphics[width=0.46\textwidth]{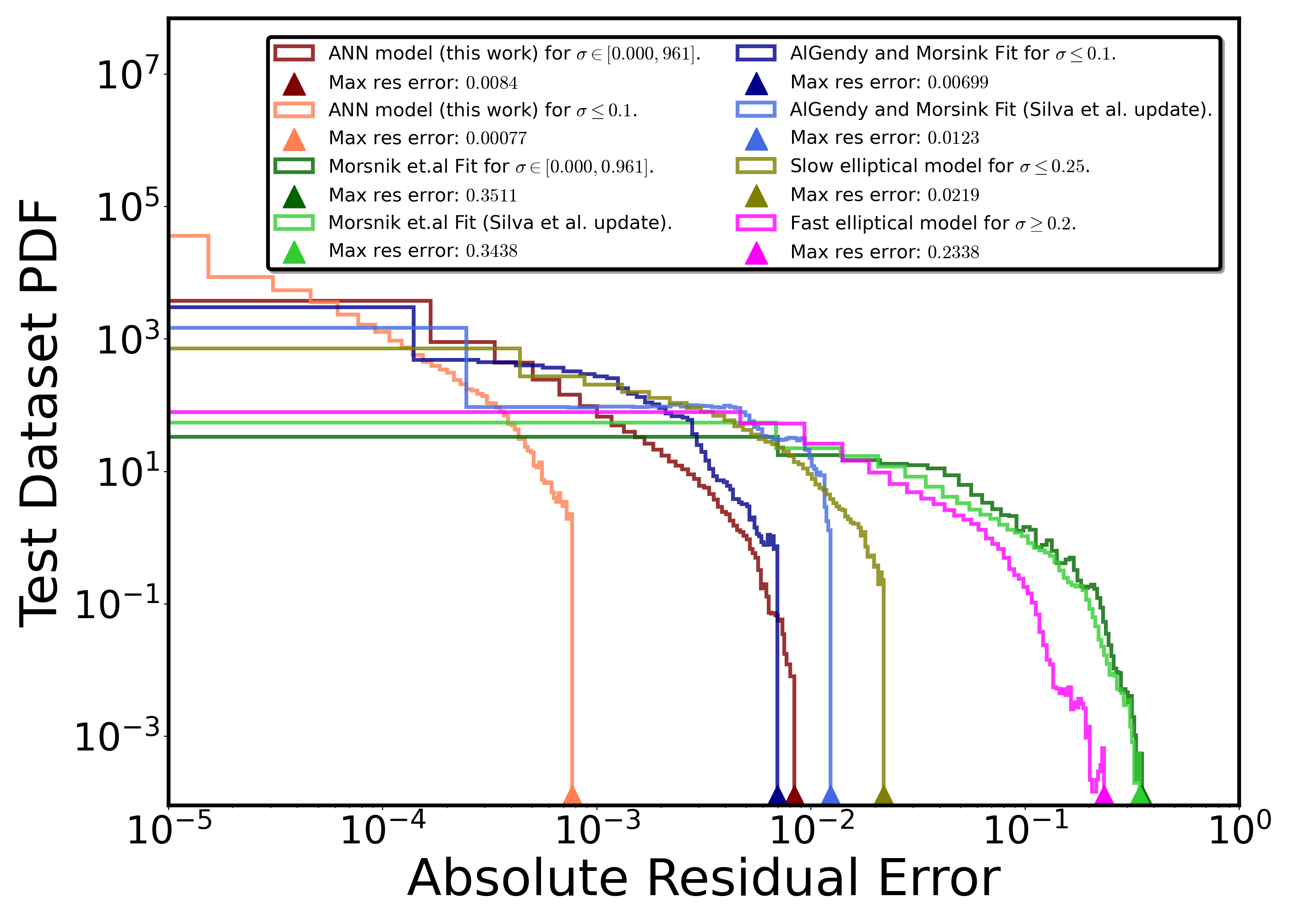}
        \caption{\label{fig:Rsiduals_plot} 
        Colored histograms depicting the distribution of absolute residual errors $\left(d\log R(\mu)/d\theta \right)_{\mathrm{fit}} - \left(d\log R(\mu)/d\theta \right)$, comparing the proposed ANN model's results against the logarithmic derivative models derived from fitting functions previously discussed in the literature using Eq.(\ref{log_der}). 
        Each histogram refers to the test set, while the colored triangles denote the absolute maximum relative error produced by each model to verify the data.}         
\end{figure}

Furthermore, Fig. \ref{fig:Rsiduals_plot}
displays a histogram distribution on the test set, illustrating the absolute residual errors between our fitting function (\ref{dlogR_reg_fit}) and the actual data.
Employing this parametrization, the regression model (\ref{dlogR_reg_fit}) accurately verifies the data on the test set, exhibiting a high precision with a residual error of less than $8.36\times 10^{-3}$. For each NS model within our ensemble, the associated formulation demonstrates perfect generalization ability and is insensitive to the star's EoS. In addition, Fig. \ref{fig:Rsiduals_plot}
demonstrates the residual error comparison between our regression model (\ref{dlogR_reg_fit}) and the models derived using Eq.(\ref{log_der}) for the surface fitting functions that are already available in the literature for both slowly and rapidly rotating stellar models \cite{silva2021surface,morsink2007oblate,algendy2014universality}. For slowly rotating NS configurations with $\sigma \leq 0.1$, the fitting curve derived from the AlGendy and Morsink fitting function produces results with a residual error of $\lesssim 7 \times 10^{-3}$, while applying the same formula with coefficients from Silva et al. results in a residual error of less than $1.23 \times 10^{-2}$. In comparison, for the same range spin case, our model achieves significantly lower residual errors, at $\leq 0.77 \times 10^{-3}$. In any case, regardless of the star's rotation, the proposed formula (\ref{dlogR_reg_fit}) surpasses the performance of the previously established fitting models employed to extract the logarithmic derivative. Hence, it provides a significant enhancement in the estimation of the measurement of the deviation from the sphericity of the star’s surface, irrespective of EoS.

A crucial aspect to investigate is the source of the maximum relative error observed in the regression model (\ref{dlogR_reg_fit}) on the test set, considering both the overall EoS categories and individual EoSs associated with a specific category. In Appendix \ref{sec:regression_violin_dlog_R_mu}, we present violin plots that display the distribution of absolute residuals for each case in the test set, enabling a thorough evaluation of the model’s performance and the variability in its predictions. The hybrid class of EoSs has the largest residuals, around $ 8.36 \times{10^{-3}}$ (Max Error), while the hadronic and hyperonic EoS-classes demonstrate residual deviations of about $ 6.3 \times{10^{-3}} $ and $5.2 \times{10^{-3}} $, respectively. The EoS models exhibiting the largest residual differences, as compared to others within the same category, are the the Holographic V-QCD model APR intermediate \cite{akmal_equation_1998, jokela2019holographic, ishii2019cool, ecker2020gravitational, jokela2021unified}, the RDF model QMC-RMF4 \cite{typel1999relativistic, xia2022unified, xia2022unified_2}, and the SU$(3)$-CMF model DNS \cite{dexheimer_proto-neutron_2008,dexheimer_reconciling_2015,dexheimer_tabulated_2017,schurhoff_neutron_2010} (see e.g., Appendix \ref{app:eos_tables} and Fig. \ref{fig:dlogR_mu_violins} for a review).

\begin{figure*}[!ht]
    \centering
    \includegraphics[width=0.9\textwidth]{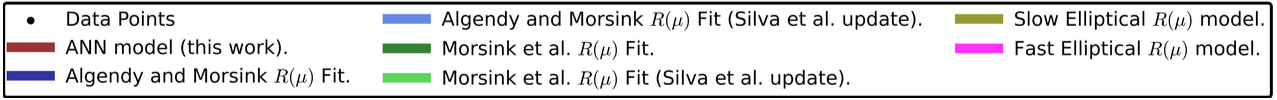}\\
    \centering
    \includegraphics[width=0.32\textwidth]{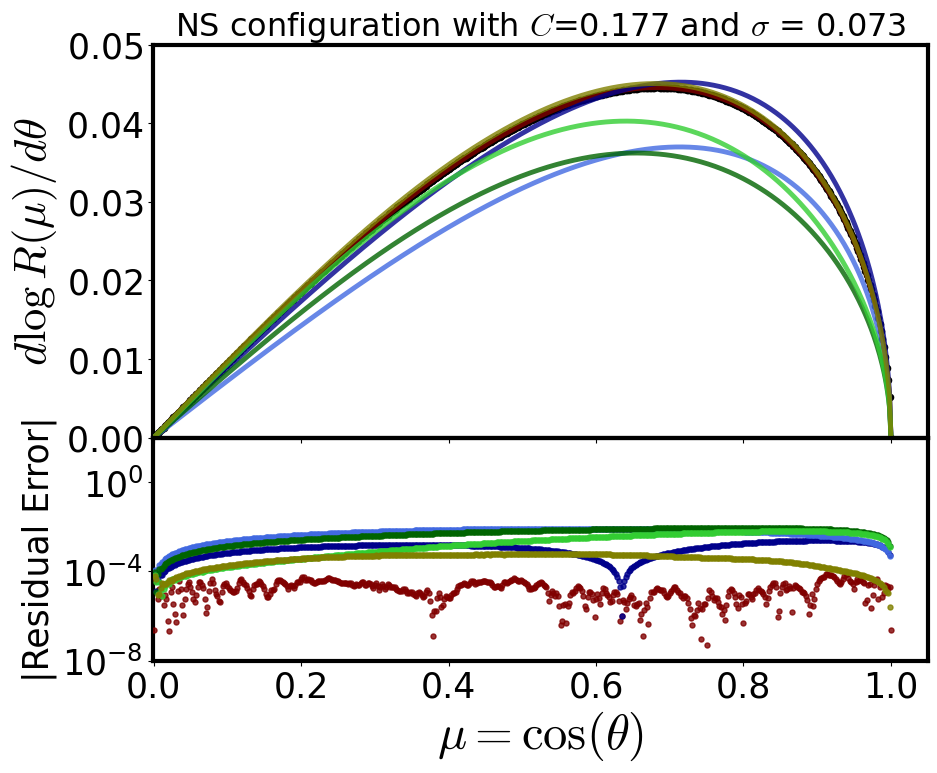}
    \includegraphics[width=0.32\textwidth]{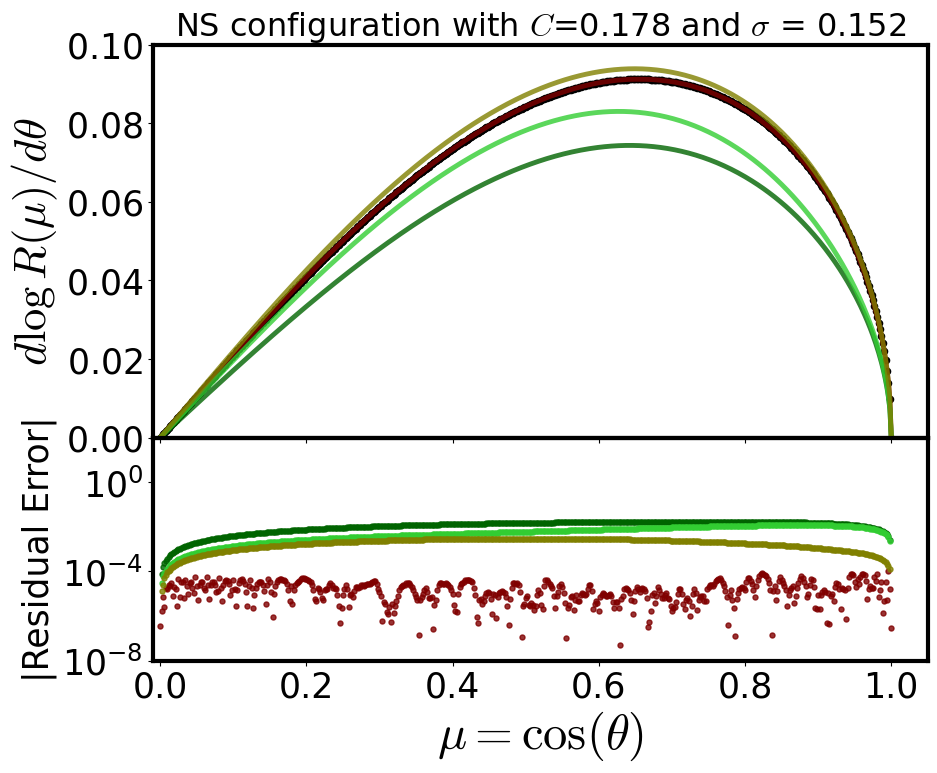}
    \includegraphics[width=0.32\textwidth]{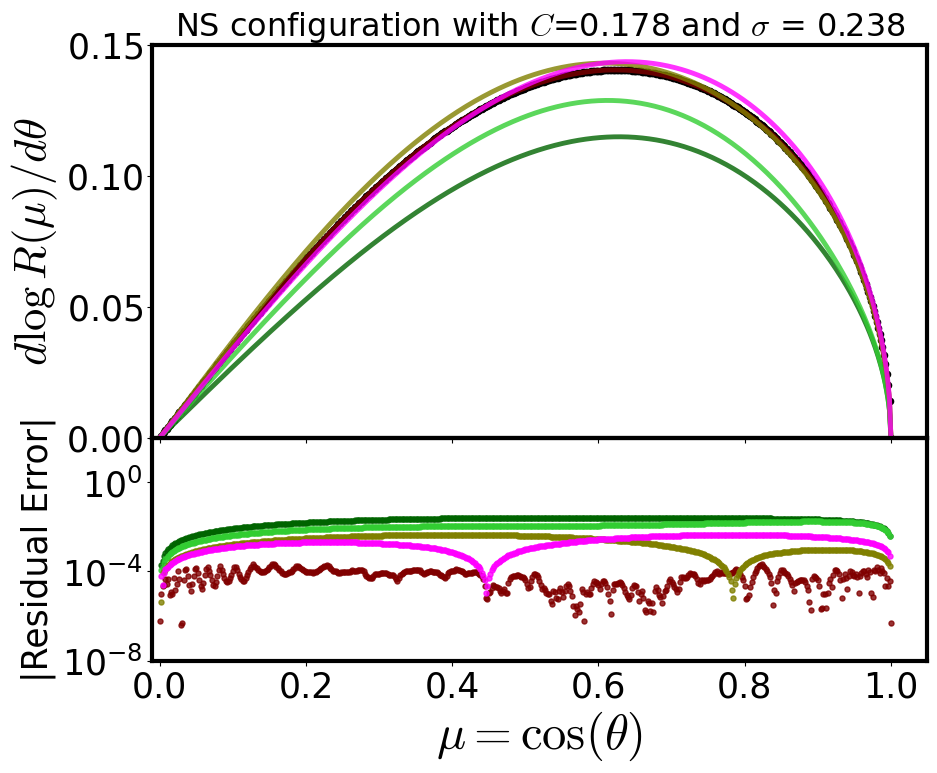}
    \includegraphics[width=0.32\textwidth]{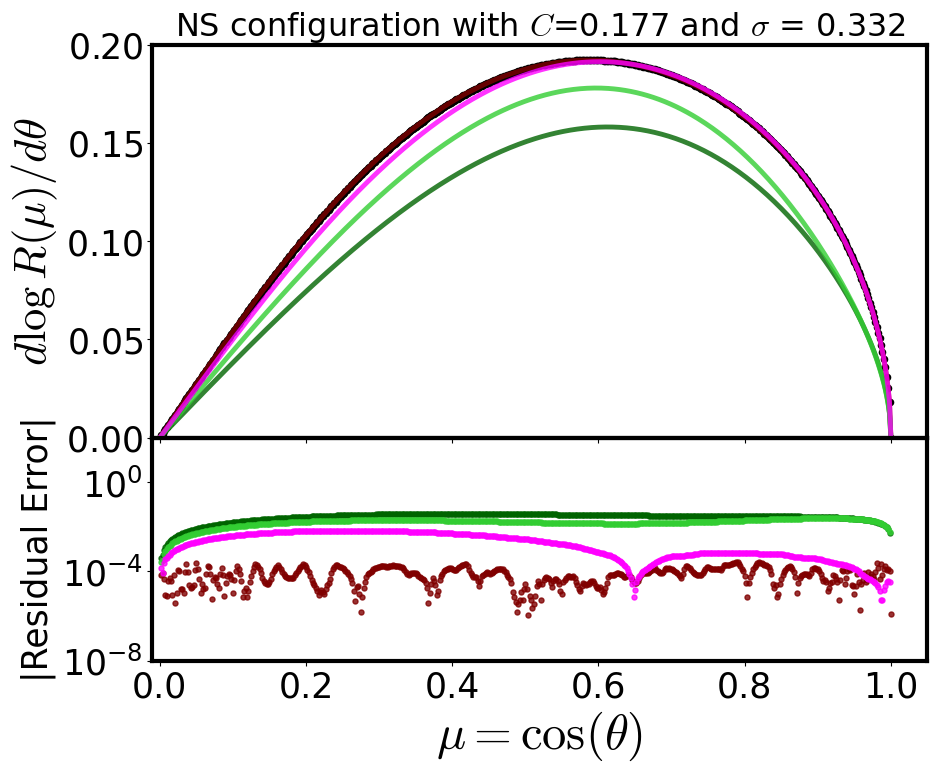}
    \includegraphics[width=0.32\textwidth]{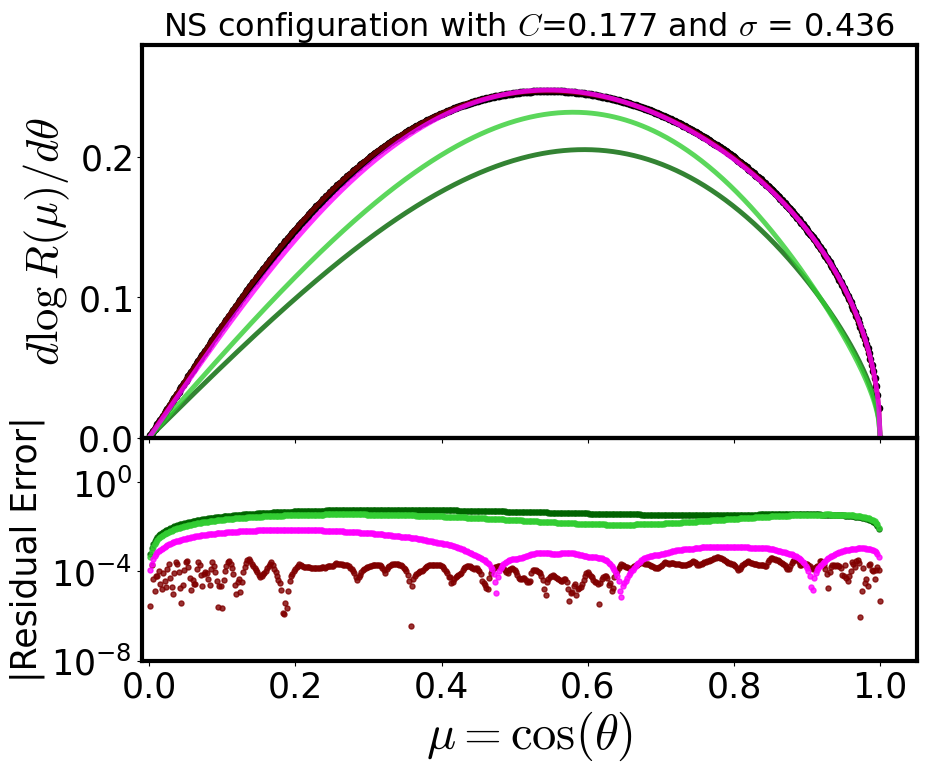}
    \includegraphics[width=0.32\textwidth]{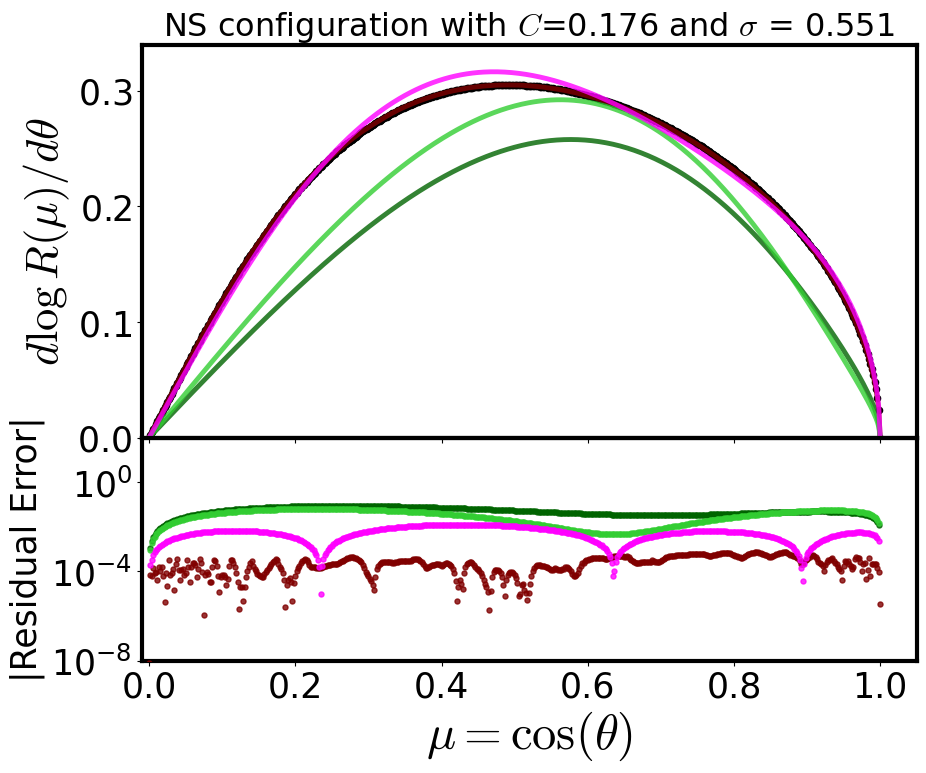}
    \includegraphics[width=0.32\textwidth]{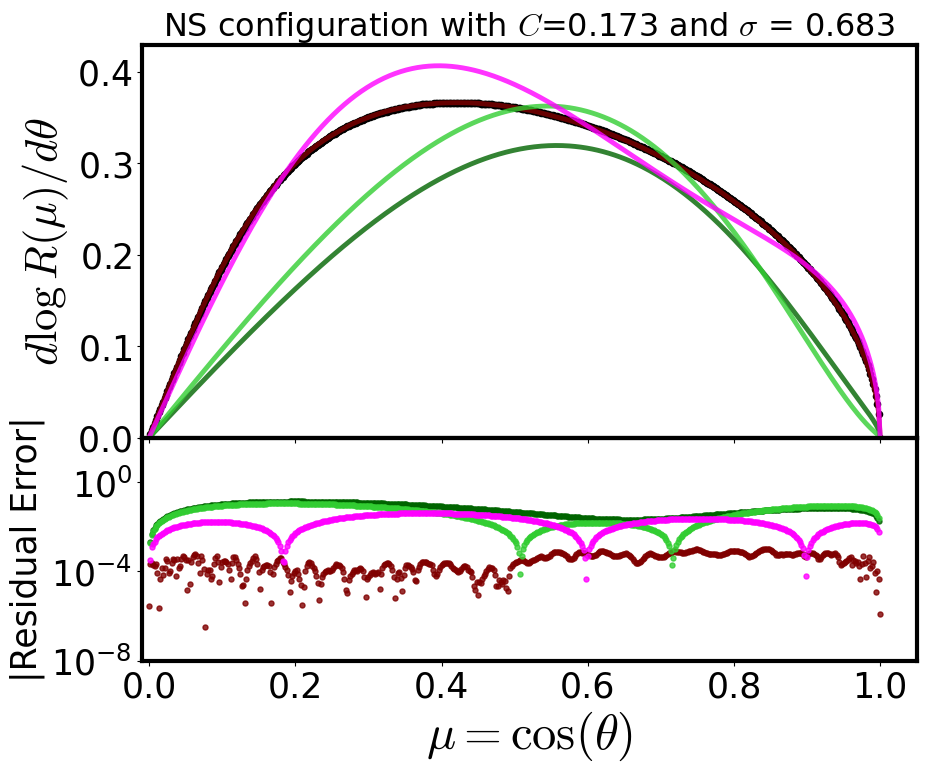}
    \includegraphics[width=0.32\textwidth]{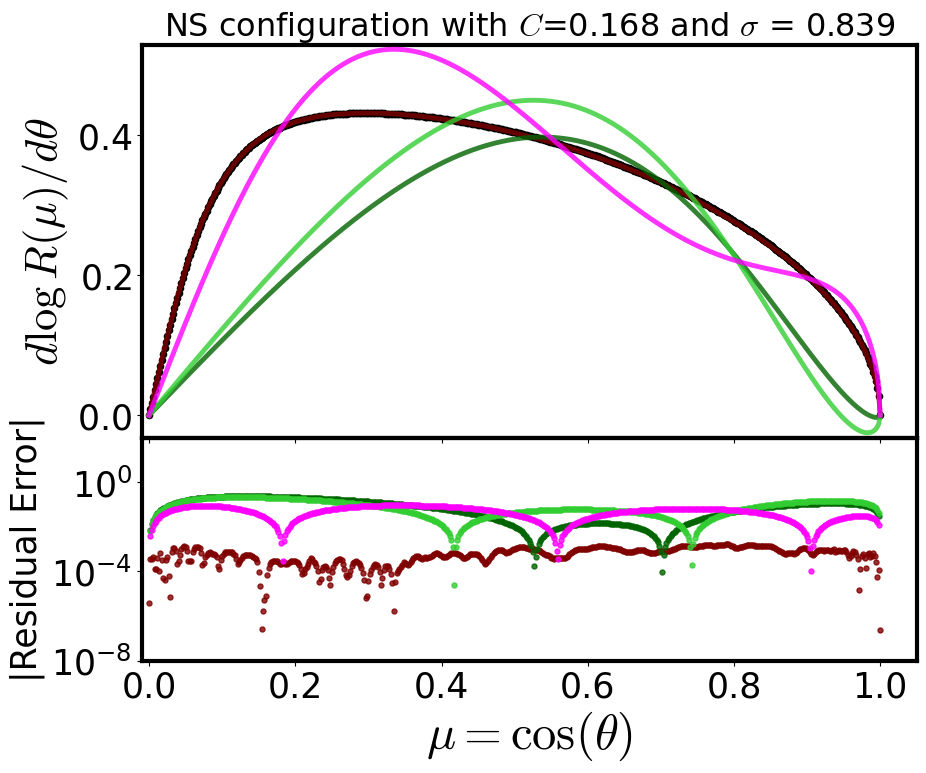}
    \caption{\label{fig:indicative_log_derivatives} Angular derivative and absolute residual errors as a function of $\mu=\cos(\theta)$. for the set of rapidly rotating NS models presented in Table (\ref{tab:indicative_propert}). The colored curves correspond to different fits. Apart from the proposed ANN model, the residual errors in the other fitting models escalate with the increase in reduced spin $\sigma$. }

\end{figure*}

As an additional presentation of the effectiveness of the regression model (\ref{dlogR_reg_fit}), Fig. \ref{fig:indicative_log_derivatives} illustrates various logarithmic derivative curves and their corresponding relative errors for the whole sample of rotating NS benchmark models outlined in Table (\ref{tab:indicative_propert}). In the same Figure, we also include the corresponding curves derived by acquiring the established fitting functions associated with the star's surface $R(\mu)$, enabling a basis for direct comparison. 
Nevertheless, the proposed ANN model excels by accurately reproducing the data. As in the case of Fig. \ref{fig:indicative_surfaces}, our results have the lowest relative errors and, moreover, do not increase for higher rotation ratios, in contrast to previously published fits.

It is important to note that the already established models used for comparison rely on the existence of a fitting function $R(\mu)$ that represents the star's surface. As a result, the corresponding logarithmic derivative is derived indirectly from Eq.(\ref{log_der}). {Except for our new fitting function (\ref{dlogR_reg_fit}), to the best of our knowledge, there is currently no other general fitting function in the existing literature dedicated to the direct universal estimation of this particular quantity.} Moreover, for the sample of NS configurations covering the parameter space, our ANN model accurately reproduces the data regardless of the star's spin parametrization. This outcome is paving the way for accurate calculations, particularly in determining the beaming angle for a photon emitted at the star's surface \cite{baubock2012ray,cadeau2007light}.

Finally, beyond the consideration of global parameters like $C, \sigma$, and $\mathcal{R}$, practical utilization of the regression function (\ref{dlogR_reg_fit}) demands knowledge of the maximum value of the logarithmic derivative for each specific NS configuration. From an observational point of view, this requirement is adeptly addressed through the utilization of the derived EoS-insensitive polynomial function (\ref{eq:dlogR_c_sigma_Rp_Re}), as highlighted in the last part of the Sec. \ref{sec:poly_methods_1}.

\subsection{\label{sec:univ_rel_g}UNIVERSAL ESTIMATION OF THE STAR'S EFFECTIVE GRAVITY USING ANN}

The hotspot models utilized in NICER data analysis are based on the assumption that the NS's atmosphere is mainly composed of Hydrogen \cite{heinke2006hydrogen}. The characteristics of Hydrogen atmospheres and other possible atmospheric compositions depend on the local acceleration due to gravity. Therefore, investigating an EoS-insensitive relation for the local star's surface gravity $g(\mu)$ is highly important.

Building upon this, our focus shifts towards leveraging the designed ANN architecture for the estimation of the effective acceleration due to gravity at the star's surface, $g(\mu)$. To achieve this objective, we employ the parameters $|\mu|, C, \sigma$, and $e$ as input features for the model.

Within a given sequence of data points representing the star's effective gravity, we subsequently proceed to calculate the scaled effective acceleration, denoted as 
\begin{equation}
    \label{g_normalization}
    \large
    \hat{z}_{3} =
    \begin{cases}
    \ \frac{g(\mu) - g_{\mathrm{pole}}}{g_{\mathrm{eq}}-g_{\mathrm{pole}}}, \ \ \sigma \neq 0 \\
    \\
    \ \frac{g(\mu)}{g_{\mathrm{eq}}},  \ \ \sigma = 0.
    \end{cases}
\end{equation}
Again, this formulation is performed individually for each star within our ensemble. In the absence of rotation, this linear transformation ensures that $g(\mu)=g_{\mathrm{eq}}$, while for the rotating case, it maps the interval $[g_{\mathrm{eq}}, g_{\mathrm{pole}}]$ into the unit interval $[1,0]$. As discussed in previous sections, we can leverage this observation to improve the ANN model's performance, thereby improving the learning process.

\begin{figure}[!htb]
    \includegraphics[width=0.46\textwidth]{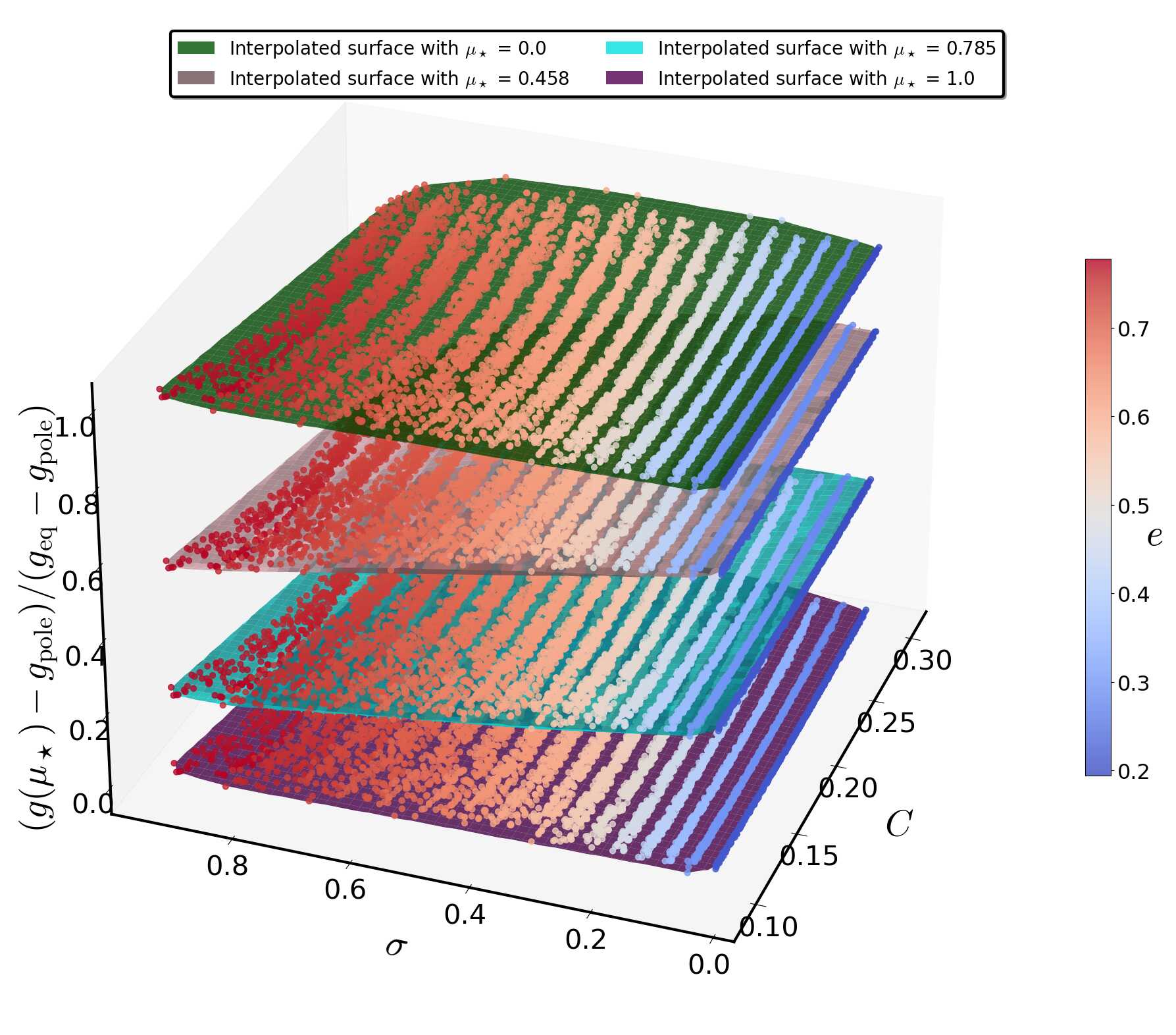}
    \caption{\label{fig:g(mu)_univ_illustration} EoS-insensitive representation: Normalized effective gravity $(g(\mu_\star)-g_{\mathrm{pole}})/(g_{\mathrm{eq}}-g_{\mathrm{pole}})$ as a function of the star's global parameters $C$, and $\sigma$ for a discrete array of $\mu\in [0,1]$ values moving from the rotating star's equator ($\mu = 0$) to the star's pole ($\mu = 1$). Each colored numerically interpolated surface corresponds to an assigned $\mu_{\star}$ value, while the vertical colored bar represents the star's eccentricity $e$. }
\end{figure}

\begin{figure}[!htb]
    \includegraphics[width=0.46\textwidth]{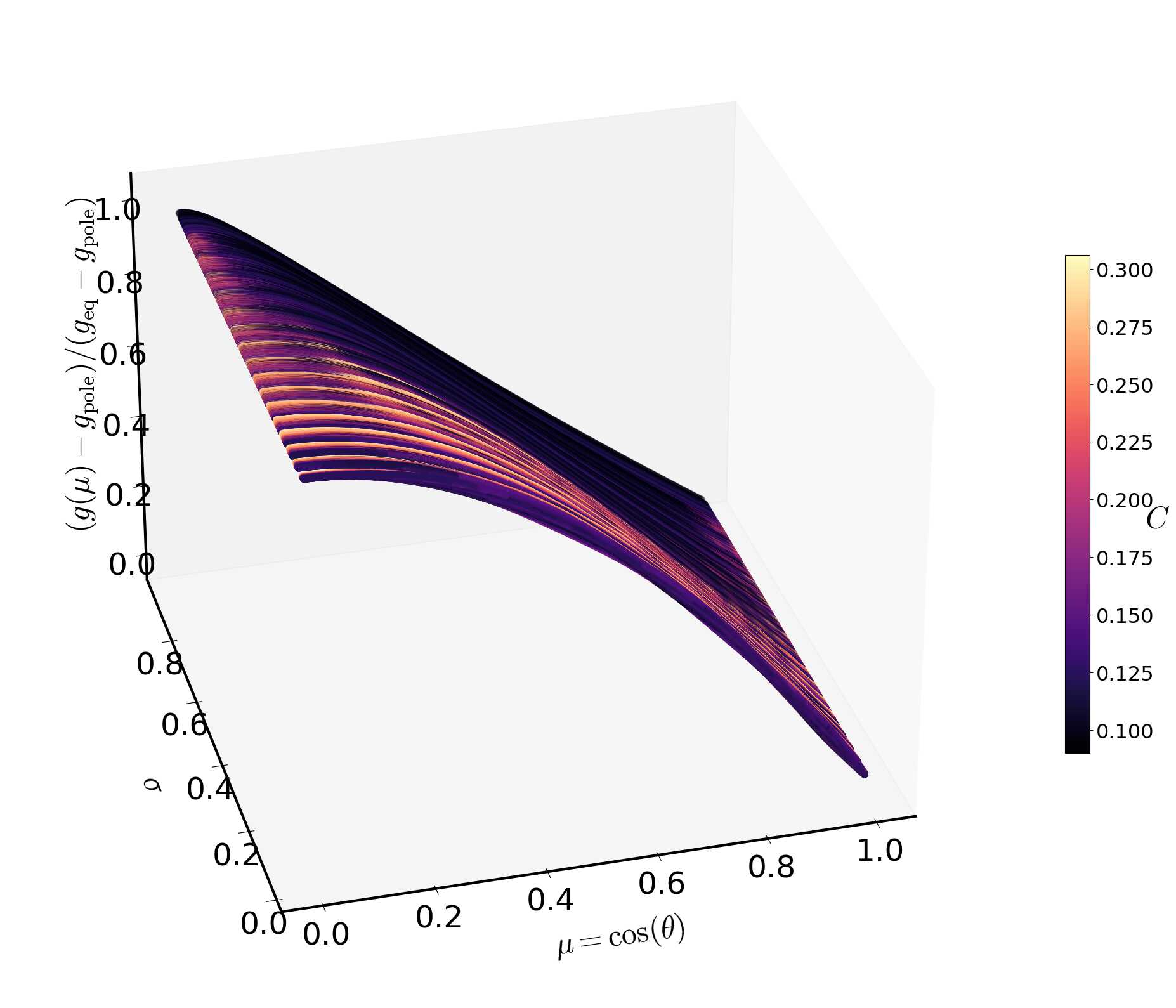}
    \caption{\label{fig:g(mu)_univ_illustration_2} EoS-insensitive parametrization: Normalized radius $(g(\mu)-g_{\mathrm{pole}})/(g_{\mathrm{eq}}-g_{\mathrm{pole}})$ as a function of 
    the angular position parameter on the star $\mu = \cos(\theta)$, and the dimensionless spin $\sigma$.
    The vertical colored bar represents the star's stellar compactness $C$.}
\end{figure}

For the rotating case, as one moves from the star's equator to its pole, the transformation (\ref{g_normalization}) remains independent of the EoS, aligning all data onto a nearly universal plane for each specific colatitude $\theta$. This is illustrated in Fig. \ref{fig:g(mu)_univ_illustration}, where the data is shown for a representative array of values $\mu \in [0,1]$ throughout the parameter space. In the figure, colored planes represent regions with fixed $\mu = \mu_\star$, while the vertical colored bar indicates the star's eccentricity. In total, the data for each EoS collectively establish an EoS-insensitive hyperstructure within the investigated parameter space. Lastly, Fig. \ref{fig:g(mu)_univ_illustration_2} demonstrates an equivalent visualization of the EoS-independent normalization (\ref{g_normalization}) for each $\mu$ within the parameter space, illustrating the data for each EoS in the test dataset over the full space of $\mu$ values across different spin rates.

\begin{figure}[!ht]
    \centering
    \includegraphics[width=0.46\textwidth]{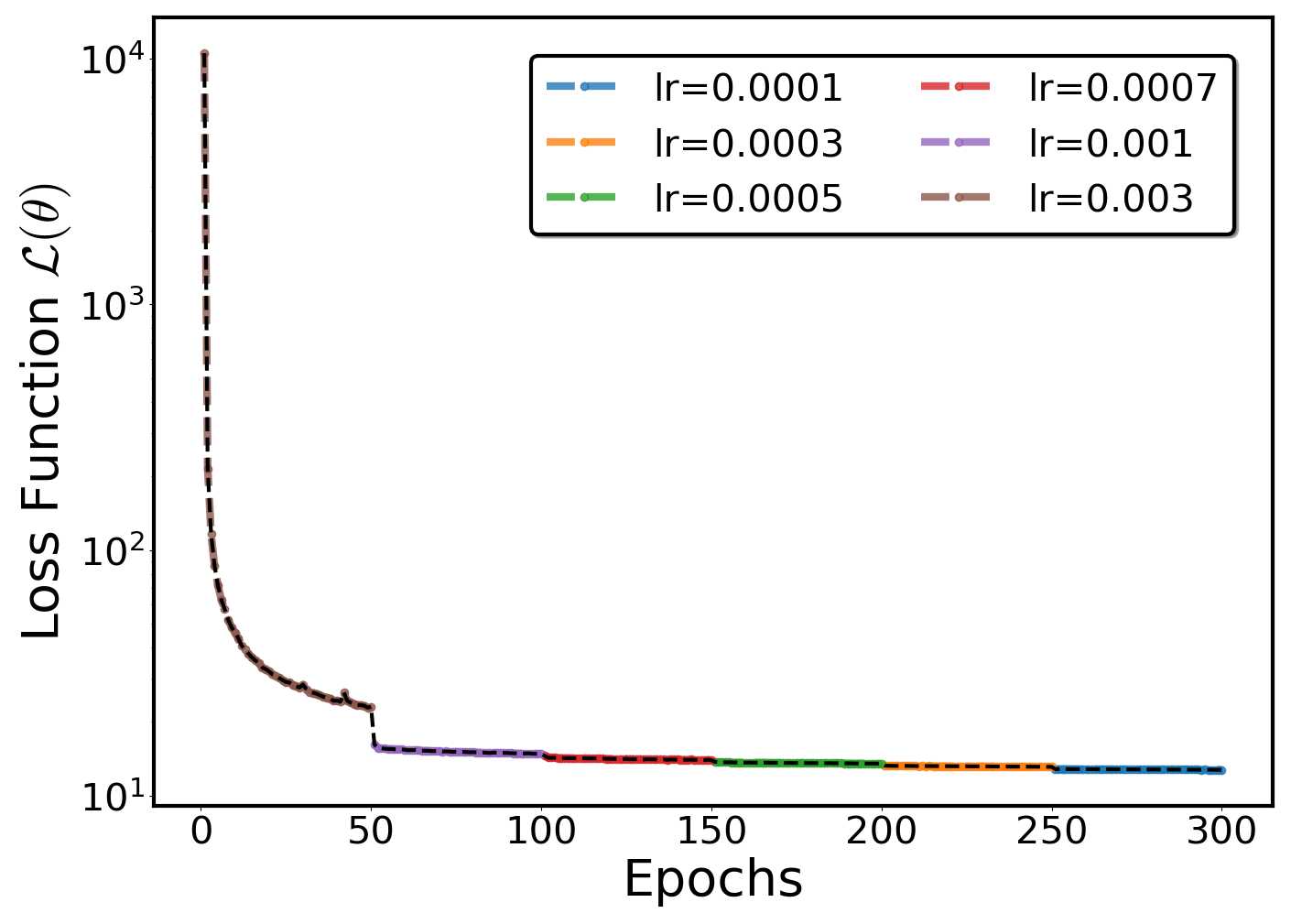}   
    \caption{\label{fig:effetive_grav_opt_process} Normalized effective gravity inference using ANN: Illustration of the Loss function minimization results derived during the ANN model training. Each colored curve corresponds to the corresponding learning rate employed. The whole training process is carried out for 300 epochs.}
\end{figure}

Following the derivation of data targets $\hat{z}_{3}$ for each NS configuration, we further advance to train the ANN model $\hat{\mathbb{F}}_{\theta}(|\mu|, C, \sigma, e)$ to extract the optimal parameters $\theta^{\star}$. In Fig.  \ref{fig:effetive_grav_opt_process}, we present the minimization of the loss function $\mathcal{L}(\theta)$ over 300 epochs of model training.

The regression formula having the optimal $\theta^{\star}$ parameters has the functional form
\begin{equation}
\label{eq:g_mu_fit}
g(\mu) = g_{\mathrm{pole}} + (g_{\mathrm{eq}} - g_{\mathrm{pole}})\hat{\mathbb{F}}_{\theta^{\star}}(|\mu|, C, \sigma, e).
\end{equation}
Incorporating $|\mu|$ as an input feature ensures the maintenance of $\mathbb{Z}_2$ symmetry across the star's surface, ensuring that $g(\mu) = g(-\mu)$ along the rotation axis. The suggested fitting formula perfectly infers the star's effective gravity at the surface for any spin-induced deformation $\sigma$. Thus, along the star's oblate topology, it offers an EoS-insensitive representation for the effective acceleration due to gravity for each $\mu$ value referring to the star's surface. It is important to highlight that in the absence of rotation ($\sigma = 0$), our formula precisely satisfies the consistency condition $g(\mu) = g_{\mathrm{pole}} = g_{\mathrm{eq}}$ (static case) for all $\mu$ values.

The corresponding evaluation measures for the fitting function (\ref{eq:g_mu_fit}) on the test set are presented in Table (\ref{tab:g_mu_eval_meas}).
\begin{table}[!ht]
    \footnotesize
    \caption{\label{tab:g_mu_eval_meas} Evaluation measures for the parametrization given by equation (\ref{eq:g_mu_fit}) on the test set.}
    \begin{ruledtabular}
        \begin{tabular}{ccccccc}
            MAE & Max Error & MSE & $d_{\text{max}}$ & MAPE & Exp Var & $R^2$ \\
            $\times 10^{-4}$ & $\times 10^{-3}$ & $\times 10^{-7}$ &  ($\%$) & $\times 10^{-4}$ ($\%$) & &   \\
            \hline
            3.380  & 7.730 & 4.512 & 0.91 & 3.436& 0.99999 & 0.99999  \\
        \end{tabular}
    \end{ruledtabular}
\end{table}

\begin{figure}[!ht]
    \centering
    \includegraphics[width=0.46\textwidth]{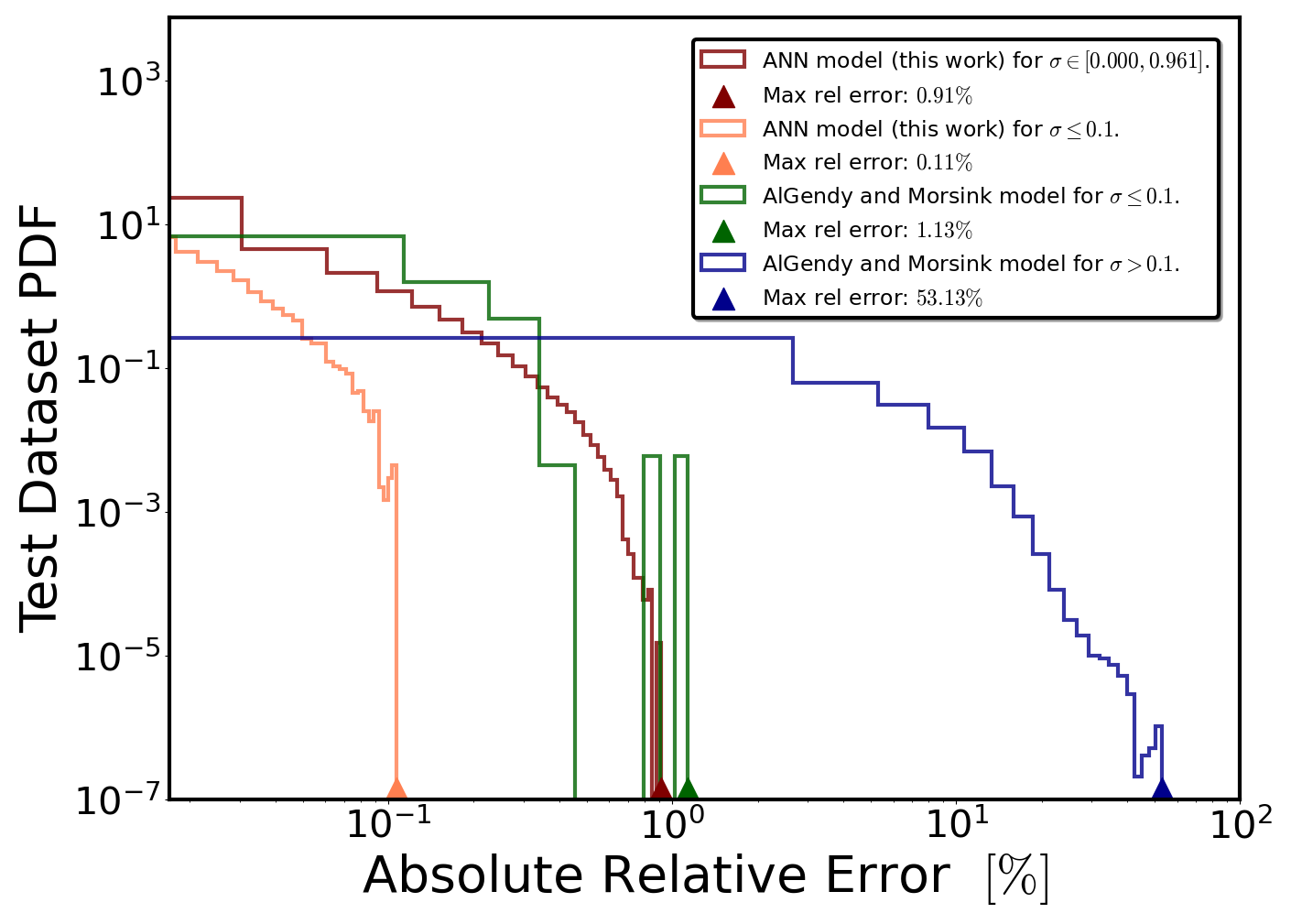}
    \caption{\label{fig:g_mu_rel_err_hist.png} Absolute value of the relative error $(100\% \times |g(\mu)_{\mathrm{fit}} - g(\mu)|/g(\mu))$ using the suggested ANN model (\ref{eq:g_mu_fit}) to estimate the star's effective gravity at surface $g(\mu)$ (maroon-colored histogram). 
    Depending on the star's rotation, the other histograms depict the distribution of absolute relative errors derived using the fitting functions provided in the literature. Each histogram is referred to the test set, while the colored triangles denote the absolute maximum relative deviation produced by each functional form used to verify the data.}
\end{figure}

Fig. \ref{fig:g_mu_rel_err_hist.png} demonstrates the associated histograms highlighting relative errors on the test set, offering a comparative analysis between our regression function and those suggested in \cite{algendy2014universality} for slowly and rapidly rotating NSs (AlGendy and Morsink fits).
We have to note that the suggested regression model (\ref{eq:g_mu_fit}) remarkably reproduces the data on the test set, showcasing relative error precision less than $0.91 \%$ for arbitrary rotation within the parameter space. When evaluated on the subset of data corresponding to slowly rotating NSs with $\sigma \leq 0.1$, our model demonstrates even higher accuracy, with relative errors not exceeding $0.11\%$. Hence, for each NS model included in our ensemble, our model demonstrates an excellent generalization ability. In addition, the proposed method surpasses the results obtained using the fitting functions proposed in \cite{algendy2014universality} for both slowly and rapidly rotating NS models.

A key aspect to examine is the origin of the maximum relative deviation exhibited by the proposed regression model (\ref{eq:g_mu_fit}) on the test set, analyzed both across EoS categories and for individual EoSs. In Appendix \ref{sec:regression_violin_g_mu}, we present the relative panels illustrating the distribution of the absolute fractional differences in the test set for each case, providing a comprehensive review of the model’s performance and its variability. The hadronic EoS class has the largest deviations of effective gravity, around $ 0.91 \%$ ($d_{\mathrm{max}}$), while the hybrid and hyperonic classes demonstrate maximum relative deviations of about $ 0.84 \%$ and $0.56 \%$, correspondingly. The EoS models associated with these maximum fractional differences are the EI-CEF-Scyrme model KDE0v1 \cite{agrawal_determination_2005,danielewicz_symmetry_2009,gulminelli_unified_2015}, the Holographic V-QCD model APR intermediate \cite{akmal_equation_1998, jokela2019holographic, ishii2019cool, ecker2020gravitational, jokela2021unified}, and the RMF DDH$\delta$ Y4 model \cite{douchin_unified_2001,gaitanos_lorentz_2004,grill_equation_2014,oertel_hyperons_2015} (see e.g., Appendix \ref{app:eos_tables} and Fig. \ref{fig:g_mu_violins} for a review).

\begin{figure*}[!ht]
    \centering
    \includegraphics[width=0.85\textwidth]{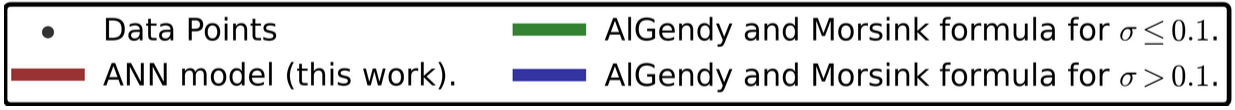}\\
    \centering
    \includegraphics[width=0.32\textwidth]{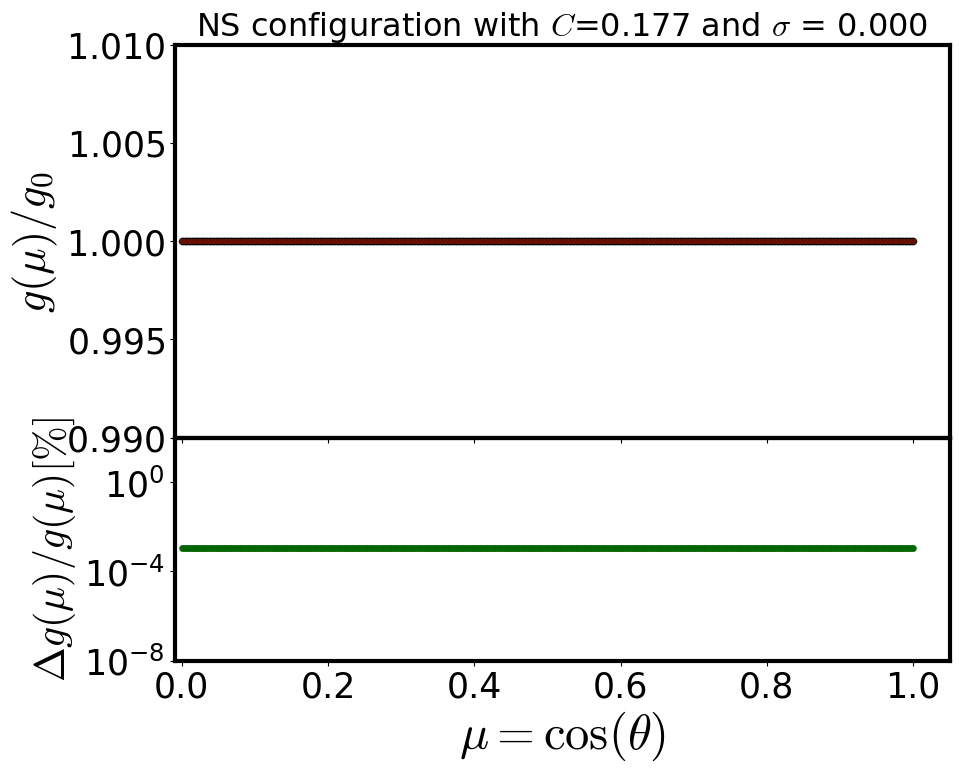}
    \includegraphics[width=0.32\textwidth]{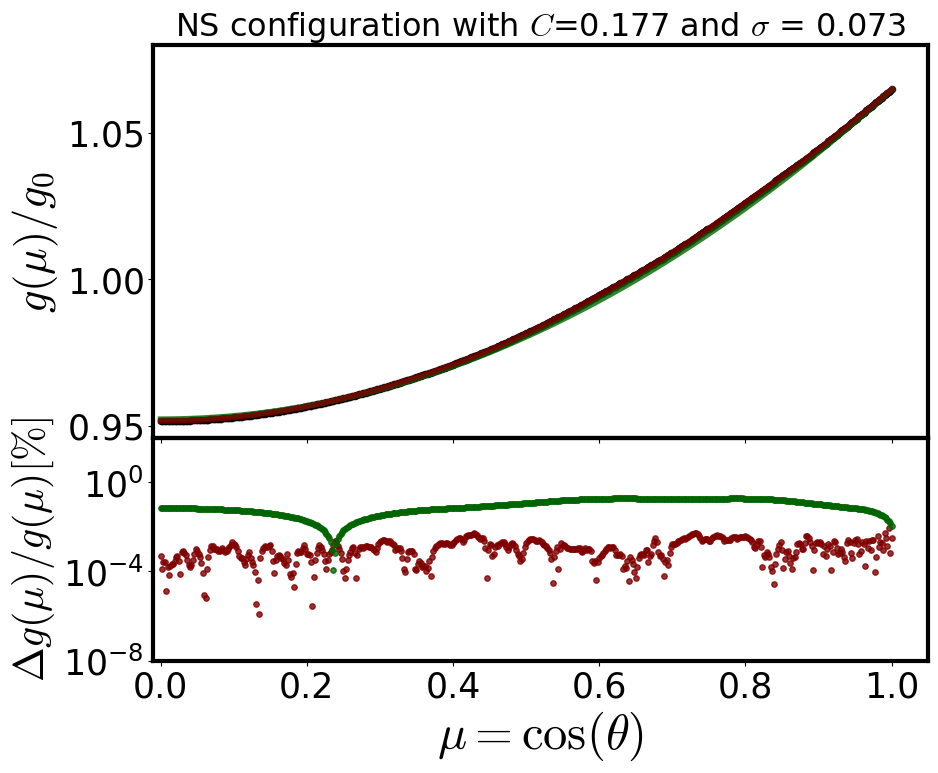}
    \includegraphics[width=0.32\textwidth]{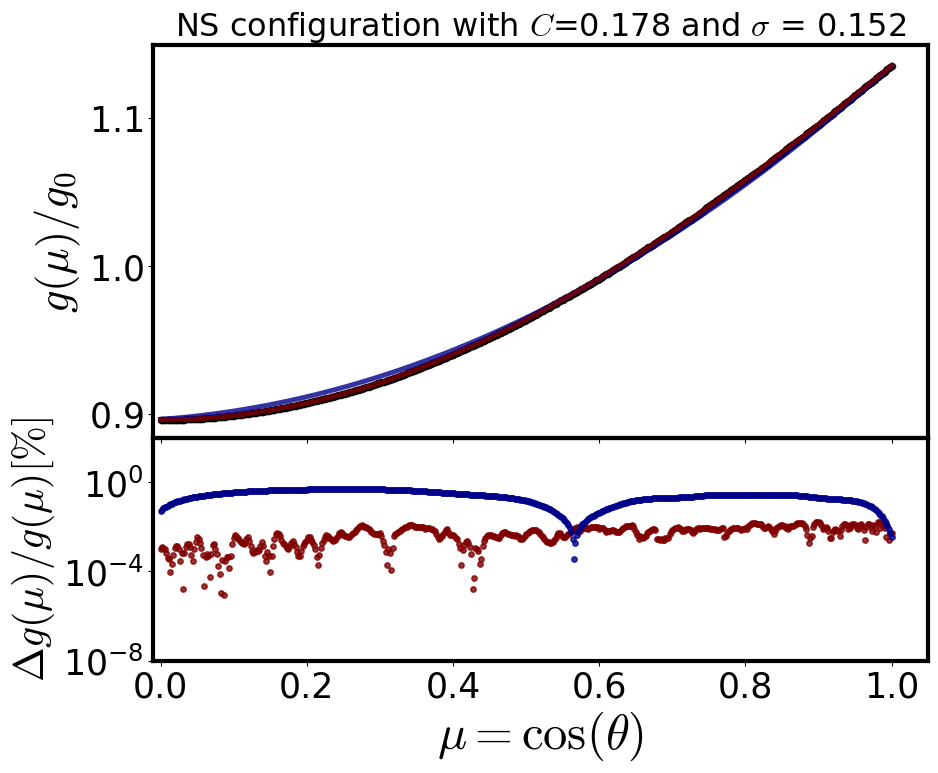}
    \includegraphics[width=0.32\textwidth]{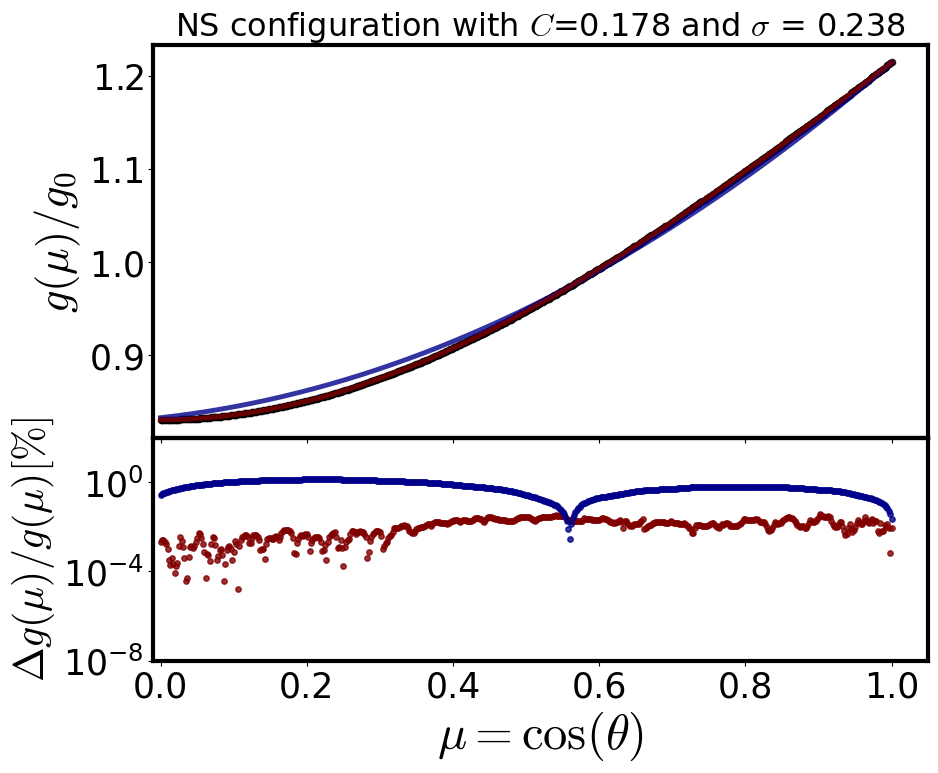}
    \includegraphics[width=0.32\textwidth]{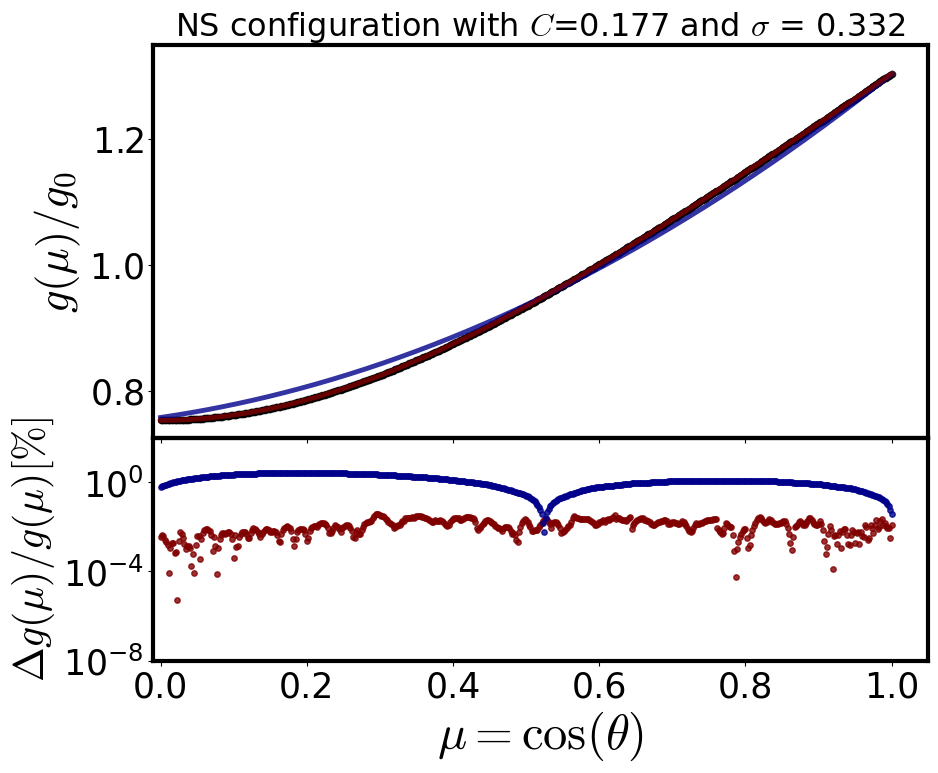}
    \includegraphics[width=0.32\textwidth]{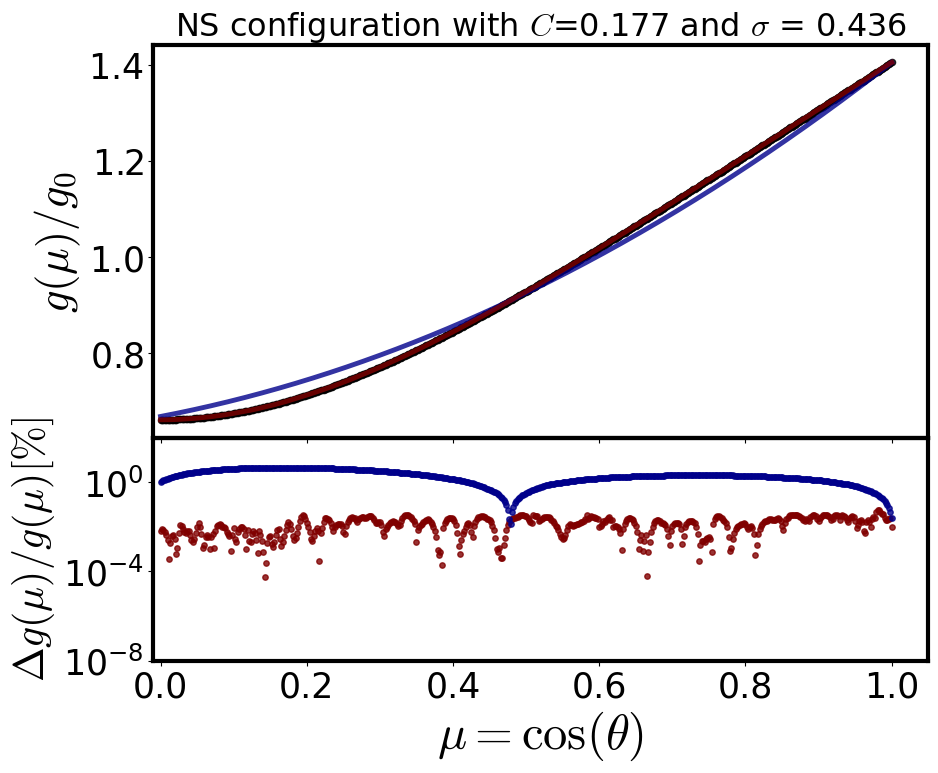}
    \includegraphics[width=0.32\textwidth]{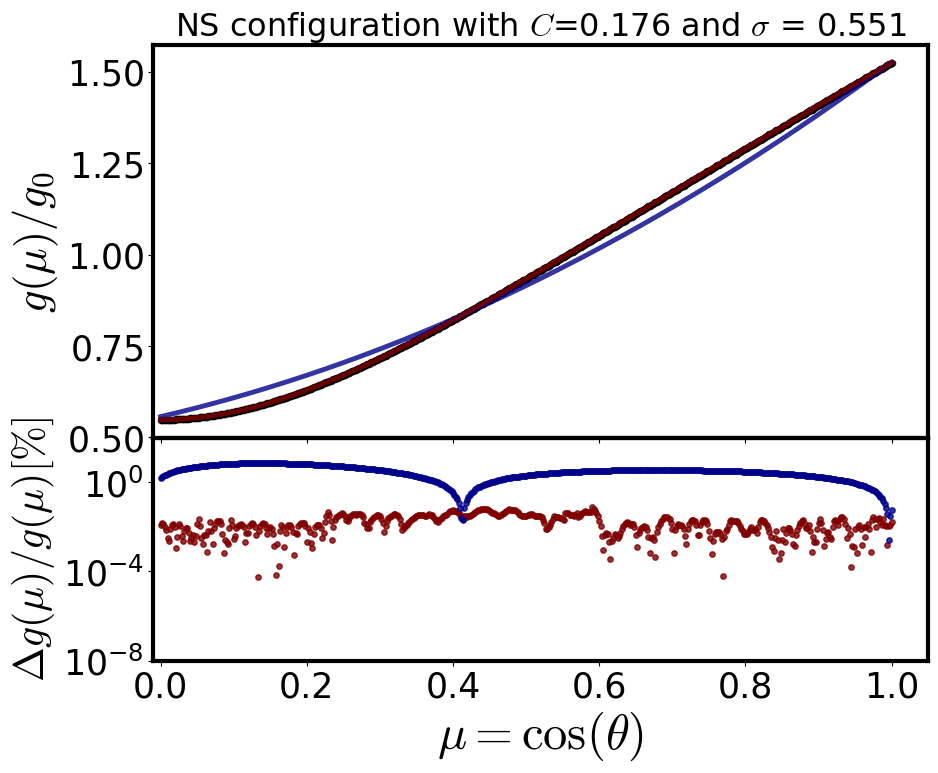}
    \includegraphics[width=0.32\textwidth]{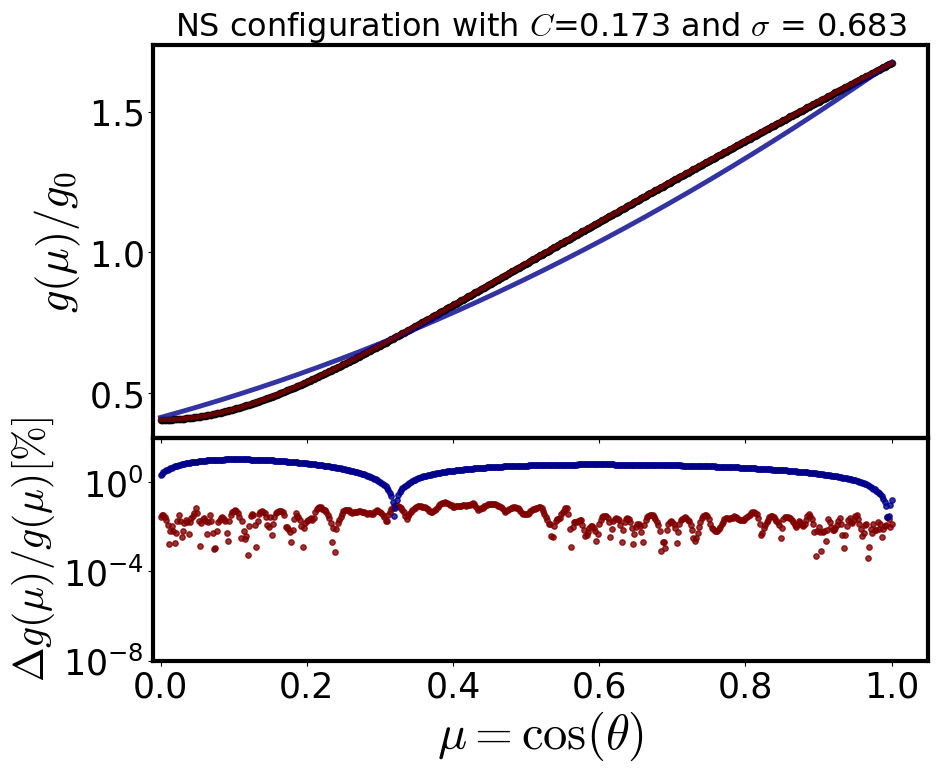}
    \includegraphics[width=0.32\textwidth]{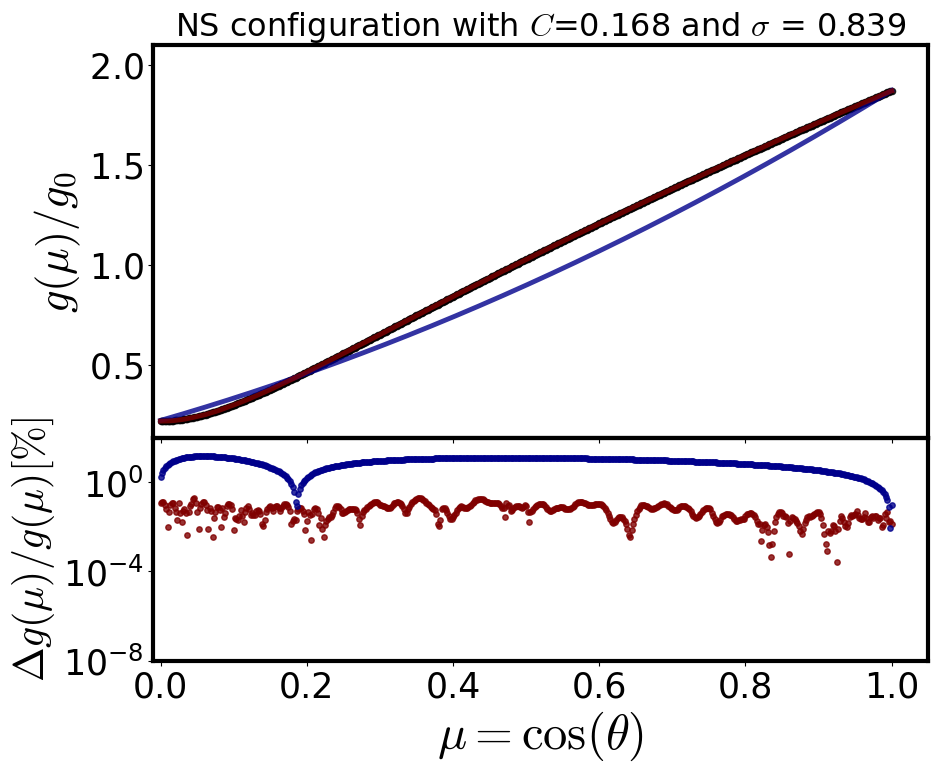}
     \caption{\label{fig:indicative_g_mu_curves} Normalized effective gravity and relative errors in logarithmic scale vs. angular position for the full set of NS models presented in Table (\ref{tab:indicative_propert}). Each plot is for a different stellar model and compares our results to previous results in the literature. The bottom panel in each plot shows the relative error for each fitting formula. Our ANN fit achieves the lowest relative errors, which are also independent of the reduced spin frequency $\sigma$, in contrast to other fits. }
    
\end{figure*}

Again, for the full set of NS models presented in Table (\ref{tab:indicative_propert}), we demonstrate the respective curves for $g(\mu)$ obtained using our regression model (\ref{eq:g_mu_fit}), along with the corresponding relative errors in Fig. \ref{fig:indicative_g_mu_curves}. For each NS model selected, we also incorporate results associated with the suggested fitting functions given by AlGendy and Morsink \cite{algendy2014universality}, providing a basis for relative comparison. Each method accurately reproduces the static case, as anticipated. Nevertheless, our regression model excels in accurately reproducing the data, whereas the other functions encounter increasing challenges in accurately representing the star's effective gravity as frequency rises. Notably, the relative errors in the AlGendy and Morsink fit intensify as the star's frequency increases, highlighting the superior performance of our approach. However, it is important to acknowledge that the limited accuracy of the AlGendy and Morsink fitting function in the context of rapid rotation can be attributed to the methodology employed. Their approach utilizes effective gravity values sampled at only three specific locations on the star: the equator, the north pole, and a point positioned $60$ degrees from the equator. In all cases, our method mitigates the error accumulation as the model's predictions move from the equator towards the star's pole, thanks to normalization (\ref{g_normalization}).

From an observational perspective, having the precise measurements for the star's global parameters $C,\sigma$, and $e$ one needs to provide accurate measurements or estimations for the $g_{\mathrm{pole}}$, and $g_{\mathrm{eq}}$ parameters to determine the star's effective acceleration due to gravity using the suggested regression model (\ref{eq:g_mu_fit}). In that direction, the previously derived universal relations for $g_{\mathrm{pole}}$ (\ref{eq:gpole_c_sigma}), and $g_{\mathrm{eq}}$ (\ref{eq:g_eq_c_sigma_e}) address this issue, thus providing the corresponding estimation for effective gravity with high accuracy.

\section{\label{sec:conclusion} SUMMARY AND DISCUSSION}
This work systematically explores EoS-insensitive relations for the NS's surface and related global properties. The whole investigation is performed for a diverse set of rotating NSs in $\beta$ equilibrium using an extended sample of 70 tabulated cold EoSs. Covering a broad spectrum of compactness ($0.0876 \leq C \leq 0.3095$) and rotation frequency values ($0.00 \leq f[\mathrm{kHz}] \leq 1.871$), the numerical solutions for NSs are obtained using the RNS code \cite{rns, stergioulas1994comparing}, with their surfaces accurately determined through an enthalpy-based method. The analysis to investigate universal relations involves utilizing supervised machine-learning methods for regression (see, Appendix \ref{sec:ML_part} for review). Throughout our systematic investigation, we employed both polynomial functions and an Artificial Neural Network (ANN). 

In exploring the global parameters of the star's surface, we employed polynomial functions within a least-squares regression framework. In this direction, a distinctive feature of our analysis involves incorporating a cross-validation evaluation procedure. To assess the generalization ability of our model functions beyond the training data, we subject them to validation sets. This approach distinguishes our fitting functions from other models in the literature that lack validation evaluation. The utilization of cross-validation ensures that the suggested universal relations possess generalization ability beyond the training data, enhancing the chance of our models performing consistently within the specified relative errors when applied to new data.

To speed up the inference for EoS-insensitive relations concerning the star's surface $R(\mu)$, its logarithmic derivative, and the associated effective acceleration due to gravity $g(\mu)$, we employed an Artificial Neural Network (ANN) architecture. We used min-max scaling to map the values of each input feature within the interval $[0,1]$. In any case, this normalization has been demonstrated to be a universal transformation. In addition, we used Hermite interpolation to augment more data points for the star's surface and its associated quantities of interest. The synthetic data points verify the condition (Eq. (\ref{surf_1})) for the star's surface efficiently. The full dataset was partitioned into an 80:20 train/test ratio. We implemented a dynamic learning rate strategy for optimization, utilizing the typical squared error as the loss function. Notably, we observed a remarkable enhancement in the network's learning ability, accompanied by a significant reduction in model errors for each case study. This improvement was especially pronounced when the data boundaries remained consistent across each case study.

Here, we briefly summarize the suggested new EoS-insensitive relations, presented in sections \ref{sec:univ_rel_surf_prop}, and \ref{sec:univ_rel_with_ANNs}.

At first, in Sec. \ref{sec:poly_methods_1}, we proposed a new EoS-insensitive relation for the star's polar-to-equatorial radius $\mathcal{R}=R_{\mathrm{pole}}/R_{\mathrm{eq}}$ in terms of the stellar compactness $C=M/R_{\mathrm{eq}}$ and reduced spin $\sigma = \Omega^2R_{\mathrm{eq}}^3/GM$. The resulting fitting function $\mathcal{R}(C,\sigma)$ (\ref{eq:R_c_sigma}) verifies the data with accuracy $\leq 2.79 \%$ for all EoSs considered, whereas it describes most data values with relative error $\leq 1 \%$. Relative deviations $> 1 \%$ correspond to a small fraction of stellar configurations covering the whole compactness space having equatorial radius $R_{\mathrm{eq}}\in [12.63, 19.41] \ \mathrm{km}$, reduced spin $\sigma \in [0.284, 0.873]$ and they are mainly associated to the Hyperonic and hybrid EoSs utilized. Additionally, when the reduced spin is constrained to $\sigma \leq 0.25$, the maximum percentage error reduces to a mere $0.96\%$. Also, in the absence of rotation, the derived formula reproduces the $R_{\mathrm{pole}}=R_{\mathrm{eq}}$ constraint with a relative deviation that is better than $0.24 \%$. Next, we explored the possibility of better estimating the star's eccentricity $e=\sqrt{1-\mathcal{R}^2}$ universally. To address this objective, we examined its dependence as a function of the parameters $C$ and $\sigma$. The evaluated new regression model $e(C,\sigma)$ (\ref{eq:e_c_sigma}) reproduces the data with accuracy $\leq 4.57 \%$ for all EoSs considered. In addition, most of the models are given with a relative error that is better than $2\%$. In contrast, for the stellar configurations exhibiting larger error variances, no distinct pattern is observed in the distribution of fractional differences with respect to the EoSs or the parameters ($R_{\mathrm{eq}}, C, \sigma$) that adequately describe the compact object's parameter space. Lastly, we looked for a universal relation between the maximum value of the logarithmic derivative ($d\log R(\mu)/d\theta$) and the parameters $C,\sigma$, and $\mathcal{R}$. The proposed new theoretical fitting function (\ref{eq:dlogR_c_sigma_Rp_Re}) is accurate with relative error $\leq 3.21\%$ for all the data, while it is better than $1\%$ for most of the data. More specifically, when applying this regression model, only an extremely small subset of the rotating NS models exhibit relative deviations exceeding $1\%$. It is important to highlight that no discernible pattern associates this subset with a specific EoS or category of EoSs. Nevertheless, it corresponds to NSs with lower to intermediate stellar compactness, characterized by parameters within the following ranges: $C\in[0.094,0.238], R_{\mathrm{eq}}\in[11.594,16.766] \ \mathrm{km},\sigma\in[0.033,0.844]$, and $\mathcal{R}\in[0.639,0.978]$.

Secondly, in Sec. \ref{sec:poly_methods_2}, we proposed a universal relation for the effective acceleration due to gravity at the star's pole, formulated as a function of the parameters $C$ and $\sigma$. The estimated regression model $g_{\mathrm{pole}}(C,\sigma)$ (\ref{eq:gpole_c_sigma}) had accuracy less than $3.07 \%$ for all EoS considered, while most of the data are reproduced with a relative error $\leq 1\%$. Regardless of rotation and compactness, only a small fraction of NS configurations out of the total, with equatorial radii in the range $R_{\mathrm{eq}}\in[11.53,18.38] \ \mathrm{km}$, exhibit relative deviations larger than $1\%$. The majority of these stellar models correspond to hybrid EoS models. Then, we turn our attention to investigating an EoS-insensitive formula related to the effective gravity at the star's equator. We have considered $g_{\mathrm{eq}}$ as a function of the parameters $C,\sigma$, and $e$. It should be noted that the dependence of $g_{\mathrm{eq}}$ from these parameters stands as an entirely new contribution to the literature. The derived fitting function $g_{\mathrm{eq}}(C,\sigma,e)$ (\ref{eq:g_eq_c_sigma_e}) has a relative error that is $\leq 4.26 \%$. Most significant deviations ($> 1\%$) correspond to a tiny subset of stellar configurations out of the full set. For these NS models, no clear pattern emerges in the variance of relative deviations with respect to either the EoSs or the $(C, \sigma, e)$ parameters that describe the examined parameter space.

After universally estimating the global parameters relating to the star's surface, our focus shifted to implementing the designed feed-forward network architecture to fully describe the star's global oblate shape and its properties. The corresponding methodology is outlined in Sec. \ref{sec:univ_rel_with_ANNs}. More specifically, in Sec. \ref{sec:univ_rel_R},  we suggested a new universal relation for the star's circumferential radius $R(\mu)$ in terms of the parameters $C,\sigma$, and $e$. The resulting new regression formula (\ref{eq:R_mu_fit}) describes the data on the test set with accuracy $\leq 0.25 \%$ for all EoS considered, showcasing a state-of-the-art generalization ability beyond the training data. A key point of evaluation is the source of maximum relative deviation of the proposed regression model (\ref{eq:R_mu_fit}) on the test set, both across EoS categories and individual EoSs. In Appendix \ref{sec:regression_violin_R_mu}, violin plots are presented to show the distribution of absolute fractional differences arising after the evaluation of the regression model for each case, offering insight into the model’s performance and variability. The hybrid EoS class exhibits the largest deviations ($\sim0.25\%$), followed by hadronic ($\sim 0.20\%$) and hyperonic ($\sim 0.16\%$) categories. Notable the EoS models associated with the reported errors originating after the model evaluation include the Holographic V-QCD model APR intermediate \cite{akmal_equation_1998, jokela2019holographic, ishii2019cool, ecker2020gravitational, jokela2021unified}, the EI-CEF-Scyrme model SKb \cite{danielewicz_symmetry_2009,gulminelli_unified_2015,kohler_skyrme_1976}, and the SU$(3)$-CMF model DS(CMF)$-1$ \cite{bennour_charge_1989,gulminelli_unified_2015,dexheimer_gw190814_2021,dexheimer_novel_2010,dexheimer_proto-neutron_2008,dexheimer_reconciling_2015,dexheimer_tabulated_2017}.

Then, in Sec. \ref{sec:univ_rel_dlogR}, we delved into the prospect of universally determining the logarithmic derivative; a measurement of the deviation from the sphericity of the rotating star’s surface. For this purpose, we examined the relation between ($d \log R(\mu)/d\theta$) and the parameters $C,\sigma$, and $\mathcal{R}$. The proposed regression function (\ref{dlogR_reg_fit}) reproduces the data on the test set with residual error $\leq 8.36 \times 10^{-3}$. The source of the maximum relative error in the regression model (\ref{dlogR_reg_fit}) on the test set was evaluated across EoS categories and individual models. Appendix \ref{sec:regression_violin_dlog_R_mu} shows violin plots of absolute residuals, revealing that the hybrid EoS class had the largest residuals ($8.36 \times 10^{-3}$), followed by hadronic ($6.3 \times 10^{-3}$) and hyperonic ($5.2 \times 10^{-3}$) classes. Notably, the Holographic V-QCD model APR intermediate \cite{akmal_equation_1998, jokela2019holographic, ishii2019cool, ecker2020gravitational, jokela2021unified}, the RDF model QMC-RMF4 \cite{typel1999relativistic, xia2022unified, xia2022unified_2}, and the SU$(3)$-CMF model DNS exhibited the largest residuals within their respective categories \cite{dexheimer_proto-neutron_2008,dexheimer_reconciling_2015,dexheimer_tabulated_2017,schurhoff_neutron_2010}. {To the best of our knowledge, the proposed new fitting formula (\ref{dlogR_reg_fit}) represents the only universal fitting model for this relation currently available.}

Lastly, in Sec. \ref{sec:univ_rel_g}, we proposed a new universal relation for the star's effective acceleration due to gravity $g(\mu)$ in terms of the parameters $C,\sigma$, and $e$. The derived new regression model (\ref{eq:g_mu_fit}) reproduces the data on the test set with relative error $\leq 0.91\%$ for all EoSs considered, highlighting remarkable generalization ability. As before, an important aspect of the analysis was identifying the source of the maximum relative deviation in the regression model (\ref{eq:g_mu_fit}) on the test set, evaluated across both EoS categories and individual models. Appendix \ref{sec:regression_violin_g_mu} presents panels showing the distribution of absolute fractional differences, offering an overview of the model's performance and variability. The largest deviations in effective gravity were found in the hadronic EoS category ($0.91\%$), with the hybrid and hyperonic categories showing deviations of $0.84\%$ and $0.56\%$, respectively. It should be noted that The largest fractional differences were found in the EI-CEF-Scyrme model KDE0v1 \cite{agrawal_determination_2005,danielewicz_symmetry_2009,gulminelli_unified_2015}, in the Holographic V-QCD model APR intermediate \cite{akmal_equation_1998, jokela2019holographic, ishii2019cool, ecker2020gravitational, jokela2021unified}, and in the RMF DDH$\delta$ Y4 \cite{douchin_unified_2001,gaitanos_lorentz_2004,grill_equation_2014,oertel_hyperons_2015} EoS models.

With the acquired new regression functions for the star's surface $R(\mu)$ (Eq.(\ref{eq:R_mu_fit})), and the corresponding effective gravity $g(\mu)$ (Eq.(\ref{eq:g_mu_fit})), one can proceed to derive the Eddington luminosity at the surface of a rotating NS as an application (see e.g. \cite{ozel2012surface} for a comprehensive review). The proposed new fitting functions, adeptly designed, allow for a highly accurate estimation of this quantity.

In summary, Table (\ref{tab:relations_table}) presents a detailed and systematic overview of the EoS-insensitive relations investigated in this work. These relations, derived from an extensive analysis of 70 tabulated EoS models, offer a robust framework for understanding the universal properties of the surface of stellar configurations across diverse astrophysical scenarios. Concerning the proposed fitting functions, the indicative evaluation notebooks with the estimations highlighted in this work can be found in the following GitHub repository:\href{https://github.com/gregoryPapi/Universal-description-of-the-NS-surface-using-ML.git}{https://github.com/gregoryPapi/Universal-description-of-the-NS-surface-using-ML.git}.
\begin{table*}[!ht]
    \footnotesize
    \caption{\label{tab:relations_table} Summary of the EoS-insensitive relations investigated in this work based on an extensive ensemble of NS configurations and $70$ tabulated EoS models of cold, and ultradense nuclear matter.}
    \begin{ruledtabular}
        \begin{tabular}{c|c|c|c}
            \multirow{2}{*}{ {\bf Universal Relation}} & \multirow{2}{*}{\bf Parameters and their respective ranges}& \multirow{2}{*} {\bf Equation} & \multirow{2}{*}{\bf Max $\%$ Error} \\
 
            &  &  &  \\ 
            \hline
           $e(C,\sigma)$ & $C\in[0.0876, 0.3075], \ \sigma \in [0.0328,0.9612]$ &Improved Fit Eq.(\ref{eq:e_c_sigma}) &$ 4.57 $   \\
            \hline
             $ g_{\mathrm{pole}}(C,\sigma)$& $C\in[0.0876,0.3095], \ \sigma \in [0.0000,0.9612]$ & Improved Fit Eq.(\ref{eq:gpole_c_sigma}) & $ 3.07$  \\
             \hline
            $\mathcal{R}(C,\sigma)$ & $C\in[0.0876,0.3095], \ \sigma \in [0.0000,0.9612]$ & New Fit Eq.(\ref{eq:R_c_sigma}) & $ 2.79 $  \\
            \hline
           \multirow{2}{*}{ $(d \log R(\mu) / d \theta)_{\mathrm{max}} (C,\sigma,\mathcal{R})$} & $C\in[0.0876, 0.3075], \ \sigma \in [0.0328,0.9612],$ & \multirow{2}{*}{New Fit Eq.(\ref{eq:dlogR_c_sigma_Rp_Re})} & \multirow{2}{*}{$ 3.21 $}  \\
            & $\ \mathcal{R}\in[0.626,0.981]$ & & \\
            
            \hline
            \multirow{2}{*}{$ g_{\mathrm{eq}}(C,\sigma,e)$} &  $C\in[0.0876,0.3095], \ \sigma \in [0.0000,0.9612],$ & \multirow{2}{*}{New Fit Eq.(\ref{eq:g_eq_c_sigma_e})}& \multirow{2}{*}{$ 4.26 $} \\
            & $\ e\in[0.000,0.780]$ & & \\
            
            \hline
            
            \multirow{3}{*}{$R(\mu;R_{\mathrm{pole}}, R_{\mathrm{eq}},C,\sigma,e)$} & $ \ R_{\mathrm{pole}} \in [8.618,14.161] \ \mathrm{km}, \ R_{\mathrm{eq}} \in [9.683,19.413] \ \mathrm{km},$ & \multirow{3}{*}{New Fit Eq.(\ref{eq:R_mu_fit})}& \multirow{3}{*}{$ 0.25 $} \\
             &  $C\in[0.0876,0.3095], \ \sigma \in [0.0000,0.9612],$& & \\

            & $e\in[0.000,0.780]$& & \\
            
            \hline
            \multirow{3}{*}{$\ g(\mu; g_{\mathrm{pole}}, g_{\mathrm{eq}},C,\sigma,e)$} & $g_{\mathrm{pole}}/g_0 \in[0.987,2.107], \ g_{\mathrm{eq}}/g_0 \in [0.069,1.000],$& \multirow{3}{*}{New Fit Eq.(\ref{eq:g_mu_fit})}& \multirow{3}{*}{$ 0.91 $} \\
            
            & $C\in[0.0876,0.3095], \ \sigma \in [0.0000,0.9612],$& & \\

            & $e\in[0.000,0.780]$& &\\
            \hline
           
            

           \multirow{3}{*}{$ \left(\frac{d \log R(\mu)}{d\theta} \right)\left(\mu; (d \log R(\mu) / d \theta)_{\mathrm{max}},C,\sigma,\mathcal{R}\right)$} & \multirow{2}{*}{$(d \log R(\mu) / d \theta)_{\mathrm{max}} \in [0.019,0.503], \ C\in[0.0876, 0.3075],$} & \multirow{3}{*}{New Fit Eq.(\ref{dlogR_reg_fit})}& \multirow{3}{*}{$  8.36 \times 10^{-3}$ $^*$}   \\
           
            \multirow{2}{*}{}& \multirow{2}{*}{$\sigma \in [0.0328,0.9612], \ \mathcal{R}\in[0.626,0.981]$}& \multirow{2}{*}{} & \multirow{2}{*}{} \\
            
            & &  &  \\

        \end{tabular}
    \end{ruledtabular}
    \begin{flushleft}
    $^*$Unlike the other cases that present the maximum percentage error, this value corresponds to the maximum residual error observed in the test set.
    \end{flushleft}
\end{table*}

From an observational perspective, the approach followed in this work relies on acquiring essential information about the star's mass, rotation frequency, and global parameters like $C, \sigma$, and $\mathcal{R}$. Leveraging the proposed highly accurate regression models, such as Eq.(\ref{eq:R_mu_fit}) for estimating the star's surface, is crucial. 
In this direction, the previously derived EoS-insensitive relations, particularly those for $R_{\mathrm{pole}}$ (Eq.(\ref{eq:R_c_sigma})) and eccentricity $e$ (Eq.(\ref{eq:e_c_sigma})) prove essential in this determination, especially when utilizing the proposed fitting function. Notably, accurate measurements or estimations of the $g_{\mathrm{pole}}$ and $g_{\mathrm{eq}}$ parameters are pivotal for determining the star's effective acceleration due to gravity, as outlined in the suggested regression model (Eq.(\ref{eq:g_mu_fit})). The utility of the derived new universal relations for $g_{\mathrm{pole}}$ (Eq.(\ref{eq:gpole_c_sigma})) and $g_{\mathrm{eq}}$ (Eq.(\ref{eq:g_eq_c_sigma_e})) is paramount in addressing this requirement, providing precise estimations for effective gravity. Beyond these considerations, practical utilization of the proposed fitting function (Eq.(\ref{dlogR_reg_fit})) necessitates knowledge of the maximum value of the logarithmic derivative, a requirement adeptly addressed through the derived EoS-insensitive polynomial theoretical function (Eq.(\ref{eq:dlogR_c_sigma_Rp_Re})) which is an entirely new universal relation. This comprehensive approach ensures accurate inferences of the star's surface, its logarithmic derivative, and the associated effective gravity across diverse NS configurations.

Ultimately, the new regression model for predicting the star's surface $R(\mu)$ (\ref{eq:R_mu_fit}) offers the potential for assessing the impact of stellar oblateness on parameter estimation using the cooling tail method \cite{suleimanov2020observational} or in studying wave propagation on the thin oceans of neutron star surfaces \cite{van2020waves}. At the same time, the newly developed regression model (\ref{eq:g_mu_fit}) enables precise predictions of the star's effective gravity at the surface, $g(\mu)$, underscoring its necessity to NICER (or future missions) parameter estimation. This approach is crucial for accurate atmospheric modeling, particularly given that NICER data analyses assume that the NS's atmosphere is mainly composed of Hydrogen. The properties of Hydrogen atmospheres depend on the local effective gravity \cite{heinke2006hydrogen}. In addition, apart for implications to NSs Hydrogen atmospheres and X-ray pulse profile observations with the NICER telescope, this work is important in the context of future large-area X-ray timing facilities \cite{watts2019dense}, such as the enhanced X-ray Timing and Polarimetry (eXTP) \cite{zhang2019enhanced}, the Spectroscopic Time-Resolving Observatory for Broadband Energy X-rays (STROBE-X) \cite{ray2018strobe,ray2019strobe}, and the Square Kilometer Array (SKA) telescope \cite{Watts:2015yU} missions. These upcoming missions are anticipated to enhance the precision of parameter estimation for the radii of NSs beyond the current capabilities of NICER. Therefore, their contribution will be crucial in the efforts to constrain the different theoretical scenarios of NS matter.

\section*{ACKNOWLEDGMENTS}

We are grateful to Sharon Morsink for a careful reading of the manuscript and for comments that improved it. GP wants to thank George Pappas for useful discussions. GP acknowledges financial through a PhD fellowship by H.F.R.I project number: 20450. GV acknowledges partial funding support from H.F.R.I project number: 15940. The EoSs used in this work come from the CompOSE database. We thank the team members of CompOSE for their development. The training process for our ANN models was conducted at the Department of Computer Science and Engineering, University of Ioannina, Greece. Especially, we would like to thank Professor Christophoros Nikou for generously providing access to the computing resources of the Impala cluster.

\appendix

\section{\label{sec:ML_part} SUPERVISED LEARNING}

In this section, we will present the tools to describe universal relations governing the NS's surface properties for rotating configurations with frequencies ranging from the static case up to the $\sim 1.87 \ \mathrm{kHz}$ limit. 

More specifically, in SubSec. \ref{sec:lern_data}, we delve into supervised learning, emphasizing the importance of defining a mapping from input features to target values. To address this, we use a learning methodology that utilizes a mathematical model with learnable parameters. The goal is to minimize a Loss function via an optimization process in order to obtain the model's learned parameters called optimizers. Then, in SubSec. \ref{sec:train_test}, we further explore training and testing processes, detailing the mathematical framework employed to adjust optimal-fit functions, whether polynomial or neural network-based. In addition, we provide insights into the selection and training of ANNs as a robust tool for capturing the complexity of data patterns. Finally, we suggest an ANN architecture designed to provide accurate estimations for the quantities of interest.

\subsection{\label{sec:lern_data}LEARNING FROM DATA}
ML has emerged as a potent tool, revolutionizing the way we analyze and interpret vast amounts of data. By facilitating the development of algorithms capable of discerning intricate patterns in complex datasets, ML  allows us to extract meaningful insights, make accurate predictions, and discover hidden data patterns that might remain concealed. Motivated by the prospect of investigating universal relations governing the NS's surface and its properties using ML methods, our emphasis lies specifically on regression. Regression represents a learning methodology within the realm of supervised learning \cite{bishop2006pattern}. Using this approach, we seek to establish a systematic way of identifying EoS-insensitive relations within the context of our study.

Given the training data $D = \{(x_i, z_i)\}_{i=1}^{N}$ consisting of $N$ pairs, the goal in supervised learning is to define a mapping that transforms input values $x$ into corresponding target values $z$. In this framework, an implicit assumption emerges, suggesting that the set of observations $D$ is obtained or derived from an underlying mapping function $z=f^{\star}(x)$. Unfortunately, in the general case, we lack any additional knowledge or insights regarding the specifics of this underlying function \cite{bishop2006pattern, murphy2012machine,Goodfellow-et-al-2016,prince2023understanding}. However, we can partially determine this function from the information incorporated into the training set $D$. Consequently, we aim to define a learning methodology to approximate, with high accuracy, the underlying function $f^{\star}(x)$ using only the information from our training data.

In that direction, we utilize a mathematical model to approximate the underlying function, $\hat{f}_\theta(x) \approx f^{\star}(x)$, with $\theta$, denoting the vector of the learnable parameters.
It is essential to acknowledge that determining the optimal mathematical model in advance is not feasible. However, we can formulate plausible hypotheses for its structure, considering any relevant prior information derived from the available data (e.g., polynomial form). We intend to identify the most concise model that adequately describes the data. In regression problems, the Squared Error is widely used as the typical Loss function $\mathcal{L}(\theta)$ \cite{bishop2006pattern, murphy2012machine,Goodfellow-et-al-2016,prince2023understanding}. Then, an optimization procedure must be employed to minimize the model’s error function and obtain the optimal parameters $\theta^\star$ given by
\begin{equation}
    \label{eq:optimization}
    \theta^{\star} = \arg\min\limits_{\theta} \mathcal{L}(\theta).
\end{equation}
It is crucial to emphasize that supervised learning extends beyond mere optimization. Minimizing the Loss function and ensuring that the selected model can capture the complexity of the training data are prerequisites for effective learning. However, the primary objective of supervised learning is the generalization ability beyond the training data \cite{bishop2006pattern, murphy2012machine,Goodfellow-et-al-2016,prince2023understanding}. A generalized model produces accurate predictions on new data that were not part of the training set. This capability sets apart a robust model from the one that is limited in describing newly observed data.

\subsection{\label{sec:train_test}TRAINING AND TESTING}
We now define the mathematical framework employed to adjust the best-fit function that describes the data. Depending on the case, this mathematical model might be either a polynomial function or a neural network.

For nonlinear polynomial functions, we use the linear least-squares regression method to identify the most accurate data fit. The Loss function optimized during the training process is formulated as the sum of squared differences between the observed values $z$ and the predicted values $\hat{z}$ generated by the regression model. For example, consider a dataset containing pairs of $(\tilde{x}_i, z_i)$ data, where $i=1,..., n$. Here, $\tilde{x}_i$ signifies the feature variables eligible for inclusion as model parameters, while $z_i$ denotes the corresponding dependent variables known as labels. The investigation for the optimal parameters (optimizers) $\theta^\star$ for the fitting function  involves the optimization of the Loss function \cite{burden_numerical_2015, ramachandran2009mathematical,bishop2006pattern, murphy2012machine,Goodfellow-et-al-2016,prince2023understanding},
\begin{equation}
  \label{sq_err_loss} 
  \mathcal{L}(\theta)=\sum_{i=1}^{n} ||z_i - \hat{z}_i||^2 
\end{equation}
where $\hat{z} = \mathcal{F}(\tilde{x};\theta)$ is the polynomial function employed to describe the data. Adjusting the model to data, the regression coefficients $\theta^\star$ are determined by minimizing the Loss function, setting the corresponding partial derivatives to zero, and then solving the resulting system of equations (normal equations) \cite{burden_numerical_2015}.

However, when exploring different polynomial functions, the central question of which model most adeptly captures the specific patterns of the dataset arises. Hence, a validation process is required to identify the best functional form that accurately verifies the data. In general, training a model to learn its parameters, while assessing its statistical performance on the same dataset (training set) can lead to overfitting. In this case, the model may reproduce patterns with duplicate labels rather than generalizing to new, unseen data (excluded from the training process). As a result, the model may lack the necessary generalization ability, potentially leading to an inability to make useful predictions on unseen data \cite{bishop2006pattern}.

To ascertain the best functional representation that describes the data, a common practice is to split the dataset into training and validation sets. Employing the Leave-One-Out (LOOCV) method for Cross-Validation, we systematically treated each of the $n$ data points as a validation set in turns. In each iteration, we applied the least-squares regression method to fit a polynomial model on a corresponding training set consisting of $n-1$ data points. The model's performance is assessed using statistical metrics on the data point excluded from the training process. This process is repeated for $n$ different training and $n$ different test sets, leveraging the entire dataset as a validation set. It is important to highlight that the LOOCV method treats all data points equally, providing exactly reproducible accurate results (although it may be computationally expensive). This iterative procedure was repeated and evaluated for various polynomial functional forms for their ability to describe the data. The best model, yielding optimal statistical scores, was then selected and refitted to the entire dataset containing all $n$ data points to determine the best-fit $\theta^{\star}$ coefficients \cite{pedregosa_scikit-learn_2011,james2013introduction}. This methodology is utilized for the universal relations extracted and outlined in Sec. \ref{sec:univ_rel_surf_prop}.

Nevertheless, investigating the universal determination of the NS's surface $R(\mu)$ for arbitrary rotation, it became evident that polynomial regression is inadequate for learning the entire spectrum ranging from nonrotating up to highly rotating NSs. The same argument also extends for the EoS-insensitive derivation for the logarithmic derivative $d \log R(\mu)/ d\theta$ (\ref{log_der}), and the effective acceleration due to gravity $g(\mu)$ (\ref{eff_grav_at_surf}) referred both at the star's surface. In order to capture the complexity of such data, one may consider constructing a polynomial model of a higher degree. However, it is essential to note that such a choice can render the model highly susceptible to overfitting. In such cases, we should re-evaluate the selection of the ML model.

The intricacy of the problem prompts us to explore the use of an ANN. It has been well-established that ANNs serve as robust learners \cite{hornik1989multilayer,cybenko1989approximation,hornik1991approximation}. In recent years, ANNs have significantly advanced the state-of-the-art in various complex learning tasks in data science \cite{lecun2015deep}. Universal Approximation Theorem \cite{hornik1989multilayer} states that a feed-forward neural network with a single hidden layer containing a finite number of neurons can approximate any continuous function on a compact input space. This statement holds to any desired degree of accuracy, given a sufficiently large number of neurons and proper activation functions \cite{hornik1989multilayer}.

To effectively capture the complexity embedded in the aforementioned datasets, we opted for the application of ANNs. In that direction, the \href{https://pytorch.org/}{PyTorch} module was used \cite{paszke2019pytorch}. For each EoS of cold, dense nuclear matter included in our ensemble, we have used a random selection for both $20\%$ of static and rotating NS configurations as a test set\footnote{{\bf random.seed(42)} method implemented in Python was used in order to reproduce the same sample of random NS models for testing. This particular choice ensures that the results are consistent across different runs.}. Similar to the previous approach, we utilized the Squared Error (\ref{sq_err_loss}) as the loss function. In this case, the model of choice corresponds to a feed-forward neural network $\hat{z} = \hat{F}_{\theta}(\tilde{x})$.

In our systematic investigation, we utilized an ANN architecture characterized by an input layer encompassing the features\footnote{The specific characteristics of these features, examined in each case, were clarified in the Sec. \ref{sec:univ_rel_with_ANNs}.} $\tilde{x}=(x_1, x_2, x_3, x_4)$, followed by five hidden layers denoted as $H_1, \ldots, H_5$, while concluding in a singular output layer represented by $\hat{z}$. The number of neurons for each hidden layer is shown in Table (\ref{tab:hidden_layers_struct}).
\begin{table}[!th]
  \caption{\label{tab:hidden_layers_struct} ANN hidden layers structure. For each neuron, we have used the non-linear $\mathrm{LeakyReLU}$ function as an activation function.}
  \begin{ruledtabular}
      \begin{tabular}{ccc} 
        \textbf{Hidden Layer} & \textbf{\# Neurons} & \textbf{ Activation Function} \\
        \hline
        $H_1$ & 200 & $\phi = \mathrm{LeakyReLU(x; \beta)}$ \\
        $H_2$ & 100 & $\phi = \mathrm{LeakyReLU(x; \beta)}$ \\
        $H_3$ & 50 & $\phi = \mathrm{LeakyReLU(x; \beta)}$ \\
        $H_4$ & 25 & $\phi = \mathrm{LeakyReLU(x; \beta)}$ \\
        $H_5$ & 10 & $\phi = \mathrm{LeakyReLU(x ; \beta)}$ \\
      \end{tabular}
  \end{ruledtabular}
\end{table}
For each neuron, we used the $\mathrm{Leaky ReLU}$ activation function \cite{maas2013rectifier},
\begin{equation}
   \label{leackyrelu} 
  \phi(\mathrm{x};\beta) =
  \begin{cases}
    \ \mathrm{x}, \ \mathrm{x} > 0 \\
    \\
    \ \beta \mathrm{x}, \ \mathrm{x} \leq 0,
  \end{cases}
\end{equation}
with hyperparameter $\beta = 0.1$. The $\mathrm{Leaky ReLU}$ activation function is used in each hidden layer to introduce the non-linearity. Through experimentation, this choice led to more stable and consistent learning for the specific characteristics of our dataset. Furthermore, it is crucial to highlight that, at the end of the fifth hidden layer, we incorporate the sigmoid function given by,
\begin{equation}
\label{sigmoid}
\boldsymbol{\sigma}(\mathrm{x}) = \frac{1}{1+e^{-\mathrm{x}}},
\end{equation}
as an activation function. This selection, elucidated in Sec. \ref{sec:univ_rel_with_ANNs}, is well-suited to the specific characteristics of our data, rendering it the best choice for this particular case study.

Due to the non-linear nature of the optimization process, we employed the Adam optimizer \cite{kingma2015adam}, based on gradient descent, to extract the optimal $\theta^{\star}$ parameters. We have to note that we utilized the Kaiming uniform initialization algorithm \cite{he2015delving} regarding the model's initial $\theta$ parameters.

As a prepossessing step, we applied min-max scaling defined as  
\begin{equation}
x^{\prime}_i= \frac{x_i - \mathrm{min}(x_i)}{\mathrm{max}(x_i) - \mathrm{mix}(x_i)}
\end{equation}
to map the values of each input feature $x_i$ within the interval $[0,1]$. Feature scaling is crucial to ensure equal contribution from all features, preventing the dominance of those with larger values. This practice enhances the convergence and overall performance of the algorithms employed.

Furthermore, a noteworthy observation led us to implement a dynamic learning rate strategy, changing the optimization step significantly. Specifically, we adjusted the learning rate hyperparameter $\eta$ for every 50 epochs during the training phase, as outlined in Table (\ref{tab:lr_per_50_ep}). The entire training process is carried out for 300 epochs in total. 

\begin{table}[!th]
  \caption{\label{tab:lr_per_50_ep} Illustration of the dynamic learning rate strategy utilized for training. For every 50 epochs, we used a distinct learning rate $\eta_i$ to accelerate the optimization process. The training process is performed for a total of 300 epochs.}
  \begin{ruledtabular}
      \begin{tabular}{ccc} 
        \textbf{Training Epochs} & \textbf{Learning Rate} \\
        \hline
        1-50 & $\eta_1 =  3\times10^{-3}$ \\
        51-100 & $\eta_2 =  1\times10^{-3}$ \\
        101-150 & $\eta_3 =  7\times10^{-4}$ \\
        151-200 & $\eta_4 =  5\times10^{-4}$ \\
        201-250 & $\eta_5 =  3\times 10^{-4}$ \\
        251-300 & $\eta_6 =  1\times 10^{-4}$ \\
      \end{tabular}
  \end{ruledtabular}
\end{table}

\begin{figure}[!htb]
    \includegraphics[width=0.46\textwidth, trim=0 0 0 0.65cm, clip]{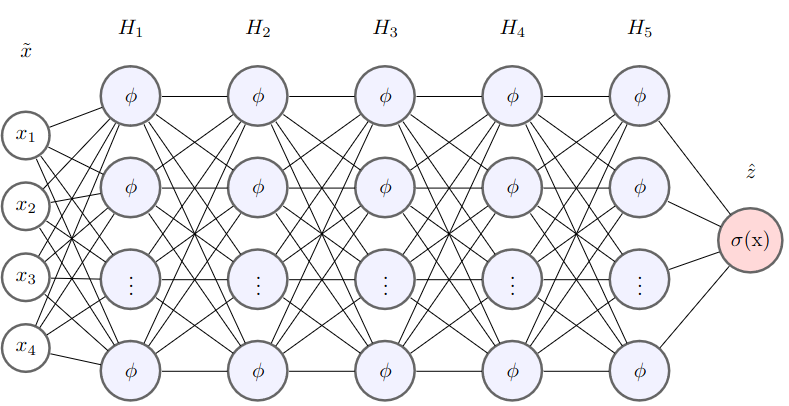}
    \caption{\label{fig:ANN_fig} Illustration of the Feed-forward ANN architecture used in this work. The representations $\tilde{x} = (x_1, x_2, x_3, x_4)$ for features, and $\hat{z}$ for labels denote the input and the output layers respectively. In addition, the $H_1,\ldots,H_5$ variables correspond to the intermediate hidden layers. For each neuron, the non-linear activation function $\phi = \mathrm{LeakyReLU(x;\beta = 0.1)}$ Eq.(\ref{leackyrelu}) was employed, while at the end of the fifth layer, we incorporated the sigmoid activation function $\boldsymbol{\sigma}(\mathrm{x})$ Eq.(\ref{sigmoid}).}
\end{figure}

The corresponding ANN model is illustrated in Fig. \ref{fig:ANN_fig}. Following the optimization process with the steps described above, the proposed feed-forward network architecture is used for each distinct case outlined in Sec. \ref{sec:univ_rel_with_ANNs}, providing the universal estimation of the neutron star's surface $R(\mu)$, the associated logarithmic derivative $d \log R(\mu)/d\theta$, and the effective gravity $g(\mu)$ at the surface, irrespectively of the star's rotation. 

Across all examined regression models, we employed evaluation measure functions to estimate and assess the model's performance in the corresponding validation set (in the context of polynomial models) or the defined test set (in the context of ANNs) \cite{james2013introduction, pedregosa_scikit-learn_2011, papigkiotis2023universal}. The aforementioned measures used for the model's evaluation are outlined as,

\begin{enumerate}[label=(\roman*)]
	\setlength\itemsep{0.09em}
	
 \item {\bf Max Error:}
        \begin{equation}
            {\textrm{Max\_Error}(z,\hat{z})=\textrm{max}\left(|z_i-\hat{z}_i|\right)},
        \end{equation}

 \item {\bf Maximum Relative Deviation}:
    \begin{equation}
        \label{dmax}
        {d_{\text{max}}(z,\hat{z}) =  \text{max} \left(\frac{|z_i-\hat{z}_i|}{\text{max}(\epsilon,|z_i|)} \right)},
    \end{equation}
    
   \item {\bf Mean Absolute Error} (MAE):
        \begin{equation}
            {\textrm{MAE}(z,\hat{z}) = \frac{1}{n}\sum_{i=1}^{n}|z_i-\hat{z_i}|},
        \end{equation}
    
    \item {\bf Mean Squared Error} (MSE):
    
        \begin{equation}
        {\textrm{MSE}(z,\hat{z}) = \frac{1}{n}\sum_{i=1}^{n}(z_i-\hat{z_i})^2},
        \end{equation}
    
	\item {\bf Mean Absolute Percentage Error} (MAPE):
        \begin{equation}
            \label{mape}
            {\textrm{MAPE}(z,\hat{z})=\frac{1}{n}\sum_{i=1}^{n}\frac{   |z_i-\hat{z}_i|}{\text{max}(\epsilon,|z_i|)}},
        \end{equation}

    \item {\bf Coefficient of Determination} ($R^2$):
        \begin{equation}
            \label{r2}
            {\textrm{$R^2$}(z,\hat{z})= 1 - \sum_{i=1}^{n}\frac{(z_i - \hat{z}_i)^2}{\left(z_i -  \langle z \rangle \right)^2},}
        \end{equation}
    
    \item and {\bf Explained Variance} (Exp Var):
        \begin{equation}
            \label{exp_var}
           {\text{Explained\_Variance}(z,\hat{z})=1-\frac{\textrm{Var}[z-\hat{z}]}{\textrm{Var}[z]}}.
       \end{equation}
\end{enumerate}
In the above definitions, $n$ is the number of data samples, $z_i$ is the corresponding actual value, and $\hat{z}_i$ is the model's predicted value for the $i\mathrm{th}$ data sample. In addition, the parameter $\epsilon$ is an arbitrarily small but positive number used to avoid undefined results when $z=0$. It should be noted that the $d_{\text{max}}$ Eq.(\ref{dmax}), and the MAPE Eq.(\ref{mape}) measures lie in the range $[0, 1]$. Furthermore, for the $R^2$ coefficient and the explained variance Eq.(\ref{exp_var}) (with Var denoting the square of the standard deviation), the optimal score for these evaluation functions is 1.0. When the prediction residuals have zero mean ($\langle z \rangle = 0$), the $R^2$ regression score is identical to the explained variance measure. At this point, it is important to note that the evaluation measure functions provided exhibit a slight increase in the validation set compared to their corresponding values in the training set.

During the LOOCV evaluation process, the selection criteria for determining the appropriate fitting function for polynomial models rely on the evaluation measure functions that yield optimal results. Specifically, the functional form demonstrating the most favorable results is chosen compared to the various forms investigated to describe the data\footnote{For the polynomial models, evaluation procedures, and evaluation measures we have utilized the \href{https://scikit-learn.org/stable/index.html}{\textsc{scikit-learn}} \cite{scikit-learn} Python library.}.

\section{\label{app:eos_tables}EQUATION OF STATE TABLES}
In this section, we summarize the EoS tables employed in this work. Also, in Fig. \ref{fig:color_band}, an EoS-color map is shown for the various fitting functions demonstrated in Sec. \ref{sec:univ_rel_surf_prop}.

\begin{widetext}

\begin{figure}[!h]
	\includegraphics[width=1.\textwidth]{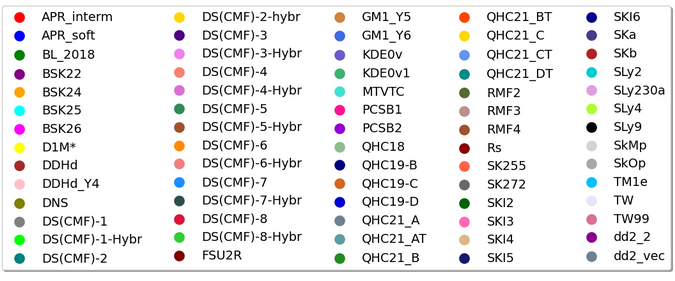}
	\caption{EoS-Color map used for the various regression models presented in Sec. \ref{sec:univ_rel_surf_prop}.}%
	\label{fig:color_band}
\end{figure}

\begin{table*}[!h]
	\caption{\label{tab:hadronic} Hadronic cold EoS models.}
    \scriptsize
    \begin{ruledtabular}
		\begin{tabular}{ccccccc}
			EoS & Model & Matter & \texttt{$M_{\max }\left[M_{\odot}\right]$} &  \texttt{$R_{M_{max}}[\mathrm{km}]$} &  \texttt{$R_{1.4 M_{\odot}}[\mathrm{km}]$}& References\\ 
			\hline
			\text{SLY2}& \text{EI-CEF-Scyrme}& \text{$n,p,e,\mu$} & 2.06& 10.06 & 11.79&\cite{chabanat_skyrme_1998,danielewicz_symmetry_2009,gulminelli_unified_2015} \\
			\hline
			\text{SKb}& \text{EI-CEF-scyrme}& \text{$n,p,e,\mu$} & 2.20& 10.58 & 12.21& \cite{danielewicz_symmetry_2009,gulminelli_unified_2015,kohler_skyrme_1976}\\
			\hline
			
			\text{SkMp}& \text{EI-CEF-scyrme}& \text{$n,p,e,\mu$} & 2.11& 10.60 & 12.50& \cite{bennour_charge_1989,danielewicz_symmetry_2009,gulminelli_unified_2015}  \\
			\hline
			\text{SLY9}& \text{EI-CEF-scyrme}& \text{$n,p,e,\mu$} & 2.16& 10.65 & 12.47& \cite{chabanat_skyrme_1998,danielewicz_symmetry_2009,gulminelli_unified_2015}  \\
			\hline
			\text{SkI3}& \text{EI-CEF-scyrme}& \text{$n,p,e,\mu$} & 2.25& 11.34 & 13.55& \cite{danielewicz_symmetry_2009,gulminelli_unified_2015,reinhard_nuclear_1995}  \\
			\hline
			\text{KDE0v}& \text{EI-CEF-scyrme}& \text{$n,p,e,\mu$} & 1.97& 9.62 & 11.42& \cite{agrawal_determination_2005,danielewicz_symmetry_2009,gulminelli_unified_2015}  \\
			\hline
			\text{SK255}& \text{EI-CEF-scyrme}& \text{$n,p,e,\mu$} & 2.15& 10.84 & 13.15& \cite{agrawal_determination_2005,danielewicz_symmetry_2009,gulminelli_unified_2015}  \\
			\hline
			\text{Rs}& \text{EI-CEF-scyrme}& \text{$n,p,e,\mu$} & 2.12& 10.76 & 12.93& \cite{danielewicz_symmetry_2009,friedrich_skyrme-force_1986,gulminelli_unified_2015}  \\
			\hline
			\text{SkI5}& \text{EI-CEF-scyrme}& \text{$n,p,e,\mu$} & 2.25& 11.47 & 14.08& \cite{danielewicz_symmetry_2009,gulminelli_unified_2015,reinhard_nuclear_1995}  \\
			\hline
			\text{SKa}& \text{EI-CEF-scyrme}& \text{$n,p,e,\mu$} & 2.22& 10.82 & 12.92& \cite{danielewicz_symmetry_2009,gulminelli_unified_2015,kohler_skyrme_1976}  \\
			\hline
			\text{SkOp}& \text{EI-CEF-scyrme}& \text{$n,p,e,\mu$} & 1.98& 10.16 & 12.13& \cite{danielewicz_symmetry_2009,gulminelli_unified_2015,reinhard_nuclear_1995}  \\
			\hline
			\text{SLY230a}& \text{EI-CEF-scyrme}& \text{$n,p,e,\mu$} & 2.11& 10.18 & 11.83& \cite{chabanat_skyrme_1998,danielewicz_symmetry_2009,gulminelli_unified_2015}  \\
			\hline
			\text{SKI2}& \text{EI-CEF-scyrme}& \text{$n,p,e,\mu$} & 2.17& 11.25 & 13.48& \cite{danielewicz_symmetry_2009,gulminelli_unified_2015,reinhard_nuclear_1995}  \\
			\hline
			\text{SkI4}& \text{EI-CEF-scyrme}& \text{$n,p,e,\mu$} & 2.18& 10.66 & 12.38& \cite{danielewicz_symmetry_2009,gulminelli_unified_2015,reinhard_nuclear_1995}\\
			\hline
			\text{SkI6}& \text{EI-CEF-scyrme}& \text{$n,p,e,\mu$} & 2.20& 10.71 & 12.49& \cite{danielewicz_symmetry_2009,gulminelli_unified_2015,reinhard_nuclear_1995}  \\
			\hline
			\text{KDE0v1}& \text{EI-CEF-scyrme}& \text{$n,p,e,\mu$} & 1.98& 9.71 & 11.63& \cite{agrawal_determination_2005,danielewicz_symmetry_2009,gulminelli_unified_2015}  \\
			\hline
			\text{SK272}& \text{EI-CEF-scyrme}& \text{$n,p,e,\mu$} & 2.24& 11.20 & 13.32& \cite{agrawal_nuclear_2003,danielewicz_symmetry_2009,gulminelli_unified_2015}  \\
			\hline
			\text{SLY4}& \text{EI-CEF-scyrme}& \text{$n,p,e,\mu$} & 2.06& 10.02 & 11.70& \cite{chabanat_skyrme_1998,danielewicz_symmetry_2009,gulminelli_unified_2015}  \\
            \hline
            \text{D1M$*$}& \text{EI}& \text{$n,p,e,\mu$} &2.00 & 10.20 &11.71 &\cite{vinas2021unified, mondal2020structure, gonzalez2018new} \\
			\hline
            \text{QMC-RMF2}& \text{RMF-EFT}& \text{$n,p,e$} &2.04 & 10.49 & 12.03&\cite{grill_equation_2014, baym1971ground,alford2022relativistic} \\
            \hline
            \text{FSU2R}& \text{RMF}& \text{$n,p,e,\mu$} & 2.048& 11.73 & 12.98&\cite{grill_equation_2014, negreiros2018cooling, providencia2019hyperonic, pearson2018unified} \\
			\hline
            \text{GPPVA(TW)}& \text{RMF}& \text{$n,p,e,\mu$} &2.07 & 10.70 & 12.33&\cite{grill_equation_2014, typel1999relativistic,pearson2018unified} \\
			\hline
			\text{DDH$\delta$}& \text{RMF} & \text{$n,p,e$} & 2.16& 11.19 & 12.58 &\cite{douchin_unified_2001,gaitanos_lorentz_2004,grill_equation_2014}  \\
			\hline
            \text{PCSB1}& \text{RMF}& \text{$n,p,e,\mu$} & 2.19& 11.73 & 13.25&\cite{hempel_statistical_2010, hornick2018relativistic, pradhan2023role} \\
             \hline
            \text{PCSB2}& \text{RMF}& \text{$n,p,e,\mu$} & 2.02& 11.41 & 13.00&\cite{hempel_statistical_2010, hornick2018relativistic, pradhan2023role} \\
			\hline
            \text{TM1e}& \text{RMF}& \text{$n,p,e,\mu$} &2.12 & 11.88 & 13.16&\cite{grill_equation_2014, shen2020effects, pearson2018unified} \\
            \hline
			\text {DS(CMF)-2}& \text{SU(3)-RMF}& \text{$n,p,e$} & 2.13& 11.96 & 13.70 &\cite{bennour_charge_1989,gulminelli_unified_2015,dexheimer_gw190814_2021,dexheimer_novel_2010,dexheimer_proto-neutron_2008,dexheimer_reconciling_2015,dexheimer_tabulated_2017} \\
			\hline
			\text {DS(CMF)-4}& \text{SU(3)-RMF}& \text{$n,p,e$} & 2.05& 11.60 & 13.26 &\cite{bennour_charge_1989,gulminelli_unified_2015,dexheimer_gw190814_2021,dexheimer_novel_2010,dexheimer_proto-neutron_2008,dexheimer_reconciling_2015,dexheimer_tabulated_2017} \\
			\hline
			\text {DS(CMF)-6}& \text{SU(3)-RMF}& \text{$n,p,e$} & 2.11& 11.58 & 13.30 &\cite{bennour_charge_1989,gulminelli_unified_2015,dexheimer_gw190814_2021,dexheimer_novel_2010,dexheimer_proto-neutron_2008,dexheimer_reconciling_2015,dexheimer_tabulated_2017}\\
			\hline
			\text {DS(CMF)-8}& \text{SU(3)-RMF}& \text{$n,p,e,\Delta^{-}$} & 2.09& 11.59 & 13.30 &\cite{bennour_charge_1989,gulminelli_unified_2015,dexheimer_gw190814_2021,dexheimer_novel_2010,dexheimer_proto-neutron_2008,dexheimer_reconciling_2015,dexheimer_tabulated_2017} \\
			\hline
            \text{BSK22}& \text{NR DF}& \text{$n,p,e,\mu$} & 2.26& 11.20 & 13.04&\cite{allard20211s0, pearson2020unified,pearson2022unified, pearson2018unified, goriely2013hartree, perot2019role, xu2013databases, welker2017binding} \\
			\hline
            \text{BSK24}& \text{NR DF }& \text{$n,p,e,\mu$} & 2.28& 11.08 & 12.57&\cite{allard20211s0, pearson2020unified,pearson2022unified, pearson2018unified, goriely2013hartree, perot2019role, xu2013databases, welker2017binding} \\
            \hline
            \text{BSK25}& \text{NR DF}& \text{$n,p,e,\mu$} & 2.22& 11.05 & 12.37&\cite{allard20211s0, pearson2020unified,pearson2022unified, pearson2018unified, goriely2013hartree, perot2019role, xu2013databases, welker2017binding} \\
            \hline
            \text{BSK26}& \text{NR DF}& \text{$n,p,e,\mu$} & 2.17& 10.20 & 11.77&\cite{allard20211s0, pearson2020unified,pearson2022unified, pearson2018unified, goriely2013hartree, perot2019role, xu2013databases, welker2017binding} \\
            \hline
            \text{QMC-RMF3}& \text{RMF-chiral EFT}& \text{$n,p,e$} & 2.15& 10.68 &12.26 &\cite{grill_equation_2014, baym1971ground,alford2022relativistic} \\
            \hline 
            \text{QMC-RMF4}& \text{RDF}& \text{$n,p,e$} &2.21 & 11.03 & 12.35&\cite{grill_equation_2014, baym1971ground,alford2022relativistic} \\
            \hline 
            \text{TW99}& \text{RDF}& \text{$n,p,e,\mu$} &2.08 &10.62  &12.27 &\cite{typel1999relativistic, xia2022unified, xia2022unified_2} \\
            \hline
            \text{MTVTC}& \text{CDF}& \text{$n,p,e,\mu$} &2.02 &10.90  &13.10 &\cite{xia2022unified, xia2022unified_2, maruyama2005nuclear} \\
            \hline
			\text{BL(chiral)\_2018}& \text{chPT-BBG-BHF}& \text{$n,p,e,\mu$} & 2.08& 10.26 & 12.31 &\cite{bombaci_equation_2018,douchin_unified_2001}  \\
		\end{tabular}
    \end{ruledtabular}
\end{table*}

\begin{table*}[!h]
	\caption{\label{tab:hyperonic} Hyperonic cold EoS models.}
    \scriptsize
	\begin{ruledtabular}
		\begin{tabular}{ccccccc}
			EoS & Model & Matter & \texttt{$M_{\max }\left[M_{\odot}\right]$} &  \texttt{$R_{M_{max}}[\mathrm{km}]$} &  \texttt{$R_{1.4 M_{\odot}}[\mathrm{km}]$}& References\\ 
			
			\hline
			\text{OPGR(DDH$\delta$ Y4)}& \text{RMF}& \text{$n,p,e,\Lambda,\Xi^{-}$} & 2.05& 11.26 & 12.58& \cite{douchin_unified_2001,gaitanos_lorentz_2004,grill_equation_2014,oertel_hyperons_2015}  \\
			\hline 
			\text{OPGR(GM1Y5)}& \text{RMF}& \text{$n,p,e,\Lambda,\Xi^{-},\Xi^0$} & 2.12& 12.31 & 13.78 &\cite{glendenning_reconciliation_1991,douchin_unified_2001,oertel_hyperons_2015}  \\
			\hline 
			\text{OPGR(GM1Y6)}& \text{RMF}& \text{$n,p,e,\Lambda,\Xi^{-},\Xi^0$} & 2.29& 12.13 & 13.78& \cite{douchin_unified_2001,glendenning_reconciliation_1991,oertel_hyperons_2015}  \\
			\hline 
			\text {DNS}& \text{SU(3)-CMF}& \text{$n,p,e,\mu,\Lambda,\Sigma^{-}$} & 2.10& 12.00 &13.58 & \cite{dexheimer_proto-neutron_2008,dexheimer_reconciling_2015,dexheimer_tabulated_2017,schurhoff_neutron_2010}  \\
			\hline
			\text {DS(CMF)-1}& \text{SU(3)-CMF}& \text{$n,p,e,\Lambda,\Sigma^{-}$} & 2.07& 11.88 & 13.57&\cite{bennour_charge_1989,gulminelli_unified_2015,dexheimer_gw190814_2021,dexheimer_novel_2010,dexheimer_proto-neutron_2008,dexheimer_reconciling_2015,dexheimer_tabulated_2017} \\
			\hline
			\text {DS(CMF)-3}& \text{SU(3)-CMF}& \text{$n,p,e,\Lambda,\Sigma^{-}$} & 2.00& 11.56 & 13.15&\cite{bennour_charge_1989,gulminelli_unified_2015,dexheimer_gw190814_2021,dexheimer_novel_2010,dexheimer_proto-neutron_2008,dexheimer_reconciling_2015,dexheimer_tabulated_2017} \\
			\hline
			\text {DS(CMF)-5}& \text{SU(3)-CMF}& \text{$n,p,e,\Lambda,\Sigma^{-}$} & 2.07& 11.43 & 13.20&\cite{bennour_charge_1989,gulminelli_unified_2015,dexheimer_gw190814_2021,dexheimer_novel_2010,dexheimer_proto-neutron_2008,dexheimer_reconciling_2015,dexheimer_tabulated_2017} \\
			\hline
			 \text{DS(CMF)-7}& \text{SU(3)-CMF}& \text{$n,p,e,\Lambda,\Sigma^{-},\Delta^{-}$} & 2.07& 11.43 & 13.20& \cite{bennour_charge_1989,gulminelli_unified_2015,dexheimer_gw190814_2021,dexheimer_novel_2010,dexheimer_proto-neutron_2008,dexheimer_reconciling_2015,dexheimer_tabulated_2017} 
		\end{tabular}
	\end{ruledtabular}
\end{table*}
\begin{table*}[!h]
	\caption{\label{tab:hybrid} Hybrid cold EoS models.}
    \scriptsize
	\begin{ruledtabular}
		\begin{tabular}{ccccccc}
			EoS & Model & Matter & \texttt{$M_{\max }\left[M_{\odot}\right]$} &  \texttt{$R_{M_{max}}[\mathrm{km}]$} &  \texttt{$R_{1.4 M_{\odot}}[\mathrm{km}]$}& References\\ 
            \hline
            \text{DS(CMF)-1 Hybr}& \text{SU(3)-CMF}& \text{$n,p,e,\mu, \Lambda,q$} & 1.96& 11.11 & 13.55& \cite{dexheimer_proto-neutron_2008, dexheimer_tabulated_2017, dexheimer_novel_2010, dexheimer_gw190814_2021, dexheimer2019we, clevinger2022hybrid}  \\
            \hline
			\text{DS(CMF)-2 Hybr}& \text{SU(3)-CMF}& \text{$n,p,e,q$} & 1.96& 11.11 & 13.67& \cite{dexheimer_proto-neutron_2008, dexheimer_tabulated_2017, dexheimer_novel_2010, dexheimer_gw190814_2021, dexheimer2019we, clevinger2022hybrid}  \\
            \hline
			\text{DS(CMF)-3 Hybr}& \text{SU(3)-CMF}& \text{$n,p,e,\mu,\Lambda, \Sigma^{-},q$} & 1.99& 11.20 & 13.15& \cite{dexheimer_proto-neutron_2008, dexheimer_tabulated_2017, dexheimer_novel_2010, dexheimer_gw190814_2021, dexheimer2019we, clevinger2022hybrid}  \\
            \hline
			\text{DS(CMF)-4 Hybr}& \text{SU(3)-CMF}& \text{$n,p,e,q$} & 1.98& 11.21 & 13.24& \cite{dexheimer_proto-neutron_2008, dexheimer_tabulated_2017, dexheimer_novel_2010, dexheimer_gw190814_2021, dexheimer2019we, clevinger2022hybrid}  \\
            \hline
			\text{DS(CMF)-5 Hybr}& \text{SU(3)-CMF}& \text{$n,p,e,\mu,\Lambda, \Sigma^{-},q$} & 2.02& 11.89 & 13.18& \cite{dexheimer_proto-neutron_2008, dexheimer_tabulated_2017, dexheimer_novel_2010, dexheimer_gw190814_2021, dexheimer2019we, clevinger2022hybrid}  \\
            \hline
			\text{DS(CMF)-6 Hybr}& \text{SU(3)-CMF}& \text{$n,p,e,q$} & 2.01& 11.94 & 13.27& \cite{dexheimer_proto-neutron_2008, dexheimer_tabulated_2017, dexheimer_novel_2010, dexheimer_gw190814_2021, dexheimer2019we, clevinger2022hybrid}  \\
            \hline
            \text{DS(CMF)-7 Hybr}& \text{SU(3)-CMF}& \text{$n,p,e,\mu,\Lambda, \Sigma^{-},q$} & 2.02& 11.90 & 13.18& \cite{dexheimer_proto-neutron_2008, dexheimer_tabulated_2017, dexheimer_novel_2010, dexheimer_gw190814_2021, dexheimer2019we, clevinger2022hybrid}  \\
            \hline
			\text{DS(CMF)-8 Hybr}& \text{SU(3)-CMF}& \text{$n,p,e,q$} & 2.01& 11.94 & 13.27& \cite{dexheimer_proto-neutron_2008, dexheimer_tabulated_2017, dexheimer_novel_2010, dexheimer_gw190814_2021, dexheimer2019we, clevinger2022hybrid}  \\   
            \hline
			\text {DD2-FRG (2) flav}& \text{NP-FRG}& \text{$n,p,e,q$} & 2.05& 12.55 & 13.20&\cite{hempel_statistical_2010,otto_hybrid_2020,typel_composition_2010}  \\
			\hline 
			\text {DD2-FRG vec int-(2) flav}& \text{NP-FRG}& \text{$n,p,e,q$} & 2.14& 12.70 & 13.20&\cite{hempel_statistical_2010,otto_hybrid_2020,otto_nonperturbative_2020,typel_composition_2010}  \\
			\hline 
			\text {QHC18}& \text{NJL-MF}& \text{$n,p,e,q$} & 2.05& 10.41 & 11.49&\cite{akmal_equation_1998,baym_hadrons_2018,togashi_nuclear_2017,yu_self-consistent_2020}  \\
			\hline 
			\text {QHC19-B}& \text{NJL-MF}& \text{$n,p,e,q$} & 2.07& 10.60 & 11.60&\cite{baym_hadrons_2018,baym_new_2019,togashi_nuclear_2017,yu_self-consistent_2020}  \\
			\hline 
			\text {QHC19-C}& \text{NJL-MF}& \text{$n,p,e,q$} & 2.18& 10.80 & 11.60&\cite{baym_hadrons_2018,baym_new_2019,togashi_nuclear_2017,yu_self-consistent_2020} \\
			\hline 
			\text {QHC19-D}& \text{NJL-MF}& \text{$n,p,e,q$} & 2.28& 10.90 & 11.60&\cite{baym_hadrons_2018,baym_new_2019,togashi_nuclear_2017,yu_self-consistent_2020} \\

            \hline 
            \text{QHC21T $A_T$} & \text{NJL-MF}& \text{$n,p,e,q$} & 2.13& 11.90 &10.80 & \cite{togashi_nuclear_2017, kojo2022implications}  \\
            \hline 
            \text{QHC21T $B_T$} & \text{NJL-MF}& \text{$n,p,e,q$} & 2.20& 11.10 &10.90 & \cite{togashi_nuclear_2017, kojo2022implications}  \\
            \hline 
            \text{QHC21T $C_T$} & \text{NJL-MF}& \text{$n,p,e,q$} & 2.26& 11.10 &10.70 & \cite{togashi_nuclear_2017, kojo2022implications}  \\
            \hline 
            \text{QHC21T $D_T$} & \text{NJL-MF}& \text{$n,p,e,q$} & 2.32& 11.30 &10.80 & \cite{togashi_nuclear_2017, kojo2022implications}  \\
            \hline
            \text{QHC21 $A_{\chi}$}& \text{NJL-MF}& \text{$n,p,e,q$} &2.19& 11.70 & 12.40 & \cite{togashi_nuclear_2017, kojo2022implications, drischler2021limiting}  \\
            \hline 
            \text{QHC21 $B_{\chi}$}& \text{NJL-MF}& \text{$n,p,e,q$} &2.25& 11.50 & 12.40 & \cite{togashi_nuclear_2017, kojo2022implications, drischler2021limiting}  \\
            \hline 
            \text{QHC21 $C_{\chi}$}& \text{NJL-MF}& \text{$n,p,e,q$} &2.31& 11.40 & 12.40 & \cite{togashi_nuclear_2017, kojo2022implications, drischler2021limiting}  \\
			\hline  
            \text{VQCD(APR), soft}& \text{Holographic V-QCD }& \text{$n,p,e,q$} &2.02& 11.90 & 12.30& \cite{akmal_equation_1998, jokela2019holographic, ishii2019cool, ecker2020gravitational, jokela2021unified}  \\
            \hline 
            \text{VQCD(APR), interm}& \text{Holographic V-QCD }& \text{$n,p,e,q$} &2.15& 11.80 & 12.40& \cite{akmal_equation_1998, jokela2019holographic, ishii2019cool, ecker2020gravitational, jokela2021unified}  \\
		\end{tabular}
	\end{ruledtabular}
\end{table*}

\end{widetext}

\section{\label{sec:errors_violin_plots} Relative deviations in the test set for each regression model proposed in Sec. \ref{sec:univ_rel_with_ANNs}.}

\subsection{\label{sec:regression_violin_R_mu} Fractional difference distributions for $R(\mu)$}

\begin{widetext}
    
  \begin{figure*}[!ht]
        \centering
    	\includegraphics[width=0.71\textwidth]{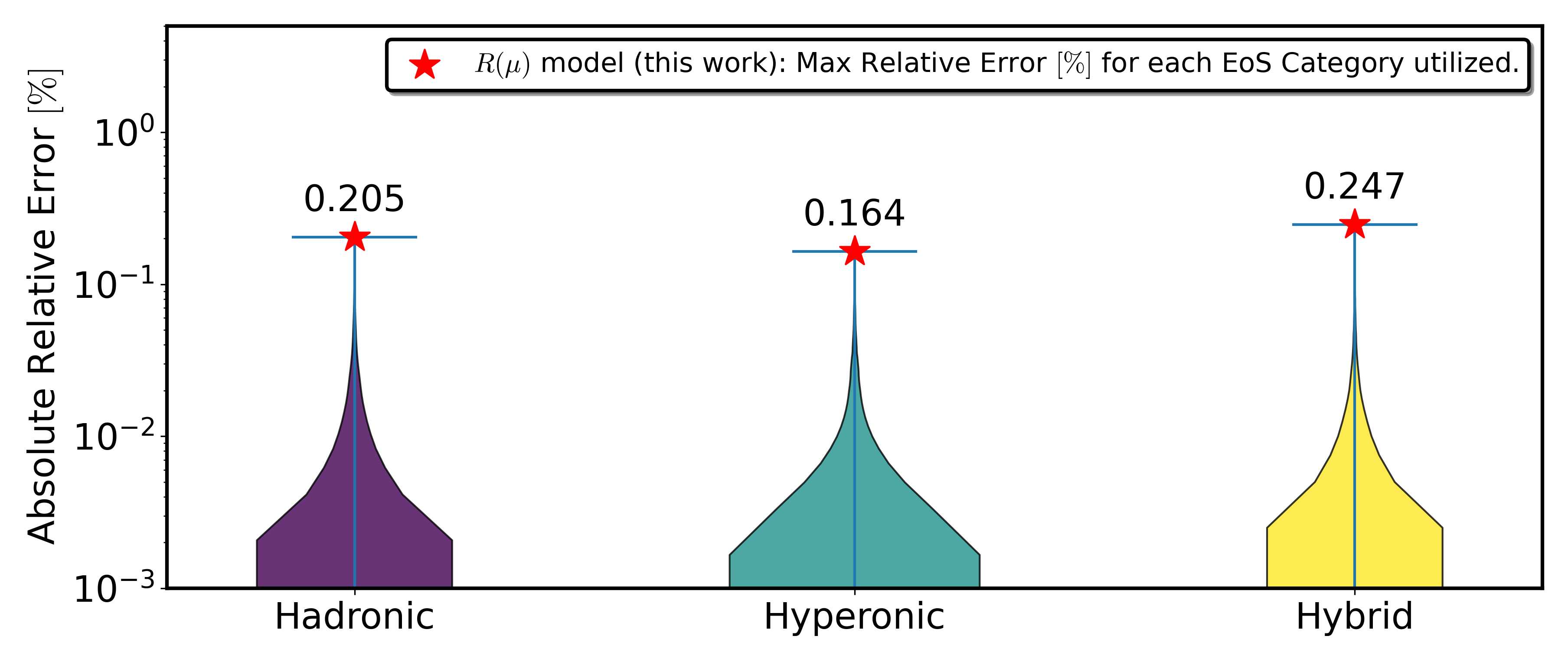}\\
        \includegraphics[width=0.71\textwidth]{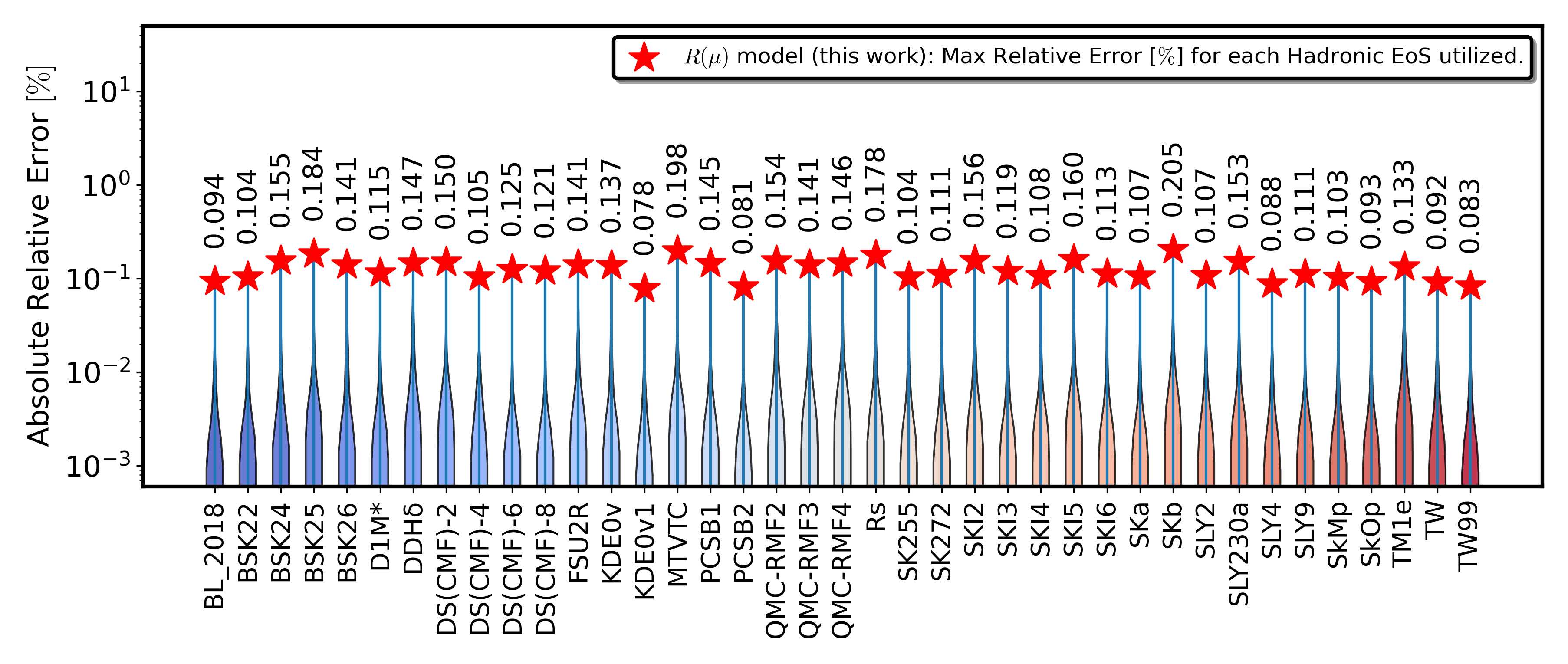}\\
        \includegraphics[width=0.71\textwidth]{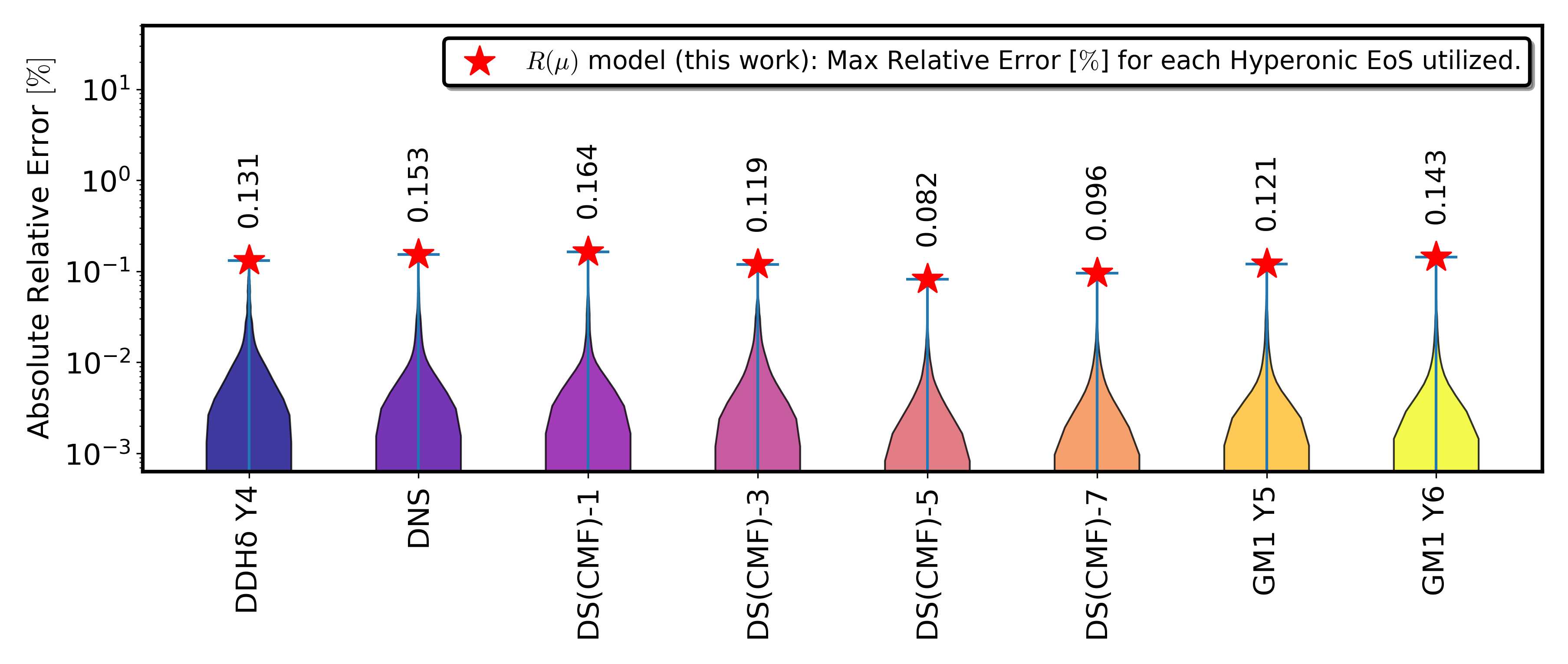}\\
        \includegraphics[width=0.71\textwidth]{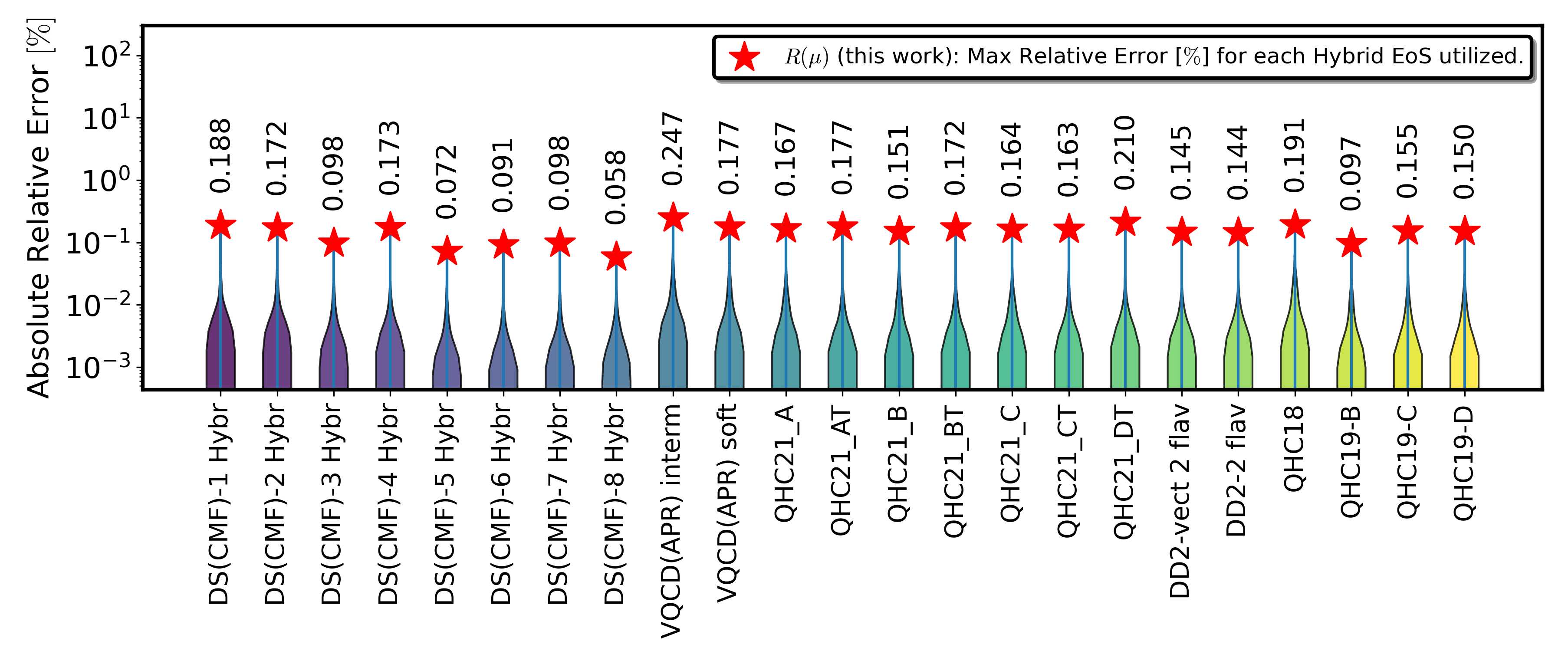}

        
        \caption{\label{fig:R_mu_violins} Violin plots illustrating the distribution of absolute fractional difference $100 \% \times (R(\mu)_{\mathrm{fit}} - R(\mu))/R(\mu)$, across the test set. The upper panel represents the variance of relative errors across EoS categories, while the subsequent ones (from top to bottom) show the relative errors for the hadronic, hyperonic, and hybrid EoSs. In addition, the absolute maximum relative error is highlighted in each distribution under investigation.}         
\end{figure*}

\end{widetext}

In this subsection, we analyze the sources of relative deviation for the regression model (\ref{eq:R_mu_fit}) related to the star's surface on the test set, considering both overall EoS categories and individual EoSs within each category. Fig. \ref{fig:R_mu_violins} concludes this analysis by presenting violin plots that depict the distribution of absolute relative variations, $100 \% \times (|R(\mu)_{\mathrm{fit}} - R(\mu)|)/R(\mu)$, for each specific case under study.

\subsection{\label{sec:regression_violin_dlog_R_mu} Residual error distributions for $d \log R(\mu)/d\theta$}

\begin{widetext}
    
  \begin{figure*}[!ht]
        \centering
    	\includegraphics[width=0.71\textwidth]{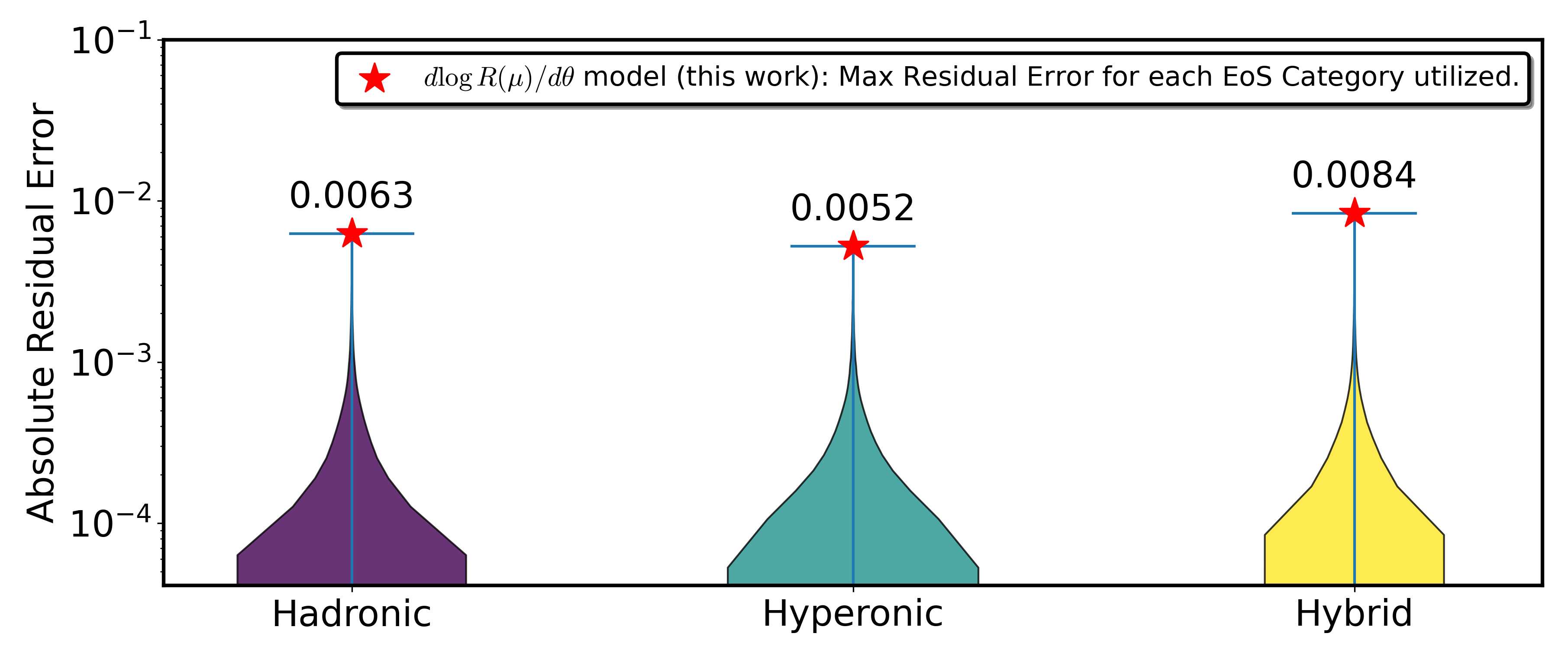}\\
        \includegraphics[width=0.71\textwidth]{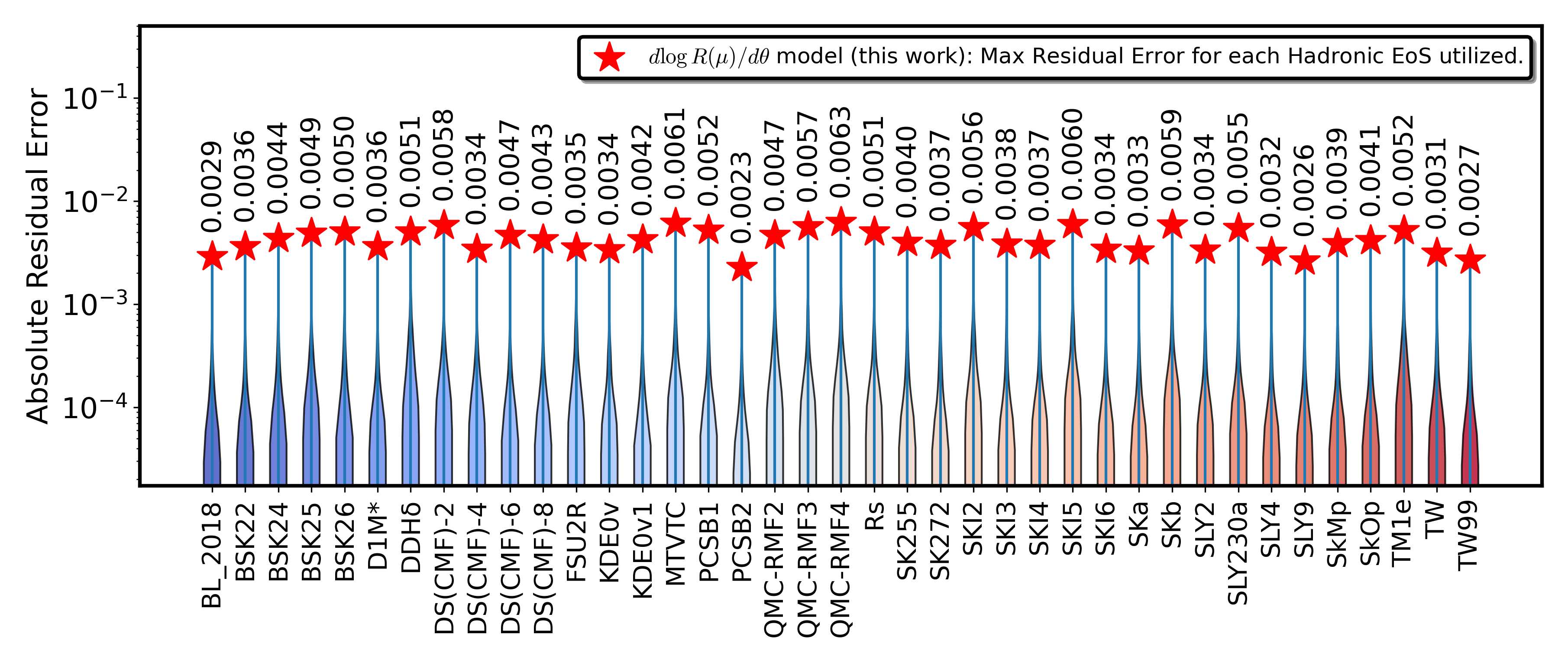}\\
        \includegraphics[width=0.71\textwidth]{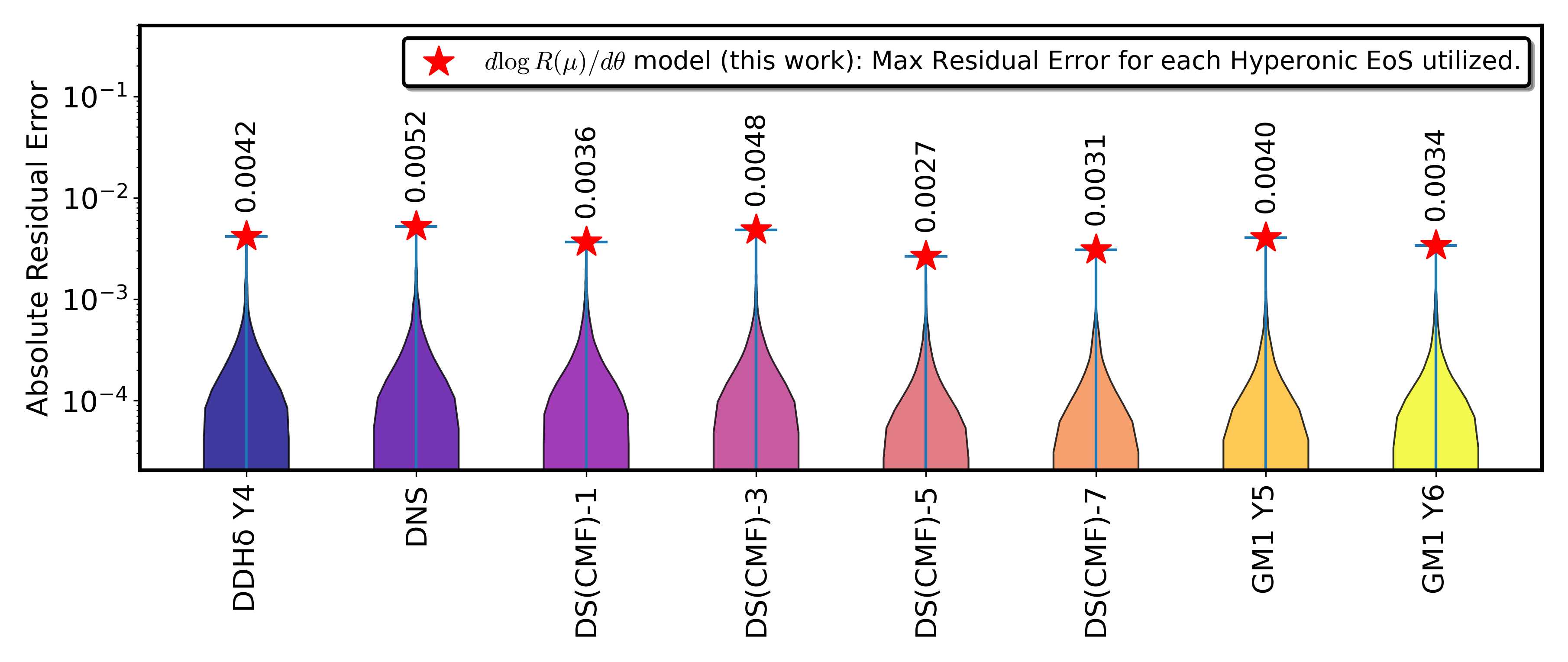}\\
        \includegraphics[width=0.71\textwidth]{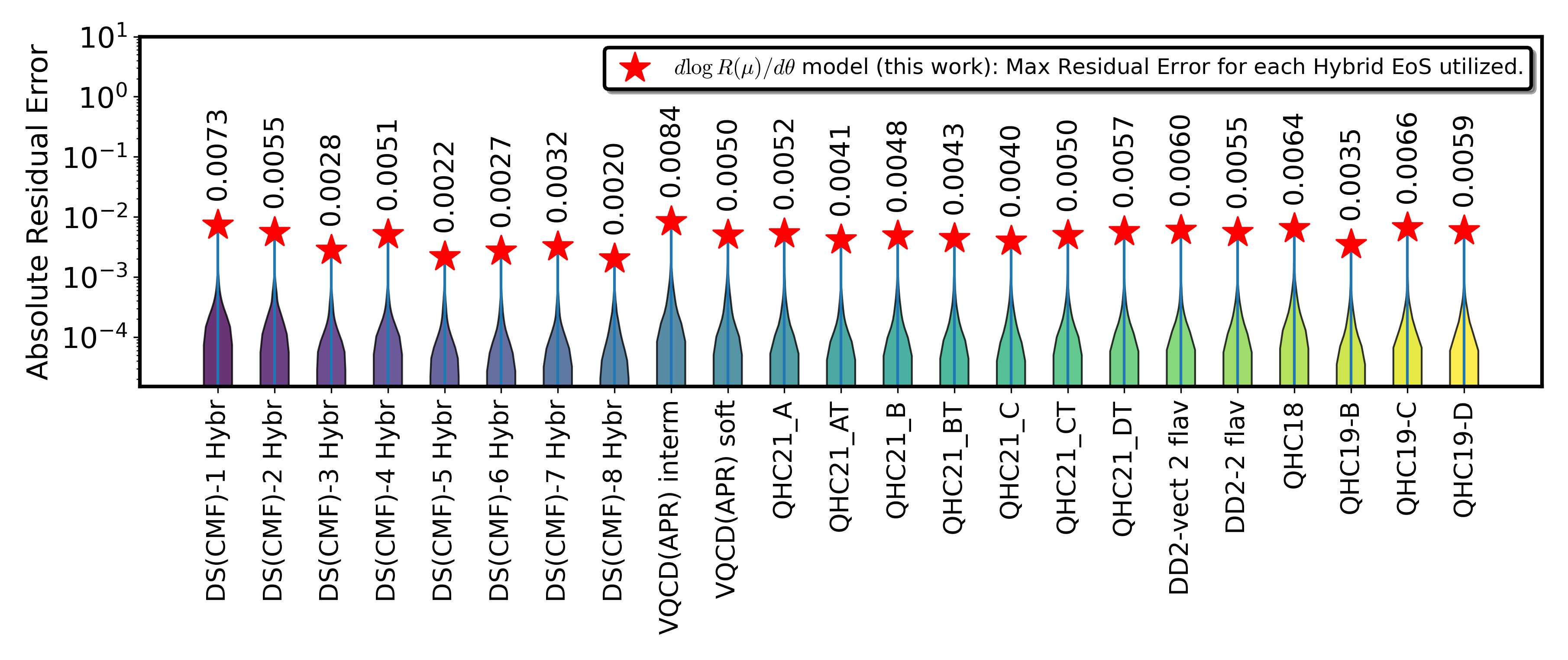}
        \caption{\label{fig:dlogR_mu_violins} Violin plots illustrating the distribution of absolute residual errors, $\left(d\log R(\mu)/d\theta \right)_{\mathrm{fit}} - \left(d\log R(\mu)/d\theta \right)$, across the test set. The upper panel represents error variance across EoS categories, while the subsequent panels (from top to bottom) show error variance for the hadronic, hyperonic, and hybrid cases, respectively.  Furthermore, the maximum residual is highlighted in each distribution for each EoS category or individual EoS of interest.}         
\end{figure*}

\end{widetext}

In this subsection, we present the sources of relative error in the regression model (\ref{dlogR_reg_fit}) on the test set, examining both overall EoS categories and individual EoSs within each category. Fig. \ref{fig:dlogR_mu_violins} concludes the analysis by presenting violin plots that illustrate the distribution of absolute residuals, $|\left(d\log R(\mu)/d\theta \right)_{\mathrm{fit}} - \left(d\log R(\mu)/d\theta \right)|$, for each specific case of interest.

\subsection{\label{sec:regression_violin_g_mu} Fractional difference distributions for $g(\mu)$}

\begin{widetext}
    
  \begin{figure*}[!ht]
        \centering
    	\includegraphics[width=0.71\textwidth]{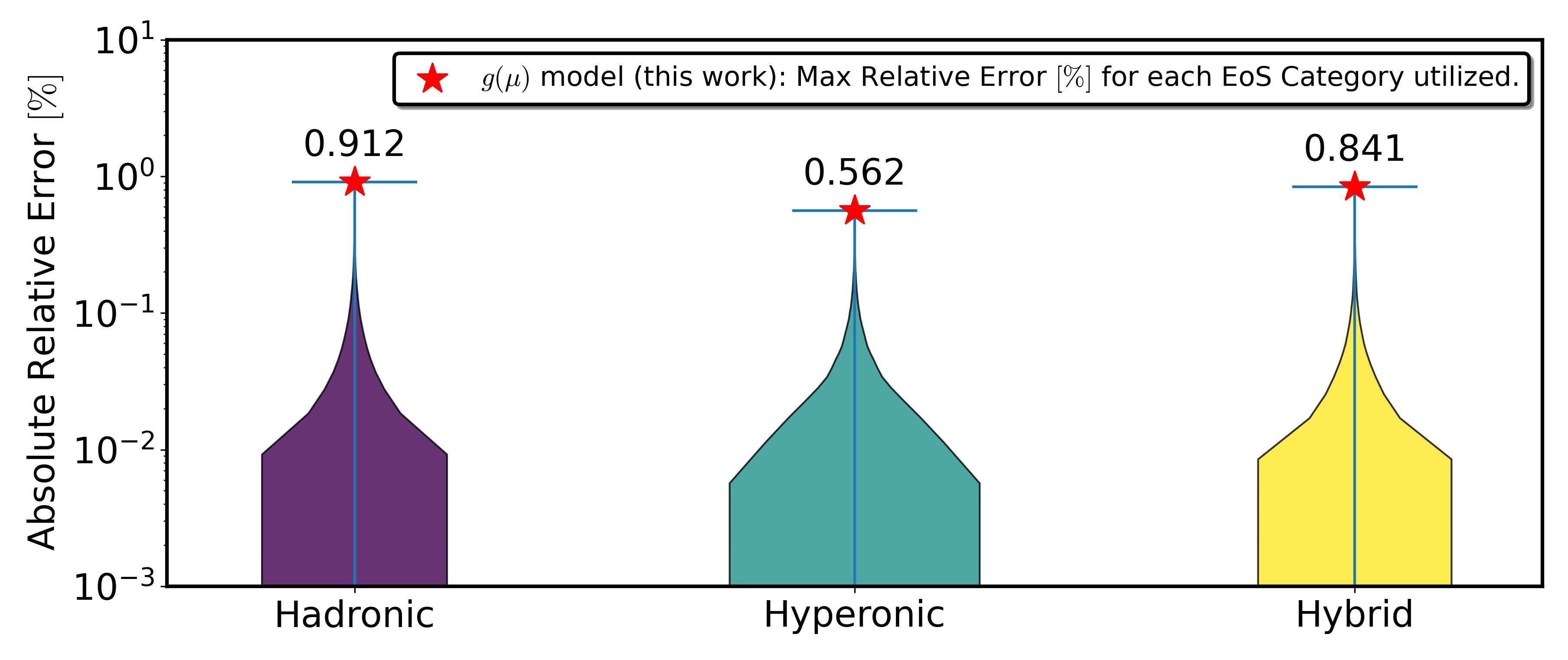}\\
        \includegraphics[width=0.71\textwidth]{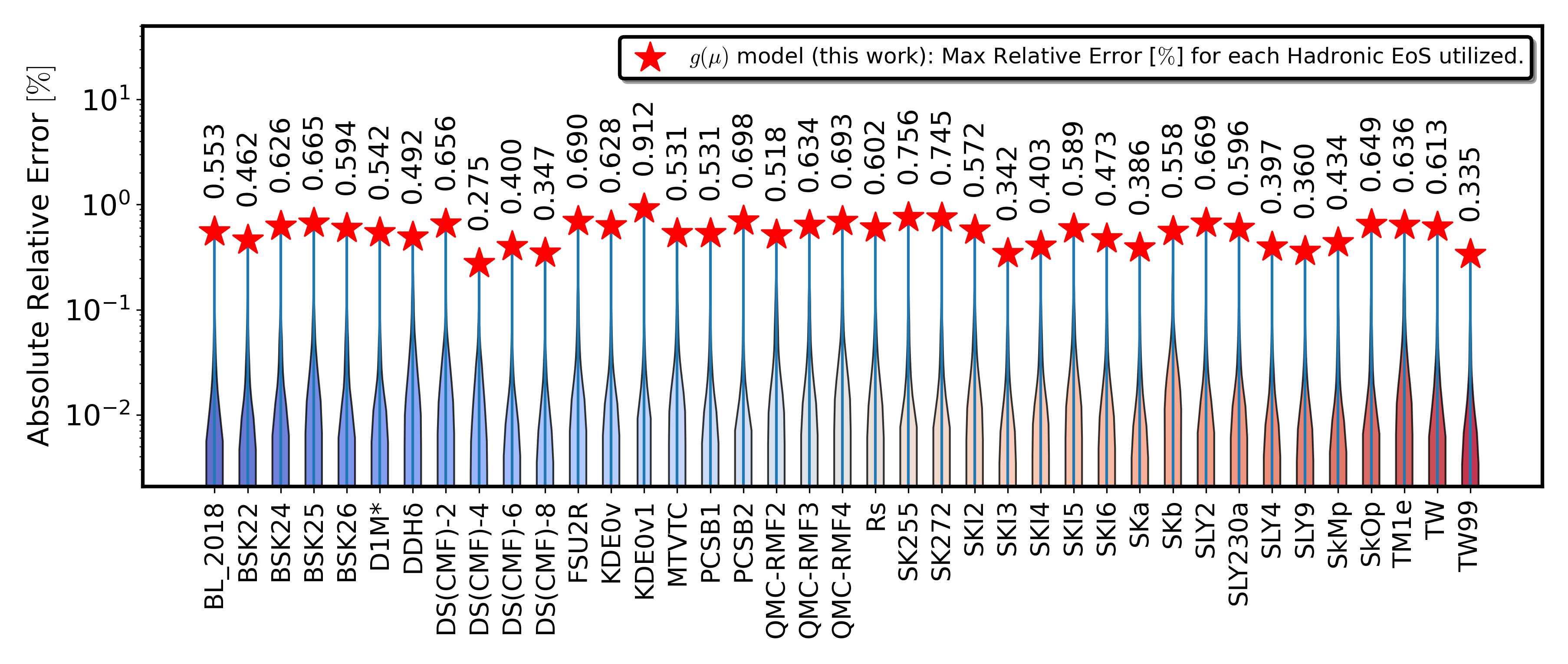}\\
        \includegraphics[width=0.71\textwidth]{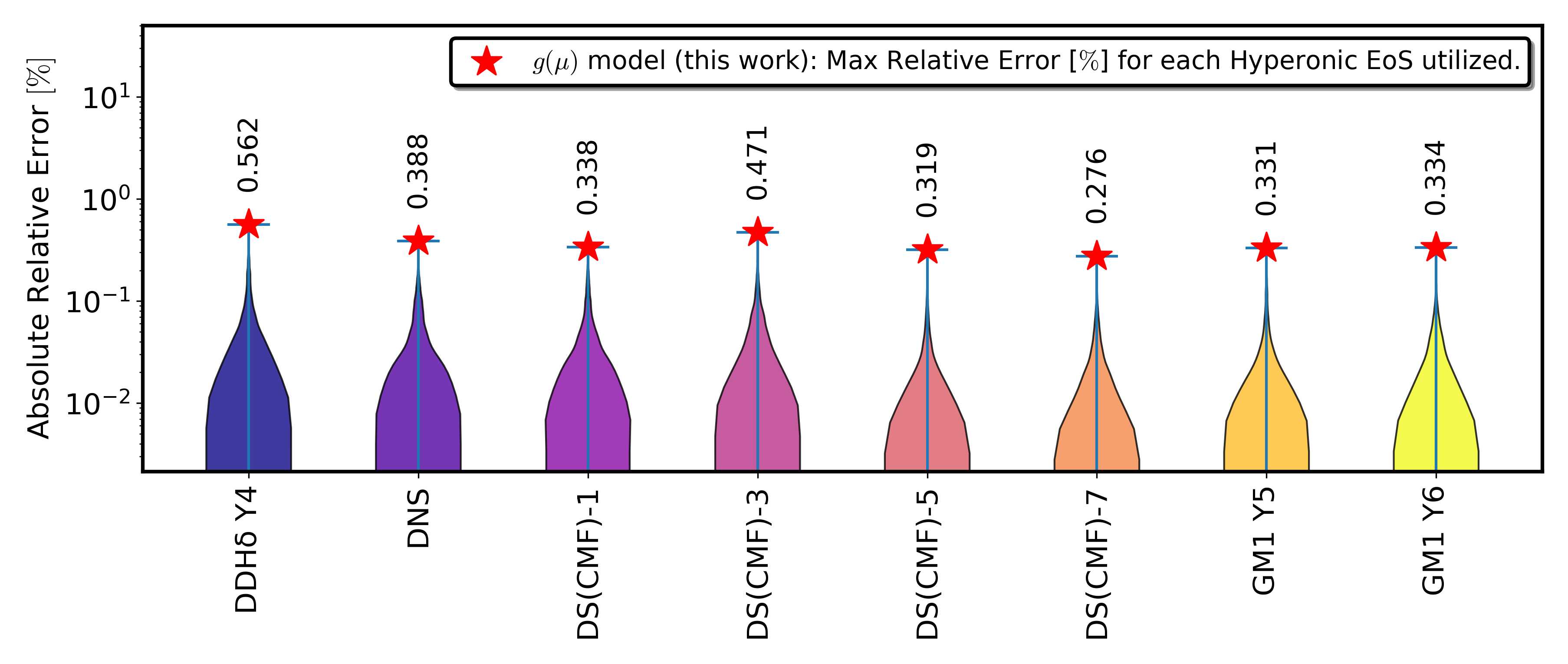}\\
        \includegraphics[width=0.71\textwidth]{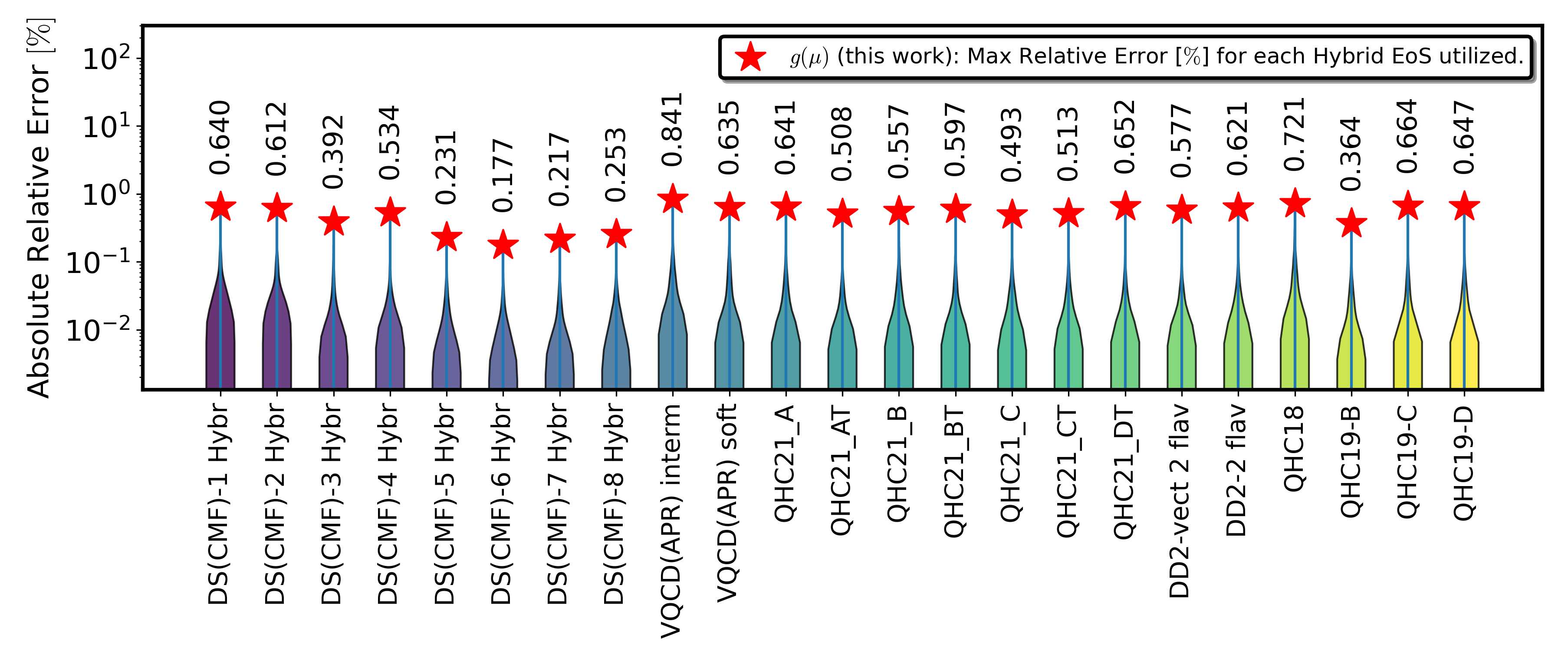}
        \caption{\label{fig:g_mu_violins} Violin plots depicting the distribution of absolute fractional differences, $100 \% \times (g(\mu)_{\mathrm{fit}} - g(\mu))/g(\mu)$, across the test set. The upper panel highlights the variance of relative errors across EoS categories, while the subsequent panels (from top to bottom) display the relative errors for hadronic, hyperonic, and hybrid EoSs. Additionally, the maximum absolute relative error is emphasized within each distribution for every EoS category and individual EoS analyzed.}         
\end{figure*}

\end{widetext}

In this subsection, we examine the sources of relative deviation in the regression model (\ref{eq:g_mu_fit}) related to the star's effective gravity on the surface, evaluating both overall EoS categories and individual EoSs within each category. Fig. \ref{fig:g_mu_violins} finalizes the analysis by showcasing violin plots that illustrate the distribution of absolute relative deviations, expressed as $100 \% \times (|g(\mu)_{\mathrm{fit}} - g(\mu)|)/g(\mu)$, for each case examined.

\bibliographystyle{ieeetr}
\bibliography{apssamp}

\providecommand{\noopsort}[1]{}\providecommand{\singleletter}[1]{#1}%
\begin{thebibliography}{100}

\bibitem{rezzolla_physics_2018}
L.~Rezzolla, P.~Pizzochero, D.~I. Jones, N.~Rea, and I.~Vida{\~n}a, {\em The physics and astrophysics of neutron stars}, vol.~457.
\newblock Springer, 2018.

\bibitem{steiner2010equation}
A.~W. Steiner, J.~M. Lattimer, and E.~F. Brown, ``The equation of state from observed masses and radii of neutron stars,'' {\em The Astrophysical Journal}, vol.~722, no.~1, p.~33, 2010.

\bibitem{lattimer2011neutron}
J.~M. Lattimer, ``Neutron stars and the dense matter equation of state,'' {\em Astrophysics and Space Science}, vol.~336, no.~1, pp.~67--74, 2011.

\bibitem{ozel_masses_2016}
F.~Ozel and P.~Freire, ``Masses, {Radii}, and {Equation} of {State} of {Neutron} {Stars},'' {\em Annual Review of Astronomy and Astrophysics}, vol.~54, no.~1, pp.~401--440, 2016.

\bibitem{burgio2021modern}
G.~F. Burgio, H.-J. Schulze, I.~Vida{\~n}a, and J.-B. Wei, ``{A modern view of the equation of state in nuclear and neutron star matter},'' {\em Symmetry}, vol.~13, no.~3, p.~400, 2021.

\bibitem{suvorov2024premerger}
A.~G. Suvorov, H.-J. Kuan, and K.~D. Kokkotas, ``{Premerger Phenomena in Neutron Star Binary Coalescences},'' {\em Universe}, vol.~10, no.~12, p.~441, 2024.

\bibitem{compose2022compose}
C.~C. Team, S.~Typel, M.~Oertel, T.~Kl{\"a}hn, D.~Chatterjee, V.~Dexheimer, C.~Ishizuka, M.~Mancini, J.~Novak, H.~Pais, {\em et~al.}, ``{CompOSE reference manual: Version 3.01, CompStar Online Supernov{\ae} Equations of State,“harmonising the concert of nuclear physics and astrophysics”, https://compose. obspm. fr},'' {\em The European Physical Journal A}, vol.~58, no.~11, p.~221, 2022.

\bibitem{oertel2017equations}
M.~Oertel, M.~Hempel, T.~Kl{\"a}hn, and S.~Typel, ``{Equations of state for supernovae and compact stars},'' {\em Reviews of Modern Physics}, vol.~89, no.~1, p.~015007, 2017.

\bibitem{lattimer2021neutron}
J.~Lattimer, ``{Neutron stars and the nuclear matter equation of state},'' {\em Annual Review of Nuclear and Particle Science}, vol.~71, no.~1, pp.~433--464, 2021.

\bibitem{chatziioannou2024neutron}
K.~Chatziioannou, H.~Cromartie, S.~Gandolfi, I.~Tews, D.~Radice, A.~W. Steiner, and A.~L. Watts, ``{Neutron stars and the dense matter equation of state: from microscopic theory to macroscopic observations},'' {\em arXiv preprint arXiv:2407.11153}, 2024.

\bibitem{miller2019psr}
M.~Miller, F.~K. Lamb, A.~Dittmann, S.~Bogdanov, Z.~Arzoumanian, K.~C. Gendreau, S.~Guillot, A.~Harding, W.~Ho, J.~Lattimer, {\em et~al.}, ``Psr j0030+ 0451 mass and radius from nicer data and implications for the properties of neutron star matter,'' {\em The Astrophysical Journal Letters}, vol.~887, no.~1, p.~L24, 2019.

\bibitem{miller2021radius}
M.~C. Miller, F.~Lamb, A.~Dittmann, S.~Bogdanov, Z.~Arzoumanian, K.~Gendreau, S.~Guillot, W.~Ho, J.~Lattimer, M.~Loewenstein, {\em et~al.}, ``The radius of psr j0740+ 6620 from nicer and xmm-newton data,'' {\em The Astrophysical Journal Letters}, vol.~918, no.~2, p.~L28, 2021.

\bibitem{baym2018hadrons}
G.~Baym, T.~Hatsuda, T.~Kojo, P.~D. Powell, Y.~Song, and T.~Takatsuka, ``From hadrons to quarks in neutron stars: a review,'' {\em Reports on Progress in Physics}, vol.~81, no.~5, p.~056902, 2018.

\bibitem{bauswein2019equation}
A.~Bauswein, ``{Equation of state constraints from multi-messenger observations of neutron star mergers},'' {\em Annals of Physics}, vol.~411, p.~167958, 2019.

\bibitem{drischler2021limiting}
C.~Drischler, S.~Han, J.~M. Lattimer, M.~Prakash, S.~Reddy, and T.~Zhao, ``Limiting masses and radii of neutron stars and their implications,'' {\em Physical Review C}, vol.~103, no.~4, p.~045808, 2021.

\bibitem{kashyap2022numerical}
R.~Kashyap, A.~Das, D.~Radice, S.~Padamata, A.~Prakash, D.~Logoteta, A.~Perego, D.~A. Godzieba, S.~Bernuzzi, I.~Bombaci, {\em et~al.}, ``{Numerical relativity simulations of prompt collapse mergers: Threshold mass and phenomenological constraints on neutron star properties after GW170817},'' {\em Physical Review D}, vol.~105, no.~10, p.~103022, 2022.

\bibitem{pang2021nuclear}
P.~T. Pang, I.~Tews, M.~W. Coughlin, M.~Bulla, C.~Van Den~Broeck, and T.~Dietrich, ``{Nuclear physics multimessenger astrophysics constraints on the neutron star equation of state: adding NICER’s PSR J0740+ 6620 measurement},'' {\em The Astrophysical Journal}, vol.~922, no.~1, p.~14, 2021.

\bibitem{imam2024implications}
S.~M.~A. Imam, T.~Malik, C.~Provid{\^e}ncia, and B.~Agrawal, ``{Implications of comprehensive nuclear and astrophysics data on the equations of state of neutron star matter},'' {\em Physical Review D}, vol.~109, no.~10, p.~103025, 2024.

\bibitem{ozel2012surface}
F.~{\"O}zel, ``Surface emission from neutron stars and implications for the physics of their interiors,'' {\em Reports on Progress in Physics}, vol.~76, no.~1, p.~016901, 2012.

\bibitem{coleman2016observational}
M.~Coleman~Miller and F.~K.~Lamb, ``Observational constraints on neutron star masses and radii,'' {\em The European Physical Journal A}, vol.~52, pp.~1--20, 2016.

\bibitem{fonseca2021refined}
E.~Fonseca, H.~T. Cromartie, T.~T. Pennucci, P.~S. Ray, A.~Y. Kirichenko, S.~M. Ransom, P.~B. Demorest, I.~H. Stairs, Z.~Arzoumanian, L.~Guillemot, {\em et~al.}, ``Refined mass and geometric measurements of the high-mass psr j0740+ 6620,'' {\em The Astrophysical Journal Letters}, vol.~915, no.~1, p.~L12, 2021.

\bibitem{heinke2006hydrogen}
C.~O. Heinke, G.~B. Rybicki, R.~Narayan, and J.~E. Grindlay, ``A hydrogen atmosphere spectral model applied to the neutron star x7 in the globular cluster 47 tucanae,'' {\em The Astrophysical Journal}, vol.~644, no.~2, p.~1090, 2006.

\bibitem{webb2007constraining}
N.~A. Webb and D.~Barret, ``Constraining the equation of state of supranuclear dense matter from xmm-newton observations of neutron stars in globular clusters,'' {\em The Astrophysical Journal}, vol.~671, no.~1, p.~727, 2007.

\bibitem{guillot2011neutron}
S.~Guillot, R.~E. Rutledge, and E.~F. Brown, ``Neutron star radius measurement with the quiescent low-mass x-ray binary u24 in ngc 6397,'' {\em The Astrophysical Journal}, vol.~732, no.~2, p.~88, 2011.

\bibitem{bogdanov2016neutron}
S.~Bogdanov, C.~O. Heinke, F.~{\"O}zel, and T.~G{\"u}ver, ``{Neutron star mass--radius constraints of the quiescent low-mass X-ray binaries X7 and X5 in the globular cluster 47 Tuc},'' {\em The Astrophysical Journal}, vol.~831, no.~2, p.~184, 2016.

\bibitem{van2024simultaneous}
M.~van~den Berg, L.~Rivera~Sandoval, C.~O. Heinke, H.~N. Cohn, P.~M. Lugger, J.~E. Grindlay, P.~D. Edmonds, J.~Anderson, and A.~Catuneanu, ``{Simultaneous Chandra and HST observations of the quiescent neutron star low-mass X-ray binaries in 47 Tucanae},'' {\em Monthly Notices of the Royal Astronomical Society}, vol.~531, no.~1, pp.~1653--1670, 2024.

\bibitem{ozel2009mass}
F.~{\"O}zel, T.~G{\"u}ver, and D.~Psaltis, ``The mass and radius of the neutron star in exo 1745- 248,'' {\em The Astrophysical Journal}, vol.~693, no.~2, p.~1775, 2009.

\bibitem{ozel2016dense}
F.~{\"O}zel, D.~Psaltis, T.~G{\"u}ver, G.~Baym, C.~Heinke, and S.~Guillot, ``{The dense matter equation of state from neutron star radius and mass measurements},'' {\em The Astrophysical Journal}, vol.~820, no.~1, p.~28, 2016.

\bibitem{parikh2021uv}
A.~Parikh, N.~Degenaar, J.~Hern{\'a}ndez~Santisteban, R.~Wijnands, I.~Psaradaki, E.~Costantini, D.~Modiano, and J.~Miller, ``{UV and X-ray observations of the neutron star LMXB EXO 0748--676 in its quiescent state},'' {\em Monthly Notices of the Royal Astronomical Society}, vol.~501, no.~1, pp.~1453--1462, 2021.

\bibitem{riley2021nicer}
T.~E. Riley, A.~L. Watts, P.~S. Ray, S.~Bogdanov, S.~Guillot, S.~M. Morsink, A.~V. Bilous, Z.~Arzoumanian, D.~Choudhury, J.~S. Deneva, {\em et~al.}, ``{A NICER view of the massive pulsar PSR J0740+ 6620 informed by radio timing and XMM-Newton spectroscopy},'' {\em The Astrophysical Journal Letters}, vol.~918, no.~2, p.~L27, 2021.

\bibitem{choudhury2024nicer}
D.~Choudhury, T.~Salmi, S.~Vinciguerra, T.~E. Riley, Y.~Kini, A.~L. Watts, B.~Dorsman, S.~Bogdanov, S.~Guillot, P.~S. Ray, {\em et~al.}, ``{A nicer view of the nearest and brightest millisecond pulsar: Psr j0437--4715},'' {\em The Astrophysical Journal Letters}, vol.~971, no.~1, p.~L20, 2024.

\bibitem{guver2016systematic}
T.~G{\"u}ver, F.~{\"O}zel, H.~Marshall, D.~Psaltis, M.~Guainazzi, and M.~D{\'\i}az-Trigo, ``{SYSTEMATIC UNCERTAINTIES IN THE SPECTROSCOPIC MEASUREMENTS OF NEUTRON STAR MASSES AND RADII FROM THERMONUCLEAR X-RAY BURSTS. III. ABSOLUTE FLUX CALIBRATION},'' {\em The Astrophysical Journal}, vol.~829, no.~1, p.~48, 2016.

\bibitem{aasi2015advanced}
J.~Aasi, B.~Abbott, R.~Abbott, T.~Abbott, M.~Abernathy, K.~Ackley, C.~Adams, T.~Adams, P.~Addesso, R.~Adhikari, {\em et~al.}, ``Advanced ligo,'' {\em Classical and quantum gravity}, vol.~32, no.~7, p.~074001, 2015.

\bibitem{acernese2014advanced}
F.~a. Acernese, M.~Agathos, K.~Agatsuma, D.~Aisa, N.~Allemandou, A.~Allocca, J.~Amarni, P.~Astone, G.~Balestri, G.~Ballardin, {\em et~al.}, ``Advanced virgo: a second-generation interferometric gravitational wave detector,'' {\em Classical and Quantum Gravity}, vol.~32, no.~2, p.~024001, 2014.

\bibitem{gendreau2012neutron}
K.~C. Gendreau, Z.~Arzoumanian, and T.~Okajima, ``The neutron star interior composition explorer (nicer): an explorer mission of opportunity for soft x-ray timing spectroscopy,'' in {\em Space Telescopes and Instrumentation 2012: Ultraviolet to Gamma Ray}, vol.~8443, pp.~322--329, SPIE, 2012.

\bibitem{arzoumanian2014neutron}
Z.~Arzoumanian, K.~Gendreau, C.~Baker, T.~Cazeau, P.~Hestnes, J.~Kellogg, S.~Kenyon, R.~Kozon, K.-C. Liu, S.~Manthripragada, {\em et~al.}, ``The neutron star interior composition explorer (nicer): mission definition,'' in {\em Space Telescopes and Instrumentation 2014: Ultraviolet to Gamma Ray}, vol.~9144, pp.~579--587, SPIE, 2014.

\bibitem{gendreau2017searching}
K.~Gendreau and Z.~Arzoumanian, ``Searching for a pulse,'' {\em Nature Astronomy}, vol.~1, no.~12, pp.~895--895, 2017.

\bibitem{watts2016colloquium}
A.~L. Watts, N.~Andersson, D.~Chakrabarty, M.~Feroci, K.~Hebeler, G.~Israel, F.~K. Lamb, M.~C. Miller, S.~Morsink, F.~{\"O}zel, {\em et~al.}, ``{Colloquium: Measuring the neutron star equation of state using x-ray timing},'' {\em Reviews of Modern Physics}, vol.~88, no.~2, p.~021001, 2016.

\bibitem{kurpas2024detection}
J.~Kurpas, A.~Schwope, A.~Pires, and F.~Haberl, ``{Detection of pulsed X-ray emission from the isolated neutron star candidate eRASSU J131716. 9--402647},'' {\em Astronomy \& Astrophysics}, vol.~683, p.~A164, 2024.

\bibitem{riley2019nicer}
T.~E. Riley, A.~L. Watts, S.~Bogdanov, P.~S. Ray, R.~M. Ludlam, S.~Guillot, Z.~Arzoumanian, C.~L. Baker, A.~V. Bilous, D.~Chakrabarty, {\em et~al.}, ``A nicer view of psr j0030+ 0451: millisecond pulsar parameter estimation,'' {\em The Astrophysical Journal Letters}, vol.~887, no.~1, p.~L21, 2019.

\bibitem{van2017upper}
E.~D. Van~Oeveren and J.~L. Friedman, ``Upper limit set by causality on the tidal deformability of a neutron star,'' {\em Physical Review D}, vol.~95, no.~8, p.~083014, 2017.

\bibitem{hinderer2008tidal}
T.~Hinderer, ``Tidal love numbers of neutron stars,'' {\em The Astrophysical Journal}, vol.~677, no.~2, p.~1216, 2008.

\bibitem{binnington2009relativistic}
T.~Binnington and E.~Poisson, ``Relativistic theory of tidal love numbers,'' {\em Physical Review D}, vol.~80, no.~8, p.~084018, 2009.

\bibitem{damour2009relativistic}
T.~Damour and A.~Nagar, ``Relativistic tidal properties of neutron stars,'' {\em Physical Review D}, vol.~80, no.~8, p.~084035, 2009.

\bibitem{chatziioannou2020neutron}
K.~Chatziioannou, ``Neutron-star tidal deformability and equation-of-state constraints,'' {\em General Relativity and Gravitation}, vol.~52, no.~11, p.~109, 2020.

\bibitem{dietrich2021interpreting}
T.~Dietrich, T.~Hinderer, and A.~Samajdar, ``Interpreting binary neutron star mergers: describing the binary neutron star dynamics, modelling gravitational waveforms, and analyzing detections,'' {\em General Relativity and Gravitation}, vol.~53, pp.~1--76, 2021.

\bibitem{gamba2023resonant}
R.~Gamba and S.~Bernuzzi, ``{Resonant tides in binary neutron star mergers: Analytical-numerical relativity study},'' {\em Physical Review D}, vol.~107, no.~4, p.~044014, 2023.

\bibitem{ripley2024constraint}
J.~L. Ripley, A.~Hegade~KR, R.~S. Chandramouli, and N.~Yunes, ``{A constraint on the dissipative tidal deformability of neutron stars},'' {\em Nature Astronomy}, pp.~1--7, 2024.

\bibitem{williams2024phenomenological}
N.~Williams, P.~Schmidt, and G.~Pratten, ``{Phenomenological model of gravitational self-force enhanced tides in inspiraling binary neutron stars},'' {\em Physical Review D}, vol.~110, no.~10, p.~104013, 2024.

\bibitem{carson2019future}
Z.~Carson, A.~W. Steiner, and K.~Yagi, ``Future prospects for constraining nuclear matter parameters with gravitational waves,'' {\em Physical Review D}, vol.~100, no.~2, p.~023012, 2019.

\bibitem{huxford2024accuracy}
R.~Huxford, R.~Kashyap, S.~Borhanian, A.~Dhani, I.~Gupta, and B.~Sathyaprakash, ``Accuracy of neutron star radius measurement with the next generation of terrestrial gravitational-wave observatories,'' {\em Physical Review D}, vol.~109, no.~10, p.~103035, 2024.

\bibitem{francesco2023nuclear}
I.~Francesco, M.~Michele, M.~Chiranjib, P.~Anna, D.~Tim, G.~Francesca, M.~Michele, and O.~Micaela, ``Nuclear physics constraints from binary neutron star mergers in the einstein telescope era,'' {\em Phys. Rev. D}, vol.~108, 2023.

\bibitem{raaijmakers2020constraining}
G.~Raaijmakers, S.~Greif, T.~Riley, T.~Hinderer, K.~Hebeler, A.~Schwenk, A.~Watts, S.~Nissanke, S.~Guillot, J.~Lattimer, {\em et~al.}, ``Constraining the dense matter equation of state with joint analysis of nicer and ligo/virgo measurements,'' {\em The Astrophysical Journal Letters}, vol.~893, no.~1, p.~L21, 2020.

\bibitem{biswas2022constraining}
B.~Biswas and S.~Datta, ``Constraining neutron star properties with a new equation of state insensitive approach,'' {\em Physical Review D}, vol.~106, no.~4, p.~043012, 2022.

\bibitem{biswas2021impact}
B.~Biswas, ``Impact of prex-ii and combined radio/nicer/xmm-newton’s mass--radius measurement of psr j0740+ 6620 on the dense-matter equation of state,'' {\em The Astrophysical Journal}, vol.~921, no.~1, p.~63, 2021.

\bibitem{biswas2022bayesian}
B.~Biswas, ``Bayesian model selection of neutron star equations of state using multi-messenger observations,'' {\em The Astrophysical Journal}, vol.~926, no.~1, p.~75, 2022.

\bibitem{traversi2020bayesian}
S.~Traversi, P.~Char, and G.~Pagliara, ``Bayesian inference of dense matter equation of state within relativistic mean field models using astrophysical measurements,'' {\em The Astrophysical Journal}, vol.~897, no.~2, p.~165, 2020.

\bibitem{xie2019bayesian}
W.-J. Xie and B.-A. Li, ``Bayesian inference of high-density nuclear symmetry energy from radii of canonical neutron stars,'' {\em The Astrophysical Journal}, vol.~883, no.~2, p.~174, 2019.

\bibitem{biswas2021towards}
B.~Biswas, P.~Char, R.~Nandi, and S.~Bose, ``Towards mitigation of apparent tension between nuclear physics and astrophysical observations by improved modeling of neutron star matter,'' {\em Physical Review D}, vol.~103, no.~10, p.~103015, 2021.

\bibitem{dietrich2020multimessenger}
T.~Dietrich, M.~W. Coughlin, P.~T. Pang, M.~Bulla, J.~Heinzel, L.~Issa, I.~Tews, and S.~Antier, ``Multimessenger constraints on the neutron-star equation of state and the hubble constant,'' {\em Science}, vol.~370, no.~6523, pp.~1450--1453, 2020.

\bibitem{landry2020nonparametric}
P.~Landry, R.~Essick, and K.~Chatziioannou, ``Nonparametric constraints on neutron star matter with existing and upcoming gravitational wave and pulsar observations,'' {\em Physical Review D}, vol.~101, no.~12, p.~123007, 2020.

\bibitem{raaijmakers2021constraints}
G.~Raaijmakers, S.~Greif, K.~Hebeler, T.~Hinderer, a.~Nissanke, A.~Schwenk, T.~Riley, A.~Watts, J.~Lattimer, and W.~Ho, ``Constraints on the dense matter equation of state and neutron star properties from nicer’s mass--radius estimate of psr j0740+ 6620 and multimessenger observations,'' {\em The Astrophysical Journal Letters}, vol.~918, no.~2, p.~L29, 2021.

\bibitem{ho2023new}
W.~C. Ho and N.~Andersson, ``{New dynamical tide constraints from current and future gravitational wave detections of inspiralling neutron stars},'' {\em Physical Review D}, vol.~108, no.~4, p.~043003, 2023.

\bibitem{abbott2017gw170817}
B.~P. Abbott, R.~Abbott, T.~Abbott, F.~Acernese, K.~Ackley, C.~Adams, T.~Adams, P.~Addesso, R.~Adhikari, V.~B. Adya, {\em et~al.}, ``Gw170817: observation of gravitational waves from a binary neutron star inspiral,'' {\em Physical review letters}, vol.~119, no.~16, p.~161101, 2017.

\bibitem{abbott2019properties}
B.~Abbott, R.~Abbott, T.~Abbott, F.~Acernese, K.~Ackley, C.~Adams, T.~Adams, P.~Addesso, R.~Adhikari, V.~Adya, {\em et~al.}, ``Properties of the binary neutron star merger gw170817,'' {\em Physical Review X}, vol.~9, no.~1, p.~011001, 2019.

\bibitem{jiang2019equation}
J.-L. Jiang, S.-P. Tang, D.-S. Shao, M.-Z. Han, Y.-J. Li, Y.-Z. Wang, Z.-P. Jin, Y.-Z. Fan, and D.-M. Wei, ``The equation of state and some key parameters of neutron stars: Constraints from gw170817, the nuclear data, and the low-mass x-ray binary data,'' {\em The Astrophysical Journal}, vol.~885, no.~1, p.~39, 2019.

\bibitem{fattoyev2018neutron}
F.~Fattoyev, J.~Piekarewicz, and C.~J. Horowitz, ``Neutron skins and neutron stars in the multimessenger era,'' {\em Physical Review Letters}, vol.~120, no.~17, p.~172702, 2018.

\bibitem{most2018new}
E.~R. Most, L.~R. Weih, L.~Rezzolla, and J.~Schaffner-Bielich, ``New constraints on radii and tidal deformabilities of neutron stars from gw170817,'' {\em Physical Review Letters}, vol.~120, no.~26, p.~261103, 2018.

\bibitem{abbott2018gw170817}
B.~P. Abbott, R.~Abbott, T.~Abbott, F.~Acernese, K.~Ackley, C.~Adams, T.~Adams, P.~Addesso, R.~X. Adhikari, V.~B. Adya, {\em et~al.}, ``Gw170817: Measurements of neutron star radii and equation of state,'' {\em Physical review letters}, vol.~121, no.~16, p.~161101, 2018.

\bibitem{landry2018constraints}
P.~Landry and B.~Kumar, ``Constraints on the moment of inertia of psr j0737-3039a from gw170817,'' {\em The Astrophysical Journal Letters}, vol.~868, no.~2, p.~L22, 2018.

\bibitem{annala2018gravitational}
E.~Annala, T.~Gorda, A.~Kurkela, and A.~Vuorinen, ``Gravitational-wave constraints on the neutron-star-matter equation of state,'' {\em Physical review letters}, vol.~120, no.~17, p.~172703, 2018.

\bibitem{lim2018neutron}
Y.~Lim and J.~W. Holt, ``Neutron star tidal deformabilities constrained by nuclear theory and experiment,'' {\em Physical review letters}, vol.~121, no.~6, p.~062701, 2018.

\bibitem{kumar2019inferring}
B.~Kumar and P.~Landry, ``Inferring neutron star properties from gw170817 with universal relations,'' {\em Physical Review D}, vol.~99, no.~12, p.~123026, 2019.

\bibitem{zdunik2013maximum}
J.~Zdunik and P.~Haensel, ``Maximum mass of neutron stars and strange neutron-star cores,'' {\em Astronomy \& Astrophysics}, vol.~551, p.~A61, 2013.

\bibitem{haensel_neutron_2007}
P.~Haensel, A.~Y. Potekhin, and D.~G. Yakovlev, {\em {Neutron stars 1: Equation of state and structure}}, vol.~326.
\newblock New York, USA: Springer, 2007.

\bibitem{al2021combining}
M.~Al-Mamun, A.~W. Steiner, J.~N{\"a}ttil{\"a}, J.~Lange, R.~O’Shaughnessy, I.~Tews, S.~Gandolfi, C.~Heinke, and S.~Han, ``Combining electromagnetic and gravitational-wave constraints on neutron-star masses and radii,'' {\em Physical Review Letters}, vol.~126, no.~6, p.~061101, 2021.

\bibitem{kruger2023rapidly}
C.~J. Kr{\"u}ger and S.~H. V{\"o}lkel, ``{Rapidly rotating neutron stars: Universal relations and EOS inference},'' {\em Physical Review D}, vol.~108, no.~12, p.~124056, 2023.

\bibitem{fujimoto2020mapping}
Y.~Fujimoto, K.~Fukushima, and K.~Murase, ``Mapping neutron star data to the equation of state using the deep neural network,'' {\em Physical Review D}, vol.~101, no.~5, p.~054016, 2020.

\bibitem{fujimoto2018methodology}
Y.~Fujimoto, K.~Fukushima, and K.~Murase, ``Methodology study of machine learning for the neutron star equation of state,'' {\em Physical Review D}, vol.~98, no.~2, p.~023019, 2018.

\bibitem{ferreira2021unveiling}
M.~Ferreira and C.~Provid{\^e}ncia, ``Unveiling the nuclear matter eos from neutron star properties: a supervised machine learning approach,'' {\em Journal of Cosmology and Astroparticle Physics}, vol.~2021, no.~07, p.~011, 2021.

\bibitem{farrell2023deducing}
D.~Farrell, P.~Baldi, J.~Ott, A.~Ghosh, A.~W. Steiner, A.~Kavitkar, L.~Lindblom, D.~Whiteson, and F.~Weber, ``{Deducing neutron star equation of state from telescope spectra with machine-learning-derived likelihoods},'' {\em Journal of Cosmology and Astroparticle Physics}, vol.~2023, no.~12, p.~022, 2023.

\bibitem{krastev2023deep}
P.~G. Krastev, ``A deep learning approach to extracting nuclear matter properties from neutron star observations,'' {\em Symmetry}, vol.~15, no.~5, p.~1123, 2023.

\bibitem{Morawski2020}
F.~Morawski and M.~Bejger, ``Neural network reconstruction of the dense matter equation of state derived from the parameters of neutron stars,'' {\em A\&A}, vol.~642, p.~A78, 2020.

\bibitem{soma2022neural}
S.~Soma, L.~Wang, S.~Shi, H.~St{\"o}cker, and K.~Zhou, ``Neural network reconstruction of the dense matter equation of state from neutron star observables,'' {\em Journal of Cosmology and Astroparticle Physics}, vol.~2022, no.~08, p.~071, 2022.

\bibitem{soma2023reconstructing}
S.~Soma, L.~Wang, S.~Shi, H.~St{\"o}cker, and K.~Zhou, ``Reconstructing the neutron star equation of state from observational data via automatic differentiation,'' {\em Physical Review D}, vol.~107, no.~8, p.~083028, 2023.

\bibitem{lobato2022cluster}
R.~V. Lobato, E.~V. Chimanski, and C.~A. Bertulani, ``Cluster structures with machine learning support in neutron star mr relations,'' in {\em Journal of Physics: Conference Series}, vol.~2340, p.~012014, IOP Publishing, 2022.

\bibitem{lobato2022unsupervised}
R.~V. Lobato, E.~V. Chimanski, and C.~A. Bertulani, ``Unsupervised machine learning correlations in eos of neutron stars,'' {\em arXiv preprint arXiv:2202.13940}, 2022.

\bibitem{ferreira2022extracting}
M.~Ferreira, V.~Carvalho, and C.~Provid{\^e}ncia, ``Extracting nuclear matter properties from the neutron star matter equation of state using deep neural networks,'' {\em Physical Review D}, vol.~106, no.~10, p.~103023, 2022.

\bibitem{ravenhall1994neutron}
D.~Ravenhall and C.~J. Pethick, ``Neutron star moments of inertia,'' {\em The Astrophysical Journal}, vol.~424, pp.~846--851, 1994.

\bibitem{lattimer2001neutron}
J.~Lattimer and M.~Prakash, ``Neutron star structure and the equation of state,'' {\em The Astrophysical Journal}, vol.~550, no.~1, p.~426, 2001.

\bibitem{bejger2002moments}
M.~Bejger and P.~Haensel, ``Moments of inertia for neutron and strange stars: Limits derived for the crab pulsar,'' {\em Astronomy \& Astrophysics}, vol.~396, no.~3, pp.~917--921, 2002.

\bibitem{lattimer2005constraining}
J.~M. Lattimer and B.~F. Schutz, ``Constraining the equation of state with moment of inertia measurements,'' {\em The Astrophysical Journal}, vol.~629, no.~2, p.~979, 2005.

\bibitem{breu_maximum_2016}
C.~Breu and L.~Rezzolla, ``Maximum mass, moment of inertia and compactness of relativistic stars,'' {\em Monthly Notices of the Royal Astronomical Society}, vol.~459, no.~1, pp.~646--656, 2016.

\bibitem{Laarakkers:1997hb}
W.~G. Laarakkers and E.~Poisson, ``Quadrupole moments of rotating neutron stars,'' {\em Astrophys. J.}, vol.~512, pp.~282--287, 1999.

\bibitem{Pappas:2012ns}
G.~Pappas and T.~A. Apostolatos, ``Revising the multipole moments of numerical spacetimes, and its consequences,'' {\em Phys. Rev. Lett.}, vol.~108, p.~231104, 2012.

\bibitem{Pappas:2012qg}
G.~Pappas and T.~A. Apostolatos, ``Multipole moments of numerical spacetimes,'' {\em arXiv preprint arXiv:1211.6299}, 2012.

\bibitem{urbanec2013quadrupole}
M.~Urbanec, J.~C. Miller, and Z.~Stuchlik, ``Quadrupole moments of rotating neutron stars and strange stars,'' {\em Monthly Notices of the Royal Astronomical Society}, vol.~433, no.~3, pp.~1903--1909, 2013.

\bibitem{yagi2017approximate}
K.~Yagi and N.~Yunes, ``Approximate universal relations for neutron stars and quark stars,'' {\em Physics Reports}, vol.~681, pp.~1--72, 2017.

\bibitem{yagi2013love}
K.~Yagi and N.~Yunes, ``{I-Love-Q}: Unexpected universal relations for neutron stars and quark stars,'' {\em Science}, vol.~341, no.~6144, pp.~365--368, 2013.

\bibitem{baubock2013relations}
M.~Baub{\"o}ck, E.~Berti, D.~Psaltis, and F.~{\"O}zel, ``Relations between neutron-star parameters in the hartle--thorne approximation,'' {\em The Astrophysical Journal}, vol.~777, no.~1, p.~68, 2013.

\bibitem{yagi_i-love-q_2013a}
K.~Yagi and N.~Yunes, ``{I-Love-Q} relations in neutron stars and their applications to astrophysics, gravitational waves, and fundamental physics,'' {\em Physical Review D}, vol.~88, no.~2, p.~023009, 2013.

\bibitem{maselli2013equation}
A.~Maselli, V.~Cardoso, V.~Ferrari, L.~Gualtieri, and P.~Pani, ``Equation-of-state-independent relations in neutron stars,'' {\em Physical Review D}, vol.~88, no.~2, p.~023007, 2013.

\bibitem{doneva2013breakdown}
D.~D. Doneva, S.~S. Yazadjiev, N.~Stergioulas, and K.~D. Kokkotas, ``Breakdown of {I-LOVE-Q} universality in rapidly rotating relativistic stars,'' {\em The Astrophysical Journal Letters}, vol.~781, no.~1, p.~L6, 2014.

\bibitem{pappas2014effectively}
G.~Pappas and T.~A. Apostolatos, ``Effectively universal behavior of rotating neutron stars in general relativity makes them even simpler than their newtonian counterparts,'' {\em Physical Review Letters}, vol.~112, no.~12, p.~121101, 2014.

\bibitem{chakrabarti2014q}
S.~Chakrabarti, T.~Delsate, N.~G{\"u}rlebeck, and J.~Steinhoff, ``{I-Q} relation for rapidly rotating neutron stars,'' {\em Physical Review Letters}, vol.~112, no.~20, p.~201102, 2014.

\bibitem{Konstantinou:2022vkr}
A.~Konstantinou and S.~M. Morsink, ``Universal relations for the increase in the mass and radius of a rotating neutron star,'' {\em The Astrophysical Journal}, vol.~934, no.~2, p.~139, 2022.

\bibitem{stein2014three}
L.~C. Stein, K.~Yagi, and N.~Yunes, ``Three-hair relations for rotating stars: Nonrelativistic limit,'' {\em The Astrophysical Journal}, vol.~788, no.~1, p.~15, 2014.

\bibitem{yagi2014effective}
K.~Yagi, K.~Kyutoku, G.~Pappas, N.~Yunes, and T.~A. Apostolatos, ``Effective no-hair relations for neutron stars and quark stars: relativistic results,'' {\em Physical Review D}, vol.~89, no.~12, p.~124013, 2014.

\bibitem{chatziioannou2014toward}
K.~Chatziioannou, K.~Yagi, and N.~Yunes, ``Toward realistic and practical no-hair relations for neutron stars in the nonrelativistic limit,'' {\em Physical Review D}, vol.~90, no.~6, p.~064030, 2014.

\bibitem{papigkiotis2023universal}
G.~Papigkiotis and G.~Pappas, ``Universal relations for rapidly rotating neutron stars using supervised machine-learning techniques,'' {\em Physical Review D}, vol.~107, no.~10, p.~103050, 2023.

\bibitem{manoharan2023finding}
P.~Manoharan and K.~D. Kokkotas, ``Finding universal relations using statistical data analysis,'' {\em arXiv preprint arXiv:2307.13063}, 2023.

\bibitem{martinon2014rotating}
G.~Martinon, A.~Maselli, L.~Gualtieri, and V.~Ferrari, ``Rotating protoneutron stars: Spin evolution, maximum mass, and {I-Love-Q} relations,'' {\em Physical Review D}, vol.~90, no.~6, p.~064026, 2014.

\bibitem{yagi2014love}
K.~Yagi, L.~C. Stein, G.~Pappas, N.~Yunes, and T.~A. Apostolatos, ``Why {I-Love-Q}: Explaining why universality emerges in compact objects,'' {\em Physical Review D}, vol.~90, no.~6, p.~063010, 2014.

\bibitem{sham2015unveiling}
Y.-H. Sham, T.~Chan, L.-M. Lin, and P.~Leung, ``Unveiling the universality of {I-Love-Q} relations,'' {\em The Astrophysical Journal}, vol.~798, no.~2, p.~121, 2015.

\bibitem{bogdanov2008thermal}
S.~Bogdanov, J.~E. Grindlay, and G.~B. Rybicki, ``{Thermal x-rays from millisecond pulsars: Constraining the fundamental properties of neutron stars},'' {\em The Astrophysical Journal}, vol.~689, no.~1, p.~407, 2008.

\bibitem{miller2015determining}
M.~C. Miller and F.~K. Lamb, ``{Determining neutron star properties by fitting oblate-star waveform models to X-ray burst oscillations},'' {\em The Astrophysical Journal}, vol.~808, no.~1, p.~31, 2015.

\bibitem{mereghetti2010x}
S.~Mereghetti, ``{X-ray emission from isolated neutron stars},'' in {\em High-Energy Emission from Pulsars and their Systems: Proceedings of the First Session of the Sant Cugat Forum on Astrophysics}, pp.~345--363, Springer, 2010.

\bibitem{silva2019neutron}
H.~O. Silva and N.~Yunes, ``{Neutron star pulse profile observations as extreme gravity probes},'' {\em Classical and Quantum Gravity}, vol.~36, no.~17, p.~17LT01, 2019.

\bibitem{yunes2022gravitational}
N.~Yunes, M.~C. Miller, and K.~Yagi, ``{Gravitational-wave and X-ray probes of the neutron star equation of state},'' {\em Nature Reviews Physics}, vol.~4, no.~4, pp.~237--246, 2022.

\bibitem{watts2019constraining}
A.~L. Watts, ``Constraining the neutron star equation of state using pulse profile modeling,'' in {\em AIP Conference Proceedings}, vol.~2127, AIP Publishing, 2019.

\bibitem{raaijmakers2019nicer}
G.~Raaijmakers, T.~E. Riley, A.~L. Watts, S.~Greif, S.~Morsink, K.~Hebeler, A.~Schwenk, T.~Hinderer, S.~Nissanke, S.~Guillot, {\em et~al.}, ``A nicer view of psr j0030+ 0451: Implications for the dense matter equation of state,'' {\em The Astrophysical Journal Letters}, vol.~887, no.~1, p.~L22, 2019.

\bibitem{jiang2020psr}
J.-L. Jiang, S.-P. Tang, Y.-Z. Wang, Y.-Z. Fan, and D.-M. Wei, ``Psr j0030+ 0451, gw170817, and the nuclear data: Joint constraints on equation of state and bulk properties of neutron stars,'' {\em The Astrophysical Journal}, vol.~892, no.~1, p.~55, 2020.

\bibitem{salmi2024radius}
T.~Salmi, D.~Choudhury, Y.~Kini, T.~E. Riley, S.~Vinciguerra, A.~L. Watts, M.~T. Wolff, Z.~Arzoumanian, S.~Bogdanov, D.~Chakrabarty, {\em et~al.}, ``{The Radius of the High Mass Pulsar PSR J0740+ 6620 With 3.6 Years of NICER Data},'' {\em arXiv preprint arXiv:2406.14466}, 2024.

\bibitem{dittmann2024more}
A.~J. Dittmann, M.~C. Miller, F.~K. Lamb, I.~M. Holt, C.~Chirenti, M.~T. Wolff, S.~Bogdanov, S.~Guillot, W.~C. Ho, S.~M. Morsink, {\em et~al.}, ``{A more precise measurement of the radius of psr j0740+ 6620 using updated nicer data},'' {\em The Astrophysical Journal}, vol.~974, no.~2, p.~295, 2024.

\bibitem{cromartie_relativistic_2020}
H.~T. Cromartie, E.~Fonseca, S.~M. Ransom, P.~B. Demorest, Z.~Arzoumanian, H.~Blumer, P.~R. Brook, M.~E. DeCesar, T.~Dolch, J.~A. Ellis, {\em et~al.}, ``Relativistic shapiro delay measurements of an extremely massive millisecond pulsar,'' {\em Nature Astronomy}, vol.~4, no.~1, pp.~72--76, 2020.

\bibitem{reardon2024neutron}
D.~J. Reardon, M.~Bailes, R.~M. Shannon, C.~Flynn, J.~Askew, N.~R. Bhat, Z.-C. Chen, M.~Cury{\l}o, Y.~Feng, G.~B. Hobbs, {\em et~al.}, ``{The neutron star mass, distance, and inclination from precision timing of the brilliant millisecond pulsar j0437-4715},'' {\em The Astrophysical Journal Letters}, vol.~971, no.~1, p.~L18, 2024.

\bibitem{salmi2024nicer}
T.~Salmi, J.~S. Deneva, P.~S. Ray, A.~L. Watts, D.~Choudhury, Y.~Kini, S.~Vinciguerra, H.~T. Cromartie, M.~T. Wolff, Z.~Arzoumanian, {\em et~al.}, ``{A NICER View of PSR J1231- 1411: A Complex Case},'' {\em The Astrophysical Journal}, vol.~976, no.~1, p.~58, 2024.

\bibitem{silva2021astrophysical}
H.~O. Silva, A.~M. Holgado, A.~C{\'a}rdenas-Avenda{\~n}o, and N.~Yunes, ``Astrophysical and theoretical physics implications from multimessenger neutron star observations,'' {\em Physical review letters}, vol.~126, no.~18, p.~181101, 2021.

\bibitem{bogdanov2019constraining}
S.~Bogdanov, S.~Guillot, P.~S. Ray, M.~T. Wolff, D.~Chakrabarty, W.~C. Ho, M.~Kerr, F.~K. Lamb, A.~Lommen, R.~M. Ludlam, {\em et~al.}, ``Constraining the neutron star mass--radius relation and dense matter equation of state with nicer. i. the millisecond pulsar x-ray data set,'' {\em The Astrophysical Journal Letters}, vol.~887, no.~1, p.~L25, 2019.

\bibitem{galloway2005discovery}
D.~K. Galloway, C.~B. Markwardt, E.~H. Morgan, D.~Chakrabarty, and T.~E. Strohmayer, ``{Discovery of the accretion-powered millisecond X-ray pulsar IGR J00291+ 5934},'' {\em The Astrophysical Journal}, vol.~622, no.~1, p.~L45, 2005.

\bibitem{Patruno2021}
A.~Patruno and A.~L. Watts, {\em Accreting Millisecond X-ray Pulsars}, pp.~143--208.
\newblock Berlin, Heidelberg: Springer Berlin Heidelberg, 2021.

\bibitem{muno2002frequency}
M.~P. Muno, D.~Chakrabarty, D.~K. Galloway, and D.~Psaltis, ``{The frequency stability of millisecond oscillations in thermonuclear X-ray bursts},'' {\em The Astrophysical Journal}, vol.~580, no.~2, p.~1048, 2002.

\bibitem{tarana2008integral}
A.~Tarana, A.~Bazzano, and P.~Ubertini, ``{INTEGRAL and BeppoSAX observations of the transient atoll source 4U 1608--522: from quiescent to hard spectral state},'' {\em The Astrophysical Journal}, vol.~688, no.~2, p.~1295, 2008.

\bibitem{morsink2007oblate}
S.~M. Morsink, D.~A. Leahy, C.~Cadeau, and J.~Braga, ``The oblate schwarzschild approximation for light curves of rapidly rotating neutron stars,'' {\em The Astrophysical Journal}, vol.~663, no.~2, p.~1244, 2007.

\bibitem{baubock2013narrow}
M.~Baub{\"o}ck, D.~Psaltis, and F.~{\"O}zel, ``Narrow atomic features from rapidly spinning neutron stars,'' {\em The Astrophysical Journal}, vol.~766, no.~2, p.~87, 2013.

\bibitem{raithel2019constraints}
C.~A. Raithel, ``Constraints on the neutron star equation of state from gw170817,'' {\em The European Physical Journal A}, vol.~55, no.~5, p.~80, 2019.

\bibitem{zimmerman2020measuring}
J.~Zimmerman, Z.~Carson, K.~Schumacher, A.~W. Steiner, and K.~Yagi, ``Measuring nuclear matter parameters with nicer and ligo/virgo,'' {\em arXiv preprint arXiv:2002.03210}, 2020.

\bibitem{essick2020direct}
R.~Essick, I.~Tews, P.~Landry, S.~Reddy, and D.~E. Holz, ``Direct astrophysical tests of chiral effective field theory at supranuclear densities,'' {\em Physical Review C}, vol.~102, no.~5, p.~055803, 2020.

\bibitem{algendy2014universality}
M.~AlGendy and S.~M. Morsink, ``Universality of the acceleration due to gravity on the surface of a rapidly rotating neutron star,'' {\em The Astrophysical Journal}, vol.~791, no.~2, p.~78, 2014.

\bibitem{silva2021surface}
H.~O. Silva, G.~Pappas, N.~Yunes, and K.~Yagi, ``Surface of rapidly-rotating neutron stars: Implications to neutron star parameter estimation,'' {\em Physical Review D}, vol.~103, no.~6, p.~063038, 2021.

\bibitem{1993ApJS...88..205L}
D.~{Lai}, F.~A. {Rasio}, and S.~L. {Shapiro}, ``{Ellipsoidal Figures of Equilibrium: Compressible Models},'' {\em Astrophys. J. Suppl.}, vol.~88, p.~205, Sept. 1993.

\bibitem{hornik1989multilayer}
K.~Hornik, M.~Stinchcombe, and H.~White, ``Multilayer feedforward networks are universal approximators,'' {\em Neural networks}, vol.~2, no.~5, pp.~359--366, 1989.

\bibitem{andersson1999relevance}
N.~Andersson, K.~D. Kokkotas, and N.~Stergioulas, ``{On the relevance of the r-mode instability for accreting neutron stars and white dwarfs},'' {\em The Astrophysical Journal}, vol.~516, no.~1, p.~307, 1999.

\bibitem{andersson2000r}
N.~Andersson, D.~I. Jones, K.~D. Kokkotas, and N.~Stergioulas, ``{r-Mode runaway and rapidly rotating neutron stars},'' {\em The Astrophysical Journal}, vol.~534, no.~1, p.~L75, 2000.

\bibitem{oppenheimer_massive_1939}
J.~R. Oppenheimer and G.~M. Volkoff, ``On massive neutron cores,'' {\em Physical Review}, vol.~55, no.~4, p.~374, 1939.

\bibitem{rns}
``https://github.com/cgca/rns..''

\bibitem{stergioulas1994comparing}
N.~Stergioulas and J.~L. Friedman, ``{Comparing models of rapidly rotating relativistic stars constructed by two numerical methods},'' {\em Astrophys. J.}, vol.~444, p.~306, 1995.

\bibitem{friedman_rotating_2013}
J.~L. Friedman and N.~Stergioulas, {\em Rotating relativistic stars}.
\newblock Cambridge University Press, 2013.

\bibitem{cook1992spin}
G.~B. Cook, S.~L. Shapiro, and S.~A. Teukolsky, ``Spin-up of a rapidly rotating star by angular momentum loss-effects of general relativity,'' {\em Astrophysical Journal, Part 1 (ISSN 0004-637X), vol. 398, no. 1, p. 203-223.}, vol.~398, pp.~203--223, 1992.

\bibitem{rezzolla2013relativistic}
L.~Rezzolla and O.~Zanotti, {\em Relativistic hydrodynamics}.
\newblock Oxford University Press, 2013.

\bibitem{dutra2014relativistic}
M.~Dutra, O.~Louren{\c{c}}o, S.~S. Avancini, B.~V. Carlson, A.~Delfino, D.~P. Menezes, C.~Provid{\^e}ncia, S.~Typel, and J.~R. Stone, ``{Relativistic mean-field hadronic models under nuclear matter constraints},'' {\em Physical Review C}, vol.~90, no.~5, p.~055203, 2014.

\bibitem{dutra2012skyrme}
M.~Dutra, O.~Louren{\c{c}}o, J.~S. SaMartins, A.~Delfino, J.~R. Stone, and P.~D. Stevenson, ``Skyrme interaction and nuclear matter constraints,'' {\em Physical Review C}, vol.~85, no.~3, p.~035201, 2012.

\bibitem{yu_self-consistent_2020}
Z.-X. Yu, T.~Zhao, and H.-S. Zong, ``Self-consistent mean field approximation and application in three-flavor njl model,'' {\em Chinese Physics C}, vol.~44, no.~7, p.~074104, 2020.

\bibitem{malik_gw170817_2018}
T.~Malik, N.~Alam, M.~Fortin, C.~Provid{\^e}ncia, B.~K. Agrawal, T.~K. Jha, B.~Kumar, and S.~K. Patra, ``{GW170817}: Constraining the nuclear matter equation of state from the neutron star tidal deformability,'' {\em Physical Review C}, vol.~98, no.~3, p.~035804, 2018.

\bibitem{compose}
``https://compose.obspm.fr/home.''

\bibitem{demorest2010two}
P.~B. Demorest, T.~Pennucci, S.~Ransom, M.~Roberts, and J.~Hessels, ``A two-solar-mass neutron star measured using shapiro delay,'' {\em nature}, vol.~467, no.~7319, pp.~1081--1083, 2010.

\bibitem{antoniadis_massive_2013}
J.~Antoniadis, P.~C. Freire, N.~Wex, T.~M. Tauris, R.~S. Lynch, M.~H. Van~Kerkwijk, M.~Kramer, C.~Bassa, V.~S. Dhillon, T.~Driebe, {\em et~al.}, ``A massive pulsar in a compact relativistic binary,'' {\em Science}, vol.~340, no.~6131, p.~1233232, 2013.

\bibitem{bauswein_neutron-star_2017}
A.~Bauswein, O.~Just, H.-T. Janka, and N.~Stergioulas, ``Neutron-star radius constraints from {GW170817} and future detections,'' {\em The Astrophysical Journal Letters}, vol.~850, no.~2, p.~L34, 2017.

\bibitem{friedman_astrophysical_2020-2}
J.~L. Friedman and N.~Stergioulas, ``Astrophysical implications of neutron star inspiral and coalescence,'' {\em International Journal of Modern Physics D}, vol.~29, no.~11, p.~2041015, 2020.

\bibitem{rezzolla_using_2018}
L.~Rezzolla, E.~R. Most, and L.~R. Weih, ``Using gravitational-wave observations and quasi-universal relations to constrain the maximum mass of neutron stars,'' {\em The Astrophysical Journal Letters}, vol.~852, no.~2, p.~L25, 2018.

\bibitem{saffer2024lower}
A.~Saffer, E.~Fonseca, S.~Ransom, I.~Stairs, R.~Lynch, D.~Good, K.~W. Masui, J.~W. McKee, B.~W. Meyers, S.~S. Patil, {\em et~al.}, ``{A Lower Mass Estimate for PSR J0348+ 0432 Based on CHIME/Pulsar Precision Timing},'' {\em arXiv preprint arXiv:2412.02850}, 2024.

\bibitem{butterworth_structure_1976}
E.~M. Butterworth and J.~R. Ipser, ``On the structure and stability of rapidly rotating fluid bodies in general relativity. {I.} the numerical method for computing structure and its application to uniformly rotating homogeneous bodies,'' {\em The Astrophysical Journal}, vol.~204, pp.~200--223, 1976.

\bibitem{paschalidis2017rotating}
V.~Paschalidis and N.~Stergioulas, ``Rotating stars in relativity,'' {\em Living Reviews in Relativity}, vol.~20, no.~1, pp.~1--169, 2017.

\bibitem{wilson1972models}
J.~R. Wilson, ``Models of differentially rotating stars.,'' {\em The Astrophysical Journal}, vol.~176, p.~195, 1972.

\bibitem{bonazzola1974exact-2}
S.~Bonazzola and J.~Schneider, ``An exact study of rigidly and rapidly rotating stars in general relativity with application to the crab pulsar,'' {\em The Astrophysical Journal}, vol.~191, pp.~273--290, 1974.

\bibitem{friedman1989implications}
J.~L. Friedman, J.~R. Ipser, and L.~Parker, ``Implications of a half-millisecond pulsar,'' {\em Phys. Rev. Lett.}, vol.~62, pp.~3015--3019, Jun 1989.

\bibitem{komatsu_rapidly_1989-2}
H.~Komatsu, Y.~Eriguchi, and I.~Hachisu, ``Rapidly rotating general relativistic stars{--I}. numerical method and its application to uniformly rotating polytropes,'' {\em Monthly Notices of the Royal Astronomical Society}, vol.~237, no.~2, pp.~355--379, 1989.

\bibitem{komatsu_rapidly_1989_II-2}
H.~Komatsu, Y.~Eriguchi, and I.~Hachisu, ``Rapidly rotating general relativistic stars{--II}. differentially rotating polytropes,'' {\em Monthly Notices of the Royal Astronomical Society}, vol.~239, no.~1, pp.~153--171, 1989.

\bibitem{cook_spin-up_1992}
G.~B. Cook, S.~L. Shapiro, and S.~A. Teukolsky, ``Spin-up of a rapidly rotating star by angular momentum loss-effects of general relativity,'' {\em Astrophysical Journal, Part 1 (ISSN 0004-637X), vol. 398, no. 1, p. 203-223.}, vol.~398, pp.~203--223, 1992.

\bibitem{2020SciPy-NMeth}
P.~Virtanen, R.~Gommers, T.~E. Oliphant, M.~Haberland, T.~Reddy, D.~Cournapeau, E.~Burovski, P.~Peterson, W.~Weckesser, J.~Bright, S.~J. {van der Walt}, M.~Brett, J.~Wilson, K.~J. Millman, N.~Mayorov, A.~R.~J. Nelson, E.~Jones, R.~Kern, E.~Larson, C.~J. Carey, {\.I}.~Polat, Y.~Feng, E.~W. Moore, J.~{VanderPlas}, D.~Laxalde, J.~Perktold, R.~Cimrman, I.~Henriksen, E.~A. Quintero, C.~R. Harris, A.~M. Archibald, A.~H. Ribeiro, F.~Pedregosa, P.~{van Mulbregt}, and {SciPy 1.0 Contributors}, ``{{SciPy} 1.0: Fundamental Algorithms for Scientific Computing in Python},'' {\em Nature Methods}, vol.~17, pp.~261--272, 2020.

\bibitem{baubock2012ray}
M.~Baub{\"o}ck, D.~Psaltis, F.~{\"O}zel, and T.~Johannsen, ``A ray-tracing algorithm for spinning compact object spacetimes with arbitrary quadrupole moments. ii. neutron stars,'' {\em The Astrophysical Journal}, vol.~753, no.~2, p.~175, 2012.

\bibitem{cadeau2007light}
C.~Cadeau, S.~M. Morsink, D.~Leahy, and S.~S. Campbell, ``Light curves for rapidly rotating neutron stars,'' {\em The Astrophysical Journal}, vol.~654, no.~1, p.~458, 2007.

\bibitem{osti_4483768}
K.~S. Thorne, ``Relativistic stellar structure and dynamics.,'' {\em pp 259-441 of High Energy Astrophysics. Vol. III. DeWitt, C. Schatzman, E. Veron, P. (eds.). New York, Gordon and Breach, Science Publishers, 1967.}, 10 1968.

\bibitem{cumming2002hydrostatic}
A.~Cumming, S.~M. Morsink, L.~Bildsten, J.~L. Friedman, and D.~E. Holz, ``Hydrostatic expansion and spin changes during type i x-ray bursts,'' {\em The Astrophysical Journal}, vol.~564, no.~1, p.~343, 2002.

\bibitem{nozawa1998construction}
T.~Nozawa, N.~Stergioulas, E.~Gourgoulhon, and Y.~Eriguchi, ``Construction of highly accurate models of rotating neutron stars--comparison of three different numerical schemes,'' {\em Astronomy and Astrophysics Supplement Series}, vol.~132, no.~3, pp.~431--454, 1998.

\bibitem{suleimanov2017direct}
V.~F. Suleimanov, J.~Poutanen, J.~N{\"a}ttil{\"a}, J.~J. Kajava, M.~G. Revnivtsev, and K.~Werner, ``The direct cooling tail method for x-ray burst analysis to constrain neutron star masses and radii,'' {\em Monthly Notices of the Royal Astronomical Society}, vol.~466, no.~1, pp.~906--913, 2017.

\bibitem{suleimanov2020observational}
V.~F. Suleimanov, J.~Poutanen, and K.~Werner, ``Observational appearance of rapidly rotating neutron stars-x-ray bursts, cooling tail method, and radius determination,'' {\em Astronomy \& Astrophysics}, vol.~639, p.~A33, 2020.

\bibitem{akmal_equation_1998}
A.~Akmal, V.~R. Pandharipande, and D.~G. Ravenhall, ``Equation of state of nucleon matter and neutron star structure,'' {\em Physical Review C}, vol.~58, no.~3, pp.~1804--1828, 1998.

\bibitem{jokela2019holographic}
N.~Jokela, M.~J{\"a}rvinen, and J.~Remes, ``Holographic qcd in the veneziano limit and neutron stars,'' {\em Journal of High Energy Physics}, vol.~2019, no.~3, 2019.

\bibitem{ishii2019cool}
T.~Ishii, M.~J{\"a}rvinen, and G.~Nijs, ``Cool baryon and quark matter in holographic qcd,'' {\em Journal of High Energy Physics}, vol.~2019, no.~7, 2019.

\bibitem{ecker2020gravitational}
C.~Ecker, M.~J{\"a}rvinen, G.~Nijs, and W.~van~der Schee, ``Gravitational waves from holographic neutron star mergers,'' {\em Physical Review D}, vol.~101, no.~10, p.~103006, 2020.

\bibitem{jokela2021unified}
N.~Jokela, M.~J{\"a}rvinen, G.~Nijs, and J.~Remes, ``Unified weak and strong coupling framework for nuclear matter and neutron stars,'' {\em Physical Review D}, vol.~103, no.~8, p.~086004, 2021.

\bibitem{danielewicz_symmetry_2009}
P.~Danielewicz and J.~Lee, ``Symmetry energy {I}: Semi-infinite matter,'' {\em Nuclear Physics A}, vol.~818, no.~1-2, pp.~36--96, 2009.

\bibitem{gulminelli_unified_2015}
F.~Gulminelli and A.~R. Raduta, ``Unified treatment of subsaturation stellar matter at zero and finite temperature,'' {\em Physical Review C}, vol.~92, no.~5, p.~055803, 2015.

\bibitem{kohler_skyrme_1976}
H.~K{\"o}hler, ``Skyrme force and the mass formula,'' {\em Nuclear Physics A}, vol.~258, no.~2, pp.~301--316, 1976.

\bibitem{bennour_charge_1989}
L.~Bennour, P.-H. Heenen, P.~Bonche, J.~Dobaczewski, and H.~Flocard, ``Charge distributions of {Pb 208}, {Pb 206}, and {Tl 205} and the mean-field approximation,'' {\em Physical Review C}, vol.~40, no.~6, p.~2834, 1989.

\bibitem{dexheimer_gw190814_2021}
V.~Dexheimer, R.~O. Gomes, T.~Kl{\"a}hn, S.~Han, and M.~Salinas, ``{GW190814} as a massive rapidly rotating neutron star with exotic degrees of freedom,'' {\em Physical Review C}, vol.~103, no.~2, p.~025808, 2021.

\bibitem{dexheimer_novel_2010}
V.~A. Dexheimer and S.~Schramm, ``Novel approach to modeling hybrid stars,'' {\em Physical Review C}, vol.~81, no.~4, p.~045201, 2010.

\bibitem{dexheimer_proto-neutron_2008}
V.~Dexheimer and S.~Schramm, ``Proto-neutron and neutron stars in a chiral {SU (3)} model,'' {\em The Astrophysical Journal}, vol.~683, no.~2, p.~943, 2008.

\bibitem{dexheimer_reconciling_2015}
V.~Dexheimer, R.~Negreiros, and S.~Schramm, ``Reconciling nuclear and astrophysical constraints,'' {\em Physical Review C}, vol.~92, no.~1, p.~012801(R), 2015.

\bibitem{dexheimer_tabulated_2017}
V.~Dexheimer, ``Tabulated neutron star equations of state modelled within the chiral mean field model,'' {\em Publications of the Astronomical Society of Australia}, vol.~34, 2017.

\bibitem{typel1999relativistic}
S.~Typel and H.~Wolter, ``Relativistic mean field calculations with density-dependent meson-nucleon coupling,'' {\em Nuclear Physics A}, vol.~656, no.~3-4, pp.~331--364, 1999.

\bibitem{xia2022unified}
C.-J. Xia, T.~Maruyama, A.~Li, B.~Y. Sun, W.-H. Long, and Y.-X. Zhang, ``Unified neutron star eoss and neutron star structures in rmf models,'' {\em Communications in Theoretical Physics}, vol.~74, no.~9, p.~095303, 2022.

\bibitem{xia2022unified_2}
C.-J. Xia, B.~Y. Sun, T.~Maruyama, W.-H. Long, and A.~Li, ``Unified nuclear matter equations of state constrained by the in-medium balance in density-dependent covariant density functionals,'' {\em Physical Review C}, vol.~105, no.~4, p.~045803, 2022.

\bibitem{schurhoff_neutron_2010}
T.~Sch{\"u}rhoff, S.~Schramm, and V.~Dexheimer, ``Neutron stars with small radii—the role of $\delta$ resonances,'' {\em The Astrophysical Journal Letters}, vol.~724, no.~1, p.~L74, 2010.

\bibitem{agrawal_determination_2005}
B.~K. Agrawal, S.~Shlomo, and V.~K. Au, ``Determination of the parameters of a skyrme type effective interaction using the simulated annealing approach,'' {\em Physical Review C}, vol.~72, no.~1, p.~014310, 2005.

\bibitem{douchin_unified_2001}
F.~Douchin and P.~Haensel, ``A unified equation of state of dense matter and neutron star structure,'' {\em Astronomy \& Astrophysics}, vol.~380, no.~1, pp.~151--167, 2001.

\bibitem{gaitanos_lorentz_2004}
T.~Gaitanos, M.~Di~Toro, S.~Typel, V.~Baran, C.~Fuchs, V.~Greco, and H.~Wolter, ``On the lorentz structure of the symmetry energy,'' {\em Nuclear Physics A}, vol.~732, pp.~24--48, 2004.

\bibitem{grill_equation_2014}
F.~Grill, H.~Pais, C.~Provid{\^e}ncia, I.~Vidana, and S.~S. Avancini, ``Equation of state and thickness of the inner crust of neutron stars,'' {\em Physical Review C}, vol.~90, no.~4, p.~045803, 2014.

\bibitem{oertel_hyperons_2015}
M.~Oertel, C.~Provid{\^e}ncia, F.~Gulminelli, and A.~R. Raduta, ``Hyperons in neutron star matter within relativistic mean-field models,'' {\em Journal of Physics G: Nuclear and Particle Physics}, vol.~42, no.~7, p.~075202, 2015.

\bibitem{van2020waves}
B.~F. van Baal, F.~R. Chambers, and A.~L. Watts, ``Waves in thin oceans on oblate neutron stars,'' {\em Monthly Notices of the Royal Astronomical Society}, vol.~496, no.~2, pp.~2098--2106, 2020.

\bibitem{watts2019dense}
A.~L. Watts, W.~Yu, J.~Poutanen, S.~Zhang, S.~Bhattacharyya, S.~Bogdanov, L.~Ji, A.~Patruno, T.~E. Riley, P.~Bakala, {\em et~al.}, ``Dense matter with extp,'' {\em Science China Physics, Mechanics \& Astronomy}, vol.~62, pp.~1--17, 2019.

\bibitem{zhang2019enhanced}
S.~Zhang, A.~Santangelo, M.~Feroci, Y.~Xu, F.~Lu, Y.~Chen, H.~Feng, S.~Zhang, S.~Brandt, M.~Hernanz, {\em et~al.}, ``The enhanced x-ray timing and polarimetry mission—extp,'' {\em SCIENCE CHINA Physics, Mechanics \& Astronomy}, vol.~62, pp.~1--25, 2019.

\bibitem{ray2018strobe}
P.~S. Ray, Z.~Arzoumanian, S.~Brandt, E.~Burns, D.~Chakrabarty, M.~Feroci, K.~C. Gendreau, O.~Gevin, M.~Hernanz, P.~Jenke, {\em et~al.}, ``Strobe-x: a probe-class mission for x-ray spectroscopy and timing on timescales from microseconds to years,'' in {\em Space Telescopes and Instrumentation 2018: Ultraviolet to Gamma Ray}, vol.~10699, pp.~249--268, SPIE, 2018.

\bibitem{ray2019strobe}
P.~S. Ray, Z.~Arzoumanian, D.~Ballantyne, E.~Bozzo, S.~Brandt, L.~Brenneman, D.~Chakrabarty, M.~Christophersen, A.~DeRosa, M.~Feroci, {\em et~al.}, ``{STROBE-X: X-ray timing and spectroscopy on dynamical timescales from microseconds to years},'' {\em arXiv preprint arXiv:1903.03035}, 2019.

\bibitem{Watts:2015yU}
A.~Watts, C.~M. Espinoza, R.~Xu, N.~Andersson, J.~Antoniadis, D.~Antonopoulou, S.~Buchner, S.~Datta, P.~Demorest, P.~Freire, J.~Hessels, J.~Margueron, M.~Oertel, A.~Patruno, A.~Possenti, S.~Ransom, I.~Stairs, and B.~Stappers, ``{{Probing the neutron star interior and the Equation of State of cold dense matter with the SKA}},'' in {\em Proceedings of Advancing Astrophysics with the Square Kilometre Array {\textemdash} PoS(AASKA14)}, vol.~215, p.~043, 2015.

\bibitem{bishop2006pattern}
C.~M. Bishop and N.~M. Nasrabadi, {\em Pattern recognition and machine learning}, vol.~4.
\newblock New York, USA: Springer, 2006.

\bibitem{murphy2012machine}
K.~P. Murphy, {\em Machine learning: a probabilistic perspective}.
\newblock MIT press, 2012.

\bibitem{Goodfellow-et-al-2016}
I.~Goodfellow, Y.~Bengio, and A.~Courville, {\em Deep Learning}.
\newblock MIT Press, 2016.

\bibitem{prince2023understanding}
S.~J. Prince, {\em Understanding Deep Learning}.
\newblock MIT Press, 2023.

\bibitem{burden_numerical_2015}
R.~Burden, J.~Faires, and A.~Burden, {\em Numerical Analysis}.
\newblock Cengage Learning, 10 edition~ed., 2015.

\bibitem{ramachandran2009mathematical}
K.~Ramachandran and C.~Tsokos, {\em Mathematical Statistics with Applications}.
\newblock Elsevier Science, 2009.

\bibitem{pedregosa_scikit-learn_2011}
F.~Pedregosa, G.~Varoquaux, A.~Gramfort, V.~Michel, B.~Thirion, O.~Grisel, M.~Blondel, P.~Prettenhofer, R.~Weiss, V.~Dubourg, {\em et~al.}, ``Scikit-learn: Machine learning in python,'' {\em the Journal of machine Learning research}, vol.~12, pp.~2825--2830, 2011.

\bibitem{james2013introduction}
G.~James, D.~Witten, T.~Hastie, R.~Tibshirani, {\em et~al.}, {\em An introduction to statistical learning}, vol.~112.
\newblock Springer, 2~ed., 2013.

\bibitem{cybenko1989approximation}
G.~Cybenko, ``Approximation by superpositions of a sigmoidal function,'' {\em Mathematics of control, signals and systems}, vol.~2, no.~4, pp.~303--314, 1989.

\bibitem{hornik1991approximation}
K.~Hornik, ``Approximation capabilities of multilayer feedforward networks,'' {\em Neural networks}, vol.~4, no.~2, pp.~251--257, 1991.

\bibitem{lecun2015deep}
Y.~LeCun, Y.~Bengio, and G.~Hinton, ``Deep learning,'' {\em nature}, vol.~521, no.~7553, pp.~436--444, 2015.

\bibitem{paszke2019pytorch}
A.~Paszke, S.~Gross, F.~Massa, A.~Lerer, J.~Bradbury, G.~Chanan, T.~Killeen, Z.~Lin, N.~Gimelshein, L.~Antiga, A.~Desmaison, A.~Kopf, E.~Yang, Z.~DeVito, M.~Raison, A.~Tejani, S.~Chilamkurthy, B.~Steiner, L.~Fang, J.~Bai, and S.~Chintala, ``Pytorch: An imperative style, high-performance deep learning library,'' in {\em Advances in Neural Information Processing Systems}, vol.~32, Curran Associates, Inc., 2019.

\bibitem{maas2013rectifier}
A.~L. Maas, A.~Y. Hannun, A.~Y. Ng, {\em et~al.}, ``Rectifier nonlinearities improve neural network acoustic models,'' in {\em Proc. icml}, vol.~30, p.~3, Atlanta, GA, 2013.

\bibitem{kingma2015adam}
D.~P. Kingma and J.~Ba, ``Adam: A method for stochastic optimization,'' in {\em International Conference on Learning Representations (ICLR)}, 2015.

\bibitem{he2015delving}
K.~He, X.~Zhang, S.~Ren, and J.~Sun, ``Delving deep into rectifiers: Surpassing human-level performance on imagenet classification,'' in {\em Proceedings of the IEEE international conference on computer vision}, pp.~1026--1034, 2015.

\bibitem{scikit-learn}
F.~Pedregosa, G.~Varoquaux, A.~Gramfort, V.~Michel, B.~Thirion, O.~Grisel, M.~Blondel, P.~Prettenhofer, R.~Weiss, V.~Dubourg, J.~Vanderplas, A.~Passos, D.~Cournapeau, M.~Brucher, M.~Perrot, and E.~Duchesnay, ``Scikit-learn: Machine learning in python,'' {\em Journal of Machine Learning Research}, vol.~12, pp.~2825--2830, 2011.

\bibitem{chabanat_skyrme_1998}
E.~Chabanat, P.~Bonche, P.~Haensel, J.~Meyer, and R.~Schaeffer, ``A skyrme parametrization from subnuclear to neutron star densities part ii. nuclei far from stabilities,'' {\em Nuclear Physics A}, vol.~635, no.~1-2, pp.~231--256, 1998.

\bibitem{reinhard_nuclear_1995}
P.-G. Reinhard and H.~Flocard, ``Nuclear effective forces and isotope shifts,'' {\em Nuclear Physics A}, vol.~584, no.~3, pp.~467--488, 1995.

\bibitem{friedrich_skyrme-force_1986}
J.~Friedrich and P.-G. Reinhard, ``Skyrme-force parametrization: Least-squares fit to nuclear ground-state properties,'' {\em Physical Review C}, vol.~33, no.~1, p.~335, 1986.

\bibitem{agrawal_nuclear_2003}
B.~K. Agrawal, S.~Shlomo, and V.~KimAu, ``Nuclear matter incompressibility coefficient in relativistic and nonrelativistic microscopic models,'' {\em Physical Review C}, vol.~68, no.~3, p.~031304(R), 2003.

\bibitem{vinas2021unified}
X.~Vi{\~n}as, C.~Gonzalez-Boquera, M.~Centelles, C.~Mondal, and L.~M. Robledo, ``Unified equation of state for neutron stars based on the gogny interaction,'' {\em Symmetry}, vol.~13, no.~9, p.~1613, 2021.

\bibitem{mondal2020structure}
C.~Mondal, X.~Vi{\~n}as, M.~Centelles, and J.~De, ``Structure and composition of the inner crust of neutron stars from gogny interactions,'' {\em Physical Review C}, vol.~102, no.~1, p.~015802, 2020.

\bibitem{gonzalez2018new}
C.~Gonzalez-Boquera, M.~Centelles, X.~Vi{\~n}as, and L.~Robledo, ``New gogny interaction suitable for astrophysical applications,'' {\em Physics Letters B}, vol.~779, pp.~195--200, 2018.

\bibitem{baym1971ground}
G.~Baym, C.~Pethick, and P.~Sutherland, ``The ground state of matter at high densities: equation of state and stellar models,'' {\em The Astrophysical Journal}, vol.~170, p.~299, 1971.

\bibitem{alford2022relativistic}
M.~G. Alford, L.~Brodie, A.~Haber, and I.~Tews, ``Relativistic mean-field theories for neutron-star physics based on chiral effective field theory,'' {\em Physical Review C}, vol.~106, no.~5, p.~055804, 2022.

\bibitem{negreiros2018cooling}
R.~Negreiros, L.~Tolos, M.~Centelles, A.~Ramos, and V.~Dexheimer, ``Cooling of small and massive hyperonic stars,'' {\em The Astrophysical Journal}, vol.~863, no.~1, p.~104, 2018.

\bibitem{providencia2019hyperonic}
C.~Provid{\^e}ncia, M.~Fortin, H.~Pais, and A.~Rabhi, ``Hyperonic stars and the nuclear symmetry energy,'' {\em Frontiers in Astronomy and Space Sciences}, vol.~6, p.~13, 2019.

\bibitem{pearson2018unified}
J.~Pearson, N.~Chamel, A.~Potekhin, A.~Fantina, C.~Ducoin, A.~K. Dutta, and S.~Goriely, ``Unified equations of state for cold non-accreting neutron stars with brussels--montreal functionals--i. role of symmetry energy,'' {\em Monthly Notices of the Royal Astronomical Society}, vol.~481, no.~3, pp.~2994--3026, 2018.

\bibitem{hempel_statistical_2010}
M.~Hempel and J.~Schaffner-Bielich, ``A statistical model for a complete supernova equation of state,'' {\em Nuclear Physics A}, vol.~837, no.~3-4, pp.~210--254, 2010.

\bibitem{hornick2018relativistic}
N.~Hornick, L.~Tolos, A.~Zacchi, J.-E. Christian, and J.~Schaffner-Bielich, ``Relativistic parameterizations of neutron matter and implications for neutron stars,'' {\em Physical Review C}, vol.~98, no.~6, p.~065804, 2018.

\bibitem{pradhan2023role}
B.~K. Pradhan, D.~Chatterjee, R.~Gandhi, and J.~Schaffner-Bielich, ``Role of vector self-interaction in neutron star properties,'' {\em Nuclear Physics A}, vol.~1030, p.~122578, 2023.

\bibitem{shen2020effects}
H.~Shen, F.~Ji, J.~Hu, and K.~Sumiyoshi, ``Effects of symmetry energy on the equation of state for simulations of core-collapse supernovae and neutron-star mergers,'' {\em The Astrophysical Journal}, vol.~891, no.~2, p.~148, 2020.

\bibitem{allard20211s0}
V.~Allard and N.~Chamel, ``$^{1}s_0$ pairing gaps, chemical potentials and entrainment matrix in superfluid neutron-star cores for the brussels--montreal functionals,'' {\em Universe}, vol.~7, no.~12, p.~470, 2021.

\bibitem{pearson2020unified}
J.~M. Pearson, N.~Chamel, and A.~Potekhin, ``Unified equations of state for cold nonaccreting neutron stars with brussels-montreal functionals. ii. pasta phases in semiclassical approximation,'' {\em Physical Review C}, vol.~101, no.~1, p.~015802, 2020.

\bibitem{pearson2022unified}
J.~M. Pearson and N.~Chamel, ``Unified equations of state for cold nonaccreting neutron stars with brussels-montreal functionals. iii. inclusion of microscopic corrections to pasta phases,'' {\em Physical Review C}, vol.~105, no.~1, p.~015803, 2022.

\bibitem{goriely2013hartree}
S.~Goriely, N.~Chamel, and J.~Pearson, ``Hartree-fock-bogoliubov nuclear mass model with 0.50 mev accuracy based on standard forms of skyrme and pairing functionals,'' {\em Physical Review C}, vol.~88, no.~6, p.~061302, 2013.

\bibitem{perot2019role}
L.~Perot, N.~Chamel, and A.~Sourie, ``Role of the symmetry energy and the neutron-matter stiffness on the tidal deformability of a neutron star with unified equations of state,'' {\em Physical Review C}, vol.~100, no.~3, p.~035801, 2019.

\bibitem{xu2013databases}
Y.~Xu, S.~Goriely, A.~Jorissen, G.~Chen, and M.~Arnould, ``Databases and tools for nuclear astrophysics applications-brussels nuclear library (bruslib), nuclear astrophysics compilation of reactions ii (nacre ii) and nuclear network generator (netgen),'' {\em Astronomy \& Astrophysics}, vol.~549, p.~A106, 2013.

\bibitem{welker2017binding}
A.~Welker, N.~Althubiti, D.~Atanasov, K.~Blaum, T.~E. Cocolios, F.~Herfurth, S.~Kreim, D.~Lunney, V.~Manea, M.~Mougeot, {\em et~al.}, ``Binding energy of cu 79: Probing the structure of the doubly magic ni 78 from only one proton away,'' {\em Physical review letters}, vol.~119, no.~19, p.~192502, 2017.

\bibitem{maruyama2005nuclear}
T.~Maruyama, T.~Tatsumi, D.~N. Voskresensky, T.~Tanigawa, and S.~Chiba, ``Nuclear “pasta” structures and the charge screening effect,'' {\em Physical Review C}, vol.~72, no.~1, p.~015802, 2005.

\bibitem{bombaci_equation_2018}
I.~Bombaci and D.~Logoteta, ``Equation of state of dense nuclear matter and neutron star structure from nuclear chiral interactions,'' {\em Astronomy \& Astrophysics}, vol.~609, p.~A128, 2018.

\bibitem{glendenning_reconciliation_1991}
N.~K. Glendenning and S.~A. Moszkowski, ``Reconciliation of neutron-star masses and binding of the $\lambda$ in hypernuclei,'' {\em Physical review letters}, vol.~67, no.~18, p.~2414, 1991.

\bibitem{dexheimer2019we}
V.~Dexheimer, R.~de~Oliveira~Gomes, S.~Schramm, and H.~Pais, ``What do we learn about vector interactions from gw170817?,'' {\em Journal of Physics G: Nuclear and Particle Physics}, vol.~46, no.~3, p.~034002, 2019.

\bibitem{clevinger2022hybrid}
A.~Clevinger, J.~Corkish, K.~Aryal, and V.~Dexheimer, ``Hybrid equations of state for neutron stars with hyperons and deltas,'' {\em The European Physical Journal A}, vol.~58, no.~5, p.~96, 2022.

\bibitem{otto_hybrid_2020}
K.~Otto, M.~Oertel, and B.-J. Schaefer, ``Hybrid and quark star matter based on a nonperturbative equation of state,'' {\em Physical Review D}, vol.~101, no.~10, p.~103021, 2020.

\bibitem{typel_composition_2010}
S.~Typel, G.~R{\"o}pke, T.~Kl{\"a}hn, D.~Blaschke, and H.~H. Wolter, ``Composition and thermodynamics of nuclear matter with light clusters,'' {\em Physical Review C}, vol.~81, no.~1, p.~015803, 2010.

\bibitem{otto_nonperturbative_2020}
K.~Otto, M.~Oertel, and B.-J. Schaefer, ``Nonperturbative quark matter equations of state with vector interactions,'' {\em The European Physical Journal Special Topics}, vol.~229, no.~22, pp.~3629--3649, 2020.

\bibitem{baym_hadrons_2018}
G.~Baym, T.~Hatsuda, T.~Kojo, P.~D. Powell, Y.~Song, and T.~Takatsuka, ``From hadrons to quarks in neutron stars: a review,'' {\em Reports on Progress in Physics}, vol.~81, no.~5, p.~056902, 2018.

\bibitem{togashi_nuclear_2017}
H.~Togashi, K.~Nakazato, Y.~Takehara, S.~Yamamuro, H.~Suzuki, and M.~Takano, ``Nuclear equation of state for core-collapse supernova simulations with realistic nuclear forces,'' {\em Nuclear Physics A}, vol.~961, pp.~78--105, 2017.

\bibitem{baym_new_2019}
G.~Baym, S.~Furusawa, T.~Hatsuda, T.~Kojo, and H.~Togashi, ``New {Neutron} {Star} {Equation} of {State} with {Quark}-{Hadron} {Crossover},'' {\em The Astrophysical Journal}, vol.~885, no.~1, 2019.

\bibitem{kojo2022implications}
T.~Kojo, G.~Baym, and T.~Hatsuda, ``Implications of nicer for neutron star matter: The qhc21 equation of state,'' {\em The Astrophysical Journal}, vol.~934, no.~1, p.~46, 2022.

\end{thebibliography}

\end{document}